\documentclass[structabstract]{aa}
\usepackage{graphicx}
\usepackage{txfonts}
\usepackage{natbib}
\bibpunct{(}{)}{;}{a}{}{,} 

\newcommand{\micron}{$\mu$m}

\begin{document}


\title{A spectral line survey of the starless and proto-stellar cores detected
by BLAST toward the Vela-D molecular cloud}   

\author{Jorge L. Morales Ortiz\inst{\ref{inst1},\ref{inst3},\dag} \and
	Luca Olmi\inst{\ref{inst2},\ref{inst3}} \and
	Michael Burton\inst{\ref{inst4}} \and
	Massimo De Luca\inst{\ref{inst5}} \and 
	Davide Elia\inst{\ref{inst6}} \and 
	Teresa Giannini\inst{\ref{inst7}} \and 
	Dario Lorenzetti\inst{\ref{inst7}} \and 
	Fabrizio Massi\inst{\ref{inst3}} \and 
	Francesco Strafella\inst{\ref{inst8}} 
}          

\institute{
UJF-Grenoble 1 / CNRS-INSU, Institut de Plan\'etologie et d'Astrophysique de Grenoble
(IPAG) UMR 5274, Grenoble, F-38041, France \email{jorge.luis379@gmail.com} \label{inst1}
\and
Osservatorio Astrofisico di Arcetri - INAF, Largo
E. Fermi 5, I-50125, Firenze, Italy \label{inst3}
\and
University of Puerto Rico, Rio Piedras Campus, Physics Department, Box 23343,
UPR station, San Juan, Puerto Rico (USA) \label{inst2}
\and
School of Physics, University of New South Wales, Sydney NSW 2052,
Australia \label{inst4}
\and
LERMA-LRA, UMR 8112 du CNRS, Observatoire de Paris, \'Ecole Normale
Sup\'erieure, UPMC \& UCP, 24 rue Lhomond, 75231 Paris Cedex 05, France \label{inst5}
\and
Istituto di Fisica dello Spazio Interplanetario - INAF,
via Fosso del Cavaliere 100, I-00133 Roma, Italy \label{inst6}
\and
Osservatorio Astronomico di Roma - INAF, Via Frascati 33, I-00040
Monteporzio Catone, Roma, Italy \label{inst7}
\and
Dipartimento di Fisica, Universit\'a del Salento, CP 193,
I-73100 Lecce, Italy \label{inst8}
}

\date{Received; accepted }

\abstract
{ Starless cores represent a very early stage of the star formation
process, before collapse results in the formation of a central protostar or a
multiple system of protostars.  }
{ We use spectral line observations
of a sample of cold dust cores, previously detected with the BLAST telescope in the
Vela-D molecular cloud, to perform a more accurate physical and kinematical analysis of the
sources.  }
{ We present a 3-mm and 1.3-cm survey conducted with the
Mopra 22-m and Parkes 64-m radio telescopes of a sample of 40 cold dust cores, including
both starless and proto-stellar sources.
20 objects were also mapped using molecular tracers of dense gas.
To trace the dense gas we used the molecular species
NH$_3$, N$_2$H$^+$, HNC, HCO$^+$, H$^{13}$CO$^+$, HCN and H$^{13}$CN,
where some of them trace the more quiescent gas,
while others are sensitive to more dynamical processes. }
{ The selected cores have a wide variety of morphological types
and also show physical and chemical
variations, which may be associated to different evolutionary phases.
We find evidence of systematic motions in both starless and proto-stellar cores and
we detect line wings in many of the proto-stellar cores. Our observations probe linear
distances in the sources $\ga 0.1\,$pc, and are thus sensitive mainly to molecular gas
in the envelope of the cores. In this region we do find that, for example, the radial
profile of the N$_2$H$^+(1-0)$ emission falls off more quickly than that of C-bearing
molecules such as HNC$(1-0)$, HCO$^+(1-0)$ and HCN$(1-0)$. We also analyze the
correlation between several physical and chemical parameters and the dynamics of the
cores. }
{ Depending on the assumptions made to estimate the virial mass,
we find that many starless cores have masses below
the self-gravitating threshold, whereas most of the proto-stellar cores have masses
which are near or above the self-gravitating critical value.
An analysis of the median properties of the starless and proto-stellar cores suggests
that  the transition from the pre- to the proto-stellar phase
is relatively fast, leaving the core envelopes with almost unchanged physical parameters.
}

\keywords{submillimeter: ISM --- stars: formation --- ISM: clouds --- ISM: molecules --- radio lines: ISM}

\maketitle

\section{Introduction }

Starless cores represent a very early stage of the star formation 
process, before collapse results in the formation of a central protostar or a 
multiple system of protostars.
The physical properties of these cores can reveal important clues about their
nature. Mass, spatial distributions, and lifetime are important
diagnostics of the main physical processes leading to the formation of
the cores from the parent molecular cloud.

In the past, (sub)millimeter continuum surveys performed with ground-based instruments have probed 
the Rayleigh-Jeans tail of the spectral energy distribution
(SED) of these cold objects, far from its peak and, at short submillimeter wavelengths,
 have been affected by low sensitivity due to atmospheric conditions.  
Therefore, these surveys have been limited by their sensitivity or by their 
relative inadequacy to measure the temperature (e.g., \citealp{motte1998}), producing large
uncertainties in the derived luminosities and masses.  Recent surveys
with the MIPS instrument of the {\it Spitzer Space Telescope}
are able to constrain the temperatures of warmer
objects \citep{car05}, but the youngest and coldest objects are
potentially not detected, even in the long-wavelength {\it Spitzer} bands. 

On the other hand, the more recent surveys with both the BLAST 
(``Balloon-borne Large-Aperture Submillimeter Telescope'', \citealp{pascale2008}) and 
Herschel telescopes (see, e.g., \citealp{netterfield2009}, \citealp{olmi2009}, 
\citealp{molinari2010}) have demonstrated the  ability to
detect and characterize cold dust emission from both starless and proto-stellar
sources, constraining the low temperatures of these objects ($T \la 25$\,K)
using their multi-band photometry near the peak of the cold core SED. The bolometric
observations alone, however, cannot fully contrain the dynamical and evolutionary status of the cores.
Spectral lines follow-ups are necessary in order to investigate the physical, 
dynamical and chemical status of each detected core.

BLAST has identified more than a thousand new starless and proto-stellar cores
during its second long duration balloon science flight in 2006. BLAST detected
these cold cores in a $\sim 50\,$deg$^2$ map of the Vela Molecular Ridge (VMR)
\citep{netterfield2009}.  The VMR (\citealp{mur:may}, \citealp{Lis92}) is a 
giant molecular cloud complex within the galactic plane, in the area
$260 \degr \la l \la 272 \degr$ and $-2 \degr \la b \la 3 \degr$,
hence located outside the solar circle. Its main properties have been recently revisited
by \citet{netterfield2009} and \citet{olmi2009}.

\begin{figure}
\centering
\includegraphics[width=8.5cm,angle=0]{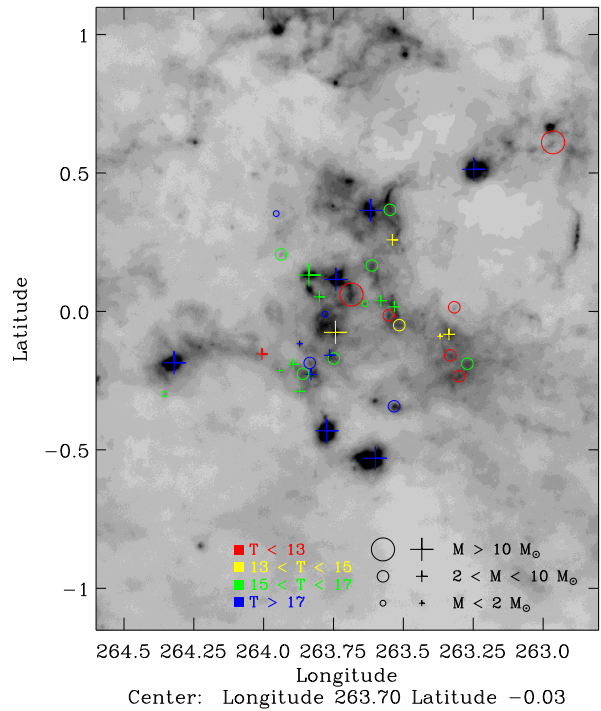}
\caption{
The gray-scale image shows the BLAST 250\,\micron\ map of Vela-D, with galactic coordinates
in degrees. Superimposed are the locations of both starless (open circles) and proto-stellar
(crosses) cores (see \citealp{olmi2009}).
The size of the symbols indicates the mass range [M$_\odot$]
of each core and color-coding indicates the core temperature [K] (see legend). 
  }
\label{fig:velaDmap}
\end{figure}

In particular, \citet{olmi2009} used both BLAST and archival data to determine the
SEDs and the physical parameters of each source detected by BLAST in the smaller region of 
Vela-D (see Figure~\ref{fig:velaDmap}), 
where observations from IR (\citealp{gianni07}) to millimeter wavelengths
(\citealp{massi07}) were already available.  \citet{olmi2010} then used the spectral line data
of \citet{elia2007} to perform a first analysis of the dynamical state of the cores toward
Vela-D. They found that $\sim 30-50$\% of the BLAST cores are gravitationally bound, 
and that a significant number of cores would need an additional source of external 
confinement. 

In this work we will use spectral line observations, both single-points 
and maps, of a sample of BLAST sources from Vela-D to perform a more accurate 
physical and kinematical analysis of the 
starless and proto-stellar cores. This paper is thus organized as follows:
in Section~\ref{sec:obs}  we describe the selected targets and molecular lines. 
In Section~\ref{sec:res} we present the results of the analysis of the observed
molecular line spectra and maps, describing the derived physical and kinematical parameters. 
We then further discuss these results in Section~\ref{sec:discussion} and conclude in 
Section~\ref{sec:concl}.

%

\begin{table*}
\caption{BLAST sources in Vela-D observed with the Mopra and Parkes telescopes}
\label{tab:srclist}
\centering
\begin{tabular}{llccccccc}
\hline\hline
\multicolumn{5}{c}{{\bf BLAST}} & & \multicolumn{1}{c}{{\bf Mopra}} & & \multicolumn{1}{c}{{\bf Parkes}} \\
\cline{1-5}
Source \#\tablefootmark{a} & Source name & $l$ & $b$ & Starless & & Observation & & Observation \\
& & [deg] & [deg] & or Proto-Stellar & & type\tablefootmark{b} & & type\tablefootmark{b}  \\
%
%
\hline
   3 & BLAST J084531-435006 & 263.6001 & $-0.5302$ & P   & & SP        &  & SP \\
   9 & BLAST J084546-432458 & 263.3005 & $-0.2341$ & S   & & SP \& M   &  & SP \\
  12 & BLAST J084552-432152 & 263.2707 & $-0.1886$ & S   & & SP        &  & -- \\
  13 & BLAST J084606-433956 & 263.5331 & $-0.3425$ & S   & & M         &  & -- \\
  14 & BLAST J084612-432337 & 263.3322 & $-0.1585$ & S   & & SP \& M   &  & SP \\      
  23 & BLAST J084633-432100 & 263.3370 & $-0.0827$ & P   & & SP        &  & -- \\
  24 & BLAST J084633-435432 & 263.7742 & $-0.4309$ & P   & & SP \& M   &  & SP \\
  26 & BLAST J084637-432245 & 263.3685 & $-0.0899$ & P   & & SP        &  & -- \\
  31 & BLAST J084654-431626 & 263.3177 & $ 0.0149$ & S   & & SP \& M   &  & SP \\
  34 & BLAST J084718-432804 & 263.5147 & $-0.0496$ & S   & & SP \& M   &  & SP \\
  38 & BLAST J084731-435344 & 263.8713 & $-0.2886$ & P   & & SP        &  & SP \\
  40 & BLAST J084735-432829 & 263.5521 & $-0.0141$ & S   & & SP \& M   &  & SP \\
  41 & BLAST J084736-434332 & 263.7488 & $-0.1699$ & S   & & SP \& M   &  & SP \\
  43 & BLAST J084738-434931 & 263.8305 & $-0.2277$ & P   & & SP        &  & SP \\
  44 & BLAST J084739-432623 & 263.5322 & $ 0.0167$ & P   & & SP        &  & -- \\
  45 & BLAST J084742-434347 & 263.7637 & $-0.1582$ & P   & & SP        &  & SP \\
  47 & BLAST J084744-435045 & 263.8591 & $-0.2249$ & S   & & SP \& M   &  & SP \\
  50 & BLAST J084749-434810 & 263.8349 & $-0.1861$ & S   & & SP \& M   &  & SP \\
  52 & BLAST J084754-432748 & 263.5802 & $ 0.0387$ & P   & & SP        &  & -- \\
  53 & BLAST J084759-433942 & 263.7428 & $-0.0757$ & P   & & SP        &  & -- \\
  54 & BLAST J084801-435108 & 263.8947 & $-0.1907$ & P   & & SP        &  & -- \\
  55 & BLAST J084803-433051 & 263.6369 & $ 0.0281$ & S   & & M         &  & -- \\
  56 & BLAST J084805-435415 & 263.9430 & $-0.2138$ & S   & & SP \& M   &  & -- \\
  57 & BLAST J084813-423730 & 262.9644 & $ 0.6103$ & S   & & SP \& M   &  & SP \\
  59 & BLAST J084815-434714 & 263.8713 & $-0.1166$ & P   & & SP        &  & SP \\
  63 & BLAST J084822-433152 & 263.6856 & $ 0.0609$ & S   & & SP \& M   &  & SP \& M \\
  65 & BLAST J084823-433858 & 263.7799 & $-0.0106$ & S   & & SP        &  & -- \\
  71 & BLAST J084834-435455 & 264.0059 & $-0.1538$ & P   & & SP        &  & -- \\
  72 & BLAST J084834-432430 & 263.6126 & $ 0.1657$ & S   & & SP        &  & -- \\
  77 & BLAST J084842-431735 & 263.5392 & $ 0.2584$ & P   & & SP        &  & SP \\
  79 & BLAST J084844-433733 & 263.8007 & $ 0.0525$ & P   & & SP        &  & -- \\
  81 & BLAST J084847-425423 & 263.2482 & $ 0.5133$ & P   & & SP \& M   &  & SP \\
  82 & BLAST J084848-433225 & 263.7415 & $ 0.1155$ & P   & & SP \& M   &  & SP \\
  88 & BLAST J084910-441636 & 264.3543 & $-0.2984$ & P   & & SP        &  & -- \\
  89 & BLAST J084912-431353 & 263.5480 & $ 0.3668$ & S   & & SP \& M   &  & SP \\
  90 & BLAST J084912-433618 & 263.8379 & $ 0.1312$ & P   & & SP        &  & SP \\
  93 & BLAST J084925-431710 & 263.6165 & $ 0.3645$ & P   & & SP        &  & SP \\
  97 & BLAST J084932-441046 & 264.3206 & $-0.1857$ & P   & & SP        &  & SP \\
 101 & BLAST J084952-433808 & 263.9381 & $ 0.2058$ & S   & & M         &  & -- \\
 109 & BLAST J085033-433318 & 263.9551 & $ 0.3532$ & S   & & M         &  & -- \\
\hline
\end{tabular}
\tablefoot{
\tablefoottext{a}{
We follow the source numbering (0 to 140) defined by \citet{olmi2009}.}
\tablefoottext{b}{ ``SP'' stands for single-point observation and ``M'' stands for map.}
}
\end{table*}

%
\begin{table}   
\caption{Molecular species observed with the Parkes and Mopra telescopes}
\label{tab:linelist}
\centering
\begin{tabular}{llcc}
\hline\hline
Spectral line & $E_{\rm u}$  & $n_{\rm eff}$\tablefootmark{a} & Rest Frequency \\
& [K] & [$\times 10^3\,$cm$^{-3}$] & [MHz]  \\ 
\hline
NH$_3(1,1)$                   & 23.4    & $-$       & 23694.496 \\
NH$_3(2,2)$                   & 64.9    & $-$       & 23722.633 \\
\hline
N$_2$H$^+(1_{21}-0_{32})$     & 4.47    & 2.5       & 93172.053 \\
HNC$(1_2-0_1)$                & 4.35    & 9.5       & 90663.574 \\
HCO$^+(1-0)$                  & 4.28    & 2.5       & 89188.526 \\
H$^{13}$CO$^+(1-0)$           & 4.16    & 2.2       & 86754.330 \\
HCN$(1_2-0_1)$                & 4.25    & 28        & 88631.847 \\
H$^{13}$CN$(1_2-0_1)$         & 4.14    & 20        & 86340.167 \\
\hline
\end{tabular}
\tablefoot{
\tablefoottext{a}{
See Section~\ref{sec:neff} for a definition of $n_{\rm eff}$.
}
}
\end{table}

\section{Observations and archival data }
\label{sec:obs}

\subsection{Targets and lines selection}
\label{sec:target}

The observations were performed in June 2009 with the Mopra and Parkes telescopes of the
Australia Telescope National Facility (ATNF). 
Although originally our aim was to observe all sources in the catalog of 
\citet{olmi2009}, because of time constraints we were able to observe
a total of 40 sources at Mopra
and 22 sources at Parkes, in either mapping or single-pointing
mode (or both, see Table~\ref{tab:srclist} and Figure~\ref{fig:velaDmap}). 
We have selected an almost equal number of starless (19) and proto-stellar (21) cores 
where, following \citet{olmi2009}, we define the BLAST cores as
proto-stellar when they turned out to be positionally associated with one or
more MIPS sources at 24$\,\mu$m, and starless otherwise. 
Each sub-sample is composed by objects that present increasing temperature 
and different morphology. This selection criterion is quite arbitrary but aimed
at identifying different evolutionary stages within each sub-sample.


In order to probe the dense and cold gas detected by BLAST toward the cores
in Vela-D we needed appropriate molecular tracers.  We thus selected 
several rotational transitions with low quantum numbers to be observed 
at Mopra in the 3-mm band, and we also observed the NH$_3$(1,1) and (2,2) 
inversion lines with the Parkes telescope. In the following sections, our analysis 
will be based on the ammonia lines and on the main spectral lines observed 
at Mopra, i.e., N$_2$H$^+(1-0)$, HCN$(1-0)$, HNC$(1-0)$ and HCO$^+(1-0)$ (and two of their
isotopologues, see Table~\ref{tab:linelist}). 
This choice of molecular tracers  ensured that we would be able to trace 
the dense as well as the more diffuse gas, and we would also be able to detect 
molecular outflows. Furthermore, these molecules should also be able to give out
clues about the different chemical states of the sources.

\subsection{Mopra}
\label{sec:mopra}

During the spectral line mapping or single-pointing mode observations with
the Mopra telescope the system temperatures were
typically comprised in the range $\sim 170 - 280\,$K
and we reached a RMS sensitivity in $T_{\rm A}^\star$ units of
about $\sim 30 - 40 \,$~mK after rebinning to a 0.22~km\,s$^{-1}$ velocity resolution.
The beam full width at half maximum (FWHM) was about $38\,$arcsec and the 
pointing was checked every hour by using 
an SiO maser as a reference source. 
Typically, pointing errors were found to be $\sim 5-10\,$arcsec.

The parameter $\eta_{\rm mb}$ to convert from antenna
temperature to main-beam brightness temperature has been assumed to be 0.44 at 100\,GHz
and 0.49 at 86\,GHz (\citealp{ladd2005}).
The Mopra spectrometer (MOPS) was used as a backend instrument in its ``zoom'' mode,
which allowed to split the 8.3~GHz instantaneous band in up to 16 zoom bands; each
sub-band was 137.5 MHz wide and had 2$\times$4096 channels.
The single-point observations were performed in position-switching mode, whereas
the spectral line maps, of size mostly $3\times 3\,$arcmin$^2$,
 were obtained using the Mopra on-the-fly
mapping mode, scanning in both right ascension and declination.

The most intense lines observed were N$_2$H$^+(1-0)$, HCN$(1-0)$, HNC$(1-0)$ and HCO$^+(1-0)$.
These lines are all good tracers of dense gas, and are also often used to 
detect velocity gradients. N$_2$H$^+$ is known to 
be  more resistant to freeze-out on grains than the carbon-bearing species. 
HNC is particularly prevalent in cold gas \citep{hirota1998}, while HCO$^+$ often 
shows infall signatures and outflow wings. These strong lines can all be optically thick
and thus two isotopologues, H$^{13}$CO$^+$ and H$^{13}$CN, were also observed to provide optical depth 
and line profile information.  Unfortunately, their intensity was in 
general too weak to allow derivation of column densities throughout the maps and other useful physical
and kinematical parameters, and thus most of our analysis will be based on the four  most intense
molecular tracers.


\subsection{Parkes}
\label{sec:parkes}

A total of 22 sources where observed with the Parkes telescope in single-pointing
mode (see Table~\ref{tab:srclist}), and only one source (BLAST063) was mapped using
the NH$_3$(1,1) and (2,2) transitions\footnote{The signal-to-noise ratio (SNR) in this 
map was very low and thus we will not consider any further this map.}.
The system temperatures were typically   $\sim 40 - 50 \,$K
and we reached a RMS sensitivity in $T_{\rm A}^\star$  units of
$ \sim 10 \,$mK after rebinning to a $\simeq 0.2 \,$km\,s$^{-1}$ velocity resolution.
The beam FWHM was about $1.3\,$arcmin and the pointing errors were found to be $< 20\,$arcsec.
The Parkes Multibeam correlator, in single-beam high resolution mode, was used as a backend.
A simultaneous band of 64\,MHz was observed, with a total of 8192 channels, which allowed
to observe the NH$_3$(1,1) and (2,2) lines within the same bandwidth.

\subsection{BLAST}
\label{sec:blast}

BLAST is a 1.8\,m Cassegrain telescope, whose under-illuminated primary
mirror has produced in-flight beams with FWHM of
36\,\arcsec, 42\,\arcsec, and 60\,\arcsec\ at 250, 350, and 500\,\micron,
respectively.  More details on the instrument, calibration methods and map-making
procedure can be found in the BLAST05 papers \citet{pascale2008},
 \citet{patanchon2008}, and \citet{truch2008}.

The Vela-D region, shown in Figure~\ref{fig:velaDmap} and
defined by \citet{olmi2009} as the area
contained within $262\fdg80 < l < 264\fdg60$ and
$-1\fdg15 < b < 1\fdg10$, was part of a larger map of the Galactic Plane  obtained by BLAST
toward the VMR \citep{netterfield2009}.
In Vela-D \citet{olmi2009} found a total of 141 sources, both starless and proto-stellar, and
in the next sections we will follow the source numbering (0 to 140) defined by these authors.

%
\begin{table*}
\caption{
Peak intensity of main detected lines or the rms Level of the spectrum. }
\label{tab:summary}
\centering
\begin{tabular}{lccccccc}
\hline\hline
Source \# & N$_2$H$^+$ & HCN & H$^{13}$CN & HNC & HC$_3$N\tablefootmark{a} & HCO$^+$ & H$^{13}$CO$^+$ \\
%
\hline
 3        & 0.17        & 0.81     & $<$0.03     & 0.58      & --       & 0.70     & 0.13  \\
 9        & 0.09        & 0.06     & $<$0.03     & 0.20      & --       & 0.22     & 0.06  \\
12        & $<$0.02     & 0.05     & $<$0.02     & $<$0.03   & --       & 0.07     & $<$0.02  \\
23        & $<$0.02     & 0.10     & $<$0.03     & 0.25      & --       & 0.38     & $<$0.03  \\
24        & 1.33        & 2.97     & 0.23        & 2.23      & --       & 3.20     & 0.32  \\
26        & $<$0.02     & 0.07     & $<$0.02     & 0.08      & --       & 0.14     & $<$0.02  \\
26\tablefootmark{b} &  --    & --       & --          & 0.10      & --       & 0.13     & --     \\
31        & $<$0.02     & 0.11     & $<$0.02     & 0.15      & --       & 0.34     & $<$0.02  \\
34        & $<$0.02     & 0.08     & $<$0.02     & 0.07      & --       & 0.23     & $<$0.02  \\
34\tablefootmark{c} & --     & --       & --          & --        & --       & 0.15     &   --   \\
38        & 0.07        & 0.11     & $<$0.02     & 0.15      & --       & 0.19     & $<$0.02  \\
40        & 0.20        & 0.31     & $<$0.02     & 0.45      & --       & 0.73     & 0.28  \\
41        & $<$0.02     & 0.27     & $<$0.03     & 0.20      & --       & 0.41     & $<$0.02  \\
43        & $<$0.05     & 0.11     & $<$0.05     & 0.26      & --       & 0.42     & $<$0.05  \\
44        & 0.21        & 0.35     & $<$0.04     & 0.57      & --       & 0.88     & 0.13  \\
45        & 0.13        & 0.24     & $<$0.02     & 0.24      & --       & 0.52     & $<$0.02  \\
47        & $<$0.02     & 0.19     & $<$0.02     & 0.12      & --       & 0.14     & $<$0.02  \\
47\tablefootmark{c} & --     & --       & --          & --        & --       & 0.13     & --     \\
50        & $<$0.02     & 0.10     & $<$0.02     & 0.07      & --       & 0.16     & $<$0.02  \\
52        & $<$0.02     & 0.11     & $<$0.02     & 0.10      & --       & 0.18     & $<$0.02  \\
53        & 0.07        & 0.20     & $<$0.02     & 0.28      & --       & 0.19     & 0.07  \\
54        & $<$0.02     & 0.14     & $<$0.02     & 0.12      & --       & 0.12     & $<$0.02  \\
56        & $<$0.02     & 0.14     & $<$0.03     & 0.08      & --       & 0.16     & $<$0.02  \\
57        & 0.12        & 0.13     & $<$0.03     & 0.21      & --       & 0.33     & 0.13  \\
59        & 0.12        & 0.18     & $<$0.03     & 0.22      & --       & 0.23     & 0.08  \\
63        & 0.11        & 0.44     & $<$0.02     & 0.55      & --       & 0.76     & 0.09  \\
65        & $<$0.02     & 0.12     & $<$0.02     & 0.09      & --       & 0.15     & $<$0.02  \\
71        & $<$0.04     & 0.11     & $<$0.04     & $<$0.05   & --       & $<$0.08  & $<$0.04  \\
72        & $<$0.02     & 0.15     & $<$0.02     & 0.14      & --       & 0.39     & $<$0.02  \\
72\tablefootmark{c} & --     & --       & --          & --        & --       & 0.19     & --     \\
77        & 0.06        & 0.26     & $<$0.02     & 0.23      & --       & 0.45     & $<$0.03  \\
79        & $<$0.05     & 0.17     & $<$0.05     & 0.23      & --       & 0.22     & $<$0.05  \\
81        & 1.49        & 4.32     & 0.39        & 2.42      & --       & 4.31     & 0.25  \\
82        & 0.50        & 0.65     & $<$0.03     & 0.74      & --       & 1.30     & 0.23  \\
88        & $<$0.02     & 0.26     & $<$0.03     & 0.18      & --       & $<$0.02  & $<$0.03  \\
89        & $<$0.02     & 0.16     & $<$0.02     & 0.09      & --       & 0.22     & $<$0.03  \\
90        & 0.50        & 1.32     & 0.12        & 1.30      & 0.16     & 2.19     & 0.38  \\
93        & 0.50        & 0.75     & 0.11        & 1.03      & 0.26     & 1.45     & 0.44  \\
97        & 0.19        & 0.54     & $<$0.03     & 0.62      & --       & 0.79     & 0.12  \\
\hline
\end{tabular}
\tablefoot{Only sources with a single-point observation are listed
(see Table~\ref{tab:srclist}). All values are in units of $T_{\rm A}^\star$ [K]. \\
\tablefoottext{a}{
The typical noise RMS in the spectra where HC$_3$N was not detected
is $\simeq 0.03\,$K.
}
\tablefoottext{b}{
This source has two separate velocity components in HNC and HCO$^+$ and
only these lines are listed for this velocity component.
}
\tablefoottext{c}{
These sources have two separate velocity components in HCO$^+$ and only
this line is listed for this velocity component.
}
}
%
%
\end{table*}

\section{ Results }
\label{sec:res}

In this section we will discuss the derivation of various physical parameters 
such as column density, mass and kinetic temperature, that
are fundamental for the analysis of the sources. We will also analyse the kinematics
of the sources, determining velocity gradients and line asymmetries. The estimate of 
the virial masses will also allow us to determine which sources are currently 
gravitationally bound and which are not.

\subsection{Morphological characteristics of spectral line maps}
\label{sec:maps}

The spectral line data and maps were reduced and analyzed using the standard IRAM
package {\sc CLASS}, and the {\sc XS} package of the Onsala Space Observatory.
The single-point spectra obtained with the Mopra telescope towards the nominal position of the 
BLAST cores are shown in Figures~\ref{fig:N2Hspectra} to \ref{fig:HC13Nspectra} of the Appendix. 
The peak intensities of the detected lines, or the spectrum RMS in case of no detection, are
listed in Table~\ref{tab:summary}. Tables~\ref{tab:N2H} to \ref{tab:HCO} in the Appendix list
several physical and kinematical parameters of the sources. 

The multi-line integrated intensity maps obtained at Mopra, with the most 
significant SNR, are shown in Figure~\ref{fig:maps1} to \ref{fig:maps4} of the Appendix, where
the white contours represent the BLAST emission at 250$\, \mu$m.
One can see that in general the line emission follows the dust continuum emission,
even when the emission is more diffuse and filamentary as observed, for example, in
source BLAST063. 
But we also note that in several sources (for example, BLAST055, BLAST056, BLAST081 and BLAST101)
the dust continuum and some of the line emission 
peak at different positions.  

One can immediately note that the three proto-stellar cores that
were mapped (BLAST024, BLAST081 and BLAST082)
have a much more regular shape than the starless ones. We also note that
the maps of sources BLAST024 and BLAST081
are partly affected by artefacts, likely caused by bad weather during the observations at Mopra,
which show up as sharp drops of the integrated intensity, particularly in the maps of the
HCO$^+(1-0)$ line.
We can also see that in source BLAST082 two separate, smaller cores have been identified.
In terms of the molecular tracers observed,
we note that there are clear differences in the spatial distribution of emission
from different molecular species. These morphological differences are more evident
in the starless cores (but also, for example, in the proto-stellar core BLAST081) and
suggest both physical and {\it chemical} differences among the observed sources.

We may summarize the wide variety of spatial morphologies observed as follows: very early and cold
starless cores appear to have an irregular shape (e.g., BLAST009) in most or all molecular tracers
mapped at Mopra. Some warmer cores (e.g., BLAST031) and cores at the transition phase from starless to
proto-stellar (BLAST063, see Section~\ref{sec:specific}) appear to be more compact.
Finally, proto-stellar cores all show a more regular shape and narrow radial intensity profiles
(see Section~\ref{sec:radprof}).  Throughout the rest
of the paper we will thus discuss how, besides to morphological differences, these cores
also show physical and chemical variations, which may be associated to different evolutionary
phases.


%
%
\begin{table}
\caption{NH$_3$(1,1) line parameters.}
\label{tab:NH311}
\centering
\begin{tabular}{lcccc}
\hline\hline
Source \# & \multicolumn{4}{c}{{\bf NH$_3$(1,1)}} \\
\cline{2-5}
& $T_A^{\star}\,\tau $ & $V_{\rm lsr}$ & $FWHM$ & $\tau$ \\
& [K] & [km\,s$^{-1}$] & [km\,s$^{-1}$] & \\
%
%
%
\hline
 3    &   0.15    &   5.3     &   1.9     &   1.40   \\
 9    &   0.04    &  10.7     &   1.1     &   0.66   \\
14    &   0.13    &   3.0     &   0.7     &   2.63   \\
24    &   0.38    &   3.5     &   2.6     &   0.51   \\
31    &   0.06    &  10.7     &   1.3     &   2.77   \\
40    &   0.07    &   1.3     &   0.9     &   0.22   \\
41    &   0.05    &  11.1     &   0.7     &   0.10   \\
47    &   0.04    &   5.5     &   1.5     &   0.77   \\
57    &   0.07    &  14.0     &   1.5     &   1.73   \\
63    &   0.11    &  12.4     &   1.7     &   0.94   \\
77    &   0.05    &   2.0     &   0.8     &   0.10   \\
81    &   0.06    &  12.3     &   3.1     &   0.10   \\
82    &   0.11    &  11.7     &   2.0     &   0.22   \\
90    &   0.10    &  11.4     &   1.7     &   0.10   \\
93    &   0.13    &   2.4     &   1.2     &   0.70   \\
97    &   0.31    &   9.4     &   1.6     &   1.28   \\
\hline
\end{tabular}
\tablefoot{
Parameters derived from method ``nh3'' of the CLASS program.
Sources where the hyperfine structure of the NH$_3$(1,1) could not be fitted are not listed here.
}
\end{table}

%
%
\begin{table}
\caption{NH$_3$(2,2) line parameters.}
\label{tab:NH322}
\centering
\begin{tabular}{lccc}
\hline\hline
Source \# & \multicolumn{3}{c}{{\bf NH$_3$(2,2)}} \\
\cline{2-4}
& $\int{T_A^{\star}\, {\rm d}V }$  & $V_{lsr}$ & FWHM \\
& [K\,km\,s$^{-1}$] & [km\,s$^{-1}$] & [km\,s$^{-1}$] \\
%
%
%
\hline
 3    &   0.06    &  5.3    &  1.1    \\
24    &   0.54    &  3.6    &  3.1    \\
47    &   0.03    &  5.3    &  2.9    \\
63    &   0.03    &  10.3   &  1.8    \\
81    &   0.08    &  12.2   &  2.7    \\
82    &   0.07    &  11.7   &  1.9    \\
90    &   0.06    &  11.4   &  2.5    \\
93    &   0.07    &  2.4    &  1.8    \\
97    &   0.14    &  9.3    &  2.1    \\
\hline
\end{tabular}
\tablefoot{
Parameters derived from a standard Gaussian fit to the main component of the NH$_3$(2,2)
hyperfine structure.
}
\end{table}

%
\begin{figure}
\centering
\includegraphics[width=9.0cm,angle=0]{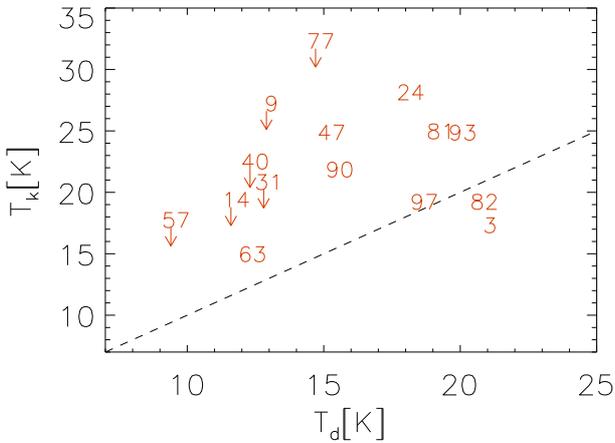}
\caption{
Kinetic temperature, $T_{\rm k}$, as derived from the
NH$_3$(1,1) and (2,2) inversion transitions, vs. the BLAST-derived dust temperature,
$T_{\rm d}$.  Sources are identified through their ID (see Table~\ref{tab:srclist}), and
the vertical arrows indicate that the corresponding $T_{\rm k}$ is an upper limit.
The dashed line indicates the $T_{\rm k} = T_{\rm d}$ locus.
  }
\label{fig:NH3temp}
\end{figure}

\subsection{Temperature and optical depth from hyperfine structure fits}
\label{sec:hfs}

The excitation temperature, $T_{\rm ex}$, and the line optical depth, $\tau$, which are required
to estimate the molecular column densities (see Sect.~\ref{sec:coldens}) could only be determined
in those sources where molecules with a hyperfine structure (hfs, hereafter) were
detected, i.e., N$_2$H$^+$, HCN and H$^{13}$CN, and/or when both
the HCO$^+(1-0)$ and H$^{13}$CO$^+(1-0)$ lines were detected.

In the molecules with a hfs, we used method ``hfs'' of the CLASS program to determine both
$T_{\rm ex}$ and $\tau$.  In those cases where  both HCO$^+$ and H$^{13}$CO$^+$ were detected,
the optical depth was estimated from their
line intensity ratio, assuming the same excitation temperature for the two
isotopic molecular species
and using the abundance ratio [HCO$^+$]/[H$^{13}$CO$^+$]$\simeq 40$ (\citealp{zinchenko2009}).
%

For the purpose of estimating the column density later,
in those sources where the hfs method could not be used for one or more molecules
(for example because of low SNR), or if only HCO$^+$ was detected but not H$^{13}$CO$^+$,
the values of both $T_{\rm ex}$ and $\tau$ for a given molecular species  were assumed
to be the averages of the parameters estimated from the detected molecular transitions
in the same source.

\subsection{Temperature and optical depth from NH$_3$}
\label{sec:nh3}


The single-point spectra of the NH$_3$(1,1) inversion line, obtained with the Parkes telescope,
are shown in Figure~\ref{fig:NH3spectra} of the Appendix.
The (2,2) line has been detected towards nine sources (mostly proto-stellar). 
We attempted to map the NH$_3$(1,1) transition toward BLAST063, but we did not detect
the line due to insufficient SNR in the individual scanning positions.

Method ``nh3'' of the CLASS program, similar to method ``hfs'' described above, was used
to fit the NH$_3$(1,1) line and resulted in the
parameters $T_A^{\star}\,\tau $, $\tau$ and FWHM
listed in Table~\ref{tab:NH311}. Due to the fact that the 
hfs components of the NH$_3$(2,2) line
were not detected (with the exception of source BLAST024), the line parameters 
were derived from a standard Gaussian fit to the main component of the NH$_3$(2,2)
hyperfine structure and are listed in Table~\ref{tab:NH322}.

%
%
%
\begin{table*}
\caption{Results of velocity gradient fitting.}
\label{tab:velfit}
\centering
\begin{tabular}{lccccccc}
\hline\hline
Source \# & \multicolumn{3}{c}{\bf HNC$(1-0)$} & & \multicolumn{3}{c}{\bf HCO$^+(1-0)$} \\
\cline{2-4}
\cline{6-8}
& $V_o$ & ${\rm d}V/{\rm d}r$ & $\theta_v$\tablefootmark{a} & & $V_o$ & ${\rm d}V/{\rm d}r$ & $\theta_v$\tablefootmark{a} \\
& [km\,s$^{-1}$] & [km\,s$^{-1}$\,pc$^{-1}$] &  [deg] & & [km\,s$^{-1}$] & [km\,s$^{-1}$\,pc$^{-1}$] & [deg] \\
%
\hline
9    & $-$    & $-$    & $-$         &      & 12.2   & 0.8    & 88.5   \\
13   & 1.0    & 2.3    & 21.1        &      & 0.9    & 1.5    & 18.1   \\
24   & 5.5    & 1.9    & 60.7        &      & 5.3    & 1.7    & 57.7   \\
34   & $-$    & $-$    & $-$         &      & 2.6    & 1.6    & 224.1  \\
40   & 2.5    & 0.2    & 16.7        &      & 2.6    & 1.2    & 71.5   \\
41   & $-$    & $-$    & $-$         &      & 12.1   & 1.2    & 289.9  \\
47   & 6.6    & 2.7    & 240.1       &      & $-$    & $-$    & $-$    \\
63   & 13.0   & 1.6    & 55.2        &      & 13.2   & 1.3    & 86.8   \\
81   & 13.5   & 1.4    & 256.9       &      & 13.3   & 1.9    & 312.6  \\
82   & 13.6   & 2.1    & 15.7        &      & 13.0   & 1.5    & 334.5  \\
%
\hline
\end{tabular}
\tablefoot{
\tablefoottext{a}{
The angle $\theta_v$ is measured positive from the axis of
positive longitude offsets toward the North.
}
}
\end{table*}

We then determined the kinetic temperature, $T_{\rm k}$, and NH$_3$ column density
using both NH$_3$(1,1) and (2,2) transitions, in those sources where
the (2,2) line was indeed detected. In these cases
we were able to determine the rotational temperature, $T_{\rm 12}$ (and thus $T_{\rm k}$),
and the column density using the method of \citet{ung1986} and \citet{bach1987}.
In this method the required data were: {\it (i)} the product
$\tau (T_{\rm ex}- T_{\rm bg})$ for the (1,1) line, where $T_{\rm ex}$ and $T_{\rm bg}=2.725\,$K are
the excitation and background temperatures, respectively, and $\tau$ is the optical depth;
{\it (ii)} the linewidth $\Delta V (1,1)$;
and {\it (iii)} the (2,2) integrated intensity.
Once $T_{\rm 12}$ has been estimated, the
kinetic temperature can be determined using the analytical
expression of \citet{tafalla2004}:
\begin{equation}
T_{\rm k} = \frac{T_{\rm 12}} {1 - \frac{T_{\rm 12}}{42} \ln [1 + 1.1 \exp(-\frac{16}{T_{\rm 12}} ) ] }
\label{eq:Tk}
\end{equation}
which is an empirical expression that fits the $T_{\rm 12} - T_{\rm k}$ relation obtained using
a radiative transfer model.
For those sources where the NH$_3$(2,2) transition was not detected, we could only determine
an upper limit to $T_{\rm k}$ (shown in Figure~\ref{fig:NH3temp}).
However, in order to obtain an estimate of the column density,
and thus of the mass, for these sources we adopted the BLAST-derived temperature, $T_{\rm d}$,
as an approximation for $T_{\rm k}$, which turns out to be a reasonably
good assumption, as discussed below.

For those sources with an independent estimate of $T_{\rm k}$ we
compared the resulting kinetic temperatures with the
BLAST-derived dust temperatures, as obtained by \citet{olmi2009}, and we show the results in
Figure~\ref{fig:NH3temp}.  
This comparison between $T_{\rm k}$ and $T_{\rm d}$, when applied separately to starless and
proto-stellar cores, is affected by the few detections of NH$_3$(2,2)
toward starless cores, and by the relatively high spread in the values of the $T_{\rm k} / T_{\rm d}$
ratio. In fact, for the two starless cores (BLAST047 and BLAST063;
see also Section~\ref{sec:specific}) where this ratio could
be measured we obtain a mean value $ \langle T_{\rm k} / T_{\rm d} \rangle = 1.4$, whereas for the
proto-stellar cores we get a median value $ T_{\rm k} / T_{\rm d}  = 1.2 \pm 0.2$. 

When all cores are considered together, we get a median value
$ T_{\rm k} / T_{\rm d} = 1.2 \pm 0.2$, i.e., the median value is dominated by the
proto-stellar cores.  A tentative 
explanation for the fact that $T_{\rm k}$ tends to be slightly higher than $T_{\rm d}$ 
could be that the NH$_3$ observations are sampling regions that
are somewhat warmer compared to the BLAST-derived $T_{\rm d}$ measurements. In fact, 
the BLAST observations are likely to average the temperature on much larger volumes of dust and,
in addition, in the less dense regions there could be a systematic difference between 
$T_{\rm d}$ and $T_{\rm k}$. In the literature both cases of systematic departures 
between dust and gas temperature, i.e. $T_{\rm k} < T_{\rm d}$ or
$T_{\rm k} > T_{\rm d}$, can be found (e.g., \citealp{krugel1984}, \citealp{pirogov1993}).
The dust and gas temperature may be affected by various effects, and a more detailed analysis 
of these effects in our sources is beyond the scopes of the present work. 

%
\begin{figure*}
\begin{centering}
%
\includegraphics[width=9.0cm,angle=0]{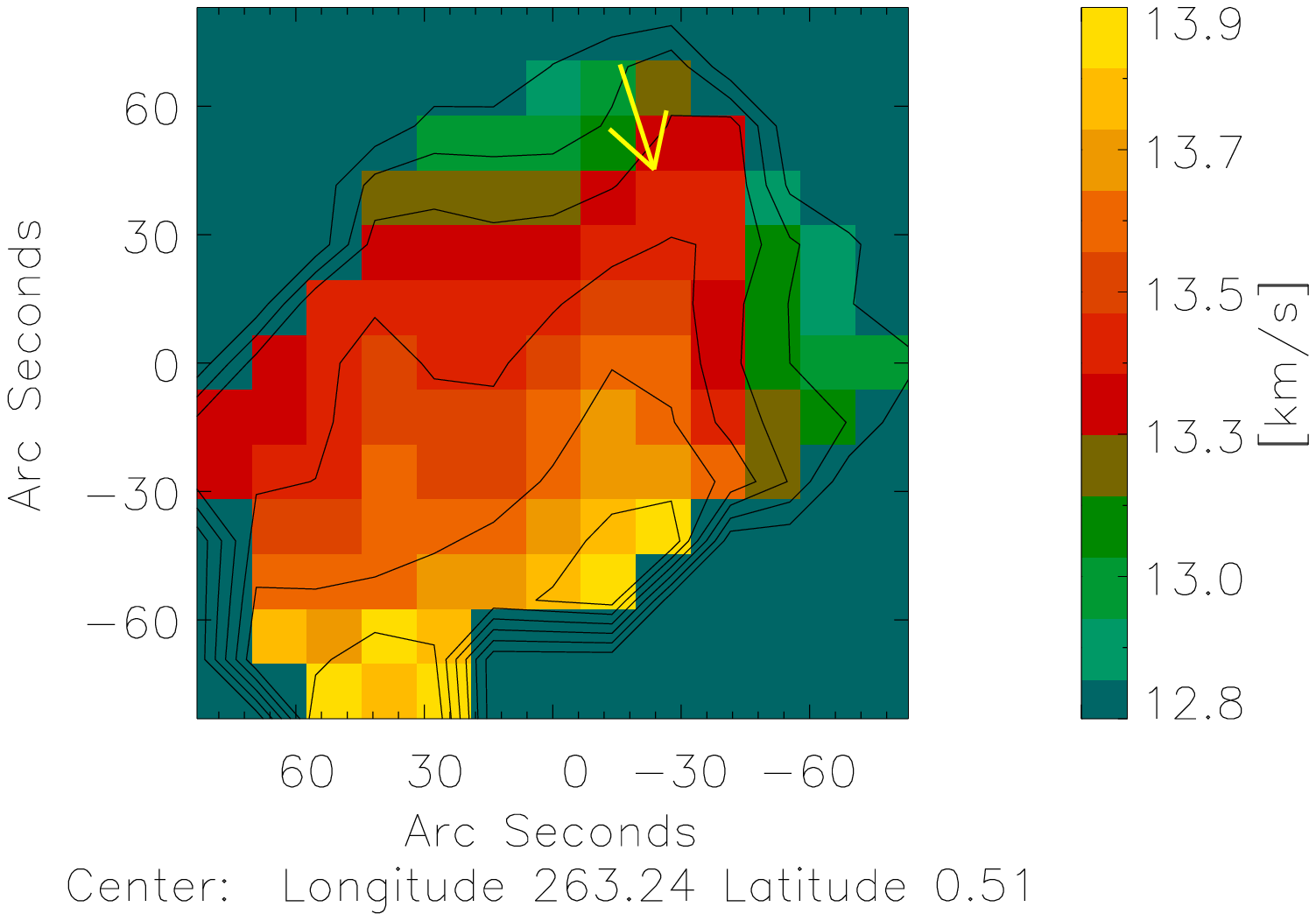} 
\includegraphics[width=9.0cm,angle=0]{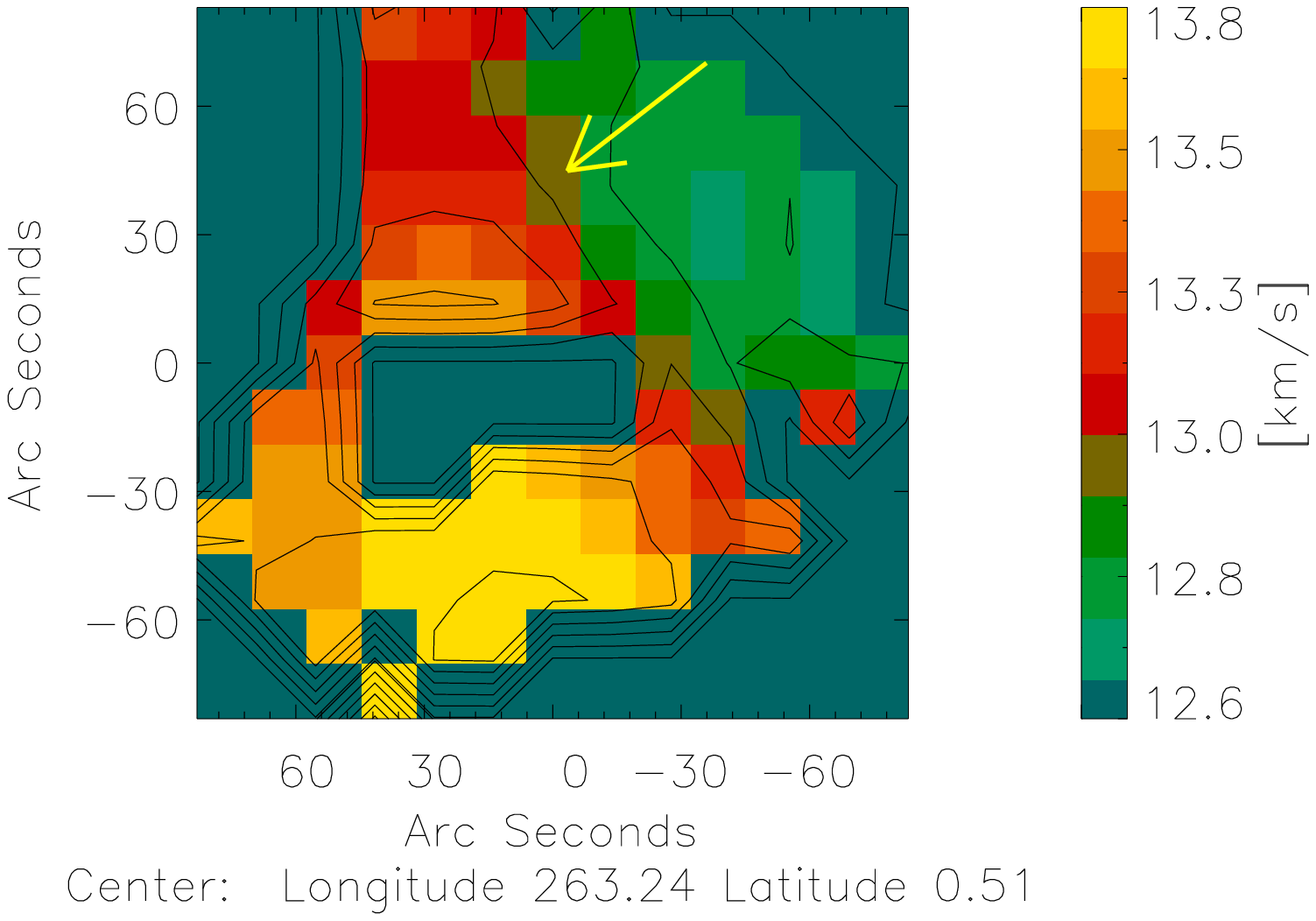} \\
\vspace{0.3cm}
\includegraphics[width=5cm,angle=270]{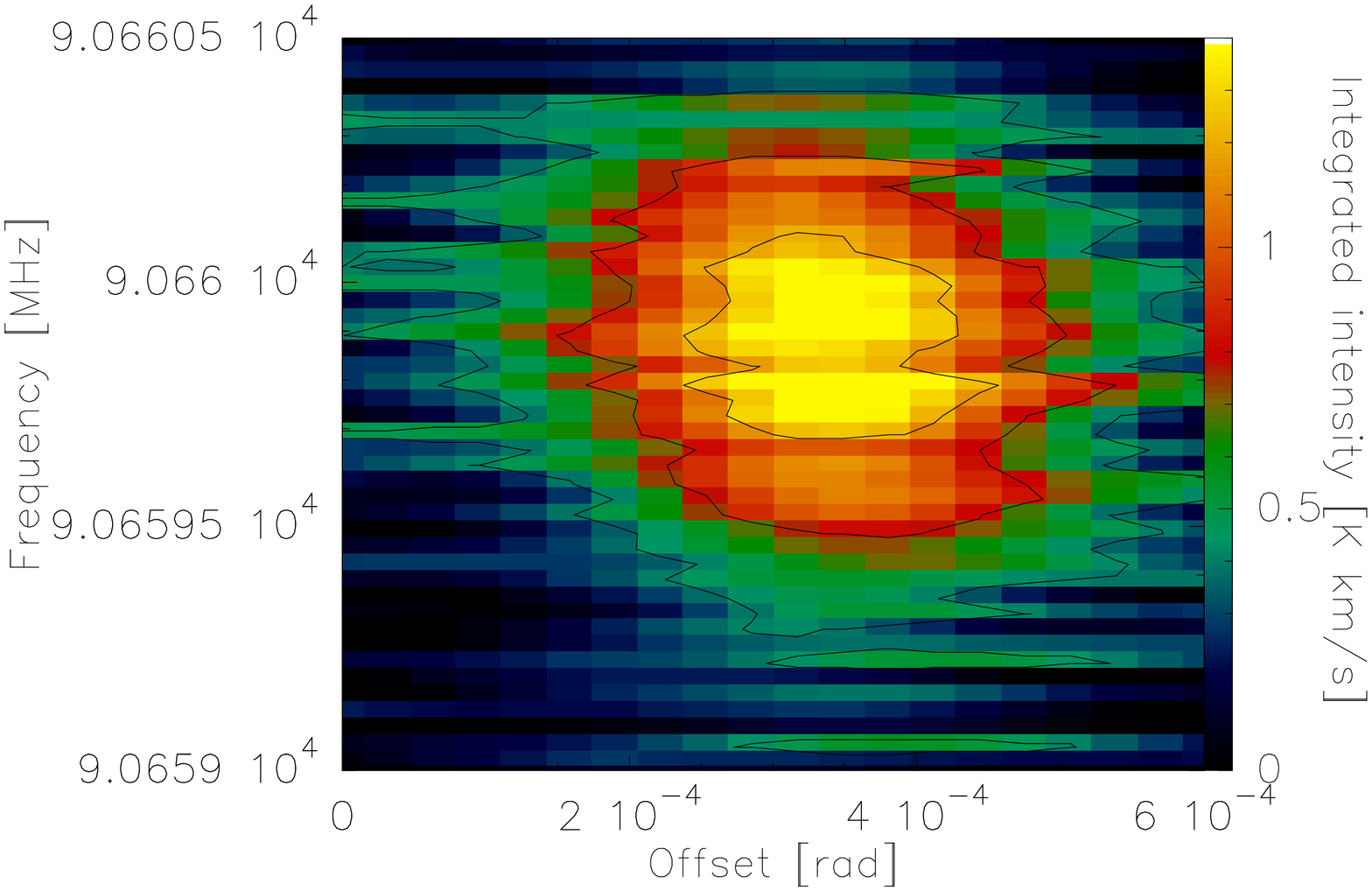} \hspace{1.0cm}
\includegraphics[width=5cm,angle=270]{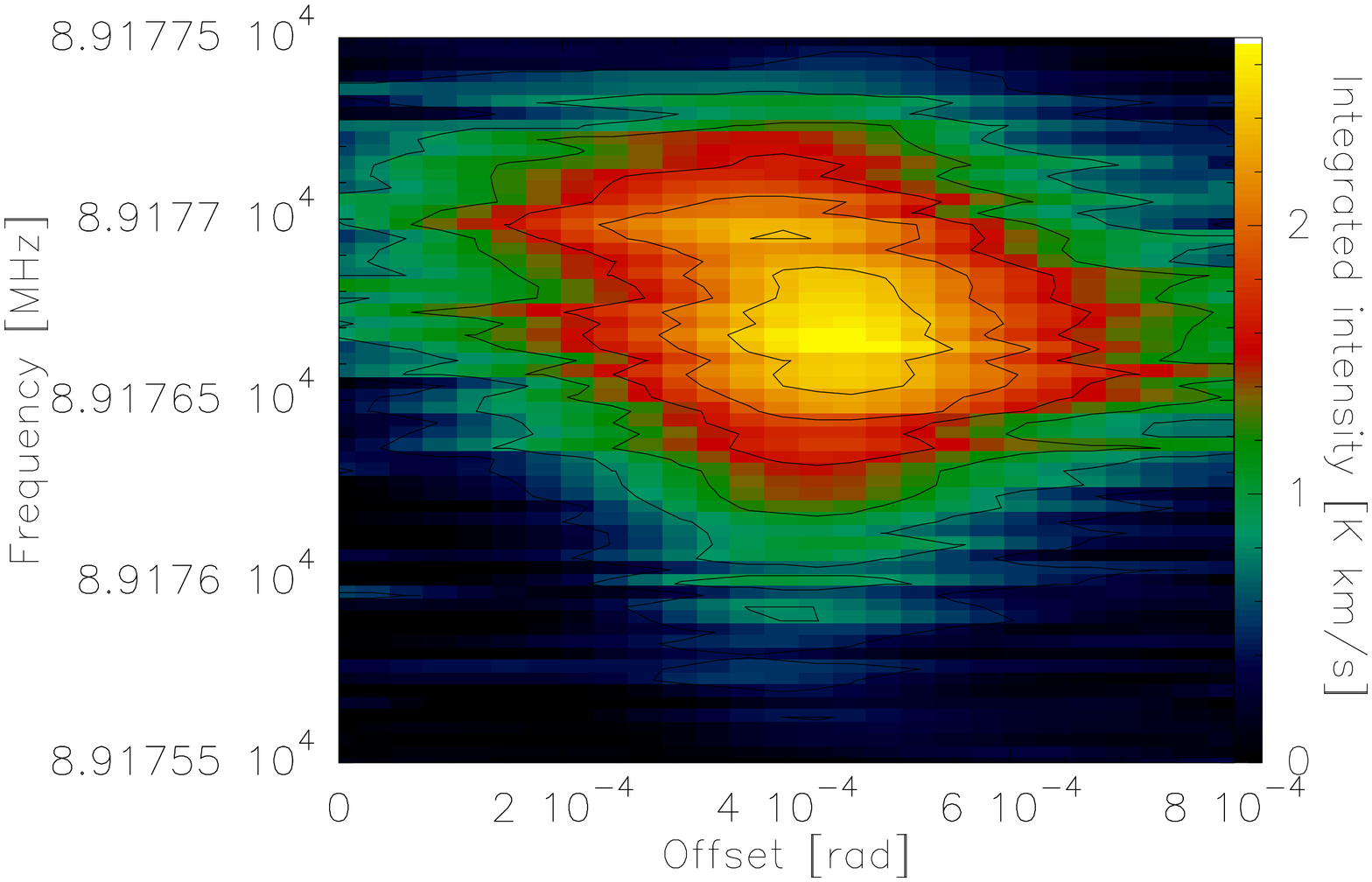}
\caption{
In the top panels we show the maps of the $V_{\rm lsr}$ for the HNC$(1-0)$ (left) and
HCO$^+(1-0)$ (right) lines toward the proto-stellar core BLAST081,
as obtained from Gaussian fits to the spectra at each map position.
The arrows show the direction of the velocity gradient, with their length
proportional to the magnitudes listed in Table~\ref{tab:velfit}.
The bottom panels show the position-velocity plots, with the
horizontal axis representing the angular offset along a direction parallel
to the velocity gradient shown by the arrows in the top panels.
Note the different direction of the velocity gradient in the two molecular tracers.
  }
\label{fig:velgrad}
\end{centering}
\end{figure*}

\subsection{Kinematics}
\label{sec:kin}

\subsubsection{Velocity gradients}
\label{sec:velgrad}

We investigated the presence of systematic velocity gradients in those sources which
could be mapped. In particular, the velocity gradient of a gas clump can be
determined by using all or most of the data in a map at once,
by least-squares fitting maps of line-center velocity for
the direction and magnitude of the best-fit velocity gradient
(see \citealp{good1993} and references therein).

A cloud undergoing solid-body rotation would exhibit a linear
gradient, ${\rm d}V/{\rm d}r$, across the face of a map, perpendicular
to the rotation axis. Thus, we have applied the procedure 
developed by \citet{good1993} and \citet{olmi2002}
to determine the magnitude and direction of the velocity
gradient in our spectral line maps. Our best results are listed in Table~\ref{tab:velfit},
and are restricted to the HCO$^+(1-0)$ and HNC$(1-0)$ lines. 
We note that for most sources with measured velocity gradients in both lines, 
the two separately estimated values of ${\rm d}V/{\rm d}r$ and $\theta_v$ are 
consistent, within a 50\,\% difference. 
However, the velocity gradient measured in the HNC map of core BLAST040 is quite small
and thus its direction can be affected by large uncertainties. Thus the largest
discrepancies between the two tracers are observed in the two proto-stellar cores
BLAST081 and BLAST082, and also in the starless core BLAST013.
%

The position-velocity plots for the proto-stellar core BLAST081 are shown in
Figure~\ref{fig:velgrad}.  The top panels show the maps of $V_{\rm lsr}$ for the HNC$(1-0)$ and
HCO$^+(1-0)$ lines, as obtained from Gaussian fits to the spectra at each observed
position. The arrows represent the direction and magnitude of the velocity gradient
as obtained from the procedure described above. We note the significant difference 
between the directions of the velocity gradients as measured in the HNC and HCO$^+$ maps.
It is not clear whether this discrepancy is real or may be caused by the presence of
some artefacts in the HCO$^+$ map of BLAST081, as noted in Section~\ref{sec:maps}. On the other hand, 
source BLAST082 has no visible artefacts, and it is thus possible that in both
proto-stellar cores the different velocity gradients in the HNC and HCO$^+$ maps are 
real and are tracing different systematic motions. One of these velocity gradients could 
be associated with a molecular outflow, as suggested by the presence of line-wings in the 
HCO$^+(1-0)$ spectra (see Table~\ref{tab:HCO}).

\subsubsection{Line asymmetries}
\label{sec:lineasymm}

The HCO$^+(1-0)$ line is generally the most intense transition observed with
the Mopra telescope in each source, and it does not have a hyperfine structure. Therefore,
it represents the ideal candidate to identify and analyze asymmetries
of the line profile. In fact, the HCO$^+(1-0)$ line shows a non-Gaussian 
profile toward several sources (see Table~\ref{tab:HCO}), 
and in some cases unambiguous wing emission is detected, indicating the likely presence
of a molecular outflow.  By inspecting Table~\ref{tab:HCO}
we also note a moderate shift, up to $\sim 1\,$km\,s$^{-1}$, between the peak 
positions of the generally optically thick HCO$^+(1-0)$ line and optically thin 
H$^{13}$CO$^+(1-0)$ transition. We can give a quantitative estimate of this asymmetry 
in order to get some indirect information about dynamical processes in the cores.

A widely used method to extract line asymmetries is based on the comparison of optically thin and 
optically thick line velocities, i.e., determining the parameter 
$\delta V = (V_{\rm thick} - V_{\rm thin}) / \Delta V_{\rm thin}$ (\citealp{mardones1997}).
In this way we can divide our cores in ``blue shifted'' cores, with $\delta V < 0$, 
and ``red shifted'' cores with $\delta V > 0$. The blue shifted values of $\delta V$ 
could be caused by infall motions and the red excess by expanding motions or outflow. 

\begin{figure}
\centering
\includegraphics[width=9.0cm,angle=0]{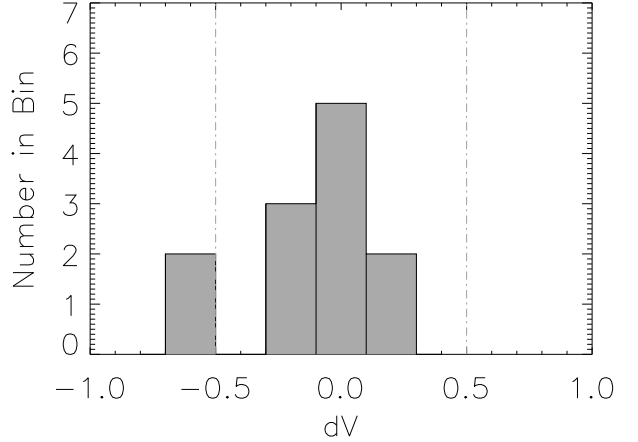}
\caption{
Histogram of $\delta V = (V_{\rm thick} - V_{\rm thin}) / \Delta V_{\rm thin}$, where
$V_{\rm thick}$ and $V_{\rm thin}$ were determined from the HCO$^+(1-0)$ and
H$^{13}$CO$^+(1-0)$ lines, respectively.
The vertical dashed lines indicate the $\pm 5 \sigma_{\delta V}$ level.
  }
\label{fig:histodV}
\end{figure}

We identified 13 sources where the HCO$^+(1-0)$ and H$^{13}$CO$^+(1-0)$ linewidths 
could be measured and in Figure~\ref{fig:histodV} we show the distribution of their
$\delta V$ parameter, with $V_{\rm thick}$ and $V_{\rm thin}$ being determined from 
the HCO$^+(1-0)$ and H$^{13}$CO$^+(1-0)$ lines, respectively. We also estimated the typical 
error on $\delta V$ and then used the same $\pm 5 \sigma_{\delta V}$ threshold as
\citet{mardones1997} to exclude those sources where the $V_{\rm lsr}$ differences were 
dominated by measurement errors. We estimated $5 \sigma_{\delta V} \simeq 0.4 - 0.6$, thus
in Figure~\ref{fig:histodV} a typical value $5 \sigma_{\delta V} \simeq 0.5$ is used.
Only three cores show a significant blue shift (BLAST044, BLAST090 and BLAST093), 
and two a red shift (BLAST003 and BLAST053).  From the red excess candidates 
we have to exclude BLAST053 which has a weak and noisy H$^{13}$CO$^+(1-0)$ spectrum.

%
%
%
\begin{table*}
\caption{Mass estimates.}
\label{tab:mass}
\centering
\begin{tabular}{lccccccc}
\hline\hline
\multicolumn{1}{c}{Source \#} & & & \multicolumn{5}{c}{$ M_{\rm cd}$\tablefootmark{b}} \\
\cline{4-8}
& $M_{\rm blast}$  &
$M_{\rm vir}$\tablefootmark{a}  &
\multicolumn{1}{c}{\bf N$_2$H$^+$} &
\multicolumn{1}{c}{\bf HNC} &
\multicolumn{1}{c}{\bf HCO$^+$} &
\multicolumn{1}{c}{\bf HCN}  &
\multicolumn{1}{c}{\bf NH$_3$\tablefootmark{c}} \\
&
[$M_\odot$]  &
[$M_\odot$]  &
[$\times 10^{-9} M_\odot$]  &
[$\times 10^{-9} M_\odot$]  &
[$\times 10^{-9} M_\odot$]  &
[$\times 10^{-9} M_\odot$]  &
[$\times 10^{-7} M_\odot$]  \\
%
\hline
   3  & 23.2  & 68.1     & $-$   & $-$    & $-$       & $-$   & $-$  \\
   9  & 3.1   & 15.9     & $-$   & 2.6    & 1.6       & 0.3   & 7.9  \\
  13  & 6.4   & $-$      & $-$   & 3.7    & 7.6       & 3.5   & $-$  \\
  14  & 7.7   & 8.9      & $-$   & $-$    & $-$       & $-$   & $-$  \\
  24  & 98.0  & 87.3     & 5.4   & 25.7   & 54.1      & 17.8  & 5.8  \\
  31  & 4.0   & 26.4     & $-$   & 0.02   & 2.3       & 0.6   & 31.8 \\
  34  & 4.6   & $-$      & $-$   & 2.1    & 1.9       & $-$   & $-$  \\
  38  & 2.2   & 32.4     & $-$   & $-$    & $-$       & $-$   & $-$  \\
  40  & 6.3   & 11.3     & 0.13  & 4.9    & 13.9      & 2.5   & 2.1  \\
  41  & 2.8   & 10.8     & $-$   & $-$    & 7.4       & $-$   & 0.8  \\
  44  & 5.8   & 19.1     & $-$   & $-$    & $-$       & $-$   & $-$  \\
  45  & 2.1   & 17.6     & $-$   & $-$    & $-$       & $-$   & $-$  \\
  47  & 2.6   & 44.7     & $-$   & 2.2    & 2.7       & 5.7   & 4.6  \\
  50  & 2.3   & $-$      & $-$   & $-$    & 6.0       & 9.9   & $-$  \\
  53  & 11.6  & 25.9     & $-$   & $-$    & $-$       & $-$   & $-$  \\
  56  & 1.8   & $-$      & $-$   & 0.1    & 1.5       & 0.6   & $-$  \\
  57  & 14.0  & 27.0     & $-$   & $-$    & 8.7       & $-$   & $-$  \\
  59  & 1.8   & 17.5     & $-$   & $-$    & $-$       & $-$   & $-$  \\
  63  & 13.4  & 64.6     & 1.5   & 4.7    & 10.3      & 8.3   & 9.0  \\
  77  & 3.7   & 12.3     & $-$   & $-$    & $-$       & $-$   & $-$  \\
  81  & 69.9  & 88.8     & 2.2   & 13.4   & 40.4      & 14.2  & 0.6  \\
  82  & 16.5  & 42.2     & 2.4   & 8.1    & 29.8      & 4.2   & 1.4  \\
  89  & 2.4   & $-$      & $-$   & $-$    & 6.7       & $-$   & $-$  \\
  90  & 22.2  & 43.9     & $-$   & $-$    & $-$       & $-$   & $-$  \\
  93  & 15.2  & 17.7     & $-$   & $-$    & $-$       & $-$   & $-$  \\
  97  & 24.4  & 54.5     & $-$   & $-$    & $-$       & $-$   & $-$  \\
 101  & 2.1   & $-$      & $-$   & 1.0    & 3.5       & 0.9   & $-$  \\
 109  & 1.0   & $-$      & $-$   & $-$    & 0.9       & $-$   & $-$  \\
\hline
\end{tabular}
\tablefoot{
\tablefoottext{a}{
$M_{\rm vir}$ is determined assuming a uniform density
spherical source.
}
\tablefoottext{b}{
$M_{\rm cd}$ is determined only when a spectral line map was
available, with the exception of NH$_3$.
}
\tablefoottext{c}{
$M[{\rm NH_3}]$ is listed only for those sources where an estimate
of the rotational temperature, $T_{\rm 12}$, was possible, and  where an estimate of the source
diameter could be obtained from at least one of the molecular tracers.
}
}
\end{table*}

\subsection{Derivation of masses}
\label{sec:gasmass}

From our spectral line data we can derive masses from the gas column density, $M_{\rm cd}$, 
as well as virial masses, $M_{\rm vir}$, which we can 
then compare with the BLAST-derived masses, $M_{\rm blast}$, estimated from the dust continuum.
Clearly, while the virial masses refer to the {\it total} gas mass, with
$M_{\rm cd}$ we can only determine the mass of a given molecular species.  The total gas
mass can only be determined by assuming specific molecular abundances; or, alternatively, we can give
a coarse estimate of the molecular abundance by taking the ratio $M_{\rm cd}/M_{\rm blast}$
(see Section~\ref{sec:molabd}).  The isotopologues observed by us, H$^{13}$CO$^+$ and H$^{13}$CN, are
generally too weak to allow derivation of column densities throughout the maps, and thus we 
use only the main molecular species, attempting to correct for first-order optical depth effects.

\subsubsection{Mass derived from column density }
\label{sec:coldens}

The calculation of masses from the column density assumes LTE and we determine the molecular gas
mass integrating the molecule column density over the extent of the source. Then
we can write:
\begin{equation}
M_{\rm cd} = d^2\, m_{\rm mol} \,
\int \, N_{\rm mol} \, {\rm d}\Omega
\label{Mcd}
\end{equation}
where $\int \, N_{\rm mol} \, {\rm d}\Omega$ is the molecule column
density integrated over the region enclosed by the chosen contour level,
$m_{\rm mol}$ is the mass of the specific molecule being considered, and $d$ is
the distance to the source.
The column density $N_{\rm mol}$ corresponding to a $J\rightarrow J-1$ rotation
transition can be calculated as:
\begin{eqnarray}
& & N_{\rm mol} \, [{\rm cm}^{-2}] = \frac{4.0\times 10^{12}}
{J^2\mu^2[{\rm D}] \,B [{\rm K}] }
Z \exp\left( \frac{E_{\rm J}}{T_{\rm ex}} \right ) \times  \nonumber \\
& & \frac{1}{\eta_{\rm mb}} \frac{\tau}{1-e^{-\tau}}
\int \, T_{\rm A}^{\star} \, {\rm d}v \, [{\rm K~ km~ s}^{-1}]
\label{eq:cd}
\end{eqnarray}
where $B$ denotes the rotational constant, $E_{\rm J}$ is the
upper state energy, $\mu$ is the dipole moment  (in Debye) and we used the
escape probability $\tau/[1-\exp(-\tau)]$ to account for first-order 
optical depth effects. The derivation of both $T_{\rm ex}$ and $\tau$ 
has been discussed in Section~\ref{sec:hfs}.
For $k T_{\rm ex} >> h B$ the partition function, $Z$, of a linear molecule is given by:
\begin{equation}
Z = \frac{k T_{\rm ex}}{h B}
\end{equation}
where $h$ and $k$ are the Planck and Boltzmann constants, respectively.

Eq.~(\ref{Mcd}) is actually implemented by writing:
\begin{equation}
M_{\rm cd} = d^2\, m_{\rm mol} \, \Delta \Omega_{\rm pix} \, \sum_{i=1}^{n_{\rm pix}}  
\, N_{\rm mol}(x_i, y_i)
\end{equation}
where $N_{\rm mol}(x_i, y_i)$ represents the column density in a single pixel $(x_i, y_i)$ of
the map, with $n_{\rm pix}$ representing the total number of pixels, and $\Delta \Omega_{\rm pix}$
represents the solid angle covered by a single pixel. The map pixels selected are those that have an
integrated intensity $I \geq \sigma_{\rm map}$, {\it and} also lie within an $1/e$ radius from the center
of a  2D Gaussian fit to the core, for each specific molecule. 
Assuming that a line is detected if at least two adjacent velocity channels lie 
above the $3 \sigma_{\rm rms}$ level of the spectrum, then we calculate
the RMS of the integrated intensity in the map simply as
$\sigma_{\rm map} = 2 (3 \sigma_{\rm rms}) \Delta v$, where $\Delta v$ is the channel width.
We thus select only those pixels that belong to the actual ``core'' of the source, and remove pixels
that come from extended diffuse envelopes.
The resulting masses are listed in Table~\ref{tab:mass}.

%

%
\begin{figure}
\centering
\includegraphics[width=7.5cm,angle=0]{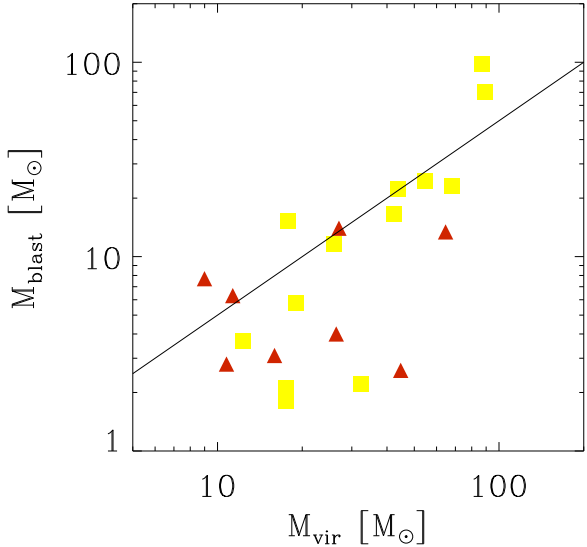}
\vspace{2mm}
\caption{
Total core mass, $M_{\rm core} = M_{\rm blast}$,  vs. the virial mass, $M_{\rm vir}$, calculated
using the velocity linewidths of the NH$_3$(1,1) and N$_2$H$^+(1-0)$ transitions (see text).
Yellow, filled squares represent proto-stellar cores and red, filled triangles represent
starless cores. The solid line indicates the minimum $M_{\rm core} = 0.5 M_{\rm vir}$ for
which the cores should be self-gravitating.
Individual values of $M_{\rm vir}$ are listed in Table~\ref{tab:mass}.
}
\label{fig:Mvir}
\end{figure}

\subsubsection{Virial mass }
\label{sec:virial}

We then estimate the virial mass of the cores assuming they are simple spherical systems
with uniform density \citep{maclaren1988}: 
%
\begin{equation}
M_{\rm vir}{\rm [M_\odot]} = 210 \, R_{\rm dec}{\rm [pc]} \, (\Delta V_{\rm av} {\rm [km~s^{-1}]})^2
\label{eq:mvir}
\end{equation}
where $R_{\rm dec}$ is the deconvolved source radius and $\Delta V_{\rm av}$ represents 
the line FWHM of the molecule of mean mass, calculated as the sum of the 
thermal and turbulent components:
%
\begin{equation}
\Delta V_{\rm av}^2 = \Delta V_{\rm mol}^2 + kT8 \ln 2 \, \left ( \frac{1}{m_{\rm av}} -
\frac{1}{m_{\rm mol}} \right )
\end{equation}
where $\Delta V_{\rm mol}$ is the line FWHM of the molecular transition being considered
and $m_{\rm av}=2.3$~amu is the mean molecular weight (with respect to the total number of
particles), assuming a mass fraction for He of 25\%. 
We note that according to \citet{maclaren1988}, Eq.~(\ref{eq:mvir}) may lead
to a mass overestimate if the density distribution is not uniform. For example, in the
case of a sphere with a density distribution $\rho \propto r^{-2}$ the numerical factor in 
Eq.~(\ref{eq:mvir}) should be replaced by 126.

As the deconvolved source radius we used the BLAST-derived values of \citet{olmi2009}. 
We decided to use the BLAST sizes, to estimate the virial masses, instead of the 
Mopra-derived sizes for two main reasons: {\it (i)} the linewidths are mostly measured 
from the single-point spectra of NH$_3$, and where NH$_3$ was not observed or not 
detected we used the observations of N$_2$H$^+$, for which only a few maps were 
available; {\it (ii)} the BLAST maps are not affected by spatial variations that may 
instead affect specific molecules and thus likely give a better representation 
of the mass distribution in each source.
%

%
%
\begin{figure*}
\centering
\includegraphics[width=4.4cm,angle=0]{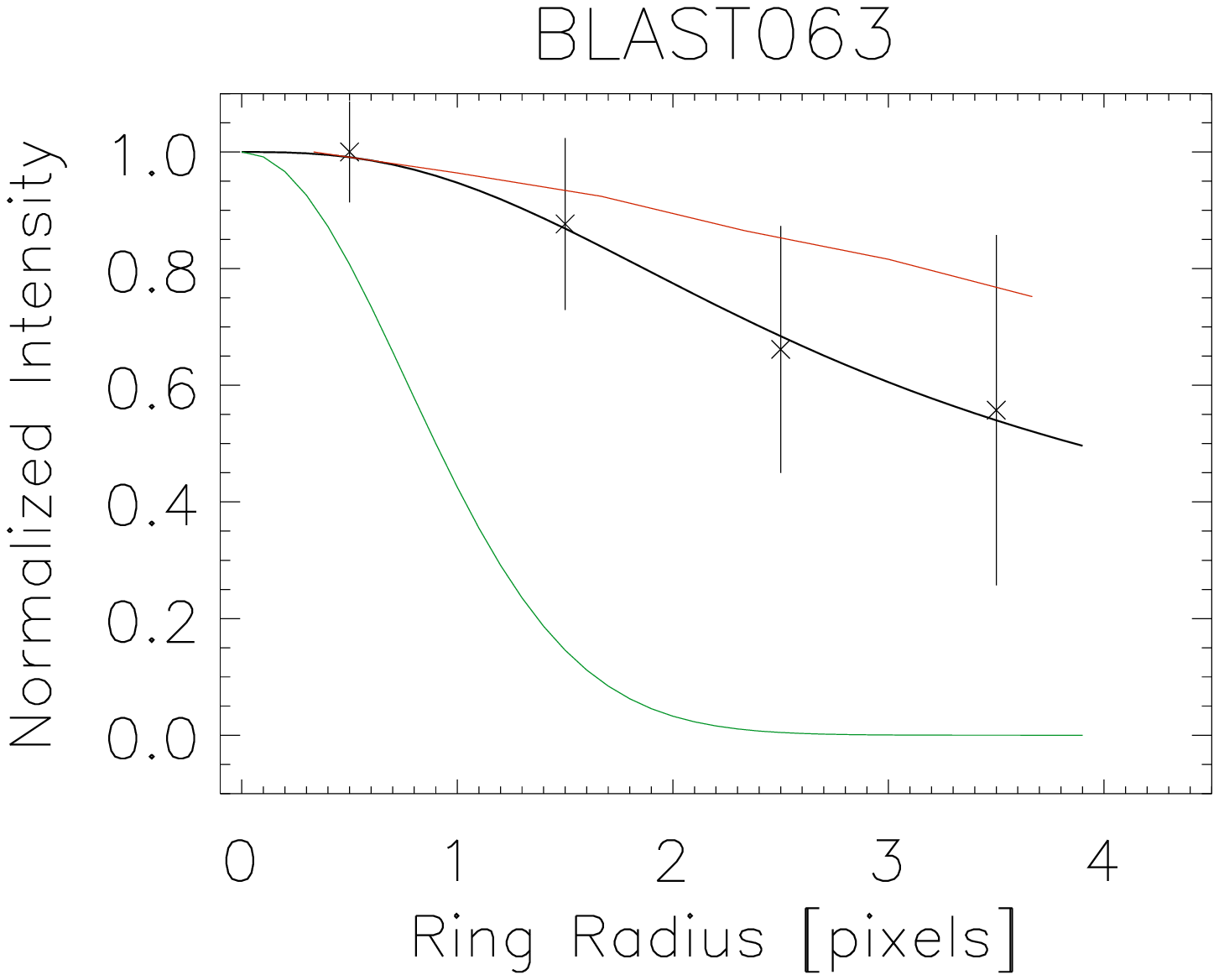}
\includegraphics[width=4.4cm,angle=0]{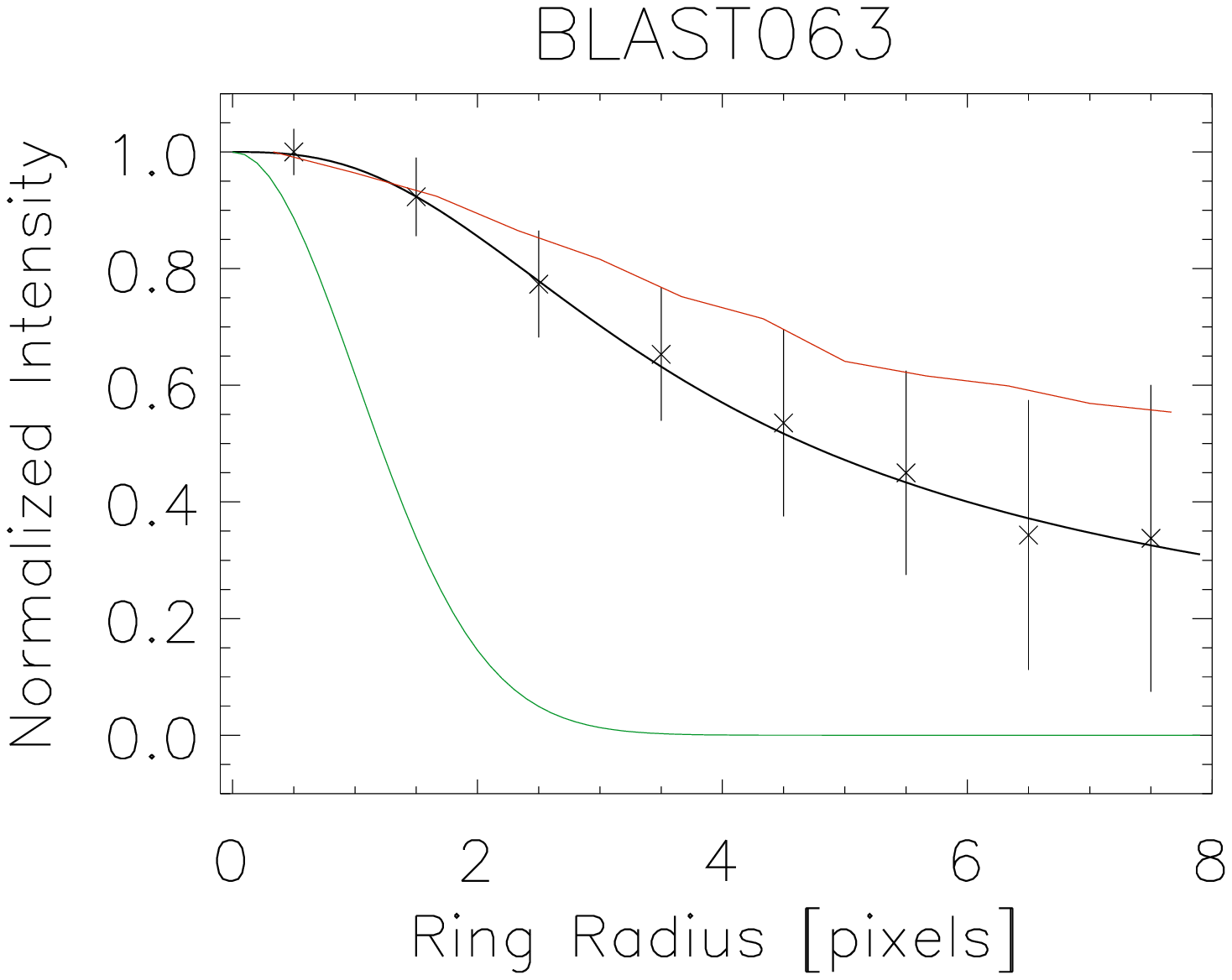}
\includegraphics[width=4.4cm,angle=0]{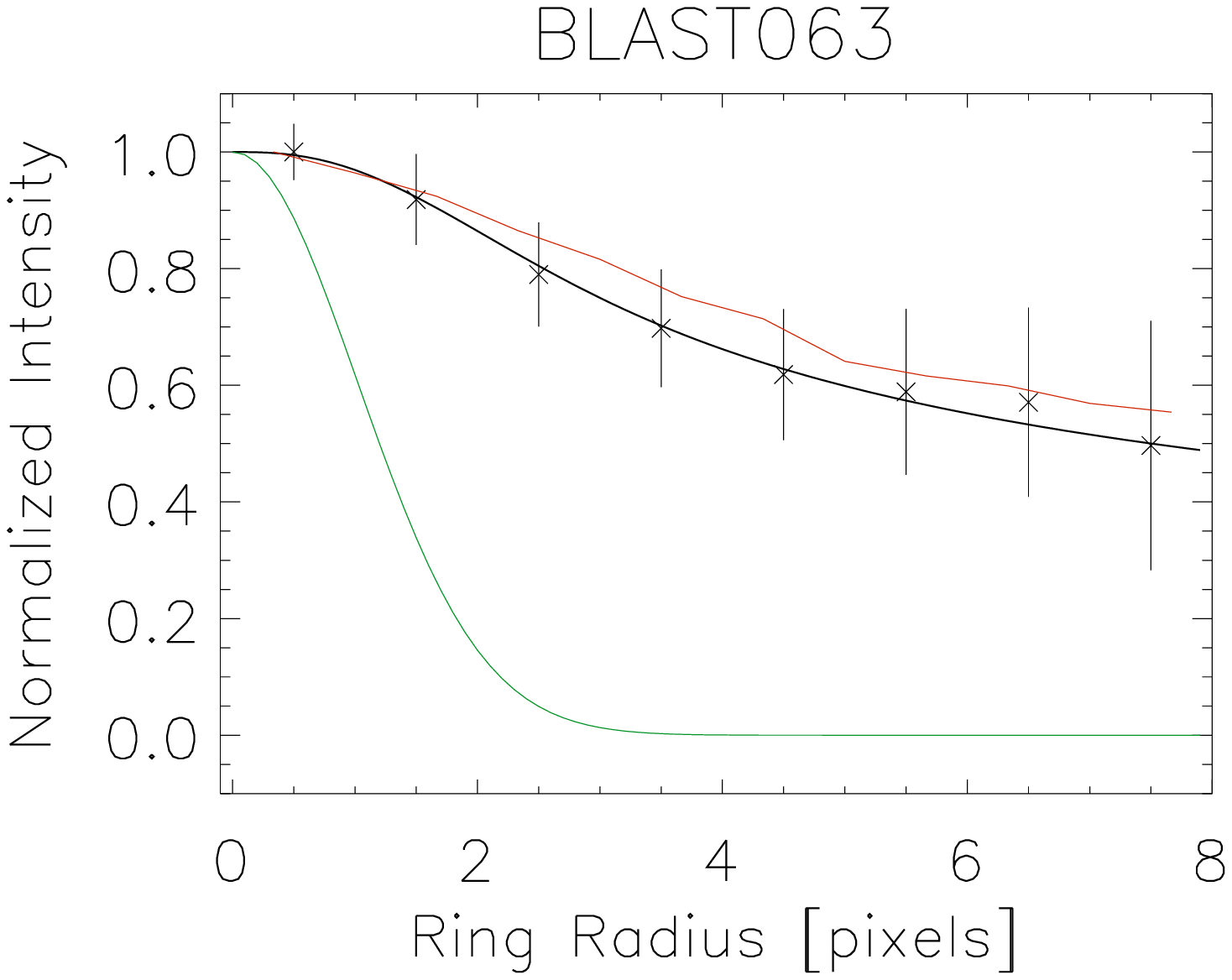}
\includegraphics[width=4.4cm,angle=0]{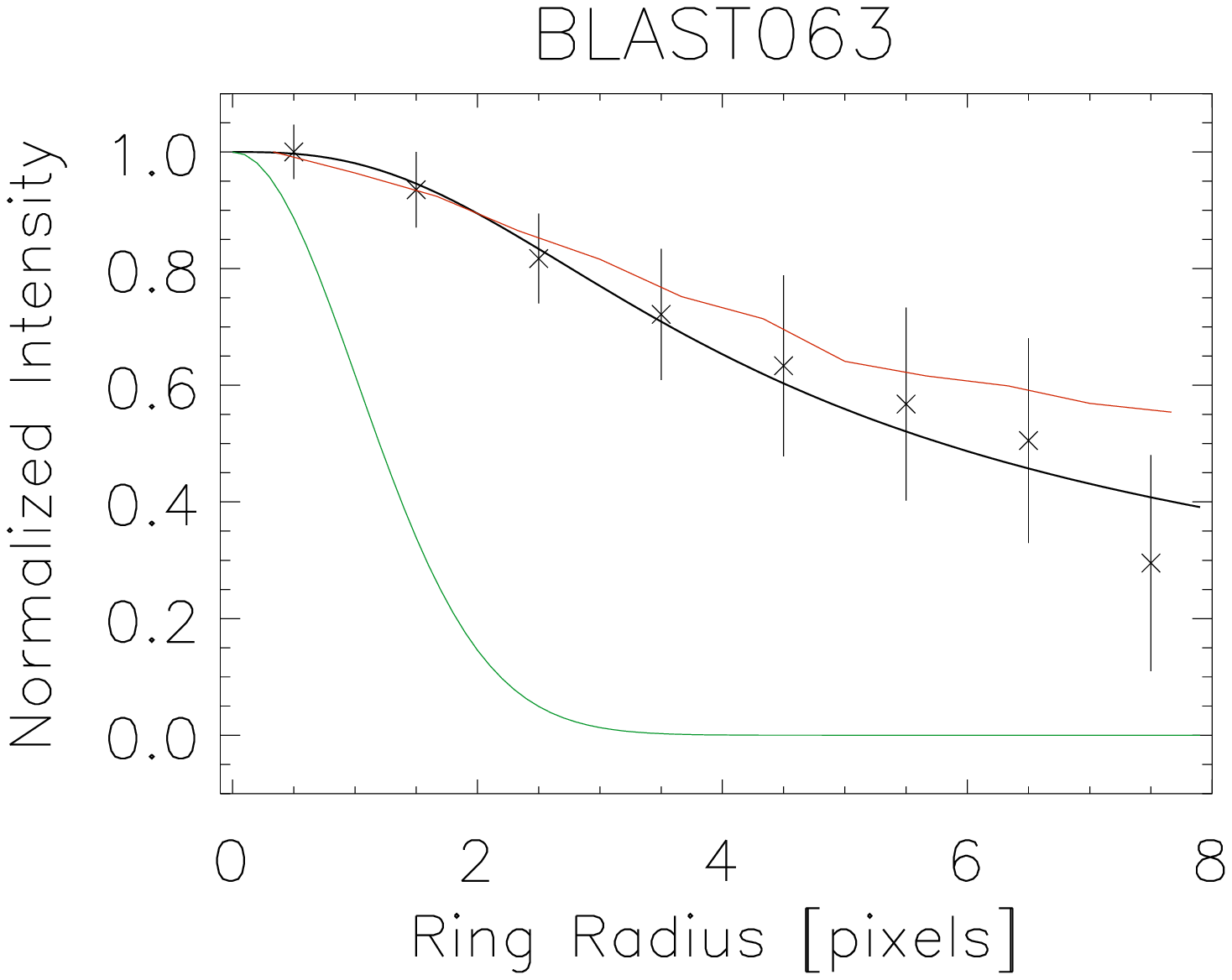} \\
\includegraphics[width=4.4cm,angle=0]{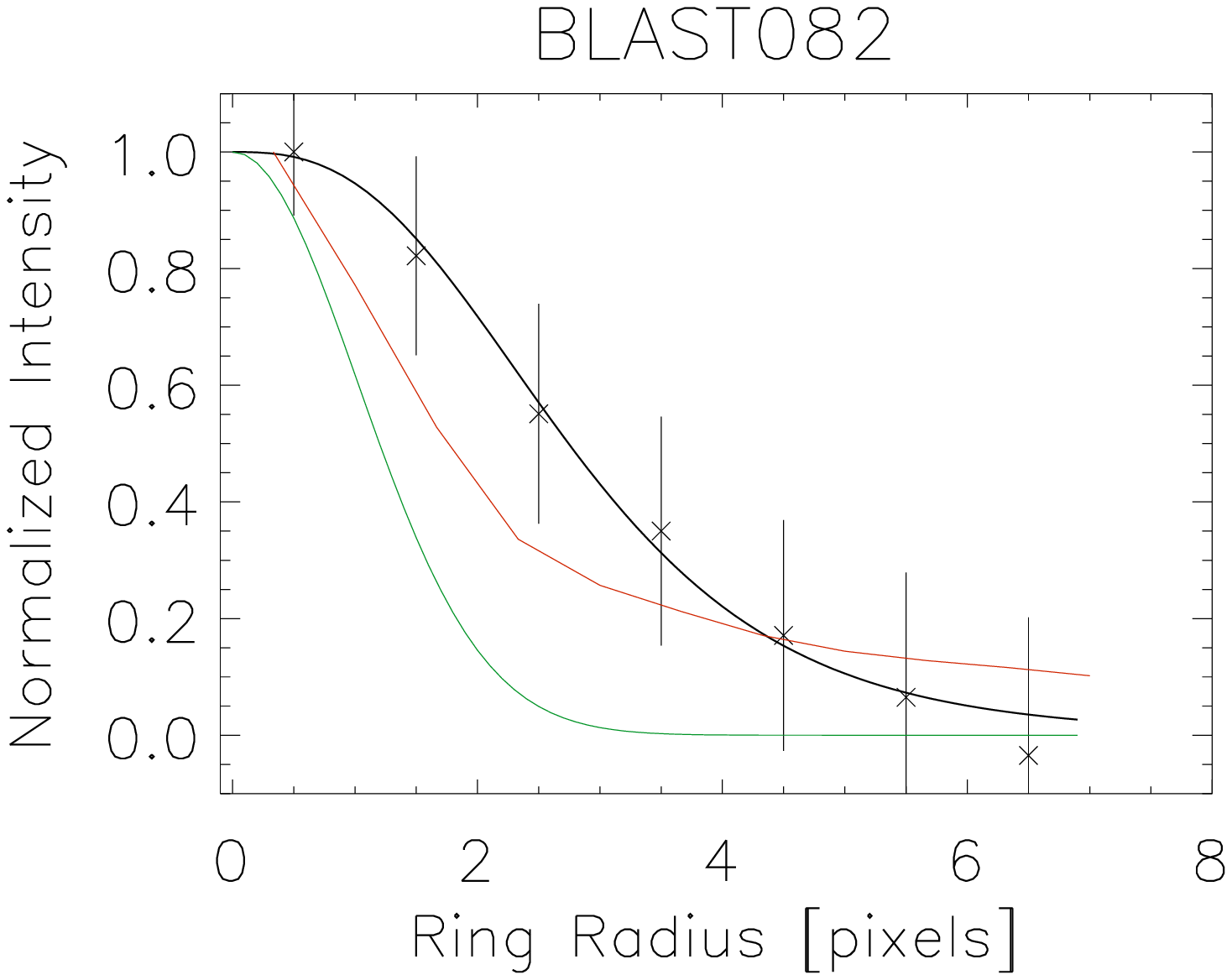}
\includegraphics[width=4.4cm,angle=0]{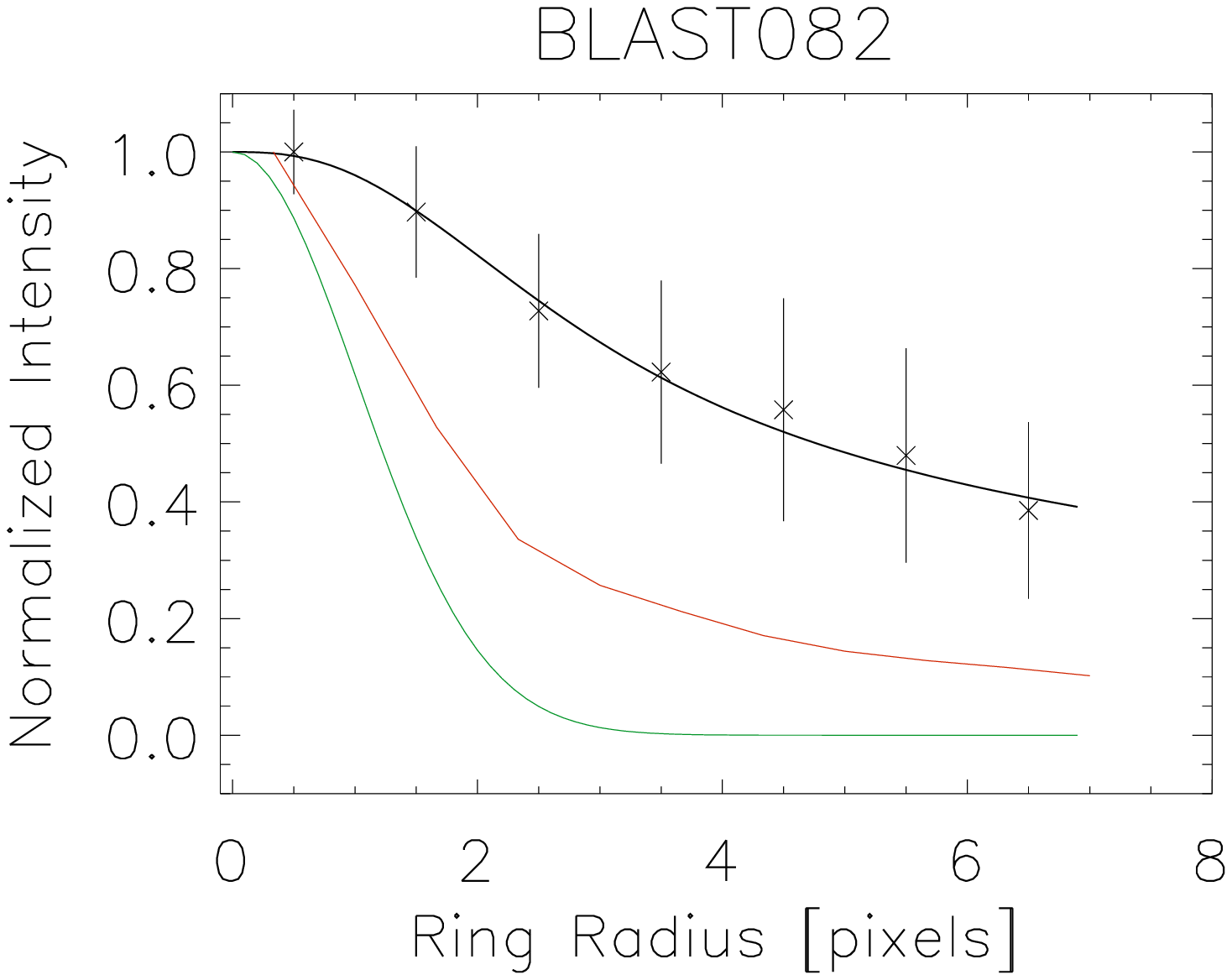}
\includegraphics[width=4.4cm,angle=0]{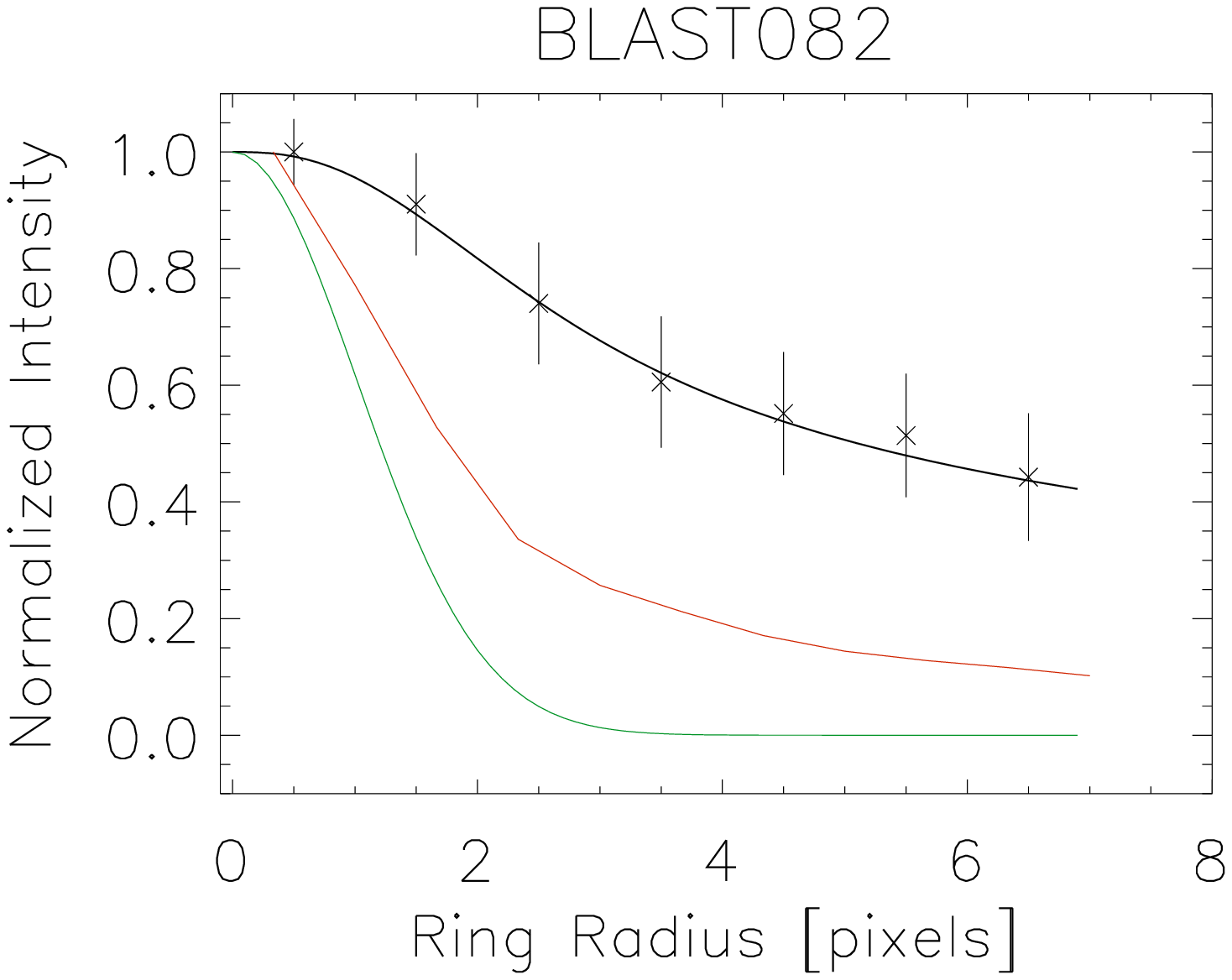}
\includegraphics[width=4.4cm,angle=0]{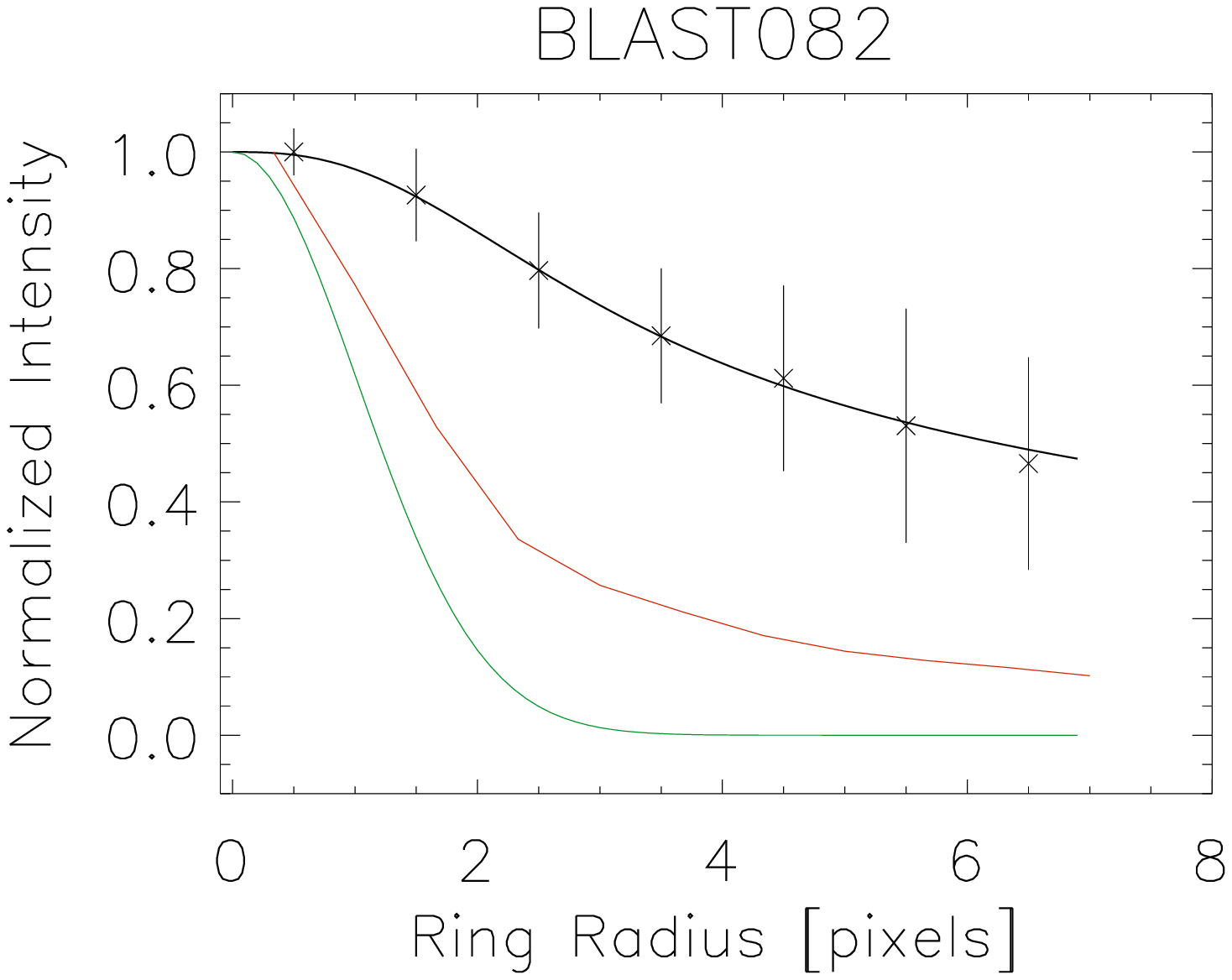} \\
\caption{
Examples of normalized radial profiles for sources BLAST063 (starless, top panels) and BLAST082
(proto-stellar, bottom panels). The points and error bars represent the ring-averaged
integrated intensity in the molecular lines (from left to right)
N$_2$H$^+(1-0)$, HNC$(1-0)$, HCO$^+(1-0)$ and HCN$(1-0)$.
The black solid line represents the fit obtained using Eq.~(\ref{eq:plummer}).
The red solid line shows the radial profile of the averaged BLAST
intensity at 250\,$\mu$m, and the green solid line represents
the 3-mm beam profile of the Mopra telescope. Each pixel corresponds to 15\,arcsec.
}
\label{fig:radprof}
\end{figure*}

As previously stated, the line FWHM is calculated from the NH$_3$(1,1) single-point 
spectra toward each source, if available, or from the N$_2$H$^+(1-0)$ line  otherwise, 
since these molecules are less likely to be affected by
depletion, they trace the denser gas and are mostly optically thin.
The linewidths determined by the fit procedure are artificially broadened by the 
velocity resolution of the observations, and thus we subtracted in quadrature the 
resolution width, $\Delta V_{\rm res}$, from the observed line FWHM, $\Delta V_{\rm obs}$, 
such that $\Delta V_{\rm mol} = \sqrt{\Delta V_{\rm obs}^2 - \Delta V_{\rm res}^2 }$.

%
%
\begin{table*}
\caption{Parameters for fit to the radial profile of column density.}
\label{tab:radprof}
\centering
\begin{tabular}{lcccccccccccc}
\hline\hline
\multicolumn{1}{c}{Source \#} &
 &
\multicolumn{2}{c}{{\bf N$_2$H$^+$}} &
 &
\multicolumn{2}{c}{{\bf HNC}} &
 &
\multicolumn{2}{c}{{\bf HCO$^+$}} &
 &
\multicolumn{2}{c}{{\bf HCN}}  \\
\cline{3-4}\cline{6-7}\cline{9-10}\cline{12-13}
& & $R_{flat}$ & $\eta$ & & $R_{flat}$ & $\eta$ & & $R_{flat}$ & $\eta$ & & $R_{flat}$ & $\eta$ \\
& & [arcsec] & & & [arcsec] & & & [arcsec] & & & [arcsec] & \\
%
\hline
   9 & & $-$        & $-$     & & 67.5       & 1.2        & & 15.0      & 0.3       & & 34.5       & 0.9  \\
  13 & & $-$        & $-$     & & 69.0       & 3.4        & & 16.5      & 0.6       & & 19.5       & 0.5  \\
  24 & & 28.5       & 1.1     & & 48.0       & 1.5        & & 15.0      & 0.8       & & 22.5       & 1.1  \\
  31 & & $-$        & $-$     & & 37.5       & 3.3        & & 27.0      & 1.8       & & 15.0       & 0.4  \\
  34 & & $-$        & $-$     & & 57.0       & 1.4        & & 16.5      & 0.7       & & $-$        & $-$  \\
  40 & & 27.0       & 2.7     & & 45.0       & 1.0        & & 16.5      & 0.3       & & 16.5       & 0.2  \\
  41 & & $-$        & $-$     & & $-$        & $-$        & & 33.0      & 0.8       & & $-$        & $-$  \\
  47 & & $-$        & $-$     & & 30.0       & 0.8        & & 16.5      & 0.4       & & 15.0       & 0.4  \\
  50 & & $-$        & $-$     & & $-$        & $-$        & & 15.0      & 0.3       & & 18.0       & 0.6  \\
  56 & & $-$        & $-$     & & 58.5       & 2.3        & & 39.0      & 1.9       & & 16.5       & 0.5  \\
  57 & & $-$        & $-$     & & $-$        & $-$        & & 20.0      & 0.6       & & $-$        & $-$  \\
  63 & & 70.0       & 2.0     & & 30.0       & 0.9        & & 18.0      & 0.4       & & 24.0       & 0.5  \\
  81 & & 24.0       & 2.6     & & 21.0       & 1.3        & & 15.0      & 0.7       & & 18.0       & 1.1  \\
 82A\tablefootmark{a} & & 31.5       & 2.4     & & 22.5       & 0.7        & & 16.5      & 0.5       & & 21.0       & 0.5  \\
 82B & & 31.5       & 2.4     & & 16.5       & 0.6        & & 16.5      & 0.5       & & 16.5       & 0.5  \\
  89 & & $-$        & $-$     & & $-$        & $-$        & & 36.0      & 1.1       & & $-$        & $-$  \\
 101 & & $-$        & $-$     & & 22.5       & 0.7        & & 27.0      & 1.2       & & 33.0       & 1.6  \\
 109 & & $-$        & $-$     & & $-$        & $-$        & & 36.0      & 2.8       & & $-$        & $-$  \\
\hline
\end{tabular}
\tablefoot{
\tablefoottext{a}{
The parameters for the two spatial components in BLAST082 are listed.
}
}
\end{table*}

The resulting virial masses are listed in Table~\ref{tab:mass} and a plot of the total 
core mass, $M_{\rm blast}$, as derived from BLAST, vs. the virial mass is shown in 
Figure~\ref{fig:Mvir}.  We note that nearly all of the starless cores lie below the 
self-gravitating line, indicating that they are unlikely to be gravitationally bound, 
whereas more than half of the proto-stellar cores lie near or above that line, confirming that 
they are gravitationally bound (about 30\% of all sources have $M_{\rm core}/M_{\rm vir} > 0.5 $). 
However, as we mentioned above, for a non-uniform 
density distribution (see Section~\ref{sec:radprof}), e.g. of type $\rho \propto r^{-2}$, 
all virial masses should be multiplied by a factor 0.6. In this case,
most of the proto-stellar cores would move above the self-gravitating line and also most
of the starless cores would be located near that line (about 60\% of all sources have 
$M_{\rm core}/M_{\rm vir} > 0.5 $ in this case).


\subsection{Radial profiles of column density}
\label{sec:radprof}

In order to identify significant differences in the distribution of both molecular gas
and dust continuum, we have derived radial profiles of the column density from both
the BLAST maps at 250$\, \mu$m and from the Mopra maps of the integrated intensity of
the different molecular transitions. 
While the BLAST 500$\,\mu$m measurements would be somewhat less affected 
by optical depth effects, we have chosen to compare our molecular data to 
the BLAST 250$\,\mu$m measurements since 
they are closely matched to the Mopra beam.

Previous work has shown that single-power-law density, or column density,
 profiles do not fit the 
emission from dense cores and that a central flattening is always needed to 
reproduce the data (e.g., \citealp{tafalla2002} and references therein). 
Among the standard analytic profiles that combine the power-law behavior 
of column density for large radii $r$ and a central flattening at small $r$, 
is the following function:
%
\begin{equation}
N(r) = N_{\rm flat} \left [ \frac{R_{\rm flat}}{(R_{\rm flat}^2 + r^2 )^{1/2} } \right ]^\eta
\label{eq:plummer}
\end{equation}
where the column density is approximately uniform, with $N \sim N_{\rm flat}$,
 for $r \ll R_{\rm flat}$,
and it falls off as $r^{-\eta}$ for $r \gg R_{\rm flat}$.

In the fitting procedure we first circularly average the maps of the integrated intensity of
selected molecular transitions around the peak, and then use a $\chi^2$ routine
to fit the column density profile, obtained from the integrated intensity
with Eq.~(\ref{eq:cd}), using the model given by Eq.~(\ref{eq:plummer}),
after convolution with a 2D $38''$ Gaussian. 
The procedure is then repeated for the BLAST250 maps. The column density profiles 
obtained from the spectral line maps and the dust continuum maps
are then compared to find any possible evidence of chemical effects such as, for example,
freezing-out on grains.


The results of the radial profile fits to individual sources are listed in Table~\ref{tab:radprof},
while in Table~\ref{tab:aveprop} we list the average values of the parameters $\eta$ and
$R_{\rm flat}$, i.e. $\langle \eta \rangle$ and $\langle R_{\rm flat} \rangle$, obtained by
selecting only fits to the radial profiles that converged. 
We can note several trends: first, apart from the large scatter observed 
in the case of the N$_2$H$^+$ and HNC molecules, the $\langle R_{\rm flat} \rangle$ values are quite consistent
within the errors. Second, the value of $\langle \eta \rangle$ for the N$_2$H$^+$ molecule
is significantly higher compared to the other molecules although, once again, the HNC molecule
shows a larger scatter. We tentatively note a trend of decreasing $\langle \eta \rangle$ values 
from HNC to HCO$^+$ and then HCN, but given the errors this cannot be considered significant.
We also note that the large uncertainty in the $\langle R_{\rm flat} \rangle$ value for 
N$_2$H$^+$ is due to the BLAST063 source. If this source is not included, then we would have
$\langle R_{\rm flat} [ {\rm N_2H^+} ] \rangle = 28.5\pm 3.2$, quite close to the values
estimated for HCN and HCO$^+$.

As a final trend, when comparing the radial profiles obtained by the BLAST dust continuum 
with those of the molecular tracers, Figure~\ref{fig:radprof} shows that, 
at least for the two cores shown in the figure, 
the agreement between the radial profiles of the molecular tracers and of the dust 
continuum is better for the starless core (BLAST063). 
In order to get a quantitative comparison, we have applied the two-sample Kolmogorov-Smirnov
test (K-S test) to the BLAST- and Mopra-derived radial profiles.
The K-S test evaluates the maximum deviation between the cumulative 
distribution functions  of the two data sets and also the associated probability, $p_{\rm ks}$ 
($ 0 < p_{\rm ks} < 1$), that two arrays of data values are drawn from the same distribution.
Small values of $p_{\rm ks}$ show that the cumulative distribution function of one data set 
is significantly different from the other. For the two sources shown in Figure~\ref{fig:radprof}
we get an average value of $p_{\rm ks} =  0.66$ for BLAST063 and $p_{\rm ks} =  0.24$ 
for BLAST082, confirming that in BLAST063 the dust and molecular gas radial profiles are more similar
than in BLAST082.
 

However, it is clear comparing the radial profiles and also
the integrated intensity maps, that the starless cores have a wider range of morphologies, with 
sources showing a good agreement between the dust continuum and the molecular radial profiles 
(although not necessarily in all molecular tracers, e.g., in sources BLAST013 and BLAST031), 
and sources where this agreement is less evident (e.g., in BLAST056 and BLAST063). These 
differences suggest that the starless sources in our sample are currently in 
different evolutionary phases.

\begin{table}
\caption{Average properties of radial profiles.}
\label{tab:aveprop}
\centering
\begin{tabular}{lccc}
\hline\hline
Spectral line &
$\langle R_{\rm flat} \rangle$ &
$\langle \eta \rangle$ &
$\langle X_{\rm mol} \rangle$ \\
& [arcsec] & &  \\
%
\hline
N$_2$H$^+(1-0)$      & $35.4 \pm 17.2$  & $2.24 \pm 0.65$    & $(7.3  \pm 5.4)\, \times 10^{-11}$  \\
HNC$(1-0)$           & $40.4 \pm 18.3$  & $1.39 \pm 1.00$    & $(4.5  \pm 2.9)\, \times 10^{-10}$  \\
HCO$^+(1-0)$         & $21.9 \pm 8.6$   & $0.87 \pm 0.67$    & $(1.3 \pm 0.8)\, \times 10^{-9}$ \\
HCN$(1-0)$           & $20.8 \pm 6.4$   & $0.68 \pm 0.39$    & $(0.8 \pm 1.2)\, \times 10^{-9}$ \\
NH$_3$\tablefootmark{a}     & $-$              & $-$                & $(1.7 \pm 3.1)\, \times 10^{-7}$ \\
\hline
\end{tabular}
\tablefoot{
\tablefoottext{a}{
Mean value estimated from NH$_3$ abundances listed in Table~\ref{tab:abd}.
}
}
\end{table}

%
%
\begin{figure*}
\centering
\includegraphics[width=5.5cm,angle=0]{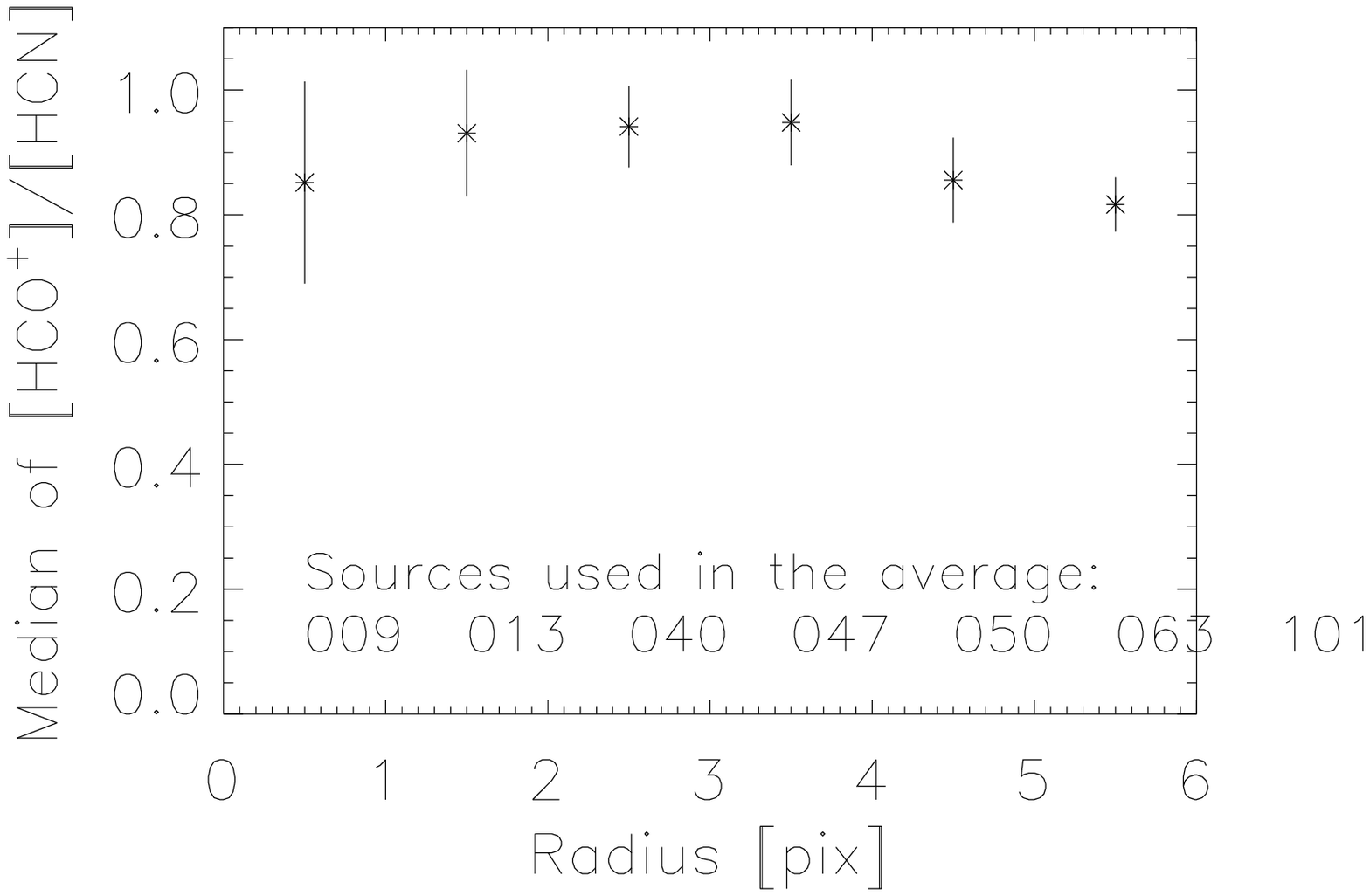}
\includegraphics[width=5.5cm,angle=0]{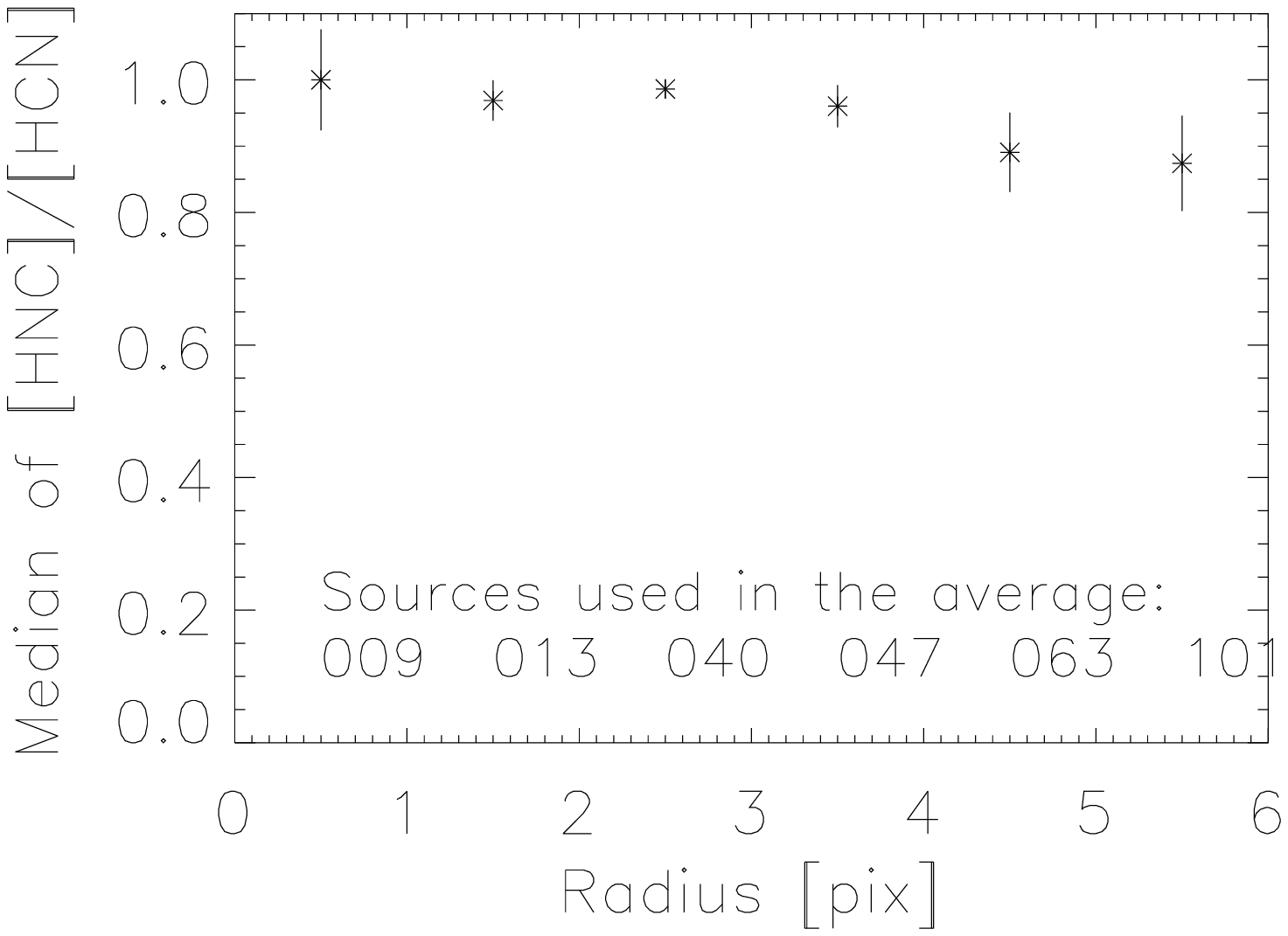}
\includegraphics[width=5.5cm,angle=0]{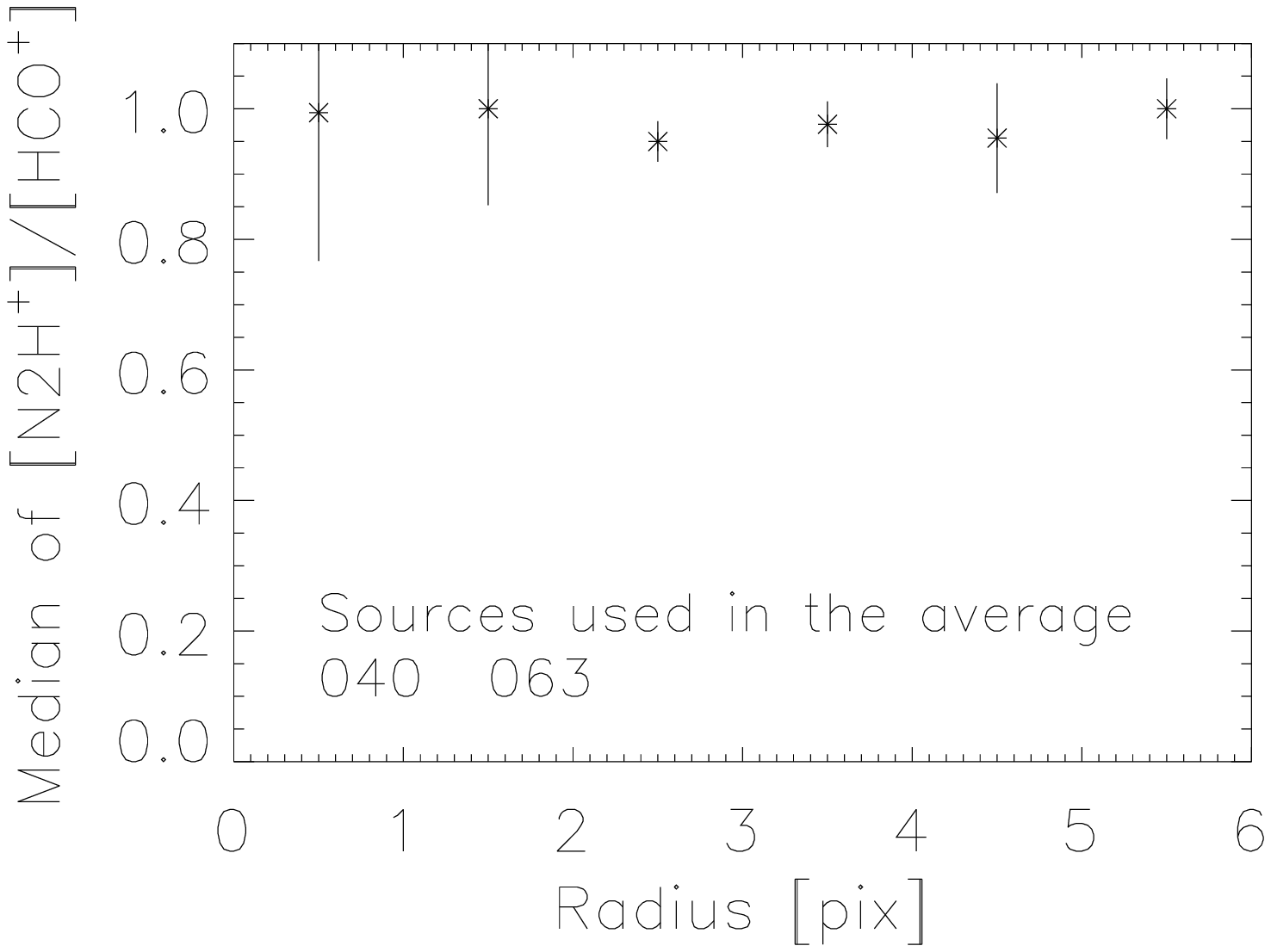} \\
\includegraphics[width=5.5cm,angle=0]{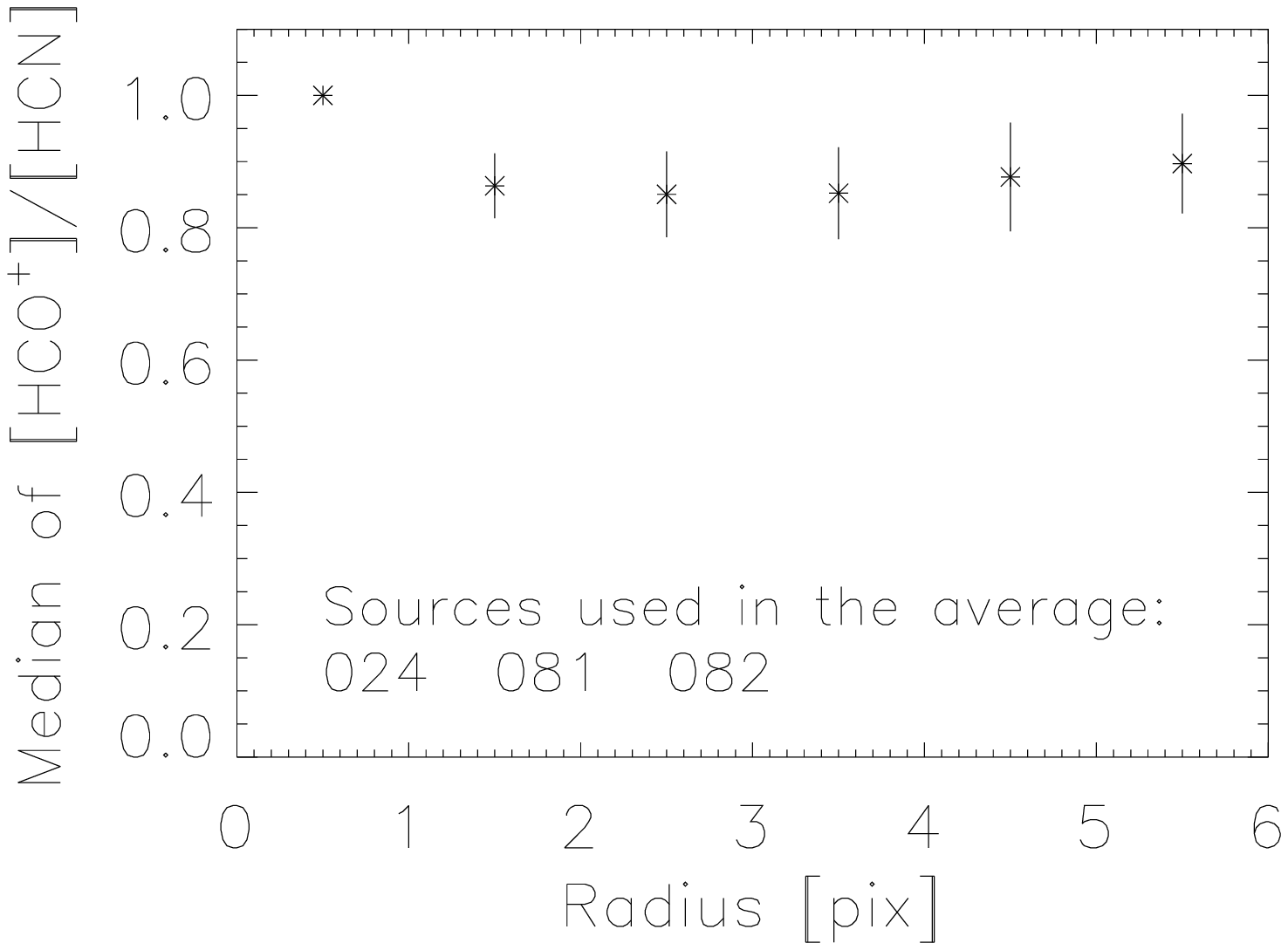}
\includegraphics[width=5.5cm,angle=0]{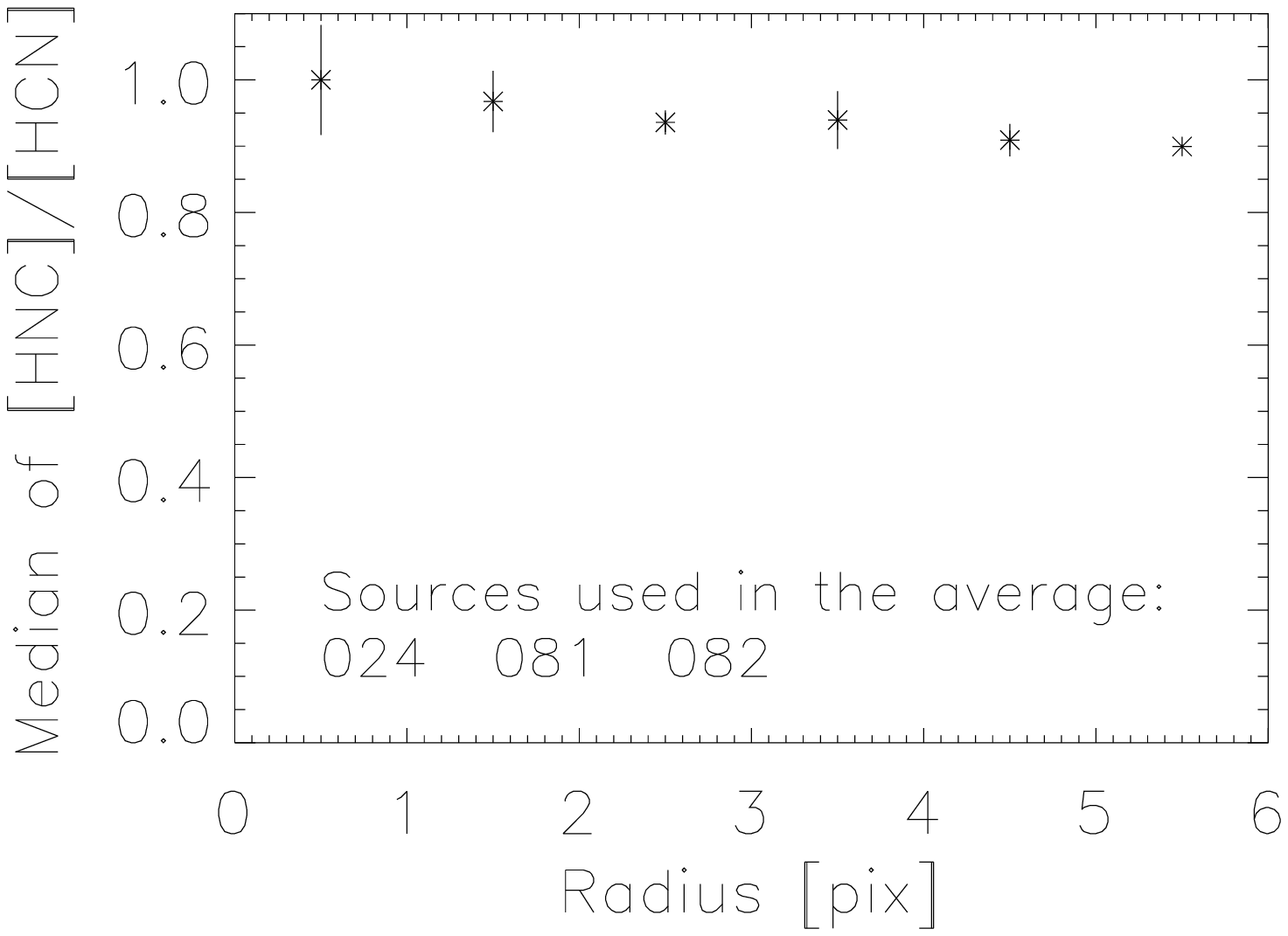}
\includegraphics[width=5.5cm,angle=0]{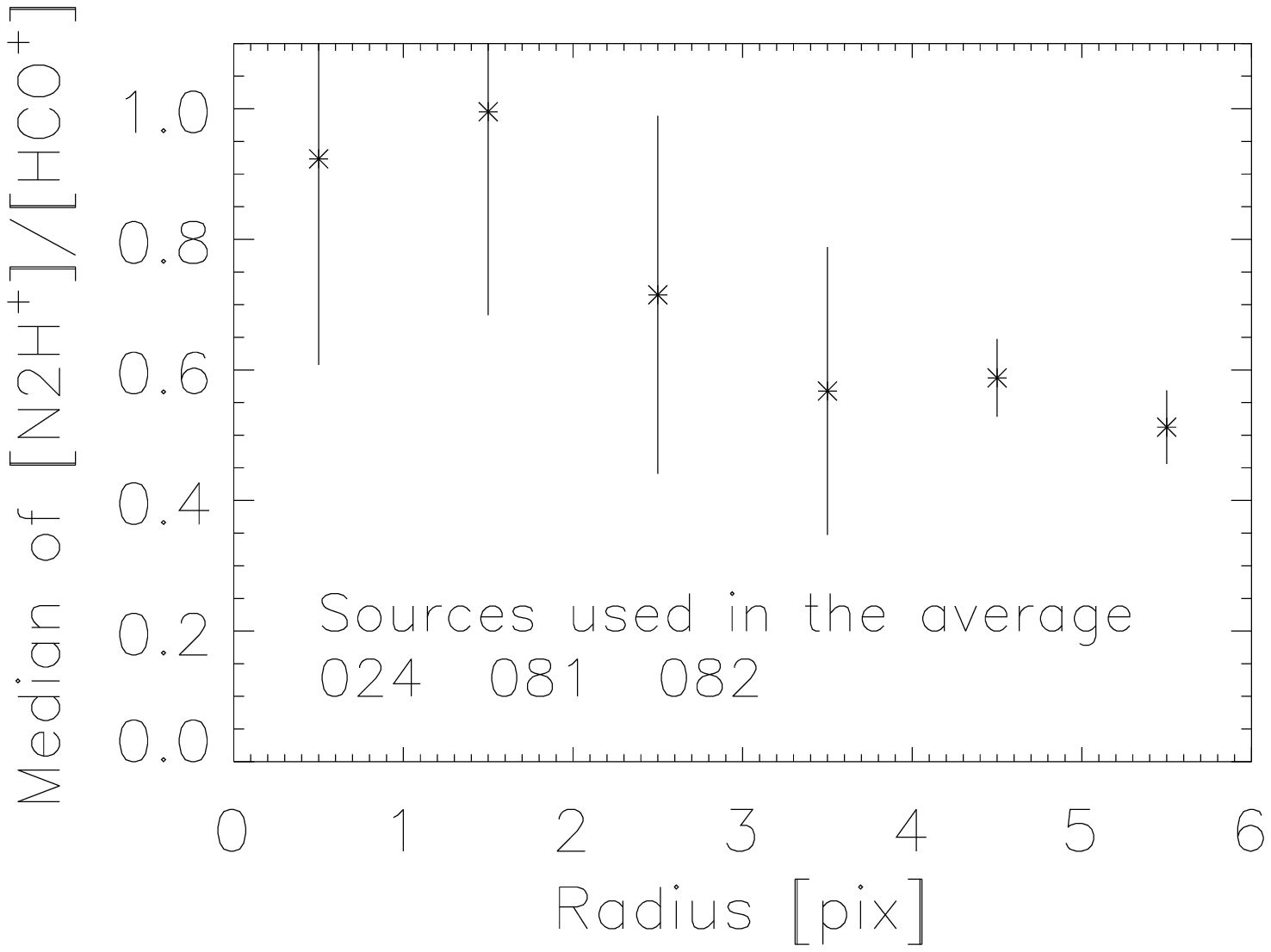} \\
\caption{Source-averaged radial profiles of the relative molecular abundances
(from left to right) of [HCO$^+$]/[HCN], [HNC]/[HCN] and [N$_2$H$^+$]/[HCO$^+$].
The abundances are normalized with respect to the peak value.
The top and bottom panels show the resulting averages for all starless and proto-stellar
cores, respectively, where data were available.
Pixel size as in Figure~\ref{fig:radprof}.
}
\label{fig:molabd}
\end{figure*}

\section{Discussion }
\label{sec:discussion}

In this section we will describe how the results obtained in Section~\ref{sec:res} 
allow us to determine the relative molecular abundances, and in particular how 
these abundances vary as a function of the radial distance in both starless and
proto-stellar cores. We will also discuss how the relative molecular abundances
can be used to roughly estimate the chemical evolutionary status of the sources.
In the last part of the section we will analyze the degree of correlations among
various physical and chemical parameters, which will also allow to highlight
further differences between starless and proto-stellar cores.

\subsection{Determining molecular abundances}
\label{sec:molabd}

As we mentioned earlier, we have estimated the molecular abundances by comparing the
molecular gas mass, $M_{\rm cd}$, with the total gas mass, $M_{\rm blast}$,
estimated by the BLAST measurements of the dust continuum, 
i.e.,  $X_{\rm mol} = M_{\rm cd}/M_{\rm blast}$.  We prefer this method to the 
alternative technique of obtaining $X_{\rm mol}$ by means of a pixel-by-pixel comparison of the
column densities in the Mopra and BLAST maps, which could be affected by differences
in the spatial distribution of the spectral line and dust continuum emission.
The estimated abundances are shown in Table~\ref{tab:abd} and their average values
are listed in Table~\ref{tab:aveprop}. The large uncertainties in the estimated
abundances are likely the consequence of uncertainties in the derivation of the 
BLAST masses, of the different spatial distribution of the molecular gas and the dust,
and possibly also of different chemical evolutionary phases (see the discussion in
Sections~\ref{sec:molabdprof} and  \ref{sec:chemevo}).

The average value of $X({\rm N_2H^+}) = (7.3 \pm 5.4)\, 10^{-11}$ listed in 
Table~\ref{tab:aveprop} is consistent with that found by, e.g., \citet{blake1995} 
in NGC1333 and also by \citet{zinchenko2009}
in S187 and W3. However, it is relatively low compared to mean values found in other cloud cores.
For example, \citet{womack1992} and \citet{benson1998} found mean values of
$X({\rm N_2H^+})$ of $ 4 \times 10^{-10}$ and $ 7 \times 10^{-10}$, respectively, from various samples.

From Table~\ref{tab:aveprop}, we can also determine the
average values of some relative abundances: 
 $X({\rm HCN})/X({\rm HNC}) \sim 1.8 $,           
$X({\rm HNC})/X({\rm HCO^+}) \sim 0.3 $, and   
 $X({\rm HCN})/X({\rm HCO^+}) \sim 0.6 $.
The value of $X({\rm HNC})/X({\rm HCO^+})$ agrees reasonable well with that estimated by 
\citet{godard2010} ($0.5 \pm 0.3$), whereas the other two relative abundances,
$X({\rm HCN})/X({\rm HNC})$ and $X({\rm HCN})/X({\rm HCO^+})$ both appear to 
be somewhat lower compared to the estimated values ($4.8 \pm 1.3$ and $1.9 \pm 0.9$, 
respectively) of \citet{godard2010}, though all estimates have large ($\ga 50$\%) uncertainties.

%
%
\begin{figure}
\includegraphics[width=7.7cm,angle=0]{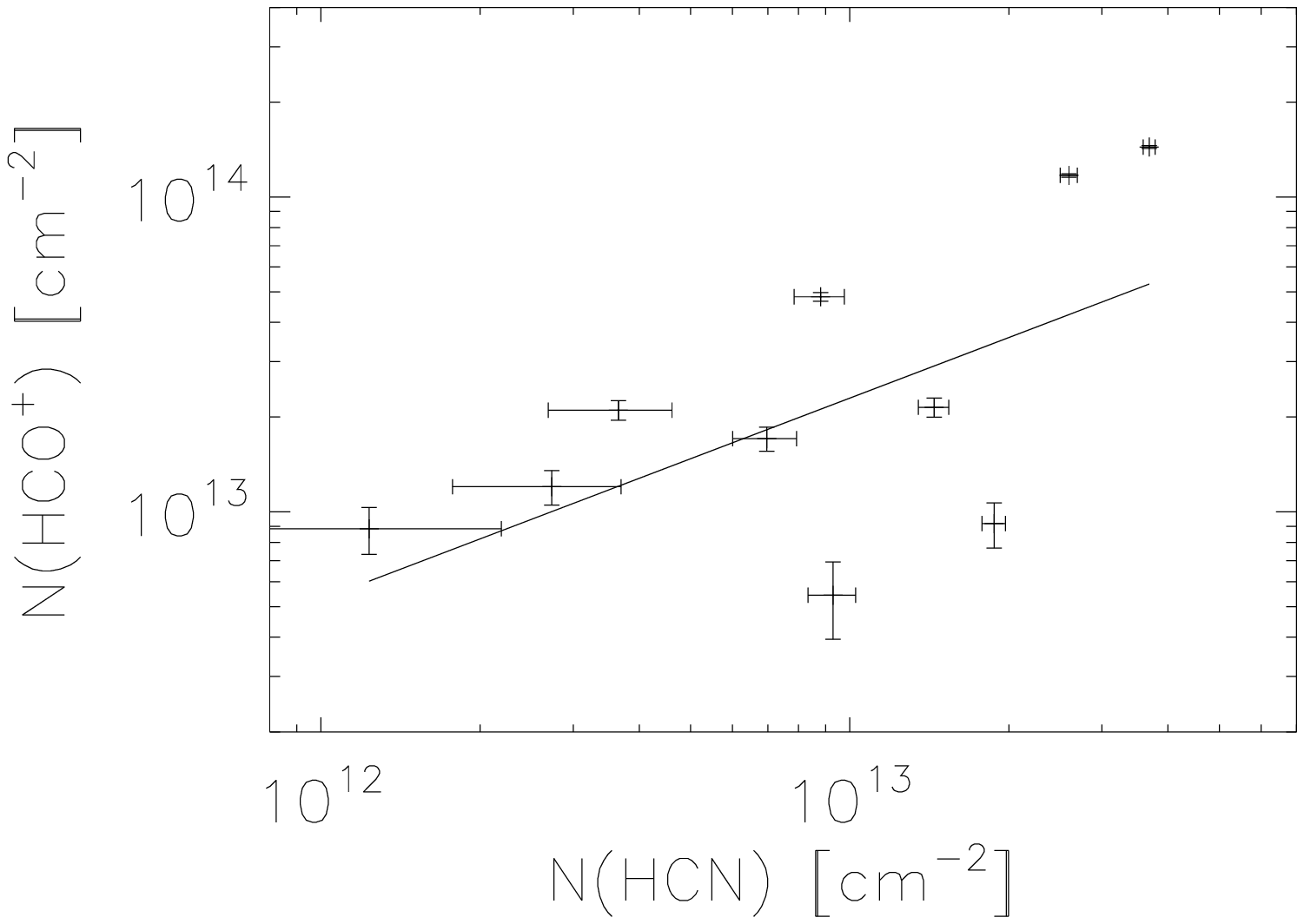}
\includegraphics[width=7.7cm,angle=0]{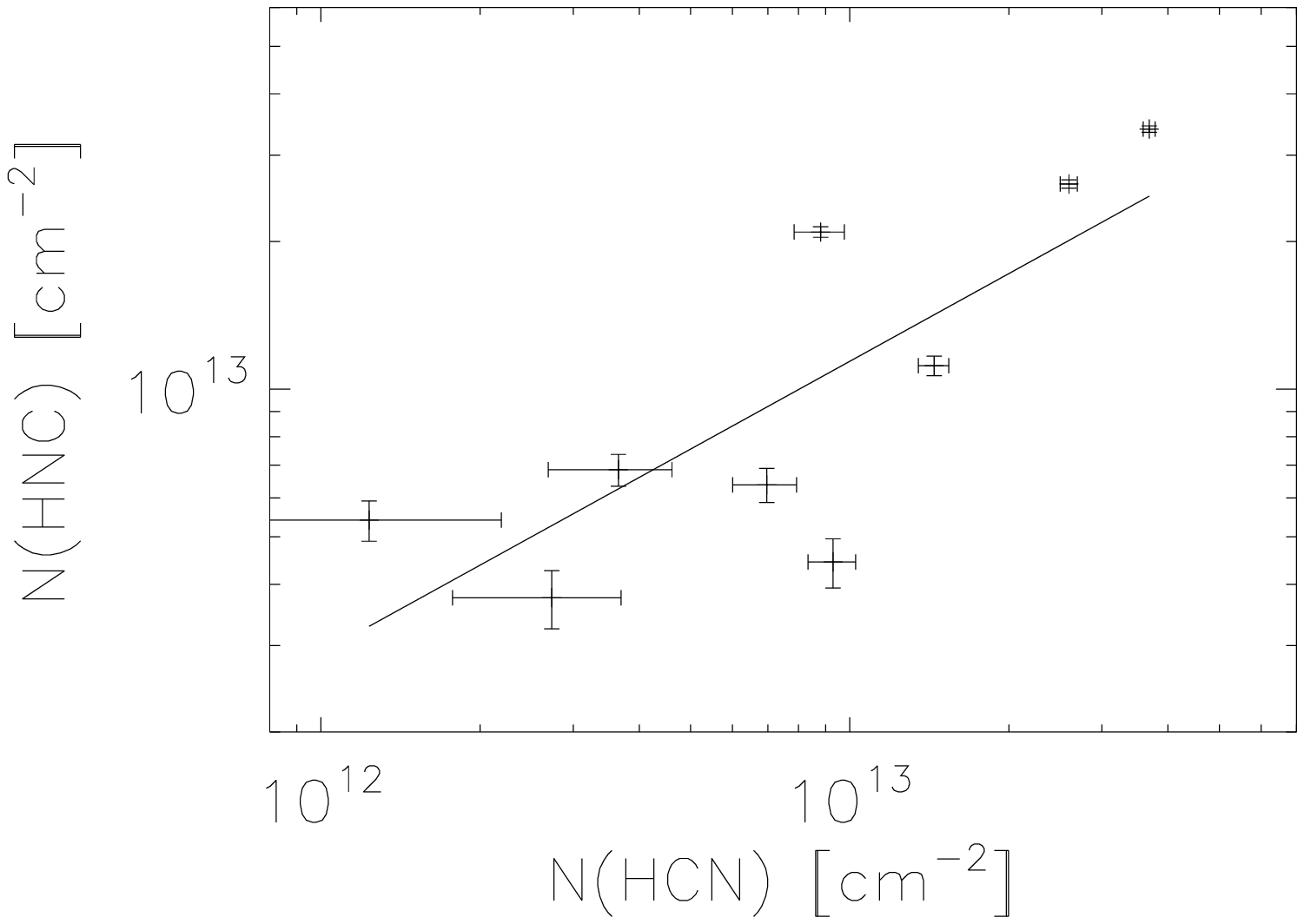}
\includegraphics[width=7.7cm,angle=0]{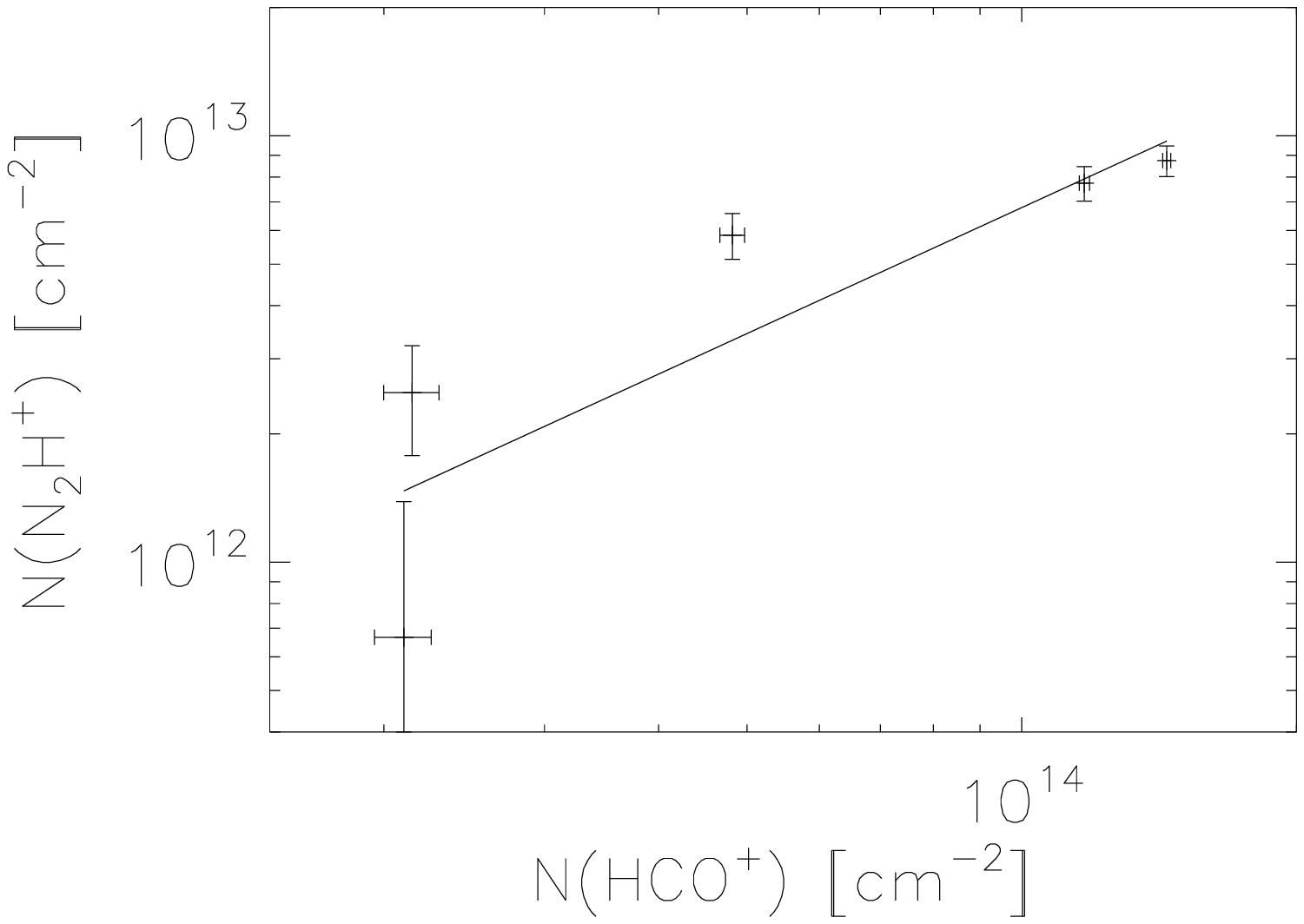}
\caption{Column density correlations for the same molecules examined in Figure~\ref{fig:molabd}.
The solid line represents the linear fit (from Bayesian statistics, see text)
to all points. The Spearman rank coefficient is $>0.5$ in all three cases, with a peak of
0.75 in the plot of $N({\rm HNC})$ vs. $N({\rm HCN})$.
}
\label{fig:cdcorr}
\end{figure}

%
%
%
%
\begin{table}
\caption{Abundance estimates.}
\label{tab:abd}
\centering
\begin{tabular}{lccccc}
\hline\hline
\multicolumn{1}{c}{Source \#} &
\multicolumn{1}{c}{{\bf N$_2$H$^+$}} &
\multicolumn{1}{c}{{\bf HNC}} &
\multicolumn{1}{c}{{\bf HCO$^+$}} &
\multicolumn{1}{c}{{\bf HCN}}  &
\multicolumn{1}{c}{\bf NH$_3$\tablefootmark{a}} \\
&
[$\times 10^{-9} $] &
[$\times 10^{-9} $] &
[$\times 10^{-9} $] &
[$\times 10^{-9} $] &
[$\times 10^{-8} $] \\
%
%
\hline
%
   9  & $-$     & 0.85      & 0.50   & 0.09   & 25.4 \\
  13  & $-$     & 0.57      & 1.18   & 0.54   & $-$ \\
  24  & 0.06    & 0.26      & 0.55   & 0.18   & 0.6 \\
  31  & $-$     & $-$       & 0.56   & 0.16   & 96.9 \\         
  34  & $-$     & 0.47      & 0.41   & $-$    & $-$ \\
  40  & 0.02    & 0.78      & 0.22   & 0.40   & 3.3 \\
  41  & $-$     & $-$       & 2.63   & $-$    & 2.7 \\
  47  & $-$     & 0.86      & 1.03   & 2.21   & 18.3 \\
  50  & $-$     & $-$       & 2.59   & 4.3    & $-$ \\
  56  & $-$     & 0.07      & 0.84   & 0.33   & $-$ \\
  57  & $-$     & $-$       & 0.62   & $-$    & $-$ \\
  63  & 0.11    & 0.35      & 0.77   & 0.62   & 6.7 \\
  81  & 0.03    & 0.19      & 0.58   & 0.20   & 0.08 \\
  82  & 0.15    & 0.49      & 1.8    & 0.26   & 0.9 \\
  89  & $-$     & $-$       & 2.80   & $-$    & $-$ \\
 101  & $-$     & 0.46      & 1.65   & 0.44   & $-$ \\
 109  & $-$     & $-$       & 0.89   & $-$    & $-$ \\
\hline
\end{tabular}
\tablefoot{The table lists the abundance values [MOL]/[H$_2$] for the main observed molecules
toward the sources that were mapped. \\
\tablefoottext{a}{
Same sources as in Table~\ref{tab:mass}.
}
}
\end{table}

\subsection{Radial profiles of relative molecular abundances}
\label{sec:molabdprof}

The computation of the absolute molecular abundances is affected by our ability to
determine the mass distributions of both a specific molecular tracer {\it and} of the
total mass.  In particular, the spatial distribution (i.e., pixel-by-pixel) of the
total mass, as determined from the dust continuum, is seriously affected  by our
ability to estimate and subtract the local background emission.

In order to avoid these difficulties, we have decided to evaluate only the spatial
distribution of the {\it relative} molecular abundances, which depend solely on our
Mopra observations and are not affected by variable levels of background in the bolometers
maps. In Figure~\ref{fig:molabd}
we show the source-averaged, radial profiles of the normalized relative abundances
[HCO$^+$]/[HCN], [HNC]/[HCN] and [N$_2$H$^+$]/[HCO$^+$] with associated standard-deviations.
We show separately the resulting averages for proto-stellar and starless
cores with available map data. For each separate source, the relative abundances
are normalized with respect to the observed maximum value, as a function of radius. We then
take the median values of these normalized relative abundances in each sub-sample of sources.
The advantage of using the normalized values is to be less affected by the variations 
of the absolute values of the abundances from source to source, and highlights instead
the behaviour of the abundances as a function of radius.

Therefore, in Figure~\ref{fig:molabd} we note that while in the proto-stellar cores 
the [HCO$^+$]/[HCN] ratio peaks at the center of the sources, in the starless
cores the peak is slightly off-center. This shows that on average the 
abundance of [HCO$^+$] relative to that of [HCN] is somewhat lower toward
the center of the starless cores. As far as the [HNC]/[HCN] ratio is
concerned, our data show that this relative abundance is quite flat in both
starless and proto-stellar cores, though a slight decrease as a function of
radius can be observed, particularly toward proto-stellar cores.
Finally, we can note the very different behaviour of the [N$_2$H$^+$]/[HCO$^+$] ratio
in starless and proto-stellar cores. In fact, while in the former this abundance
ratio is quite flat, in proto-stellar cores the [N$_2$H$^+$]/[HCO$^+$] ratio
can be observed to decrease toward larger radii.

A detailed comparison of these radial profiles with simulations obtained using 
recent chemical models (e.g., \citealp{lee2004}, \citealp{aikawa2008}),
which follow the chemical evolution of a source both as a function of time and of
radial position, is difficult because they usually model the chemistry of
the source only out to a radius $\sim 1 - 2 \times 10^4\,$AU, a size lower than
that (26600\,AU) covered by the Mopra beam at the distance of Vela-D.

However, we note that some features present in Figure~\ref{fig:molabd} are consistent
with, e.g., the model of \citet{lee2004}. The faster drop-off at large angular radii 
of the radial profile of the [N$_2$H$^+$]/[HCO$^+$] ratio in proto-stellar cores
is consistent with the molecular abundance profiles after collapse discussed
by \citet{lee2004}, though they are not consistent with similar radial profiles 
discussed by \citet{aikawa2008}. 
Also, the fact that in Figure~\ref{fig:molabd} 
the peak of the [HCO$^+$]/[HCN] ratio in starless cores is slightly off-center
could be a consequence of HCO$^+$ depletion, which is more effective in decreasing
the column density of HCO$^+$ toward the center of the sources and is more effective 
at earlier times. In fact, \citet{lee2004} have shown that depletion initiates
during the pre-collapse phase and proceeds through the collapse of the source.

\subsection{Chemical status of the cores}
\label{sec:chemevo}

In this section we further discuss some chemical implications of the relative molecular abundances.
It is often found in the literature that some molecules are classified either as ``early-time'' 
($10^4 - 10^6$ years after the onset of chemistry) or ``late-time'' species (maximum abundances 
reached at steady state, after $10^6 - 10^8$ years of chemical evolution), 
based on the production pathways via ion$-$molecule or neutral$-$neutral reactions.
Of the molecules discussed here HCO$^+$ and NH$_3$ are usually considered as late-time molecules,
but the situation of HCN and HNC is less clear in this picture.  

However, we can attempt to use the [HNC]/[HCN] abundance ratio as a measure of the 
evolutionary phase and temperature in our sources.  In fact, converting Figure~\ref{fig:molabd}
into absolute values (instead of the normalized values shown in the figure) 
for the relative molecular abundances we find that:
{\it (i)} the abundances of HCN and HNC are not significantly different between the proto-stellar 
and the starless cores, thus a systematic change in the HCN abundance due to chemical 
evolution is not apparent in this study; 
{\it (ii)} the average [HNC]/[HCN] ratio is in agreement with the values measured in 19 low-mass
starless and proto-stellar cores by \citet{hirota1998}.

%
%
\begin{figure*}
\begin{centering}
\includegraphics[width=8.1cm,angle=0]{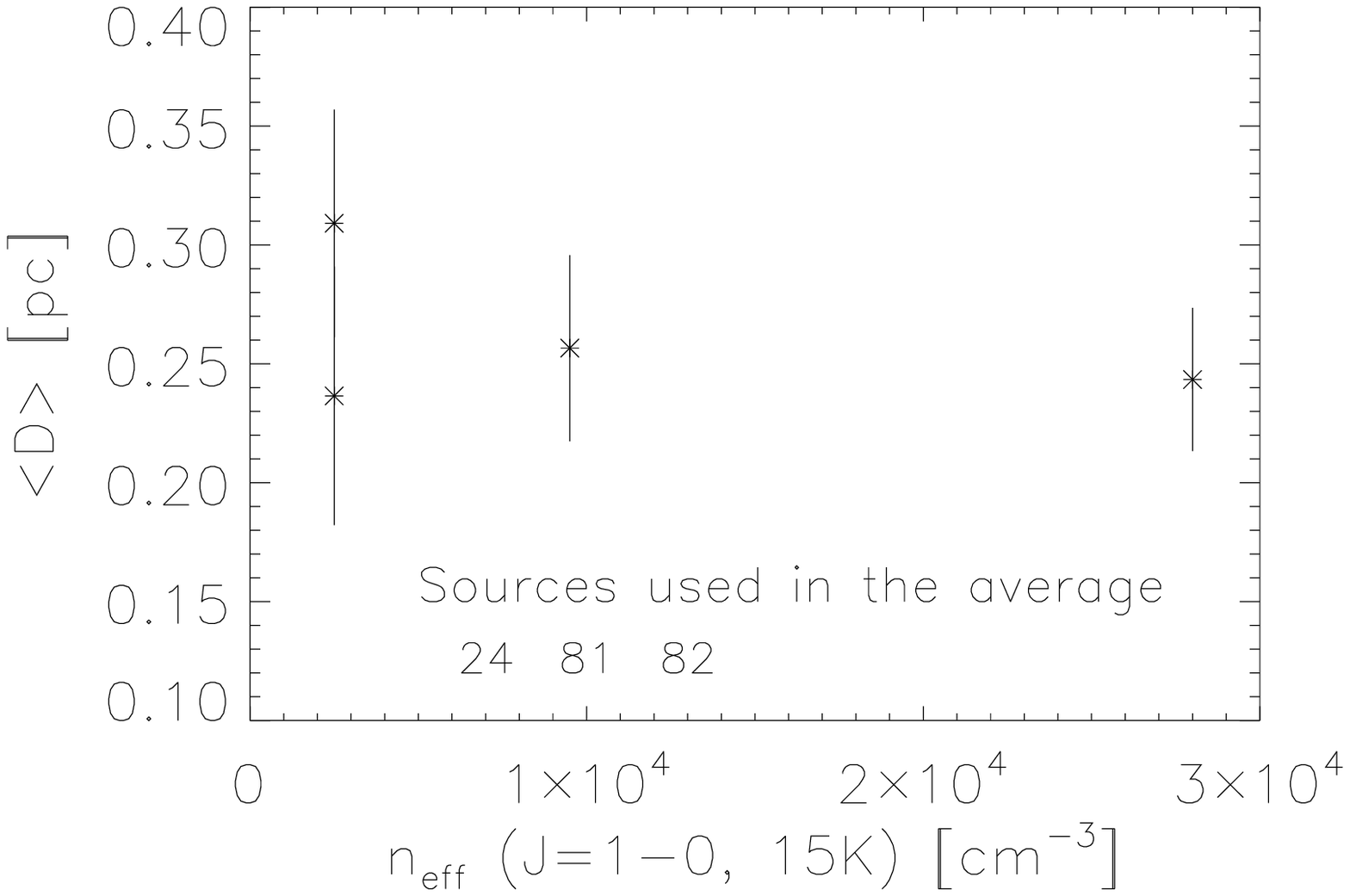}
\includegraphics[width=8.1cm,angle=0]{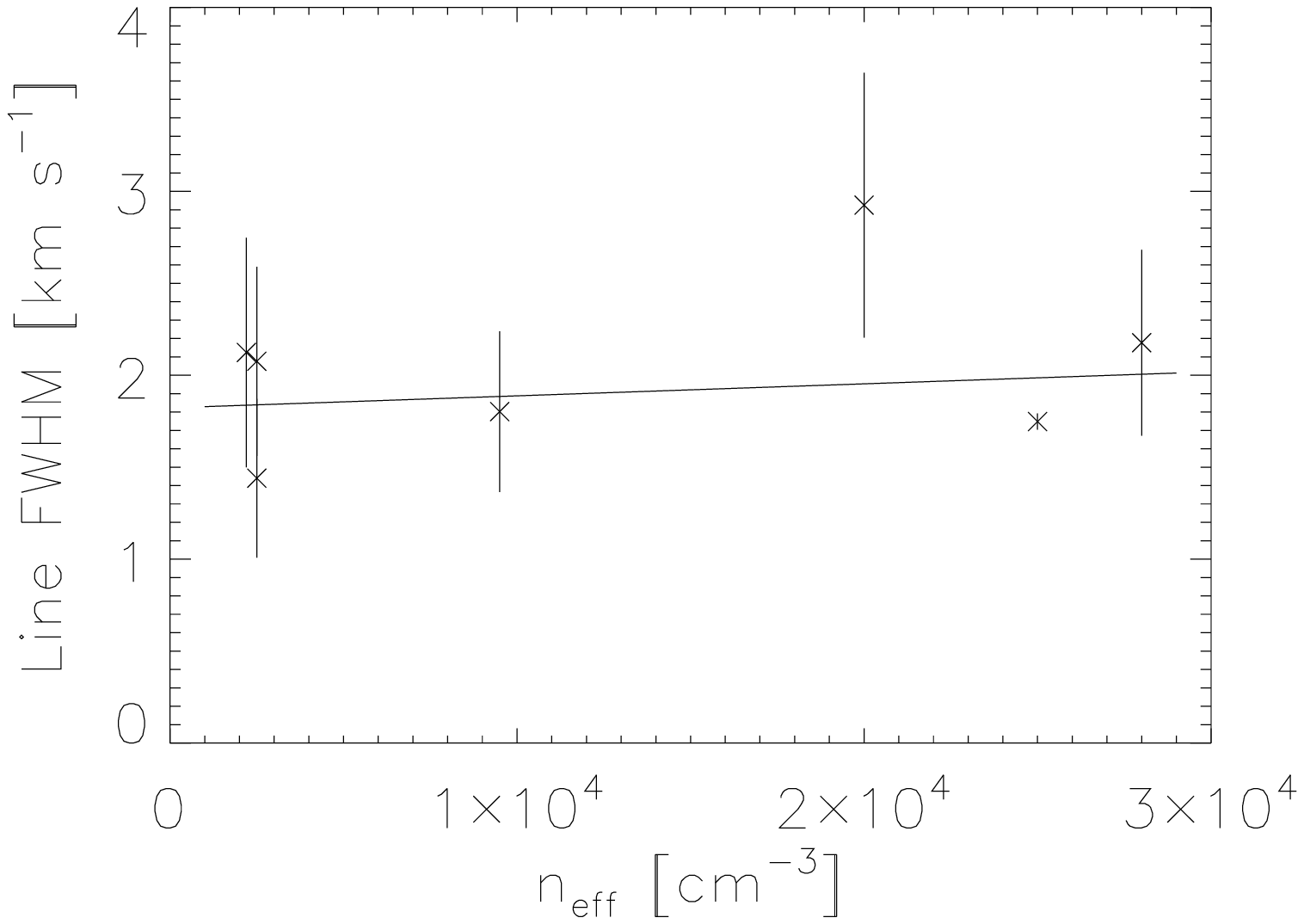} \\
\includegraphics[width=8.1cm,angle=0]{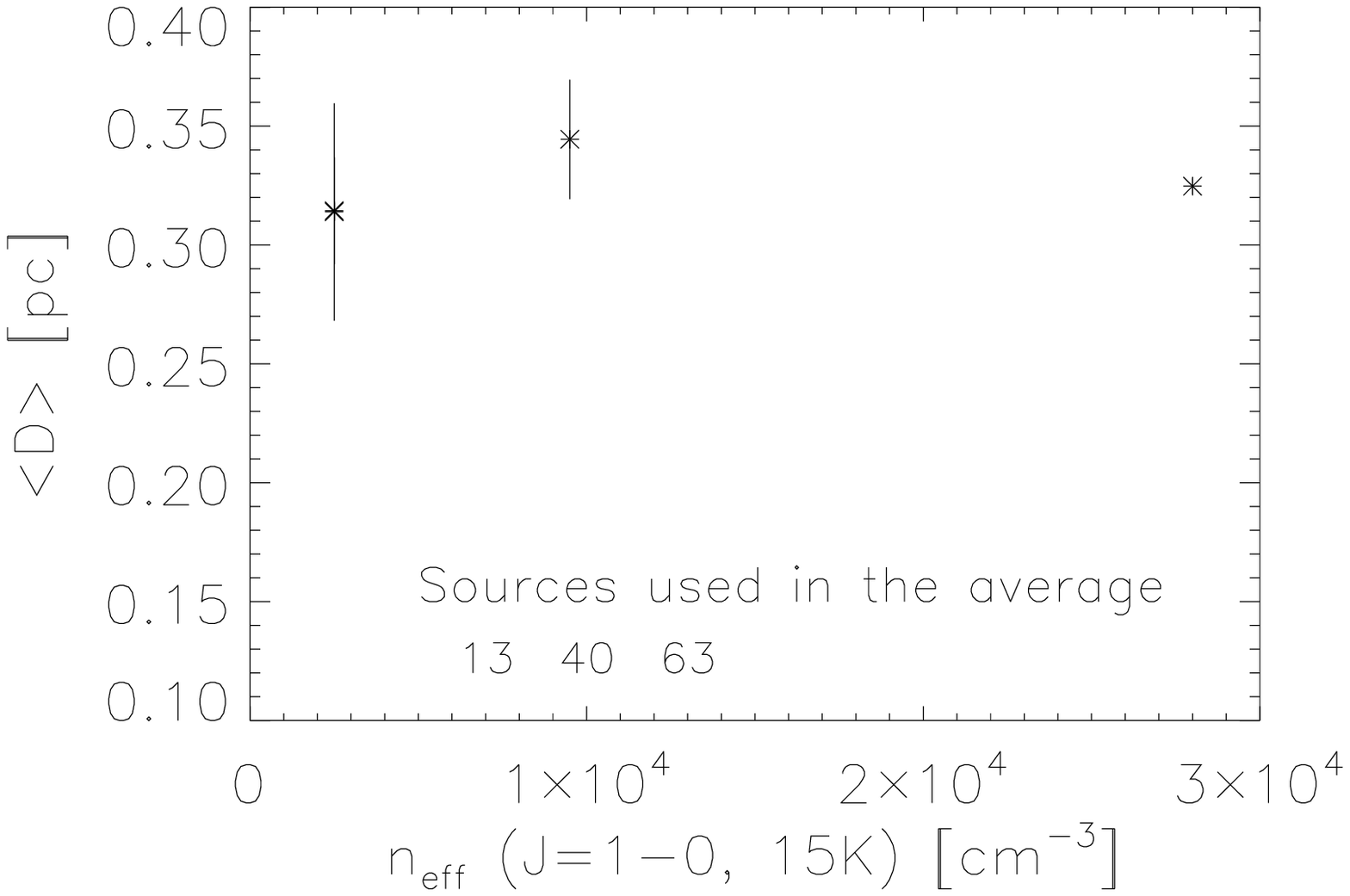}
\includegraphics[width=8.1cm,angle=0]{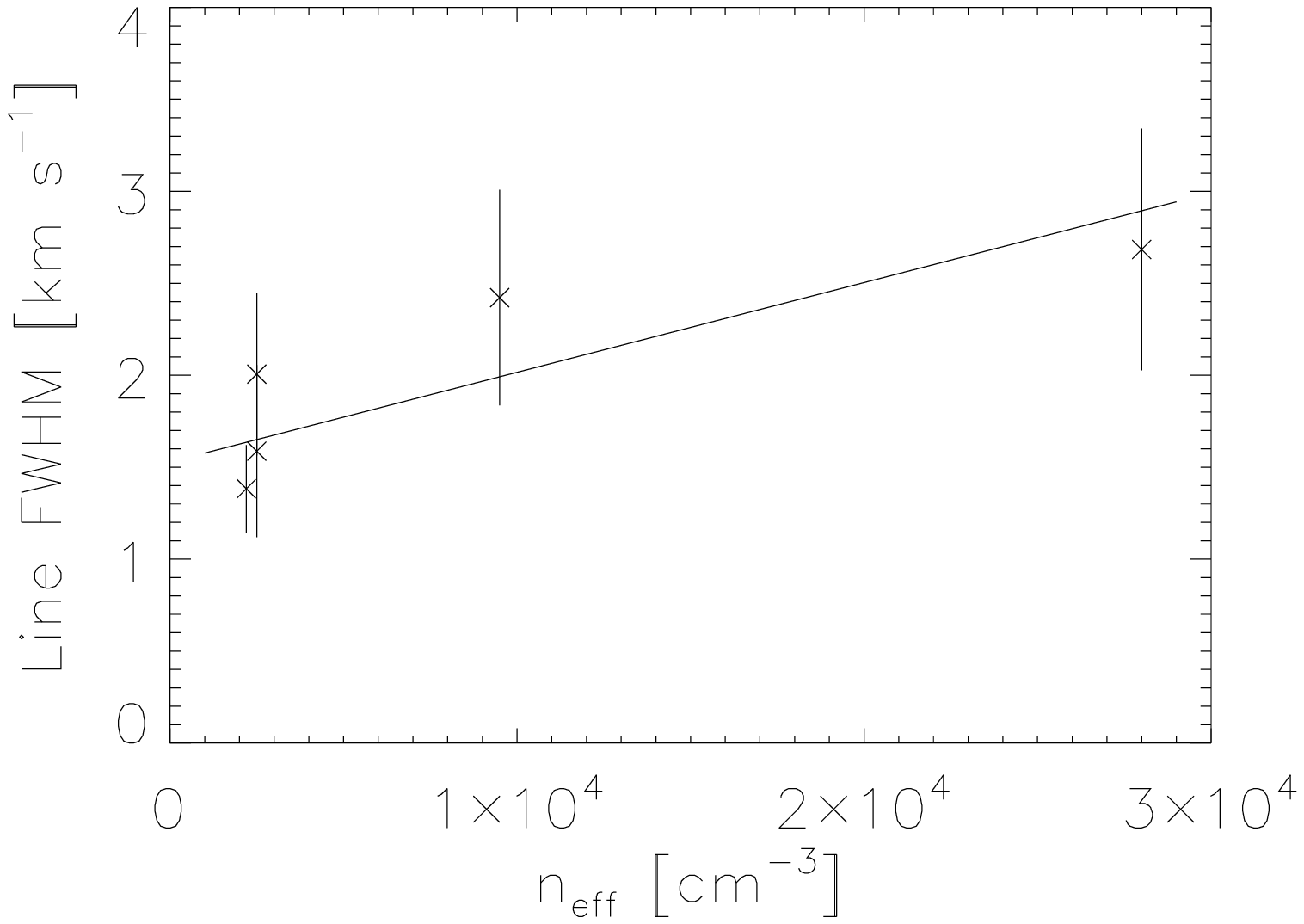}
\caption{
Median core size (left panels) and median linewidth, $\Delta V$ (right panels)
vs. $n_{\rm eff}$ (at 15\,K), shown separately for a sub-sample of proto-stellar (top row) and
starless (bottom row) cores. The plot of $\Delta V$ vs. $n_{\rm eff}$ makes use of
the single-point observations, instead of spectral line maps as in the left panels,
allowing for more molecular lines to be used.
The effective densities are listed in Table~\ref{tab:linelist}. The plot also
includes some detections of the HC$_3$N$(1-0)$ line, with $n_{\rm eff} = 2.5 \times 10^4$.
The solid line represents the linear fit (from Bayesian statistics)
to all points.
}
\label{fig:neffcorr}
\end{centering}
\end{figure*}

According to gas-phase chemical models, HCN and HNC are mainly produced by a dissociative 
recombination reaction of the HCNH$^+$ ion with an electron (see \citealp{hirota1998}
and references therein):
\begin{equation}
{\rm HCNH^+ + e^-} \rightarrow
\left \{ \begin{array}{l}
{\rm HCN+H} \\
{\rm HNC+H } \\
\end{array}
\right.
\label{eq:chem}
\end{equation}
The branching production ratio of HCN, $\alpha$, is defined as 
[HCN]/[HNC]$ = \alpha / (1 - \alpha)$, if $T_k \ll T_c$, where	$T_c \simeq 24\,$K is the 
threshold temperature above which neutral$-$neutral reactions dominate the [HCN]/[HNC] ratio.
If we take a median value of [HCN]/[HNC]$\simeq 1$ in our cores, the branching production ratio of 
HCN is estimated to be $\alpha \simeq 0.5$, somewhat higher than the value reported by
 \citet{hirota1998} ($\alpha \simeq 0.4$), but in agreement with the range of values found 
by \citet{nikolic2003} toward L1251 ($\alpha \simeq 0.2 - 0.8$).
The relatively low value of $\alpha$ suggests that the more energetic paths for destruction 
of HNC, once the temperature of the gas exceeds the critical temperature, are less favorable,
and thus it constitues a further indirect evidence that our cores are cold.

Another parameter to assess the chemical status of our cores is the [N$_2$H$^+$]/[H$^{13}$CO$^+$] 
abundance ratio. \citet{fuente2005} have proposed this ratio as a measure of
the evolutionary phase of a source. According to these authors, since the abundance 
of N$_2$H$^+$ tends to remain constant in starless clumps, while
H$^{13}$CO$^+$  could suffer from depletion in 
the densest part, then the [N$_2$H$^+$]/[H$^{13}$CO$^+$] ratio could be used as an
indication of the chemical evolutionary phase of the cores. 
%
%
%
If we use the single-point observations of both N$_2$H$^+$ and H$^{13}$CO$^+$ we find a 
median value of their relative abundance, [N$_2$H$^+$]/[H$^{13}$CO$^+$]$= 5.7\pm1.9$,
which is intermediate in the range of values proposed by \citet{fuente2005} spanning from
Class 0 objects (where molecular depletion is significant and the [N$_2$H$^+$]/[H$^{13}$CO$^+$] 
ratio is maximum, $\sim 15$) to more evolved objects where the ratio is minimum ($\sim 3$).
Note, however, that the objects studied by \citet{fuente2005} are on average more evolved
than those in our sample of cores in Vela-D.


\subsection{Correlations between parameters}
\label{sec:corr}

\subsubsection{Column density correlations}
\label{sec:virialmass}

We can use the molecular column densities estimated in the previous sections to
compare the behavior of the molecules. If we analyze molecules with different 
critical densities, we would expect their molecular transitions to be emitted
in different volumes of gas, in the presence of density gradients. Likewise, 
molecules with similar critical densities will be excited in the same volume of gas, 
in the absence of strong abundance gradients.  

We performed this test only in those sources where spectral line maps were
available, to ensure that we integrate over all molecular emission.
Thus, in these correlation plots all information about the specific spatial distribution
of the different molecular tracers is lost. In addition, the inclusion in this
analysis of spectral line maps only, dramatically decreases the size of the sample,
which also represents a range of masses, chemistries, and evolutionary states. 

%
%
\begin{figure}
\centering
\includegraphics[width=8.8cm,angle=0]{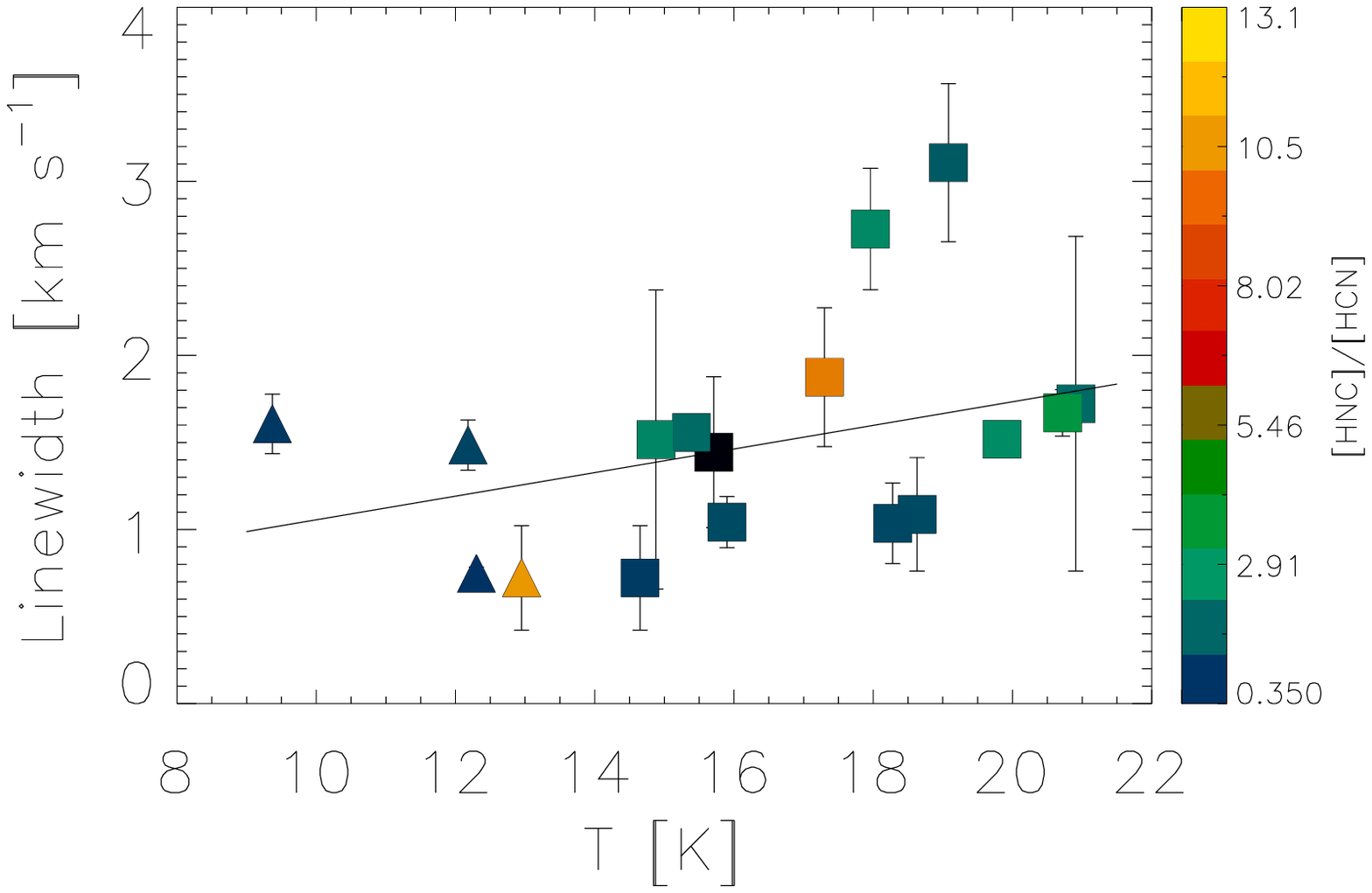}
\includegraphics[width=8.8cm,angle=0]{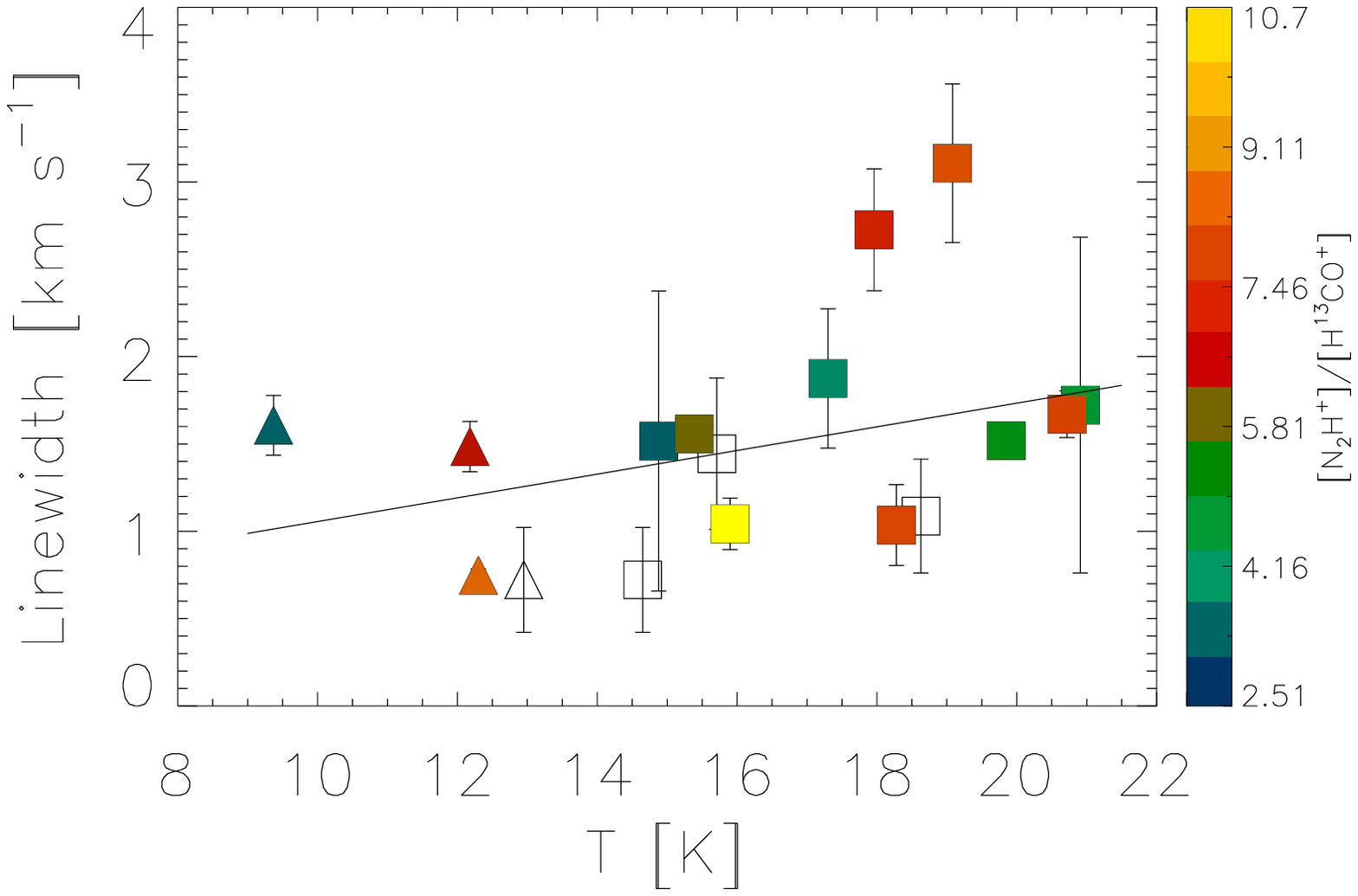} \\
%
\caption{
Plot of average linewidth vs. core temperature. The BLAST-derived dust temperature
has been used, except in those sources where $T_{\rm k}$ could be
determined from NH$_3$ (see Section~\ref{sec:nh3}).
In estimating the averages, only the linewidths of the optically thin molecular tracers
(N$_2$H$^+$, H$^{13}$CO$^+$ and H$^{13}$CN) have been considered.
The triangles and squares represent starless and proto-stellar
cores, respectively. The filling color code represents the relative
abundance [HNC]/[HCN] (top panel) and [N$_2$H$^+$]/[H$^{13}$CO$^+$] (bottom panel).
The solid line represents the linear fit (from Bayesian statistics)
to all points.
}
\label{fig:FWHMcorr}
\end{figure}

Our results are presented in  Figure~\ref{fig:cdcorr}, where
we have used the Bayesian IDL routine LINMIX\_ERR to perform a linear regression,
to find the slope of the best fit line in each case.
In the top panel we plot the column density of HCO$^+$ vs. the column density of HCN. 
HCN should be a higher-density gas tracer, compared to HCO$^+$; however, despite a 
considerable scatter, the two column densities show a modest correlation, with a 
Spearman rank coefficient $> 0.5$.
The column densities of HCN and its isomer HNC also appear to be well correlated 
(correlation coefficient $= 0.75$) in the middle 
panel of Figure~\ref{fig:cdcorr}. Finally, in the bottom panel, we also compare the
column densities of the chemically opposite species N$_2$H$^+$ and HCO$^+$. The few 
points in this panel are a consequence of the fewer available maps for N$_2$H$^+$ compared
to the other molecular tracers. However, we still see a correlation between the two
column densities (correlation coefficient $> 0.5$).

The variable degree of correlation and the scatter in Figure~\ref{fig:cdcorr} is 
an indication of source-to-source variations in the physical and chemical
conditions. However, these qualitative trends are remarkable given the
diversity of our sources and relatively small sample size.


\subsubsection{Correlations with $n_{\rm eff}$}
\label{sec:neff}

The molecular transitions observed by us are characterized by a range of critical densities,
thus allowing us to examine how specific physical parameters change as a function of the
gas traced by a specific line. However, instead of using 
the critical density, $n_{\rm crit}$, in our
analysis we prefer to use the {\it effective density}, $n_{\rm eff}$, to characterize 
the volume of gas where a given spectral line is generated. 
As discussed in \citet{evans1999}, the ability to detect a transition is usually 
associated with a number density larger than $n_{\rm crit}$, 
but lines are also easily excited in subthermally populated gas with densities more than an 
order of magnitude lower than $n_{\rm crit}$. 

To account for effects such as optical depth, 
multilevel excitation effects, and trapping, we use the definition
of \citet{evans1999}, where $n_{\rm eff}$ represents the density needed to excite 
a $T_R = 1\,$K line (where $T_R$ is the observed radiation temperature) 
for a given kinetic temperature, as calculated with a non-LTE 
radiative transfer code. We specifically calculated the effective densities of our 
molecular tracers by using the online version of RADEX (\citealp{vanderTak2007}), 
a non-LTE radiative transfer code, assuming
$\log(N/\Delta v) = 13.5\,$cm$^{-2}/($km\,s$^{-1})$  (see \citealp{evans1999})
and a kinetic temperature of 15\,K for all species (see Table~\ref{tab:linelist}).

Figure~\ref{fig:neffcorr} shows the (deconvolved) core diameter, as determined from the 
molecular intensity maps, as a function of $n_{\rm eff}$. The observed core diameter
is estimated using the area enclosed within the half-peak intensity contour and then 
deconvolved using the standard Gaussian deconvolution with the Mopra beam. 
We then take the median value of the size (using the selected sources listed in 
Figure~\ref{fig:neffcorr}) and plot it as a function of $n_{\rm eff}$ 
in Figure~\ref{fig:neffcorr}, where the error bars represent 
the median absolute deviations. 
By comparing the values of the measured core diameter in starless and proto-stellar
cores we note that for the same
value of $n_{\rm eff}$ the starless cores have on average a larger size.

%
%
%
%
%
\begin{table*}
\caption{Median properties of starless and proto-stellar cores.}
\label{tab:averages}
\centering
\begin{tabular}{lcccccccc}
\hline\hline
Source Type &
 $R_{\rm flat}$ &
 $\eta $ &
FWHM\tablefootmark{a} &
$T$ &
$\frac{ M_{\rm core} } { M_{\rm vir}}$\tablefootmark{b}  &
 $\frac{ P_{\rm tot} } {k}$  &
 $ \frac{ \rm [NH_3]}{\rm [N_2H^+]}$\tablefootmark{c}  &
 $ \frac{ \rm [N_2H^+]}{\rm [H^{13}CO^+]} $  \\
 & [arcsec] & & [km\,s$^{-1}$] & [K] & & [$\times 10^6\,$cm$^{-3}$\,K] & & \\
%
%
\hline
Starless      & $27.0 \pm 12.0$          & $0.8 \pm 0.7$           & $1.4 \pm 0.4$  & $15.1 \pm 1.6$  & $0.5 \pm 0.3$   & $\sim 5$        & $\sim 10^3$    & $5.7  \pm 1.7$  \\
BLAST063      & $24.0 \pm 6.0$\tablefootmark{d} & $0.6 \pm 0.3$\tablefootmark{d} & $1.5 \pm 0.1$  & 12.2            & 0.3             & $\sim 10$       & $\sim 600$              & $5.7$  \\
Proto-stellar & $21.0 \pm 5.9$           & $1.1 \pm 0.5$           & $1.6 \pm 0.6$  & $16.8 \pm 1.8$  & $0.8 \pm 0.4$   & $\sim 19$       & $\sim 60$             & $6.4 \pm 1.9$ \\
\hline
\end{tabular}
\tablefoot{
\tablefoottext{a}{
From optically thin tracers only.
}
\tablefoottext{b}{
$M_{\rm vir}$ estimated assuming a density distribution $\rho \propto r^{-2}$.
}
\tablefoottext{c}{
Mean values based on two (starless) and three (proto-stellar) sources.
}
\tablefoottext{d}{
Excluding N$_2$H$^+$.
}
}
\end{table*}


Figure~\ref{fig:neffcorr} also shows the source linewidth, $\Delta V$,  measured
from single-point spectra, as a function of $n_{\rm eff}$.  Because in this case we 
are using the single-point observations, we could include in the plot 
almost all molecular transitions listed in Table~\ref{tab:linelist}.
Furthermore, since linewidths may be broadened by optical depth effects,
for Figure~\ref{fig:neffcorr} we selected only those source/line cases with 
$\tau < 1$. 
We note that no significant difference in $\Delta V$ is observed, for the 
same value of $n_{\rm eff}$, between proto-stellar and starless cores, but we 
find that in starless cores $\Delta V$ and $n_{\rm eff}$ are correlated. In fact,
as for the case of Figure~\ref{fig:cdcorr}
we have used the IDL routine LINMIX\_ERR to perform a linear regression,
and we find a Spearman rank coefficient
of 0.97 for the starless cores. However, this correlation coefficient drops to
0.23 for the proto-stellar cores.

\subsubsection{Linewidth-temperature correlation}
\label{sec:FWHMcorr}

Since there is no single measurement that can determine the ``exact'' evolutionary 
stage of each source, as a further test to help estimate the status of the cores
in our sample we plot in Figure~\ref{fig:FWHMcorr} the linewidth vs. the 
core temperature. In order to avoid possible optical thickness effects, the 
figure only includes the linewidths of the (mostly) optically thin molecular tracers,
i.e., N$_2$H$^+$ and the two isotopologues H$^{13}$CO$^+$ and H$^{13}$CN. 
From this plot we also had to eliminate those sources with a poor determination of
the linewidths because of weak lines and/or noisy spectra.

Figure~\ref{fig:FWHMcorr} thus shows that there is a clear segregation in
temperature between starless and proto-stellar cores. The figure also shows that
there is a modest correlation between linewidth and temperature. 
As for the case of Figure~\ref{fig:cdcorr} 
we have used the IDL routine LINMIX\_ERR to perform a linear regression, 
and we find a Spearman rank coefficient 
of 0.47. This modest correlation is confirmed by the fact that the
median values of the linewidth determined for the starless and proto-stellar cores are
$1.4 \pm 0.4$ and $1.6 \pm 0.6$\,km\,s$^{-1}$, respectively (see Table~\ref{tab:averages}), 
which are consistent within the errors.

Figure~\ref{fig:FWHMcorr} also encodes the information about the relative 
molecular abundances [HNC]/[HCN] and [N$_2$H$^+$]/[H$^{13}$CO$^+$],
 discussed in Section~\ref{sec:chemevo}.  No apparent trend of [HNC]/[HCN]  
or [N$_2$H$^+$]/[H$^{13}$CO$^+$] with either linewidth or temperature can
be seen. 

\subsection{Comments to specific sources mapped with the Mopra telescope}
\label{sec:specific}

In this section we further discuss some starless and  proto-stellar cores that have been
mapped at Mopra, including the ``transition'' source BLAST063. 

{\bf BLAST024}. This source is a compact proto-stellar core, where the spatial
distribution of the emission of all main tracers observed by us is centered on 
the BLAST dust core. Wing emission is detected in the single-point 
HCO$^+(1-0)$ spectrum, suggesting the presence of a molecular outflow.
However, due to the low SNR in our maps, we could not produce an image of the outflow,
as it is the case also for sources BLAST081 and BLAST082.

{\bf BLAST031}. This is a moderately compact starless core, where the HCN$(1-0)$ and  HCO$^+(1-0)$
emission trace quite well the BLAST continuum emission, but the HNC$(1-0)$ emission 
is seen to peak at an offset position compared to the BLAST emission and the other two
molecular probes. Such offsets are also found toward  other sources (see below).

{\bf BLAST034}. This is an irregularly shaped, starless core which shows two well 
separated velocity components in the single-point HCO$^+(1-0)$ spectrum, with the main 
(mapped) component being the low-velocity one ($\simeq 2\,$km\,s$^{-1}$). The two 
velocity components are  also detected in the HNC$(1-0)$ spectrum. 

{\bf BLAST047}. This source also shows a double-peaked HCO$^+(1-0)$ spectrum. However, 
both the channel map of the HCO$^+(1-0)$ emission and the spatial distribution
of the HNC and HCN$(1-0)$ emission (see Figure~\ref{fig:maps2}) suggest that 
HCO$^+$ is self-absorbed at the position of the BLAST peak. One
can also note that HCN, and especially HNC, trace very well the BLAST contour plots.

{\bf BLAST063}. This source is the most intense starless core mapped by us at Mopra. 
This core has been classified throughout this paper and was originally classified as 
starless by \citet{olmi2009} because no 
MIPS compact source at 24$\,\mu m$ could be found at its location. However, we have
found that the single-point HCO$^+(1-0)$ spectrum shows line-wing emission, indicating 
the presence of a molecular outflow, which should actually change the source type 
from starless to proto-stellar. 
The fact that no 24$\,\mu m$ emission is detected toward this core,
though a molecular outflow is already active, suggests that the proto-stellar 
core has just formed and is thus in an earlier stage compared to BLAST024, BLAST081 and BLAST082.
Therefore, this core could tentatively be classified as being in a ``transition phase'' from
starless to proto-stellar.  We also note that this source has an elongated 
shape, and we can see that the HCN$(1-0)$ and HNC$(1-0)$ 
emission trace very well the BLAST continuum emission, whereas there is an offset between 
the HCO$^+(1-0)$ emission and the dust emission.

{\bf BLAST081}. This source is another compact proto-stellar core mapped with the Mopra
telescope. Despite some artefacts visible in the HCO$^+(1-0)$ map (see Figure~\ref{fig:maps4})
we note that while the N$_2$H$^+(1-0)$ and (possibly) HCO$^+(1-0)$ emission follow closely the BLAST
continuum emission, the emission of the two isomers HCN and HNC peak at an (identical) 
offset position compared to the BLAST maximum. This is the only example of such an offset
observed among the three proto-stellar cores mapped with the Mopra telescope, though
it is also partly observed toward the early proto-stellar core BLAST063.
Since our sensitivity was not high enough to map the optically thin isotopologues, such
as H$^{13}$CO$^+$ and H$^{13}$CN, we cannot determine at present whether this offset is
due to optical depth effects or chemical variations. As in BLAST024, the HCO$^+(1-0)$ spectrum 
shows line-wing emission, indicating the presence of a molecular outflow.

{\bf BLAST082}. This source is the third compact proto-stellar core mapped at Mopra. 
Contrary to BLAST081, all mapped molecular tracers are well centered on the BLAST
continuum emission and also follows quite well the less dense material. However, 
the HCN$(1-0)$ and HNC$(1-0)$ maps show an additional smaller core, 
located at $(\Delta l, \Delta b) \simeq (-60'', -60'')$  (see Figure~\ref{fig:maps4}),
which is not visible in the HCO$^+(1-0)$ map. 
This sub-structure may be a consequence of core fragmentation and the difference
observed in the various molecular tracers may indicate a different chemical status
compared to the main core.
Then, like BLAST034 and BLAST047, the single-point HCO$^+(1-0)$ spectrum is also 
double-peaked, but the channel map suggests this is the consequence of two velocity
components aligned along the line of sight. Like BLAST024 and BLAST081
the HCO$^+(1-0)$ spectrum also suggests the presence of a molecular outflow. 
Line wings in the HCO$^+(1-0)$ spectrum are also observed toward other sources, as 
listed in Table~\ref{tab:HCO}.


\subsection{Median properties of starless and proto-stellar sources and
comparison with other surveys}
\label{sec:averages}

In this section we summarize the main differences between starless and
proto-stellar cores, by presenting a list of their median properties
in Table~\ref{tab:averages}.  For comparison, we have also included source BLAST063
which we have defined as a transition source between the two classes of
objects. One can note that the starless cores appear to be moderately colder and less 
turbulent compared to the proto-stellar sources, though the differences are within
the uncertainties. The linewidths have been averaged among all optically thin tracers
(N$_2$H$^+(1-0)$, H$^{13}$CO$^+(1-0)$ and H$^{13}$CN$^+(1-0)$) and all sources.
The lower temperature of the starless cores is simply a
consequence of the analysis carried out by  \citet{olmi2009}, but we now find
that also their linewidths do not strongly differ from those of the
proto-stellar cores.

We do find a more significant difference between starless and proto-stellar cores
when we consider the radial profiles of the column densities. By using the results
of Section~\ref{sec:radprof} the median values of the parameters $R_{\rm flat}$
confirm that the proto-stellar cores are more compact on average.
The parameter $\eta$, on the other hand, is subject to larger uncertainties, also due to our 
relatively low spatial resolution at the distance of Vela-D, and no definitive statement
can thus be made. We note that the larger $R_{\rm flat}$ measured toward the starless
cores agrees well with the conclusion of \citet{caselli2002}. In fact, these authors found that 
while proto-stellar cores in their sample (which includes sources from various 
star forming regions) 
were better modeled with single power-law density 
profiles, the starless cores presented a central flattening in the integrated intensity profile.

In terms of the kinematics of the cores, our findings are also very similar to those 
of \citet{caselli2002}. In fact, we find an average value for the velocity gradient
in our sample (from the HCO$^+$ results listed in Table~\ref{tab:velfit}) of 
$1.4 \pm 0.3\,$km\,s$^{-1}$\,pc$^{-1}$, and we do not find significant differences
between starless and proto-stellar cores. This is consistent with the typical 
value of $2 \pm 1\,$km\,s$^{-1}$\,pc$^{-1}$ found by \citet{caselli2002} in their sample.
In terms of the equilibrium status of the cores, \citet{caselli2002} found that their 
``excitation mass'' to virial mass ratio is typically $M_{\rm ex} / M_{\rm vir} \simeq 1.3$,
 for both classes of cores.  From Table~\ref{tab:averages}, on the other hand, we can see
that both starless and proto-stellar cores have $0.5 \la M_{\rm core} / M_{\rm vir} \la 1$.
Therefore, while the $M_{\rm core} / M_{\rm vir}$ ratio is somewhat lower compared to 
the results of \citet{caselli2002} (but this may also be a consequence of the different
methods to calculate the core masses) it is still very near or beyond the self-gravitating 
threshold of 0.5. We also find that in proto-stellar cores  the $M_{\rm core} / M_{\rm vir}$ ratio
is slightly larger than in starless cores, though they are still consistent within the errors.

In Table~\ref{tab:averages} we have also included the median values of the
{\it average} total internal pressure (not the edge pressure), $P_{\rm tot}$, 
of each core in Vela-D for which we have measured (optically thin) linewidths. 
The values of $P_{\rm tot}$ have been calculated following the method
described by \citet{olmi2010} and are given in the usual cm$^{-3}$\,K units for $P_{\rm tot}/k$.
We note that in proto-stellar cores the median internal pressure is much higher
than in starless cores, as expected. However, the confinement of the cores is of 
concern only for those sources (mostly starless) that are gravitationally {\it unbound}.
Therefore, the starless cores which are gravitationally unbound according to the
$M_{\rm core} / M_{\rm vir}$ ratio can be stable (against expansion) only if their edge
pressure is less than the local ambient pressure 
(of order $P_{\rm ext}/k \sim 5 \times 10^5\,$cm$^{-3}$\,K in Vela-D, see \citealp{olmi2010}). 
Pressure confined cores have been previously discussed, for example, by 
\citet{lada2008} and \citet{saito2008}.
Alternative sources of support could come from the magnetic field (see \citealp{olmi2010}
and references therein), or the gravitationally unbound cores must otherwise be transient structures.

As far as chemical properties are concerned, Table~\ref{tab:averages} shows 
that the values of the relative abundance [N$_2$H$^+$]/[H$^{13}$CO$^+$]
for the starless and proto-stellar cores are quite similar. Thus, as noted in 
Section~\ref{sec:chemevo} we do not find any trend based on this ratio, as it
has been suggested by \citet{fuente2005}. However, in Table~\ref{tab:averages} we note
that the [NH$_3$]/[N$_2$H$^+$] abundance ratio is significantly larger
toward starless cores, though we had only two starless cores where
we could effectively measure this ratio. This trend, if not the quantitative difference between
the two classes of objects (because of the low statistics), is consistent with the findings
of \citet{friesen2010}, who concluded that the relative fractional abundance of 
NH$_3$ to N$_2$H$^+$ remains larger toward starless cores than toward protostellar 
cores by a factor of $\sim 2-6$.  

These properties, and in particular the fact that many of the starless cores are 
gravitationally bound (hence, {\it pre-stellar}),  suggest that at least most of the 
pre-stellar cores observed by us are indeed the precursors of the proto-stellar cores and not just
transient objects following a different evolutionary path. Their similar median temperature 
and linewidths 
indicate that the envelopes of the sources have not
had enough time to ``register'' the emergence of a protostar inside the core.
In turn, this also means that most of the turbulence in the proto-stellar cores
observed by us is generated prior to the appearance of a warm object inside the core.

Therefore, it appears that the transition from the pre- to the proto-stellar phase
is relatively fast, leaving the core envelopes with almost unchanged physical parameters. 
Alternatively, if most proto-stellar cores observed had not had much time to evolve after 
the appearance of a protostar, this might 
also explain the similar median temperatures and linewidths of the two source types.
However, the latter hypothesis seems less likely given that among our selected 
proto-stellar cores are also sources relatively evolved (see \citealp{olmi2009} and references
therein). 

In this respect, despite the uncertainties it is of interest to note that the 
values of the parameters $R_{\rm flat}$, $\eta$ and FWHM of source BLAST063 are 
somewhat intermediate between those 
of starless and proto-stellar cores. But, if the transition phase from pre- to 
proto-stellar is indeed fast, then one should not expect to detect many such
transition objects, and their properties may be indistinguishable from those
of either class of sources. More systematic observations of larger samples of 
sources are required to further study this issue.

\section{Summary and conclusions }
\label{sec:concl}

We presented 3-mm molecular line observations of the Vela-D region
 obtained with the 22-m Mopra radio telescope
and 1.3\,cm NH$_3$ observations carried out with the Parkes antenna. 
In total 8 molecular lines were analyzed in 40 sources, both starless and proto-stellar cores,
previously detected by BLAST. A total of 20 spectral line maps
were obtained.  The results of our study can be summarized as follows:

\begin{enumerate}

\item Our spectral line maps show a wide variety of morphological types: very early and cold
starless cores appear to have an irregular shape (e.g., BLAST009) in most or all molecular tracers
mapped at Mopra. Some warmer cores (e.g., BLAST031) and cores at the transition phase from starless to
proto-stellar (BLAST063) appear to be more regularly shaped and more compact. 
Finally, proto-stellar cores all show a more rounded
shape and narrow radial intensity profiles.

\item We compared the kinetic temperature, $T_{\rm k}$, 
derived from the NH$_3$(1,1) and (2,2) observations, 
with the BLAST-derived dust temperature, $T_{\rm d}$, and found that 
the median value of their ratio is
$ T_{\rm k} / T_{\rm d} = 1.2 \pm 0.2$.

\item We found that all proto-stellar sources, and at least 7 starless cores 
mapped at Mopra, show velocity gradients. In two of the proto-stellar cores the 
direction of the velocity gradients measured in HNC and HCO$^+$ are significantly 
different, suggesting the possibility that different systematic motions are 
simultaneously present (one of which likely is a molecular outflow).

\item The analysis of the virial masses showed that  nearly all of the starless cores 
have masses below the self-gravitating threshold, indicating that they are unlikely 
to be gravitationally bound, whereas more than half of the proto-stellar cores have masses which 
are near or above the self-gravitating critical value. If a density profile closer
to the one actually observed is considered in the calculation of the virial masses,
then also most starless cores turn out to have masses near or above the self-gravitating threshold.

\item The average internal pressure of proto-stellar cores is higher than in starless
cores, and both are higher than the typical ambient pressure in Vela-D. 
Cores (mostly starless) {\it not} gravitationally bound cannot thus be pressure-confined
like, for example, the cores in the Pipe nebula.

\item The radial profile of the ring-averaged integrated spectral line intensity 
confirmed that proto-stellar cores have on average a more 
compact structure. 
We also found that the radial
profile of the N$_2$H$^+(1-0)$ emission falls-off more quickly, on average, than that 
of the C-bearing molecular lines HNC$(1-0)$, HCO$^+(1-0)$ and HCN$(1-0)$.

\item 
We find a variable degree of correlation, with a significant scatter,
 between the column densities of chemically different molecular species, 
and also between molecular tracers with different effective densities. 
We also find that in starless cores the linewidth and effective density are well correlated.
Instead, linewidth and temperature are not strongly correlated.

\item 
The branching production ratio of HCN and also 
the relatively low [N$_2$H$^+$]/[H$^{13}$CO$^+$] abundance ratio confirm that all 
cores in our sample are cold and are on average in early evolutionary phases.

\item
An analysis of the median properties of the starless and proto-stellar cores suggests
that  the transition from the pre- to the proto-stellar phase
is relatively fast, leaving the core envelopes with almost unchanged physical parameters.

\end{enumerate}

\begin{acknowledgements}
The authors wish to thank the ATNF staff for support during and after the observations with
the Mopra and Parkes telescopes.
JMO acknowledges the support of NASA through the PR NASA Space Grant Doctoral Fellowship.
\end{acknowledgements}

\appendix

\section{Spectral line maps} 
\label{sec:specmaps}

\nopagebreak[4]

%
%
\begin{figure*}[h]
%
\hspace*{4.4cm}
\includegraphics[width=4.4cm,angle=0]{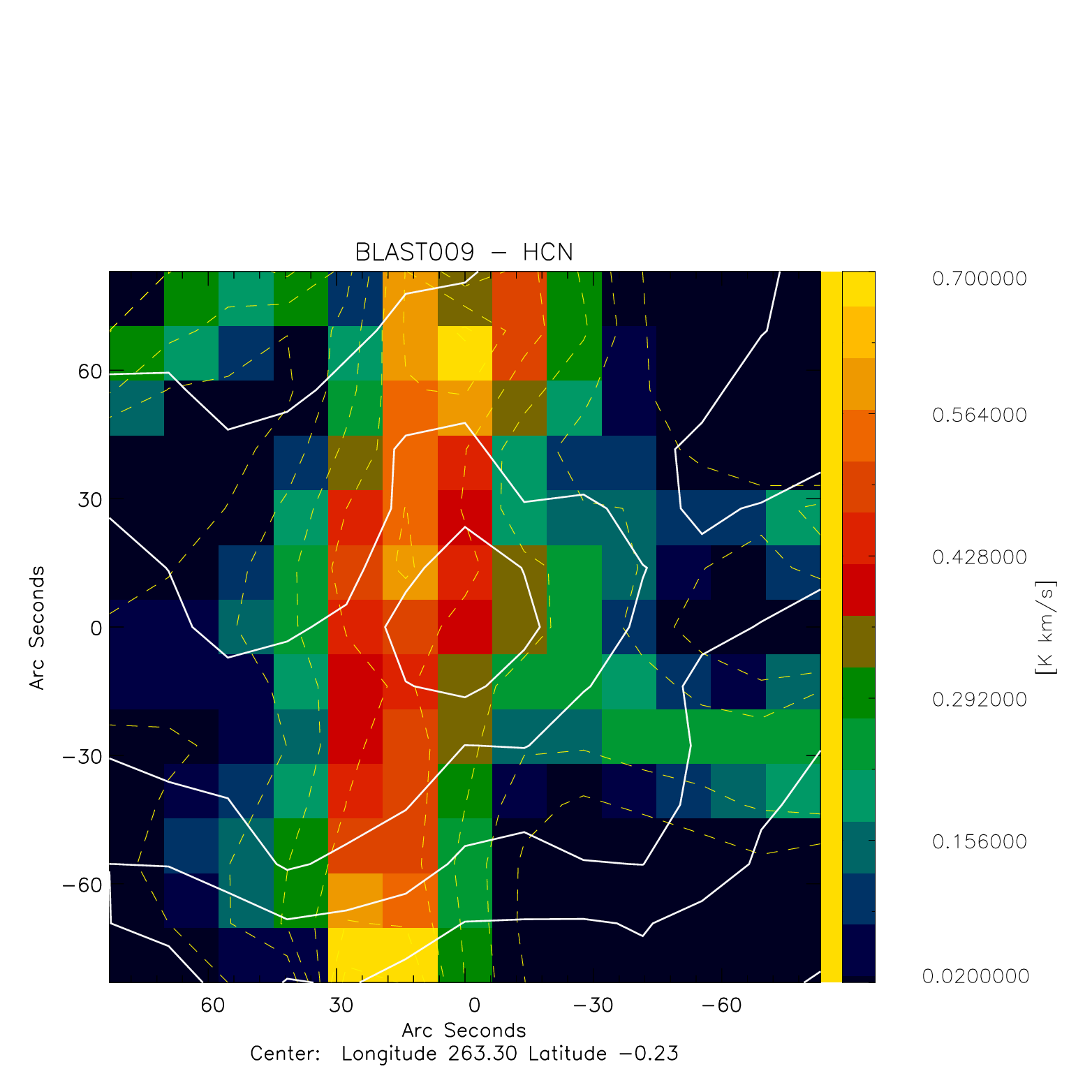}
\includegraphics[width=4.4cm,angle=0]{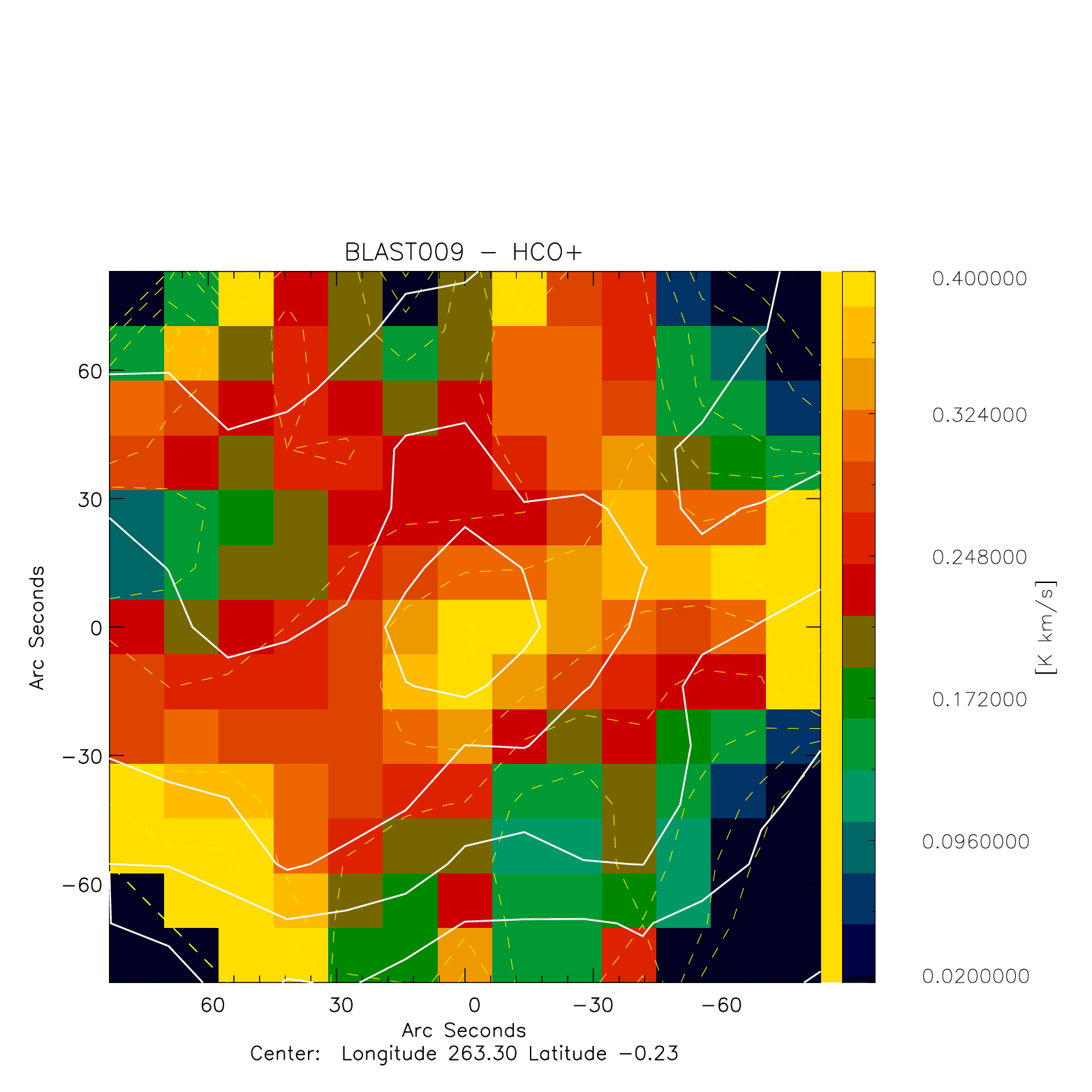}
\includegraphics[width=4.4cm,angle=0]{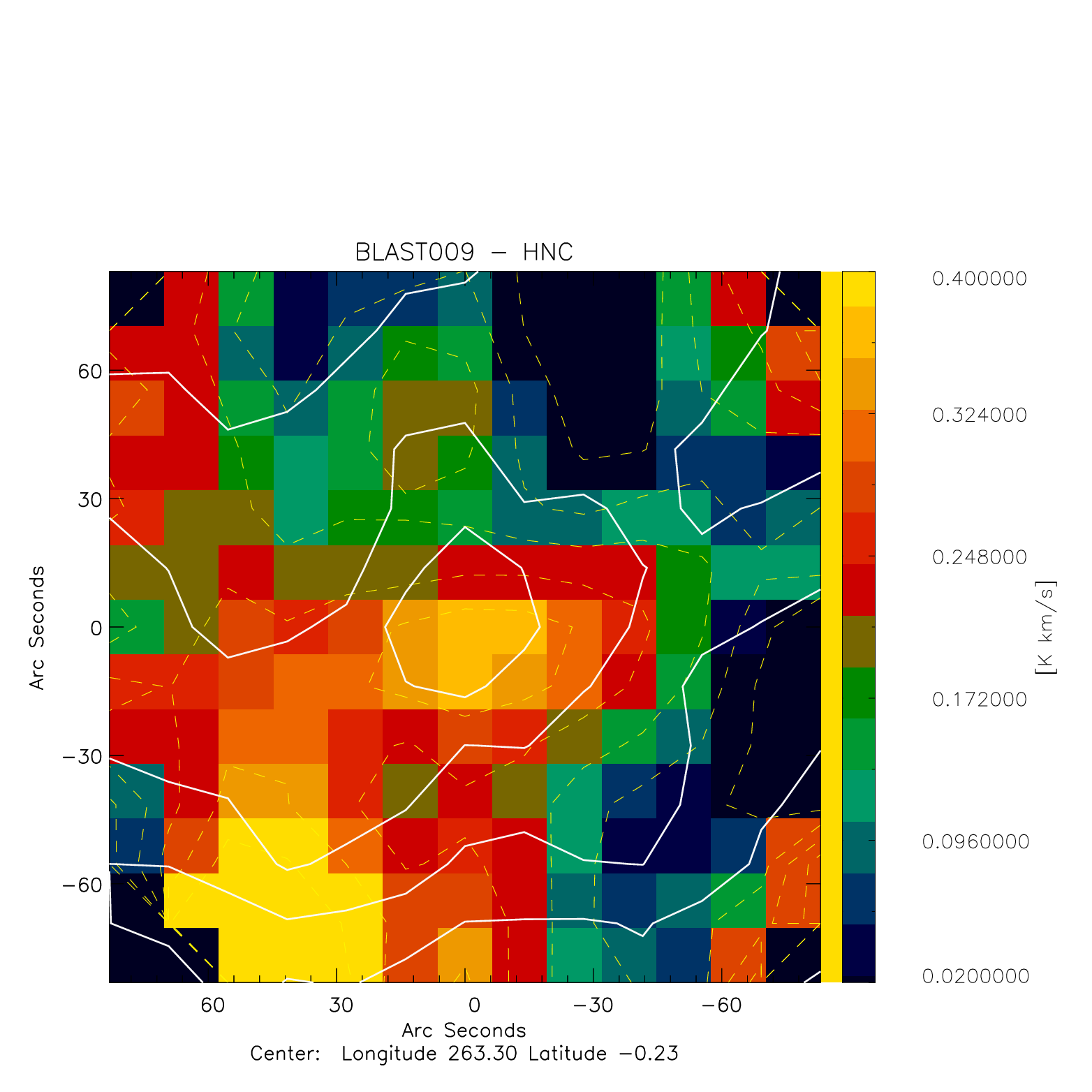} \\
\hspace*{4.4cm}
\includegraphics[width=4.4cm,angle=0]{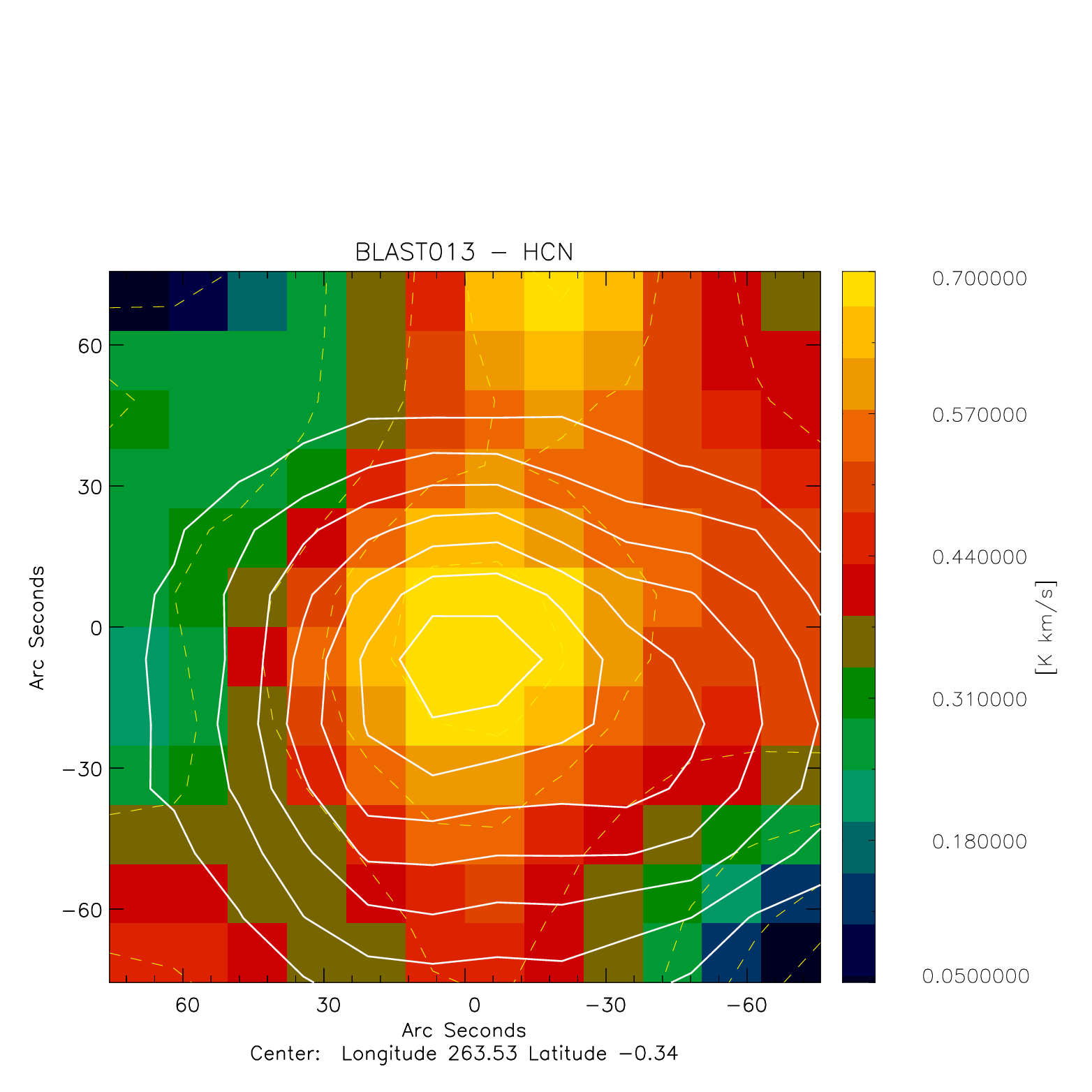}
\includegraphics[width=4.4cm,angle=0]{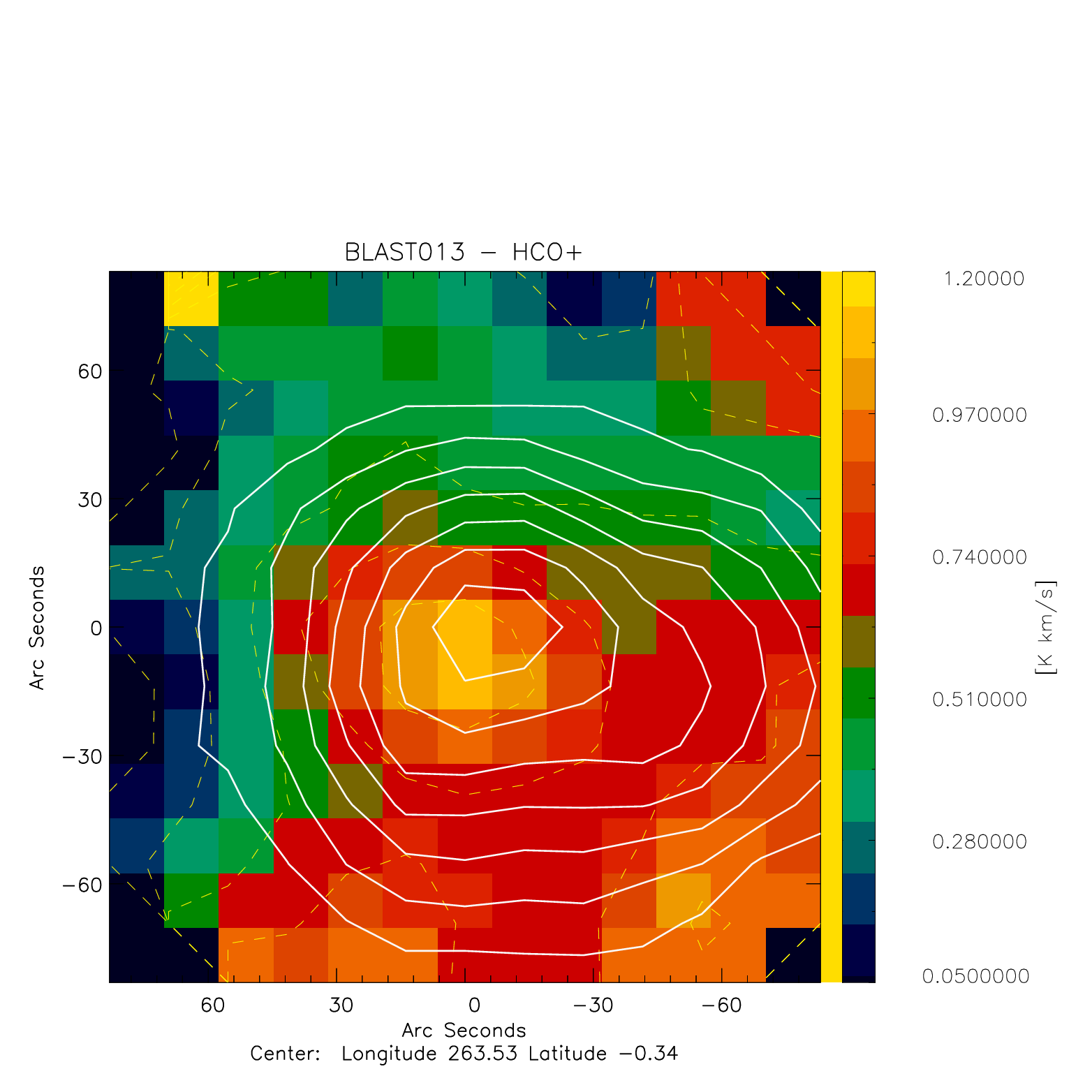}
\includegraphics[width=4.4cm,angle=0]{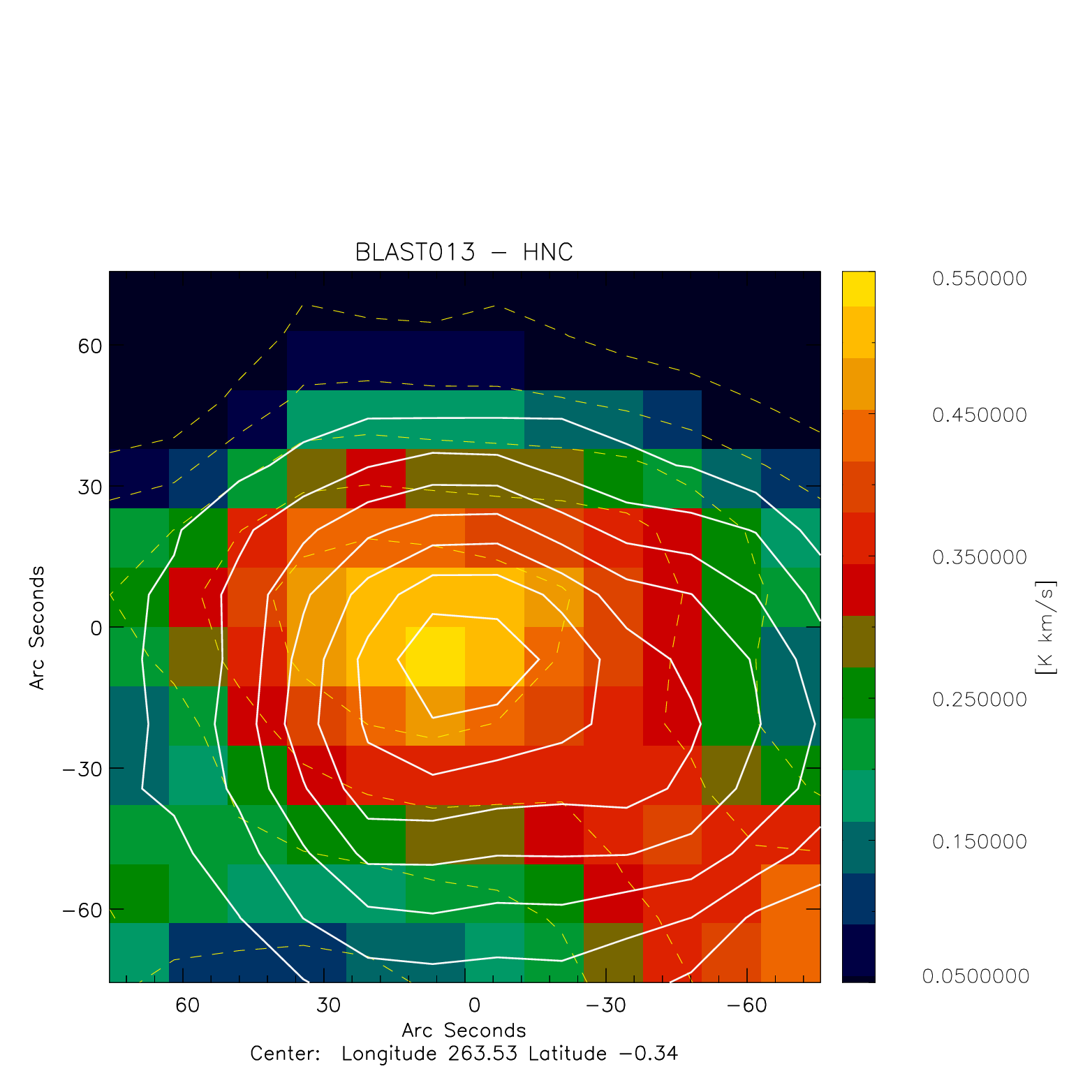} \\
\includegraphics[width=4.4cm,angle=0]{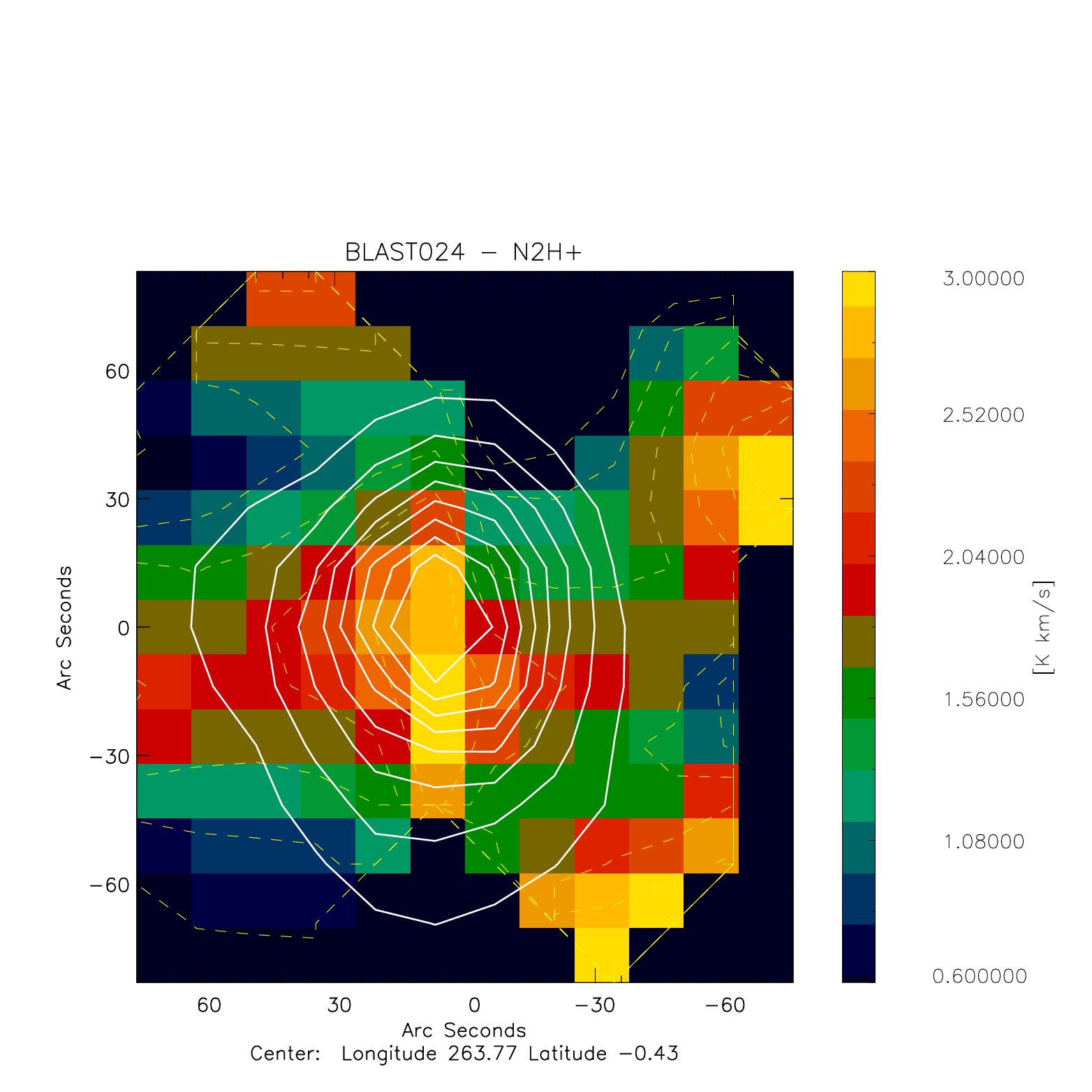}
\includegraphics[width=4.4cm,angle=0]{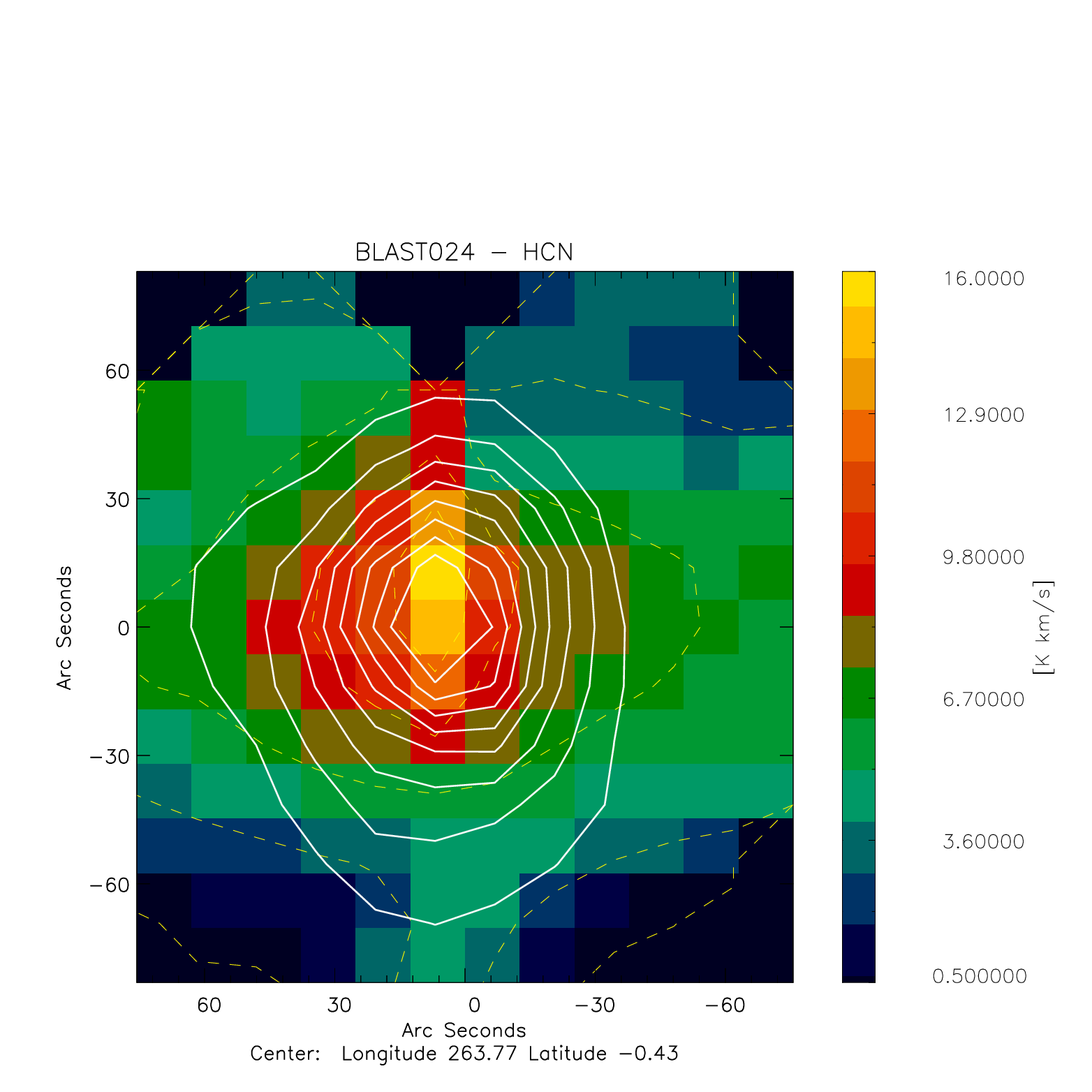}
\includegraphics[width=4.4cm,angle=0]{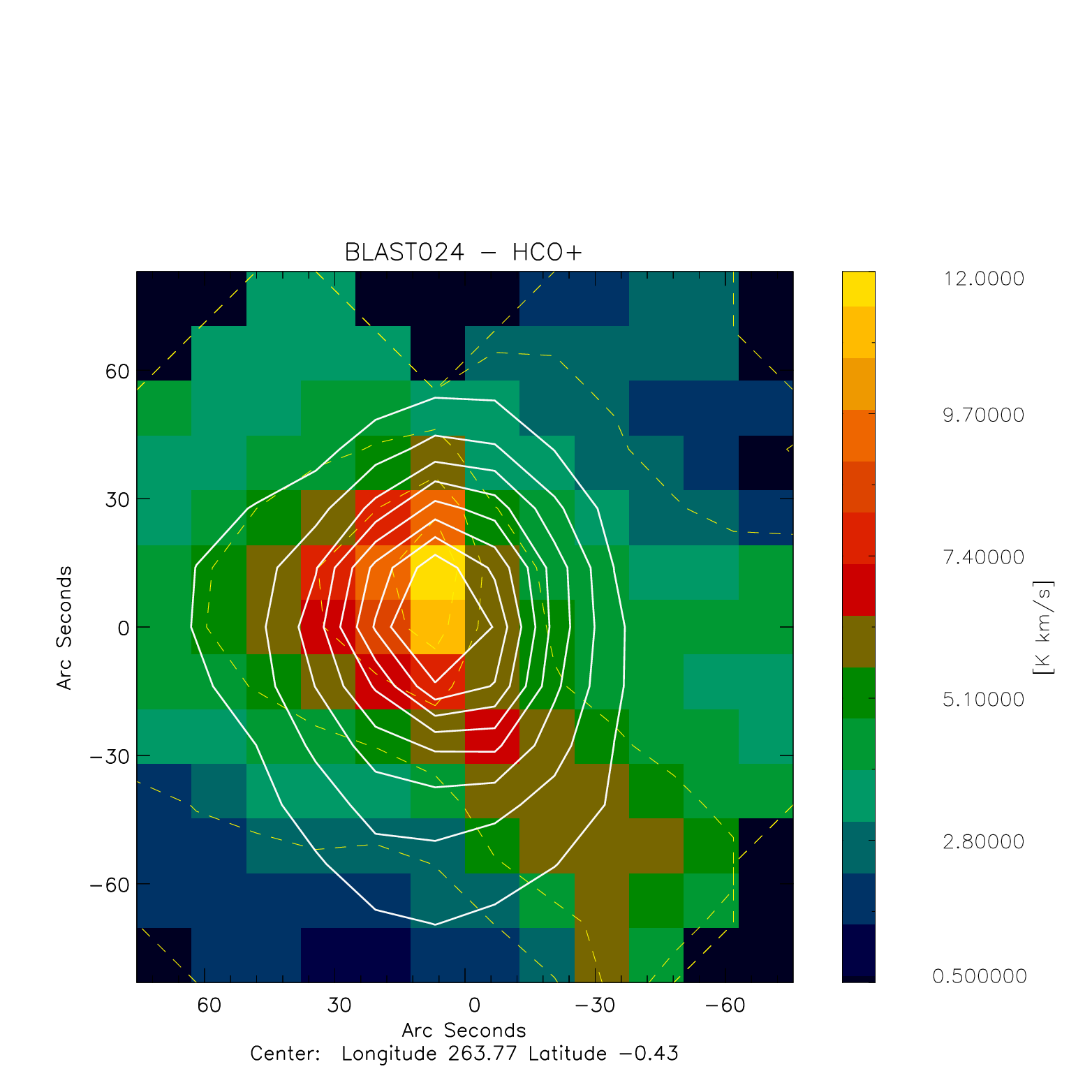}
\includegraphics[width=4.4cm,angle=0]{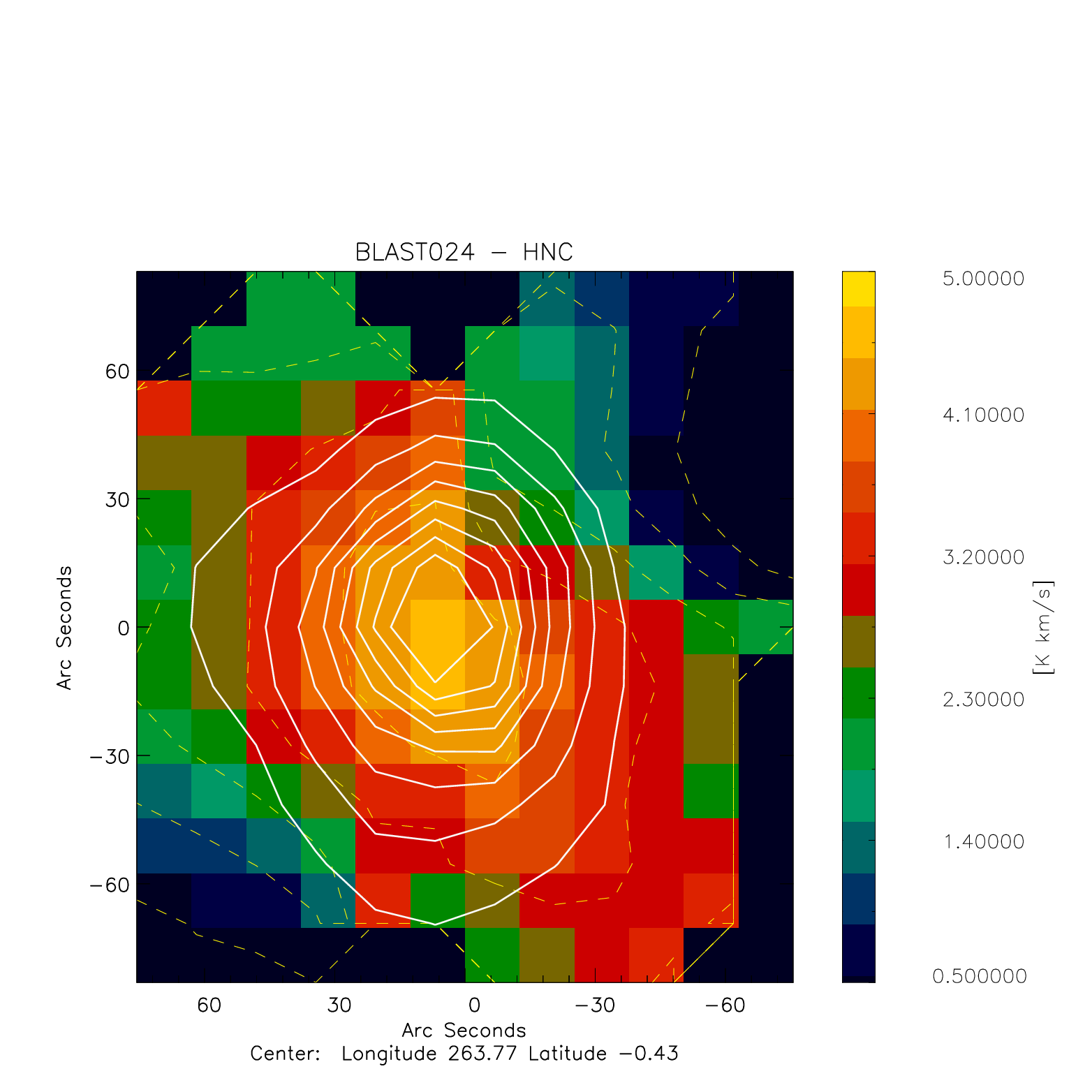} \\
\hspace*{4.4cm}
\includegraphics[width=4.4cm,angle=0]{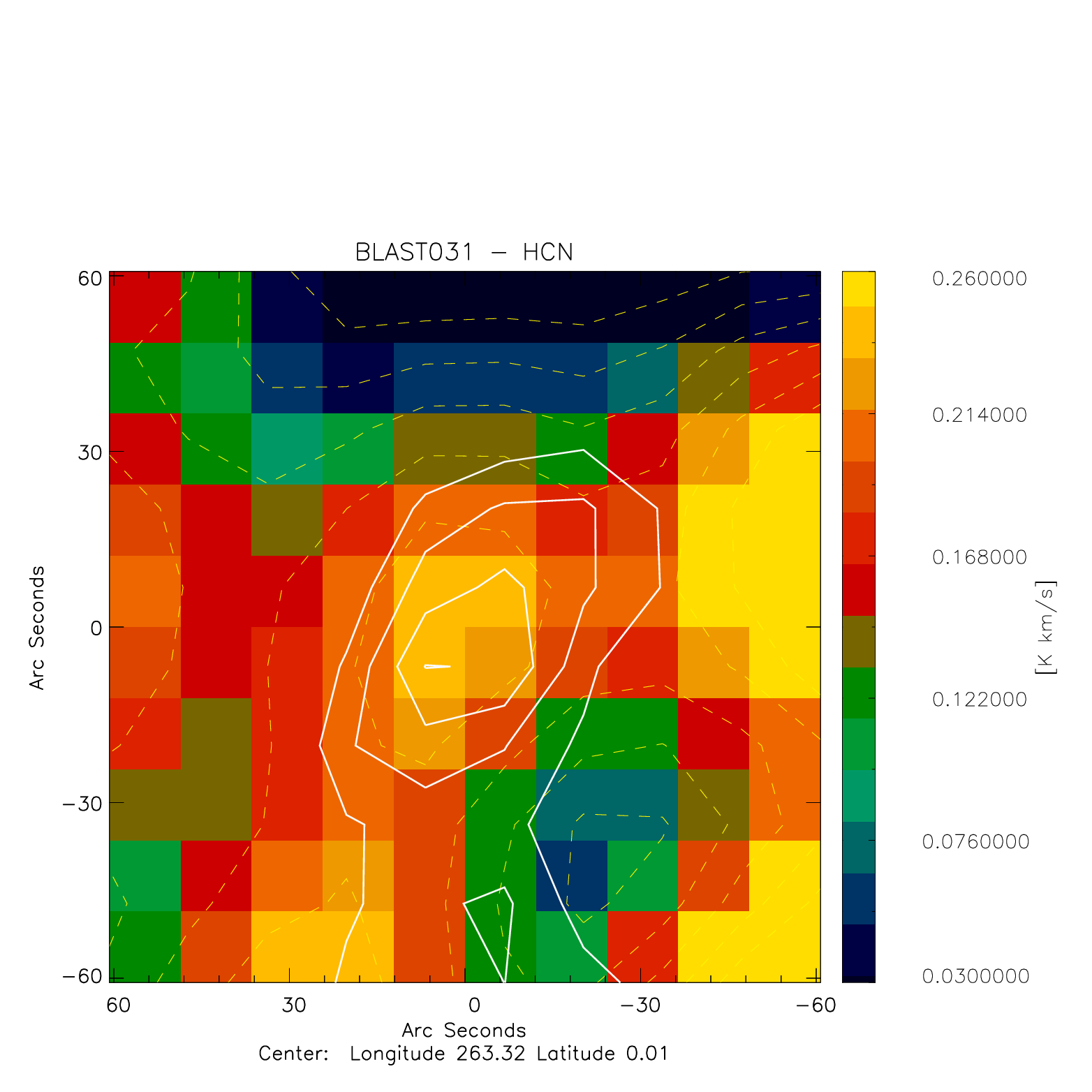}
\includegraphics[width=4.4cm,angle=0]{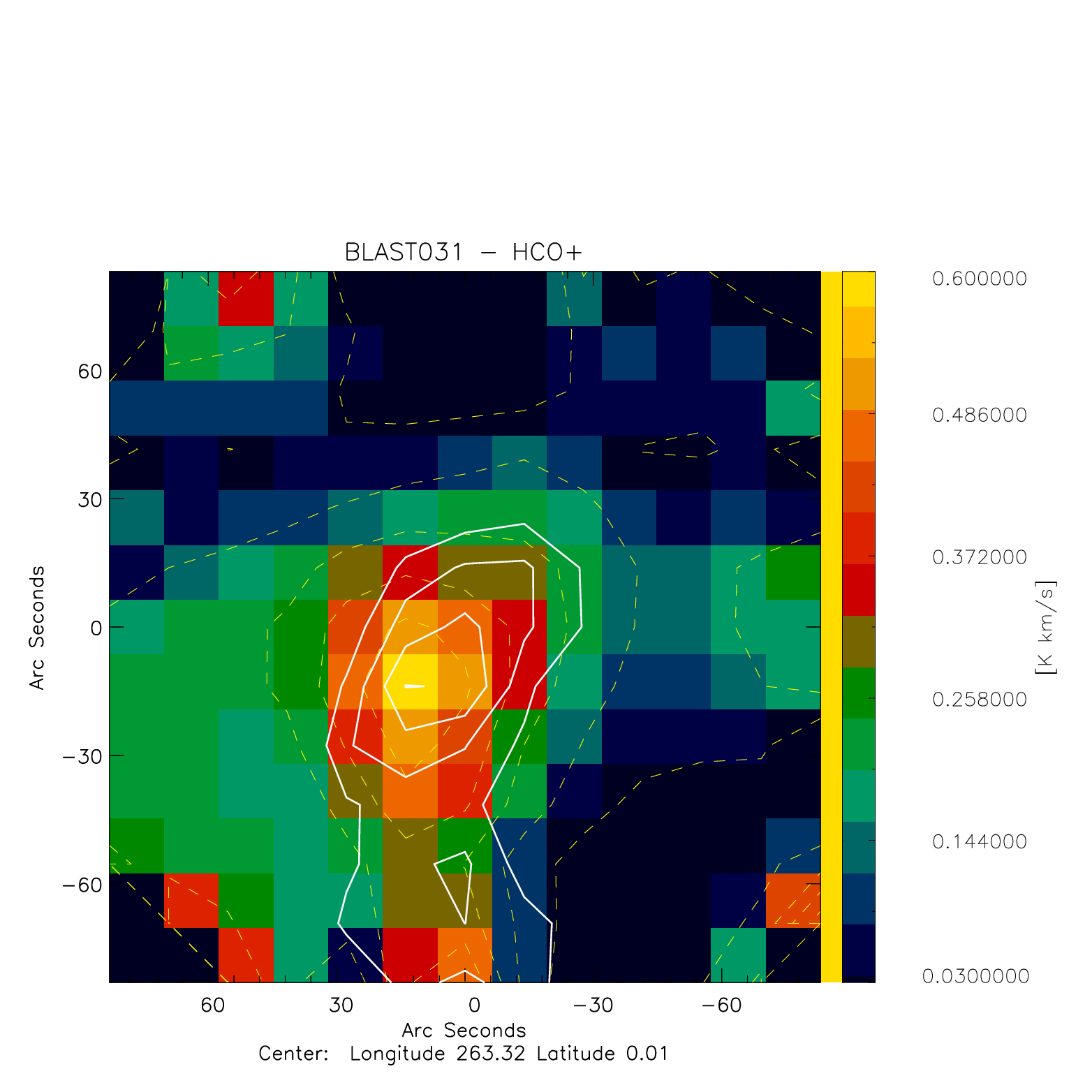}
\includegraphics[width=4.4cm,angle=0]{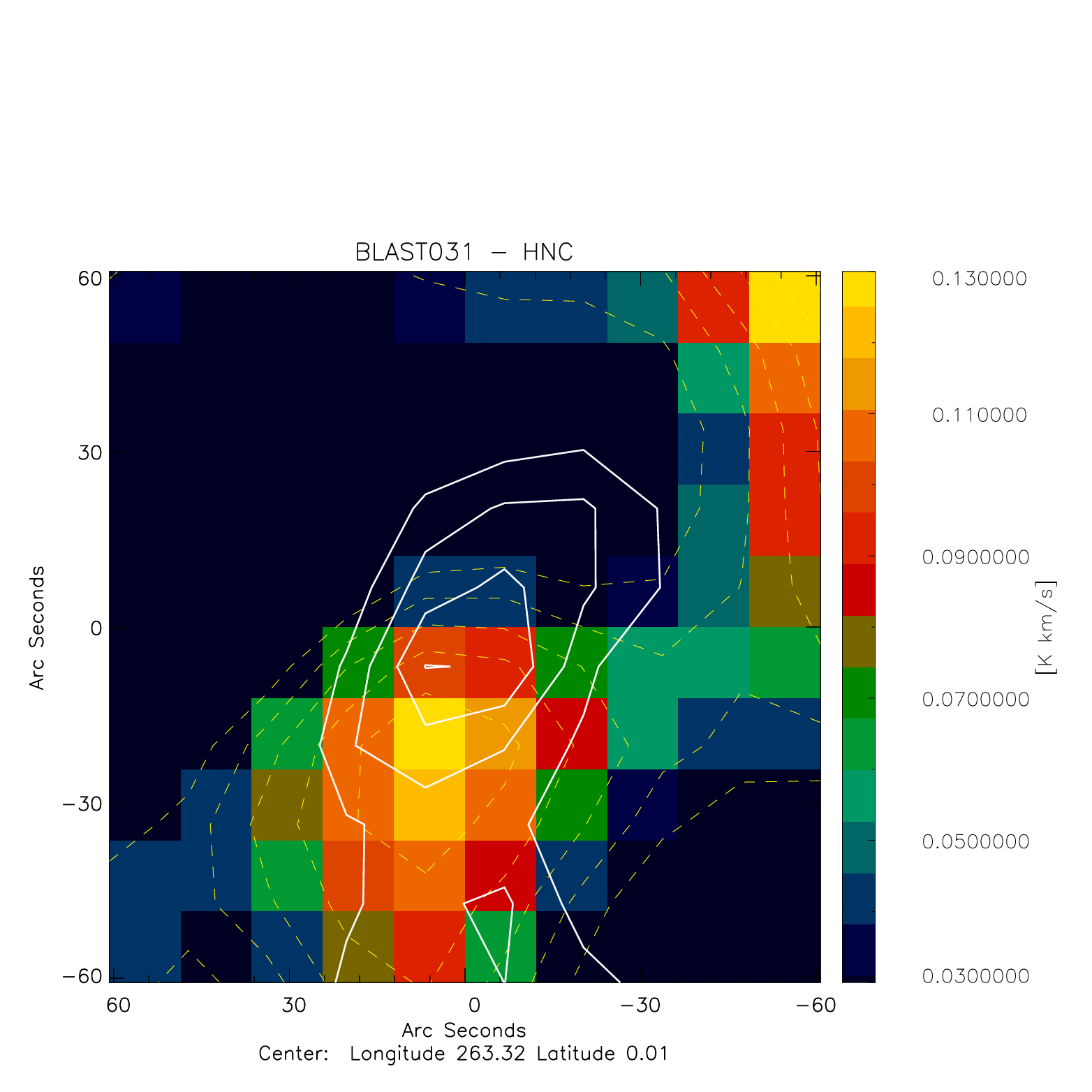} \\
\caption{
Maps of the line integrated intensities  (from left to right,
N$_2$H$^+(1-0)$, HCN$(1-0)$, HCO$^+(1-0)$ and HNC$(1-0)$,
in units of $\int T_{\rm A}^{\star} \, {\rm d}v$ [K\,km\,s$^{-1}]$) of
selected sources toward Vela-D. In this figure we show, from top to bottom, 
maps of BLAST009, BLAST013, BLAST024 and BLAST031.
The white dashed contours represent the BLAST flux density at 250\,\micron.
  }
\label{fig:maps1}
\end{figure*}

\clearpage

%
\begin{figure*}
\hspace*{4.4cm}
\includegraphics[width=4.4cm,angle=0]{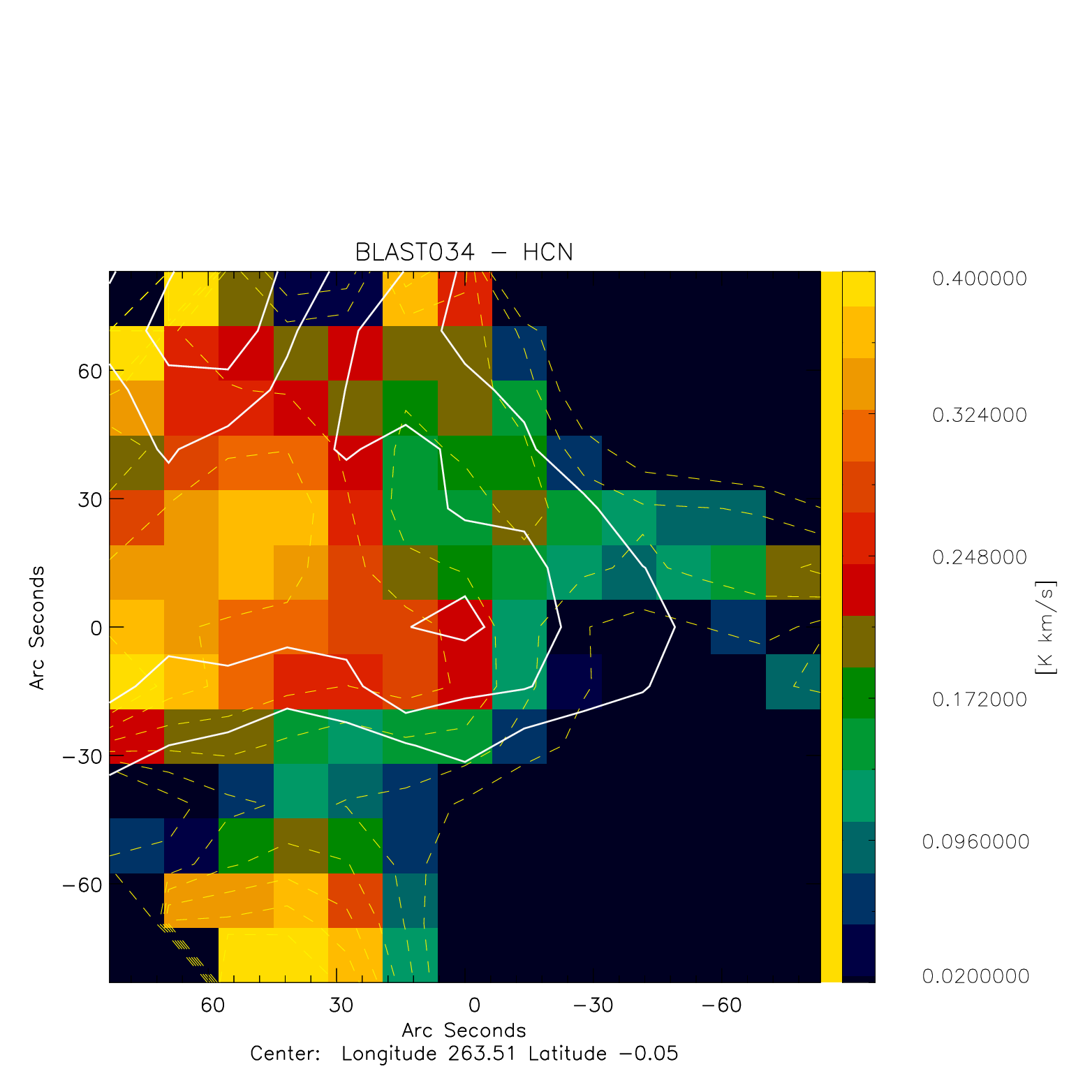}
\includegraphics[width=4.4cm,angle=0]{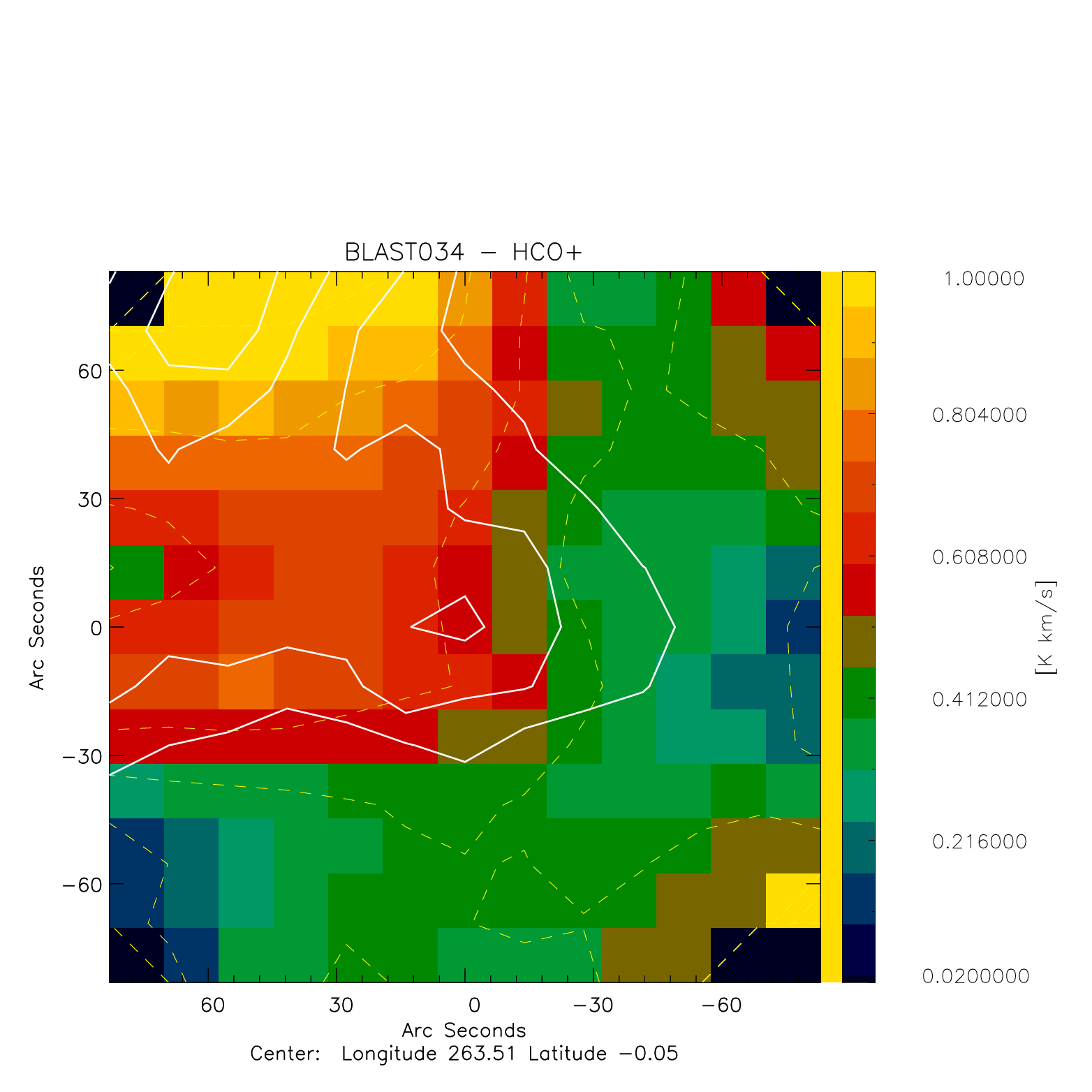}
\includegraphics[width=4.4cm,angle=0]{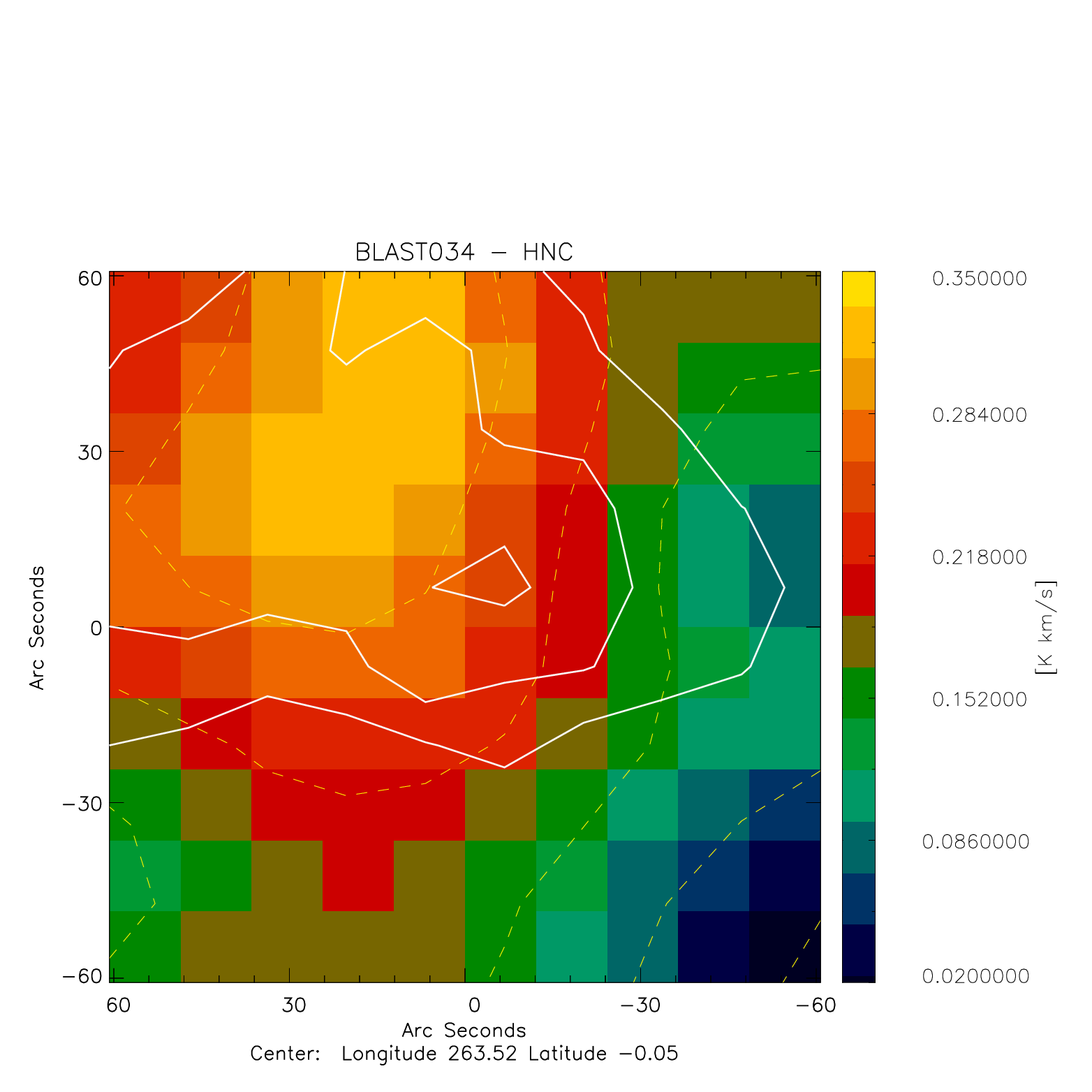} \\
\includegraphics[width=4.4cm,angle=0]{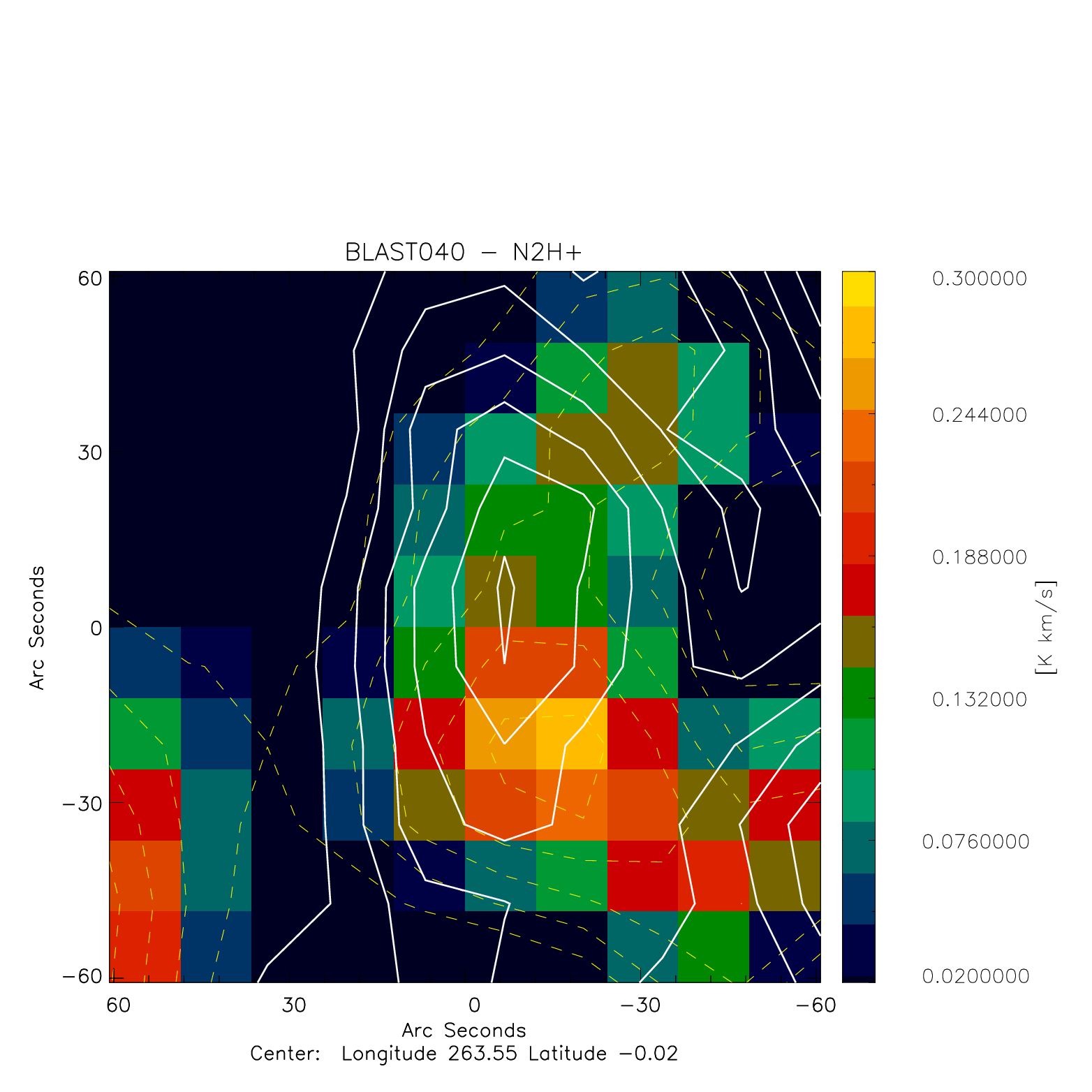}
\includegraphics[width=4.4cm,angle=0]{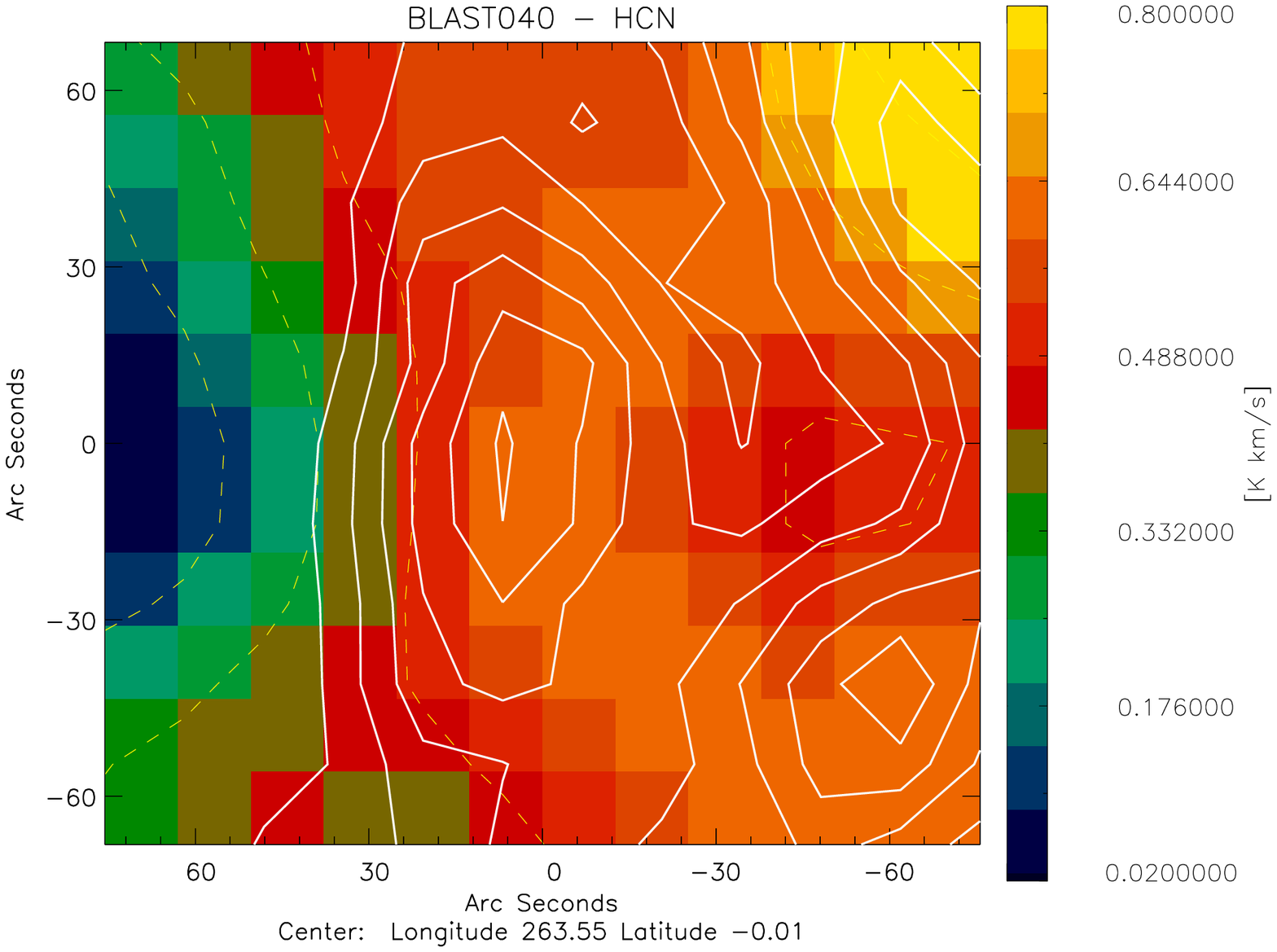}
\includegraphics[width=4.4cm,angle=0]{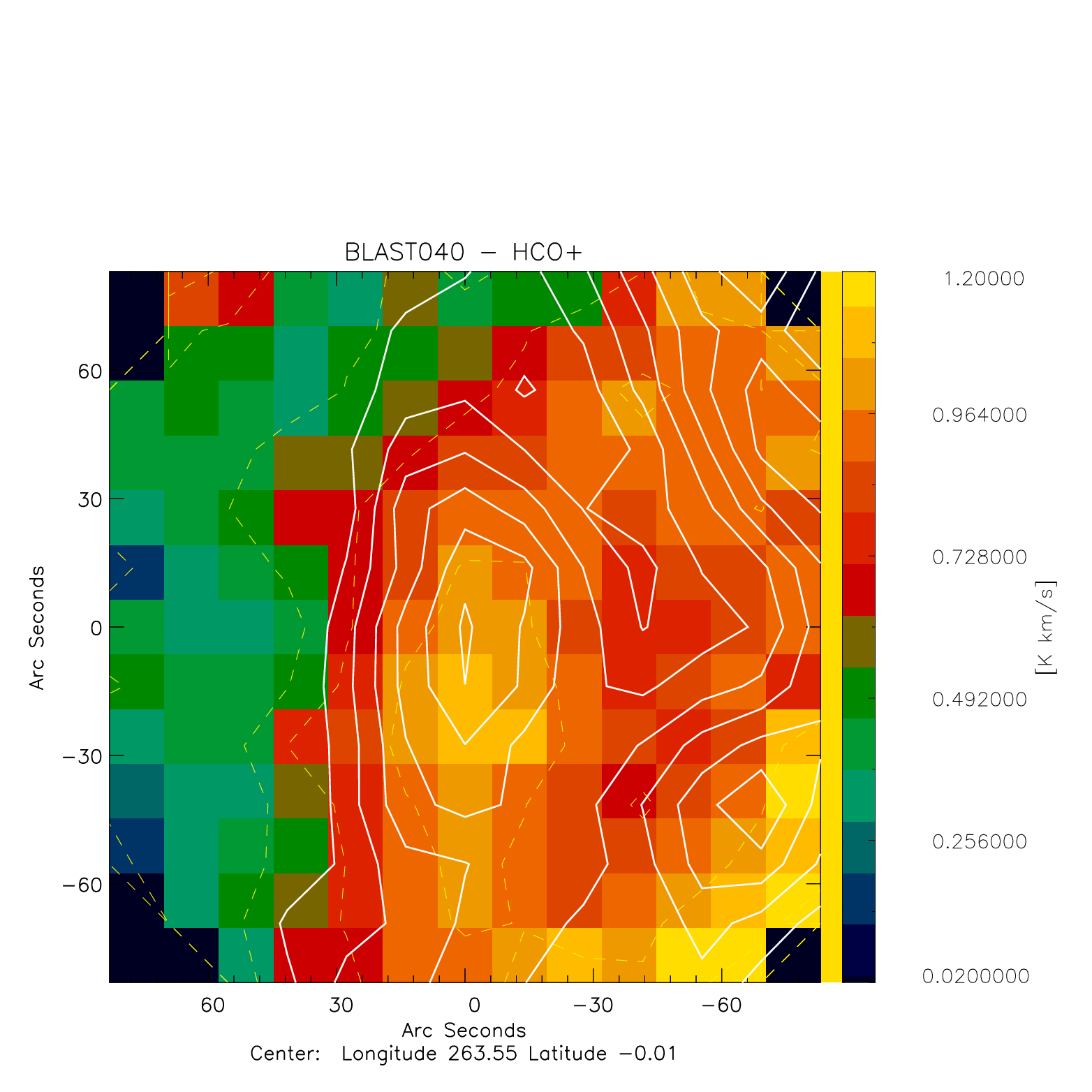}
\includegraphics[width=4.4cm,angle=0]{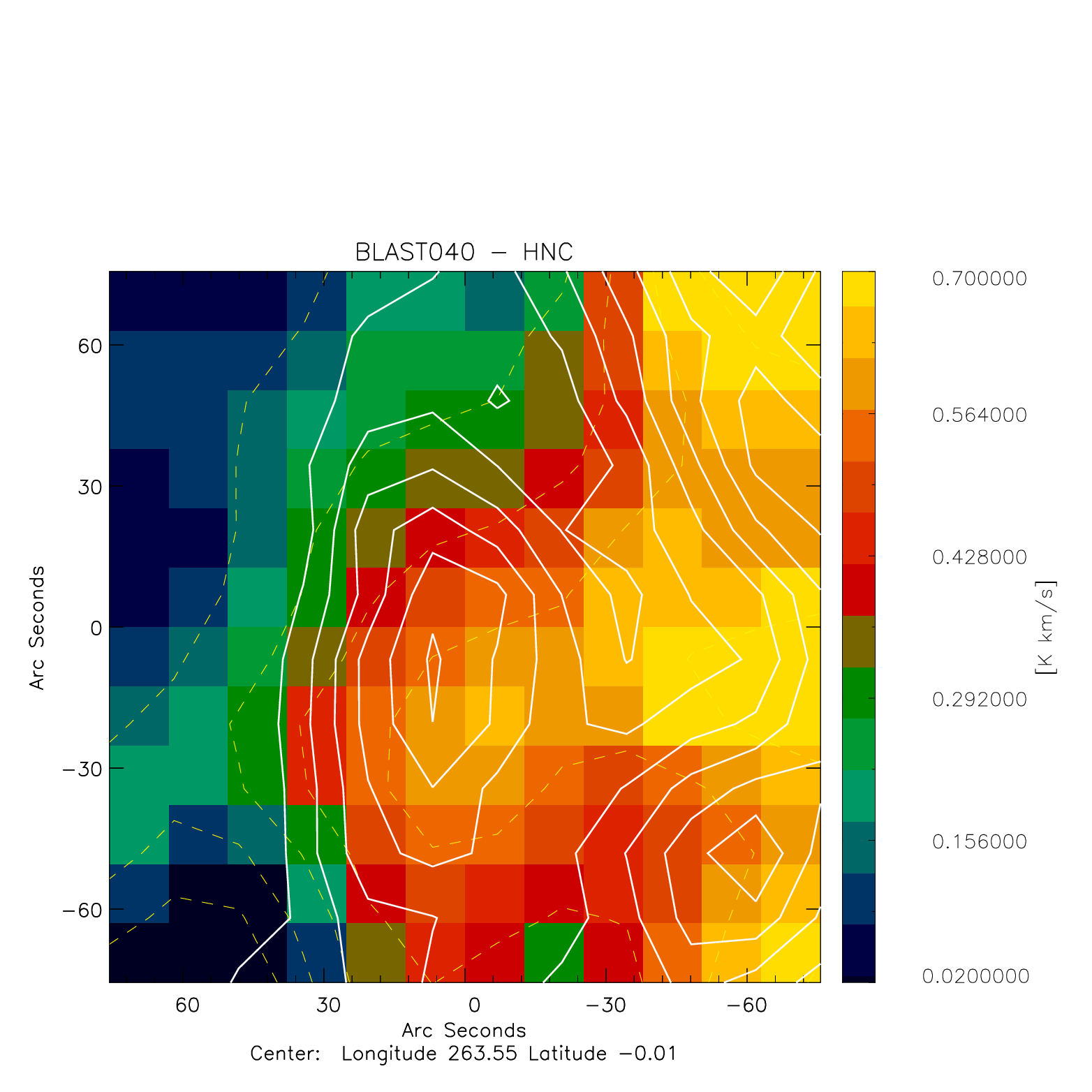} \\
\hspace*{4.4cm}
\includegraphics[width=4.4cm,angle=0]{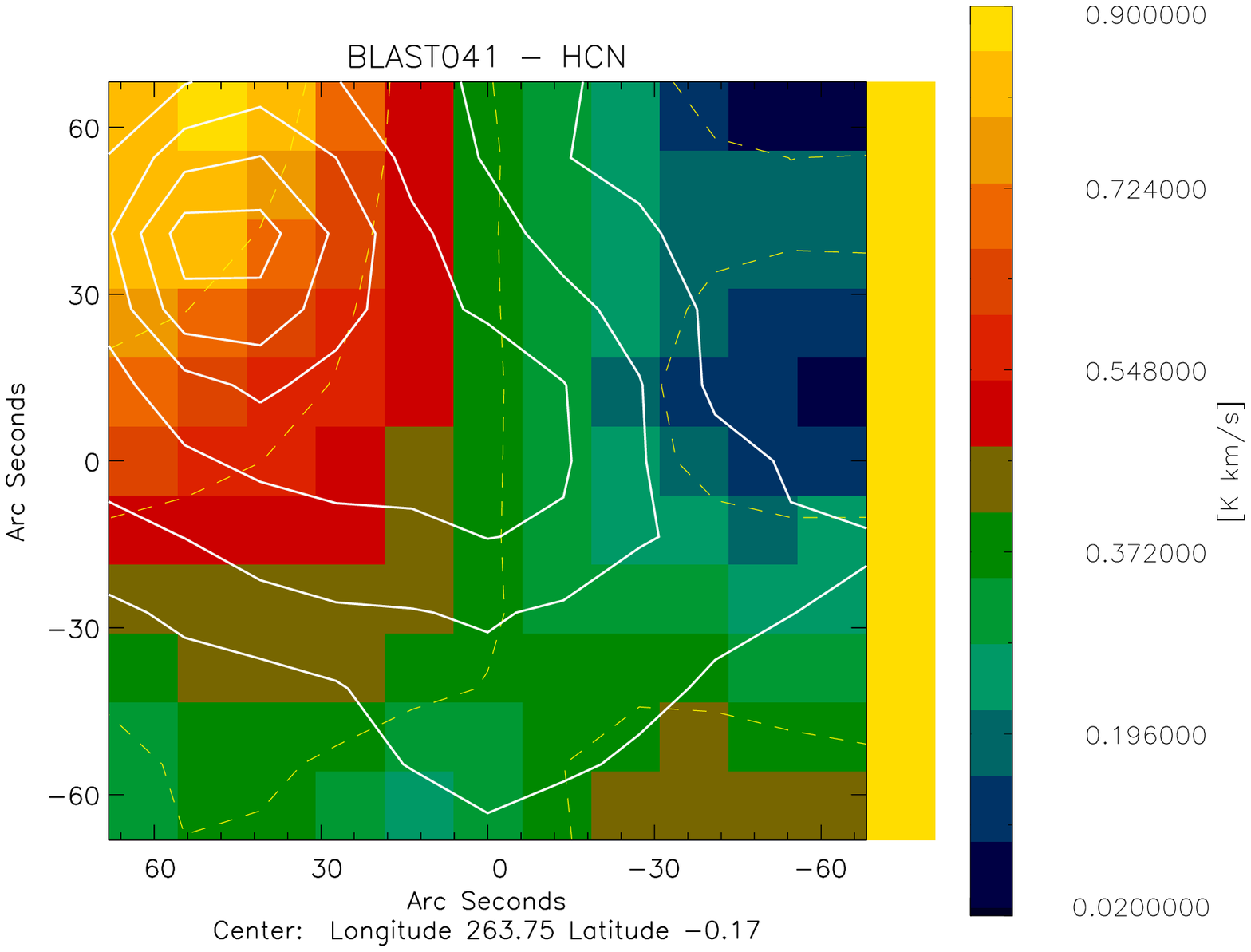}
\includegraphics[width=4.4cm,angle=0]{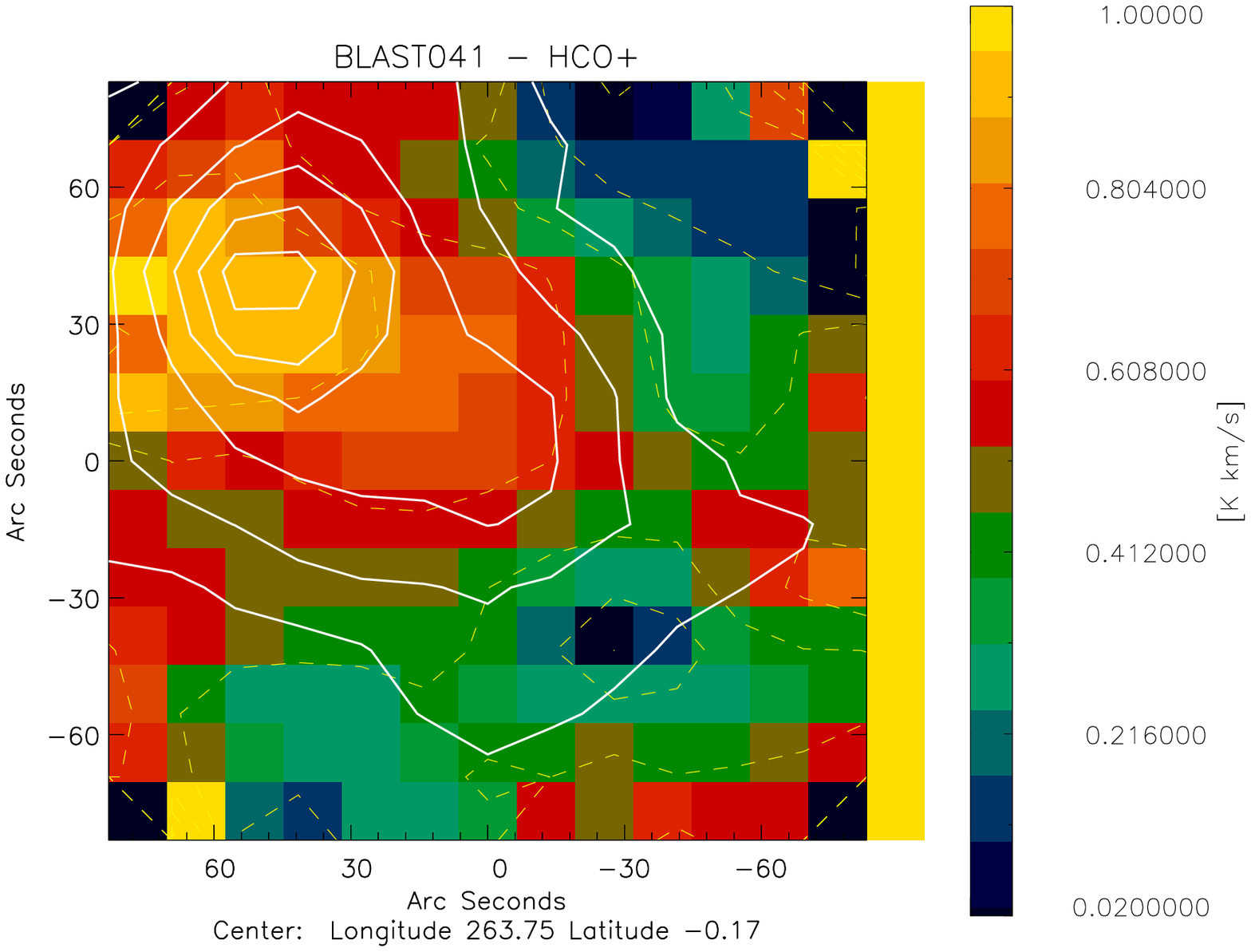}
\includegraphics[width=4.4cm,angle=0]{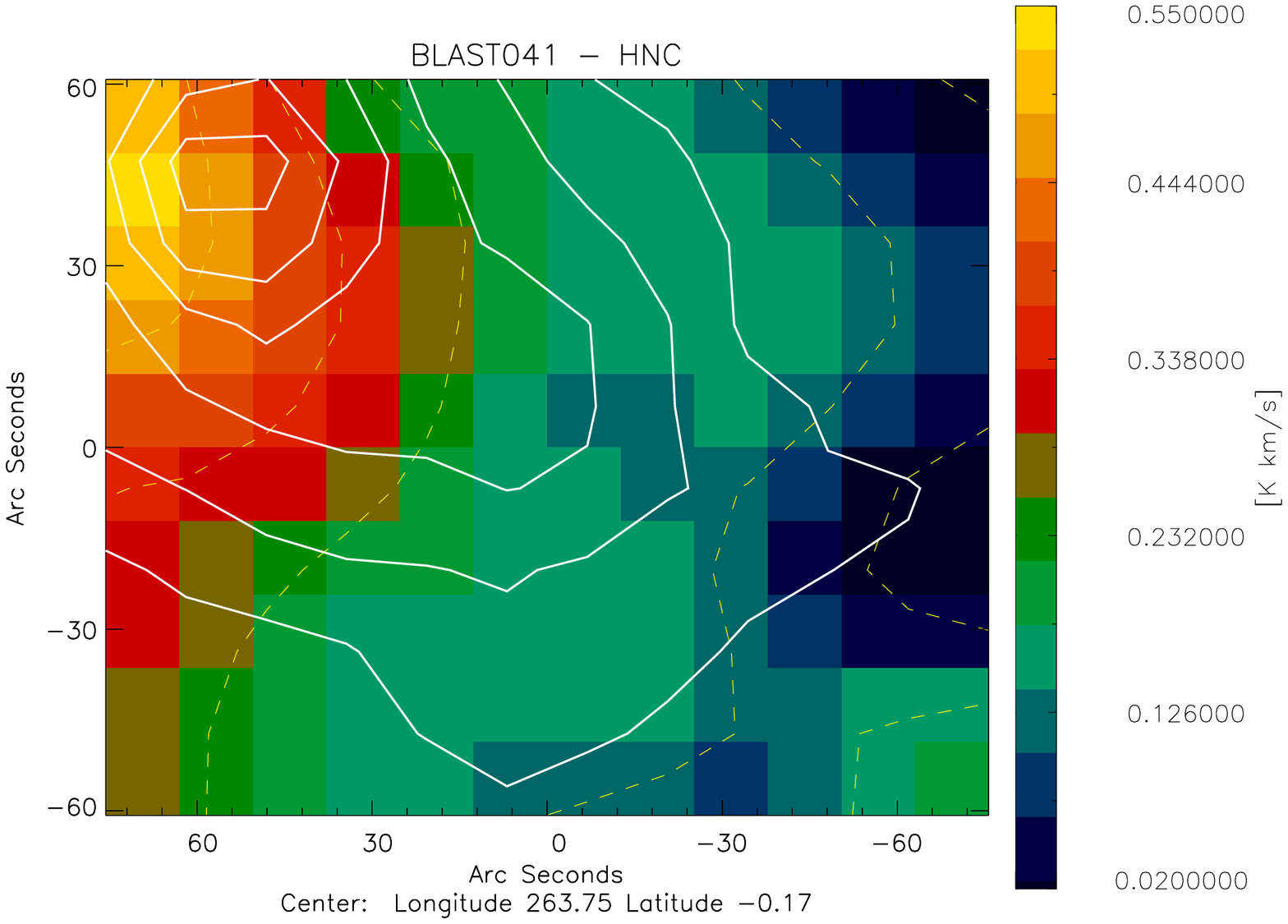} \\
\hspace*{4.4cm}
\includegraphics[width=4.4cm,angle=0]{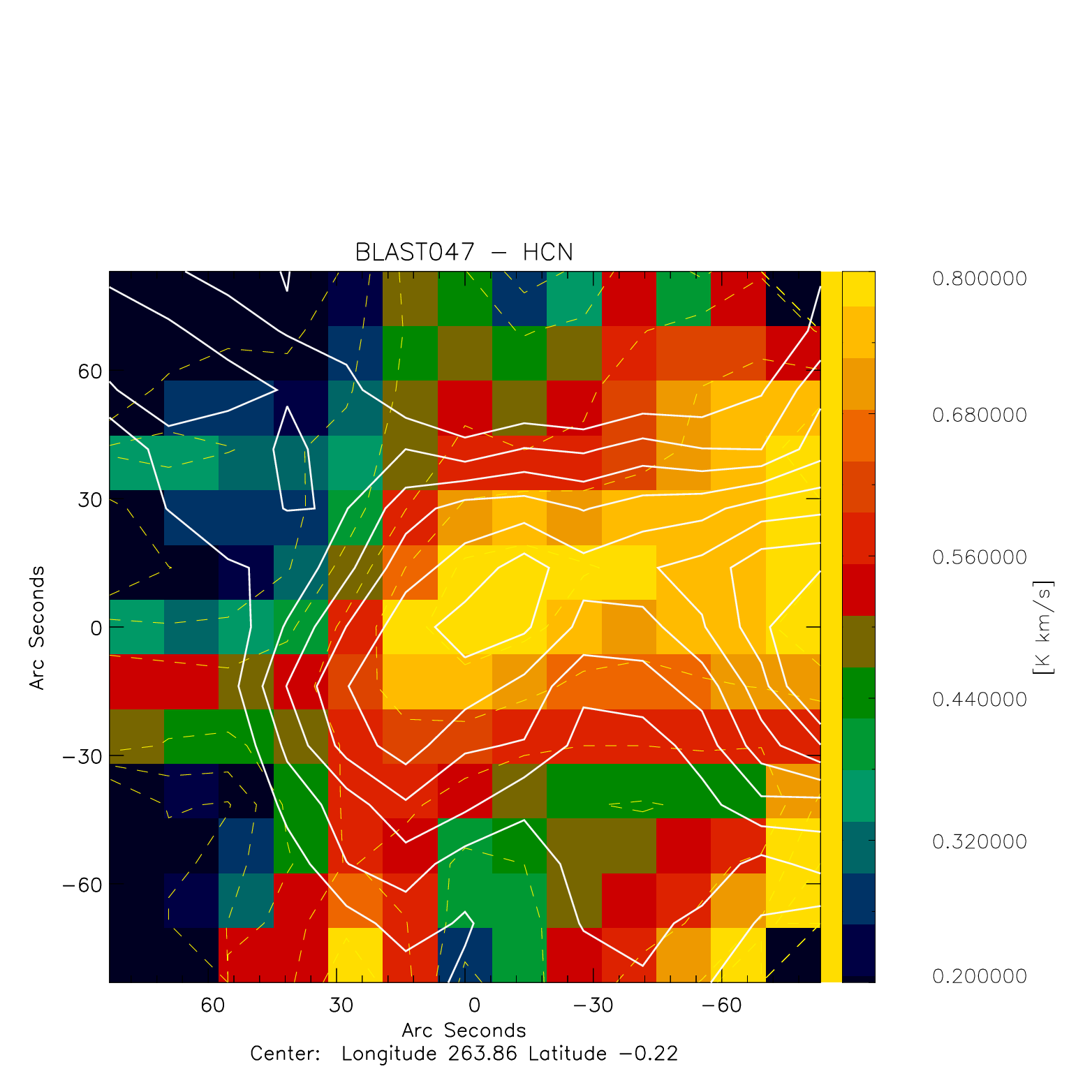}
\includegraphics[width=4.4cm,angle=0]{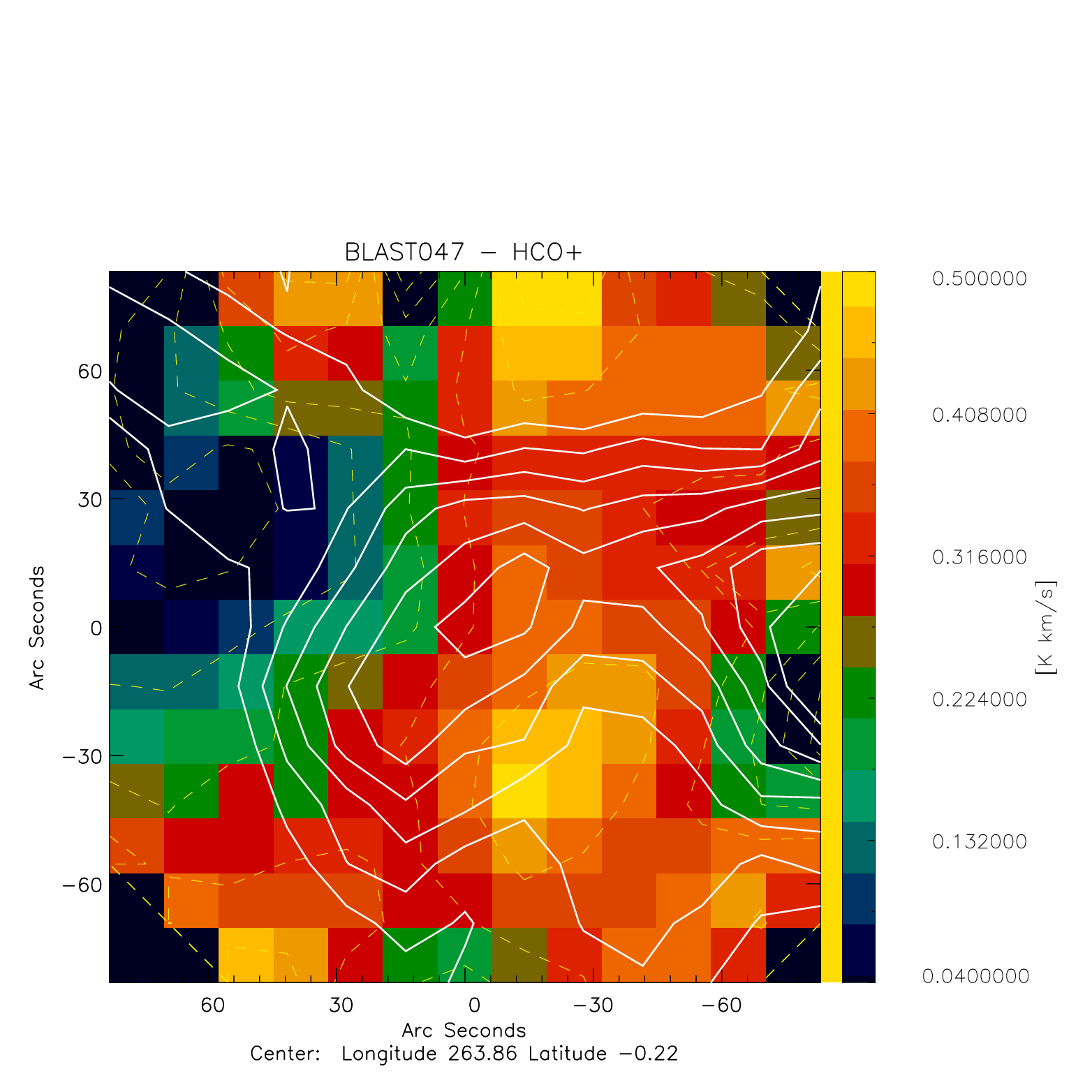}
\includegraphics[width=4.4cm,angle=0]{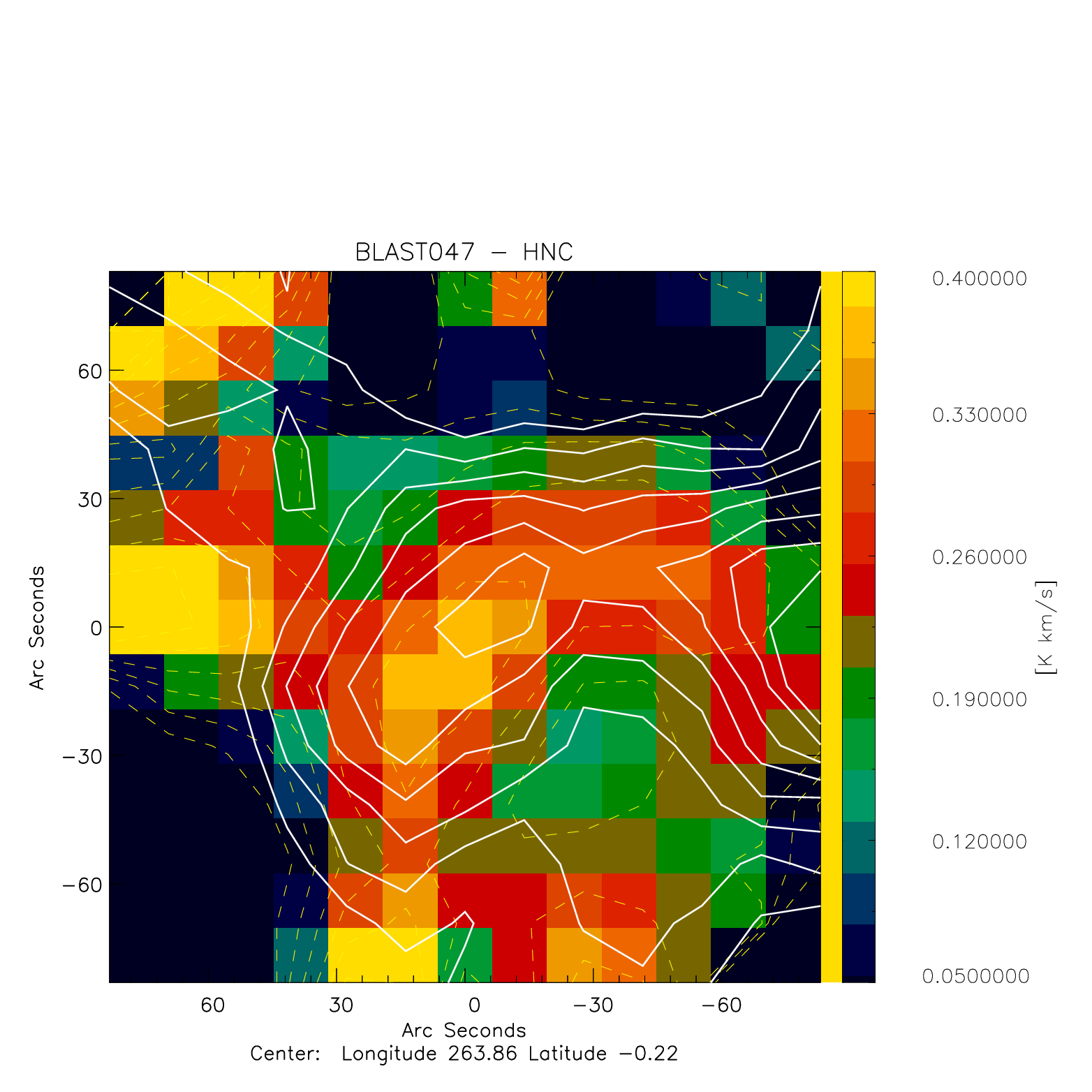} \\
\caption{
Same as Figure~\ref{fig:maps1} 
for sources BLAST034, BLAST040, BLAST041 and BLAST047.
  }
\label{fig:maps2}
\end{figure*}

\clearpage                 

%
\begin{figure*}
\hspace*{4.4cm}
\includegraphics[width=4.4cm,angle=0]{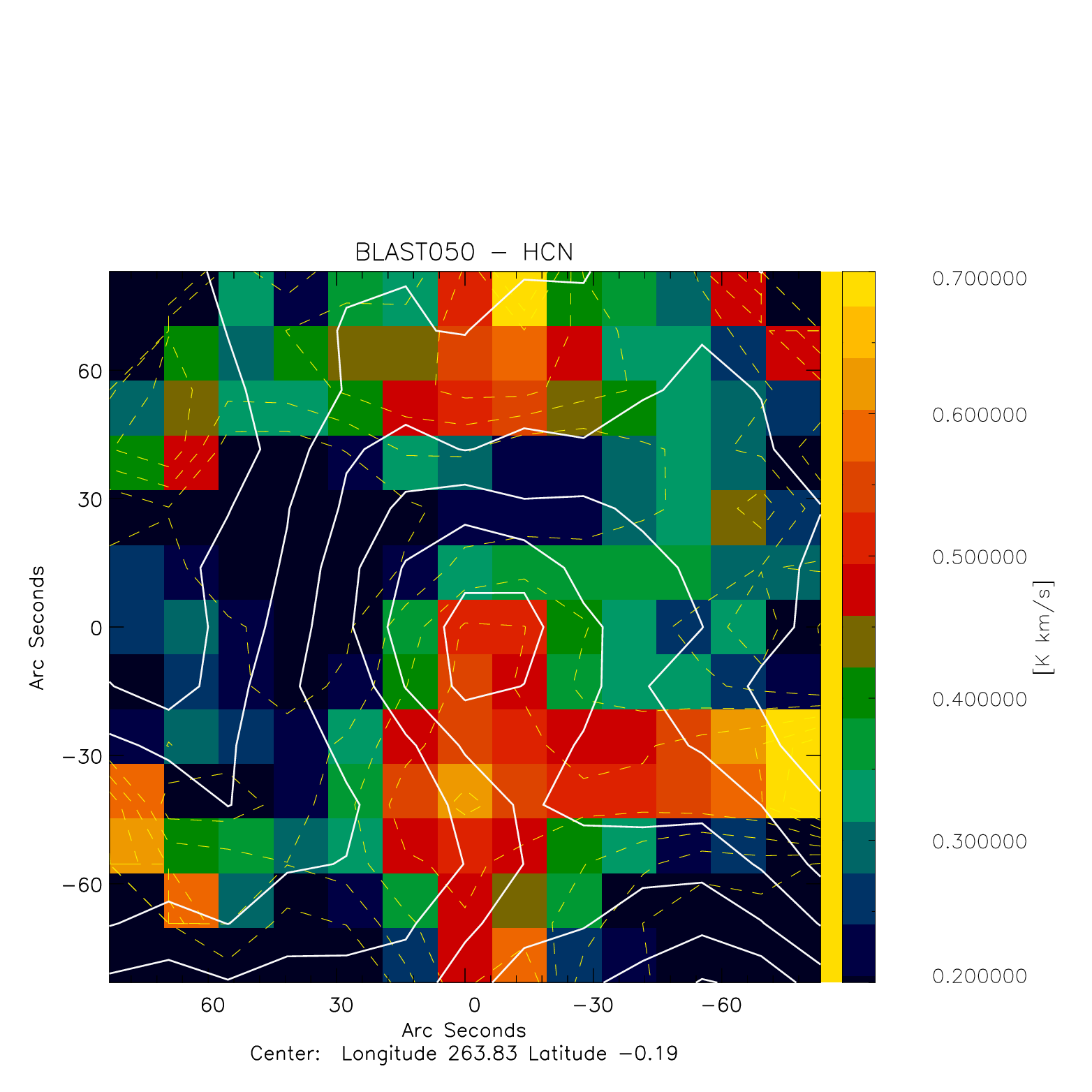}
\includegraphics[width=4.4cm,angle=0]{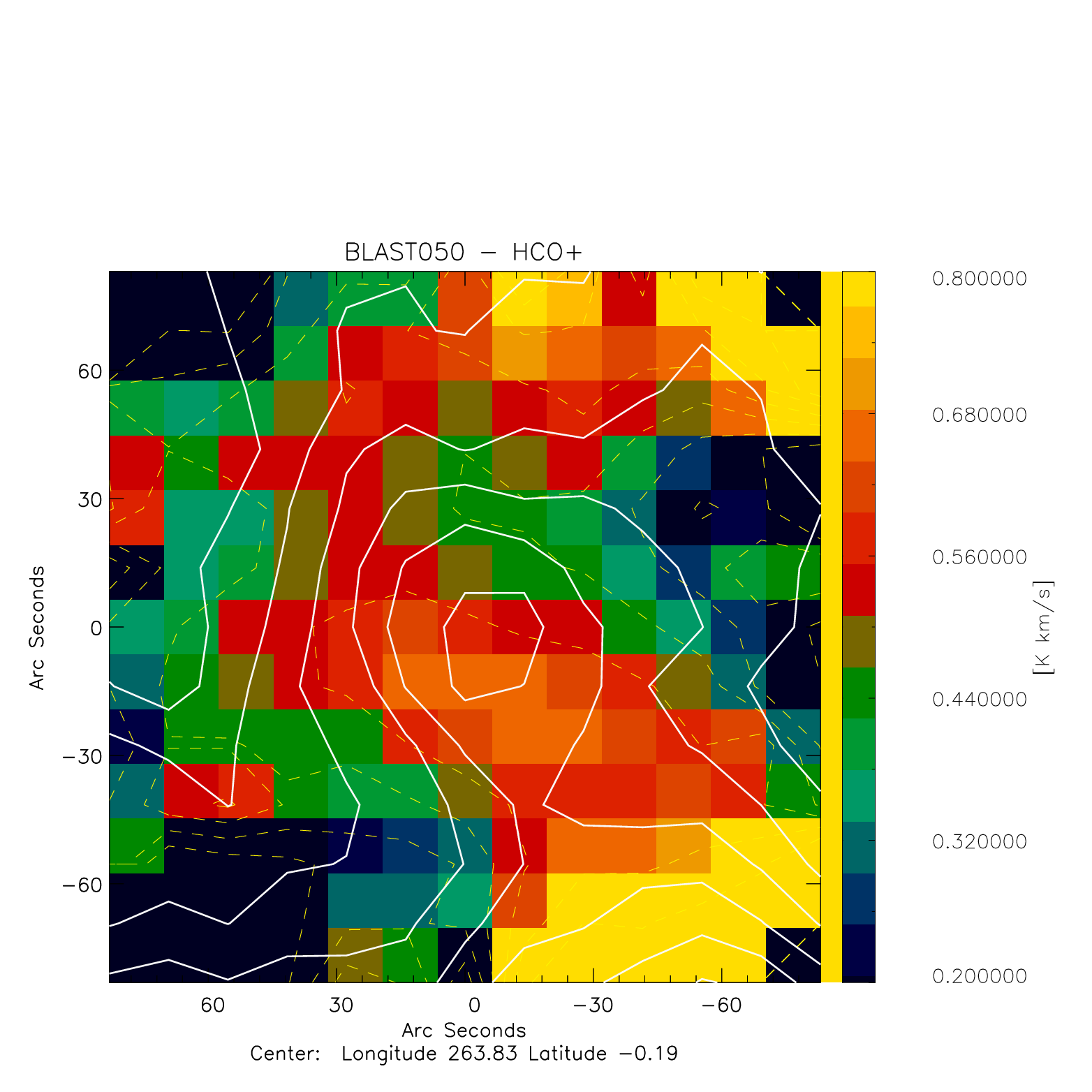}
\includegraphics[width=4.4cm,angle=0]{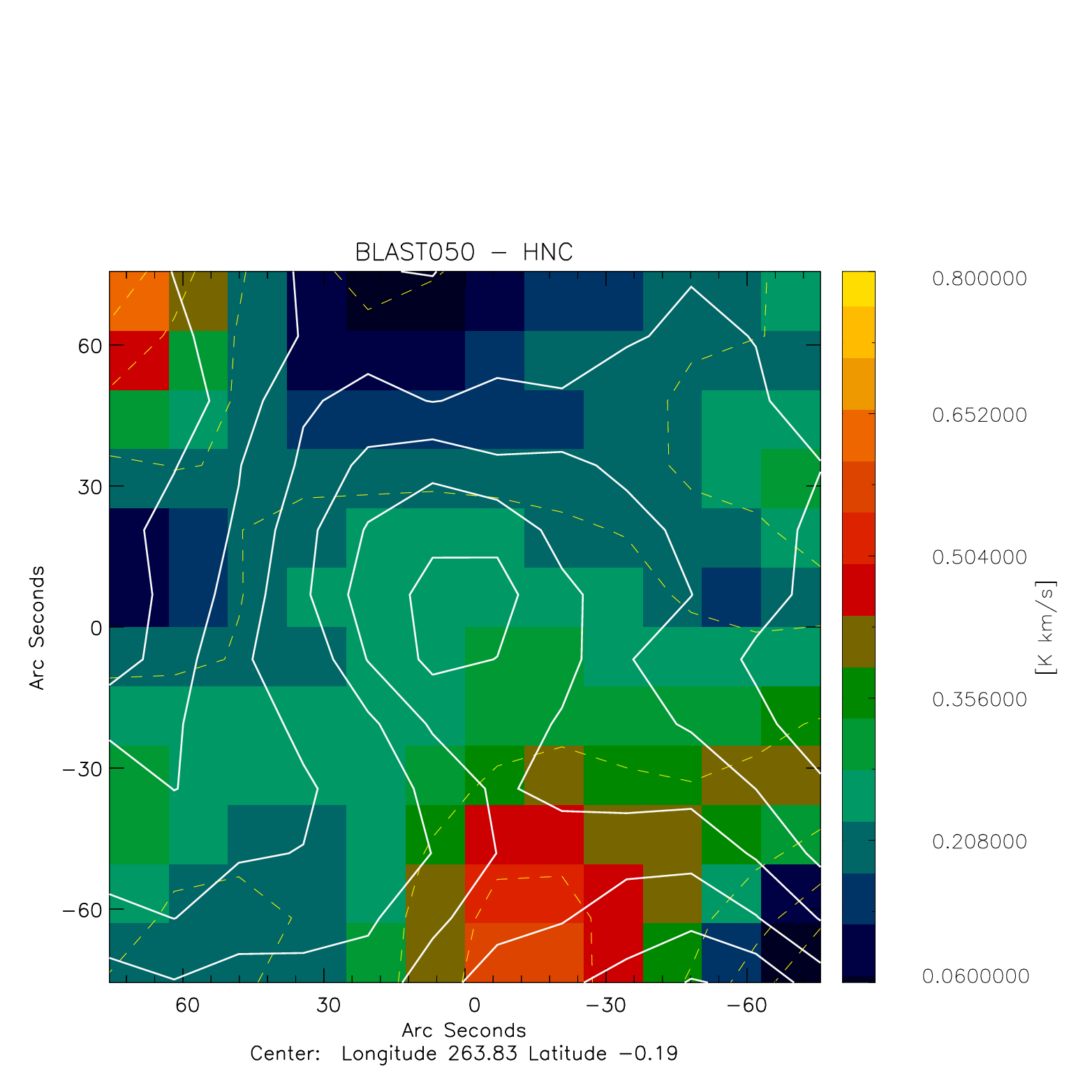} \\
\hspace*{4.4cm}
\includegraphics[width=4.4cm,angle=0]{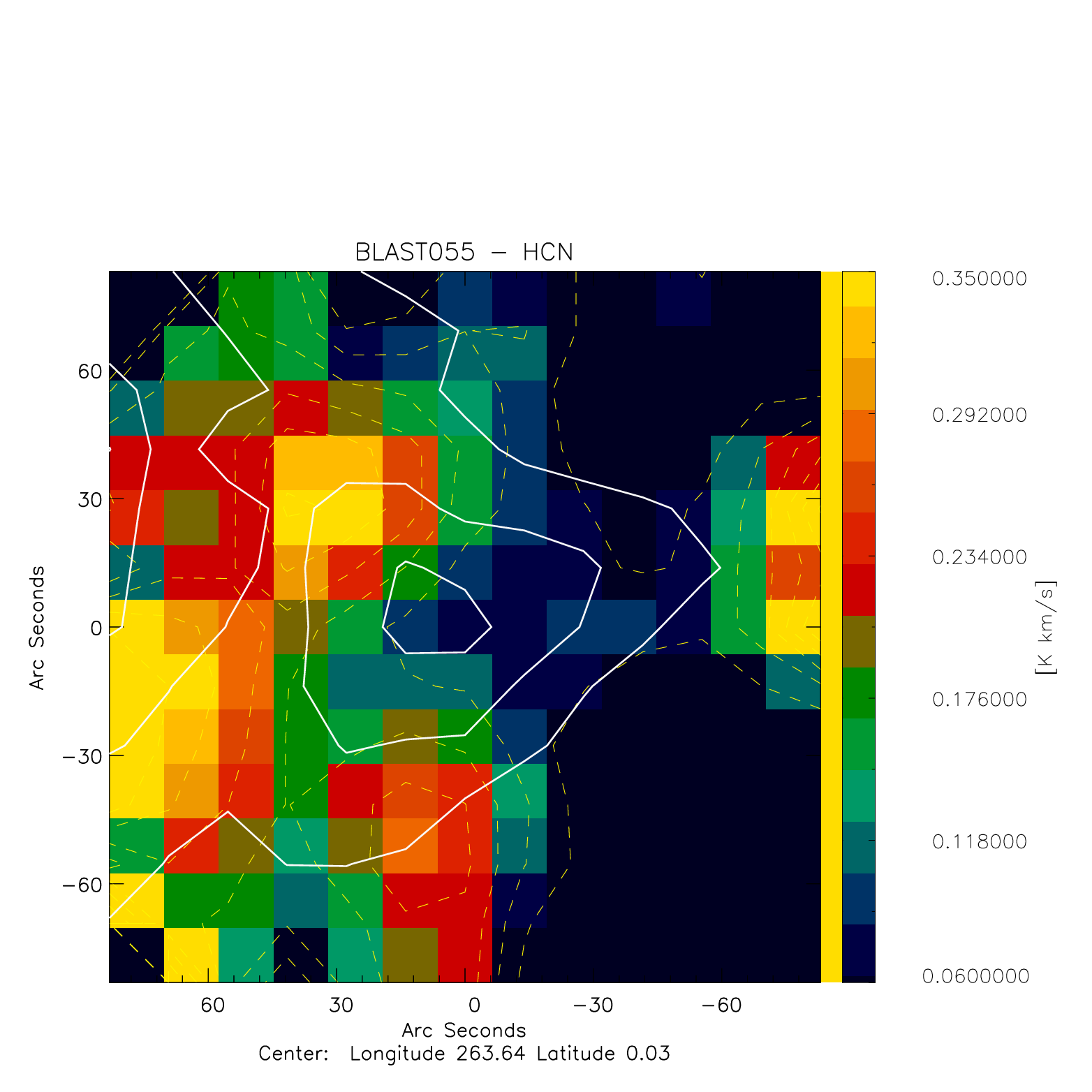}
\includegraphics[width=4.4cm,angle=0]{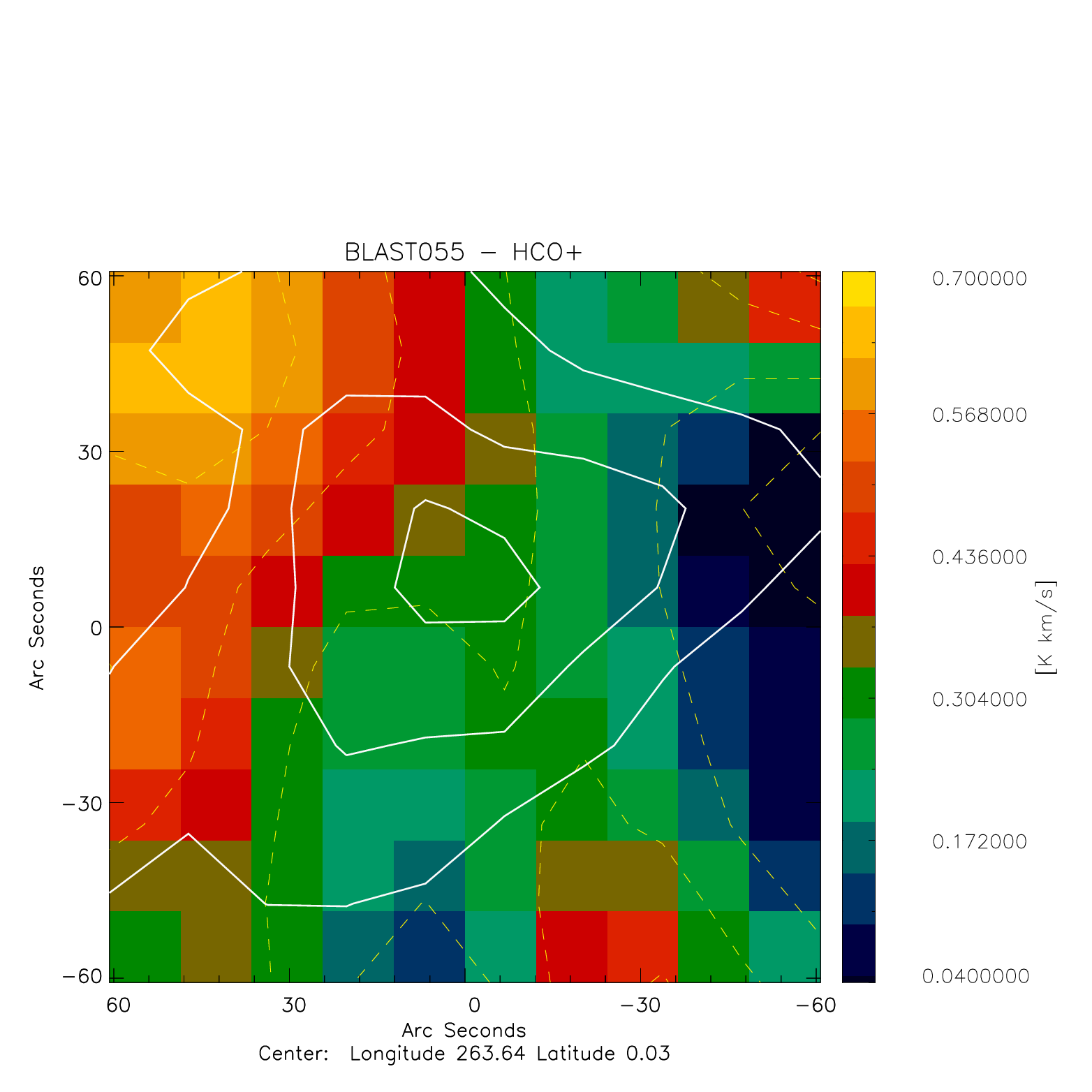}
\hspace*{4.4cm} \\
\hspace*{4.4cm}
\includegraphics[width=4.4cm,angle=0]{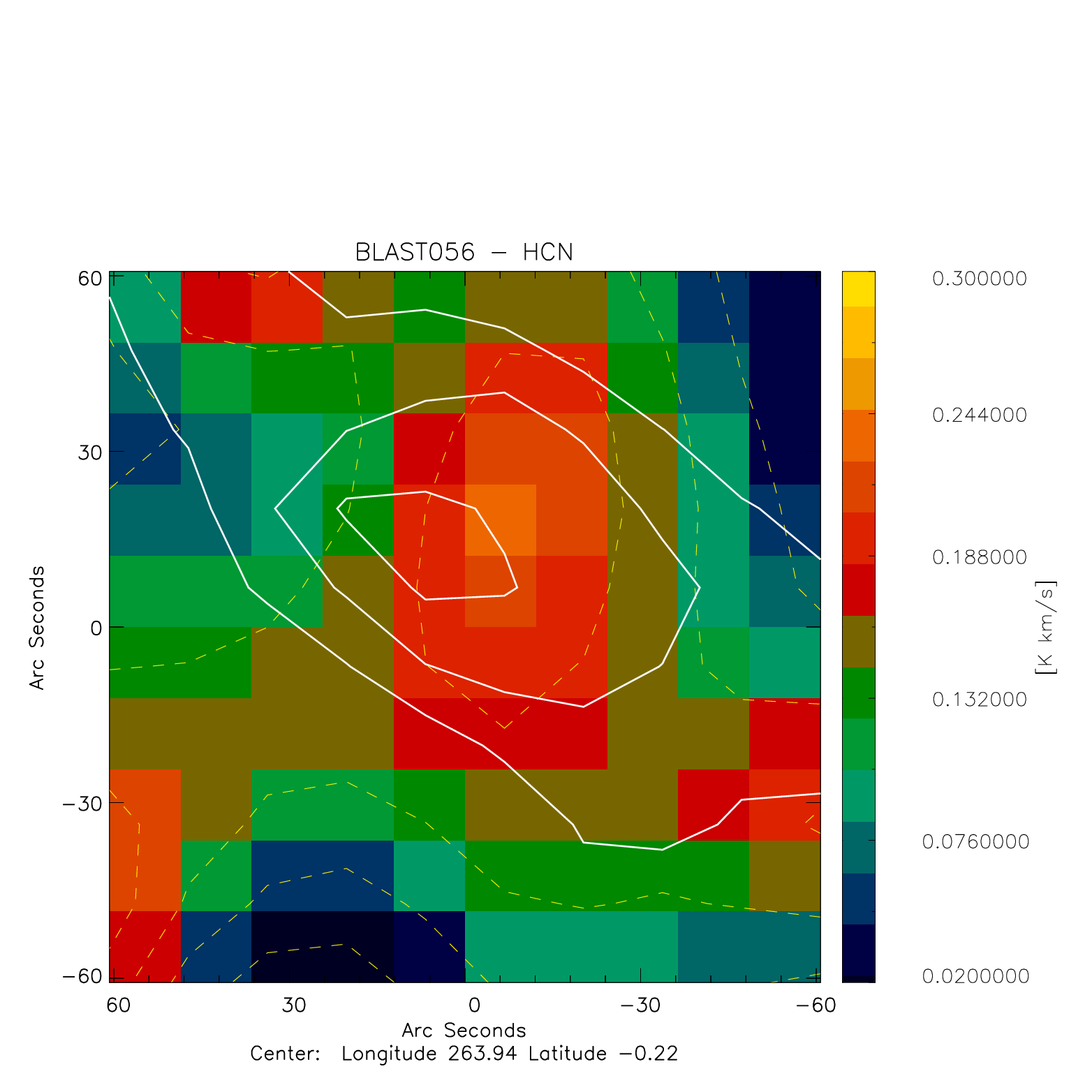}
\includegraphics[width=4.4cm,angle=0]{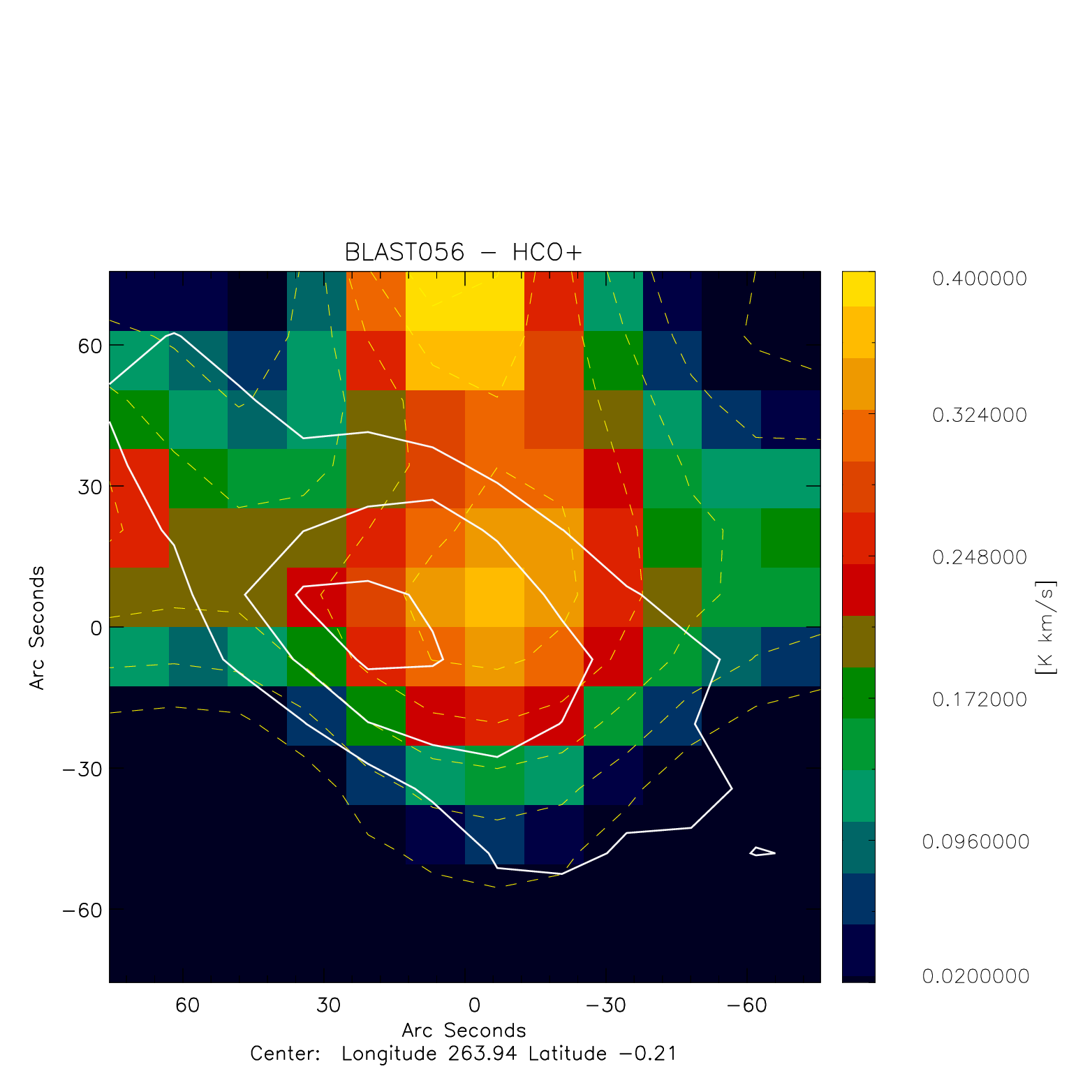}
\includegraphics[width=4.4cm,angle=0]{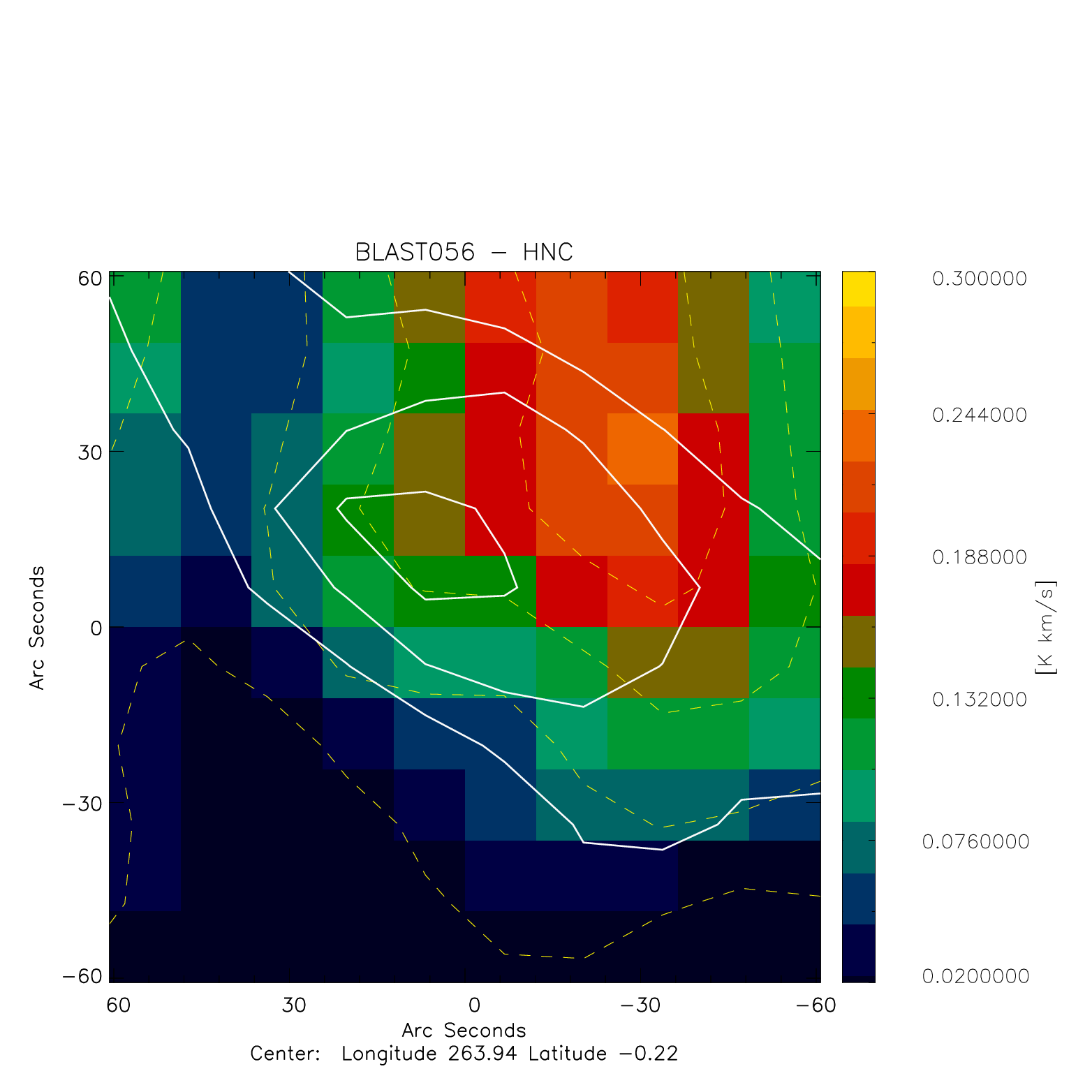} \\
\includegraphics[width=4.4cm,angle=0]{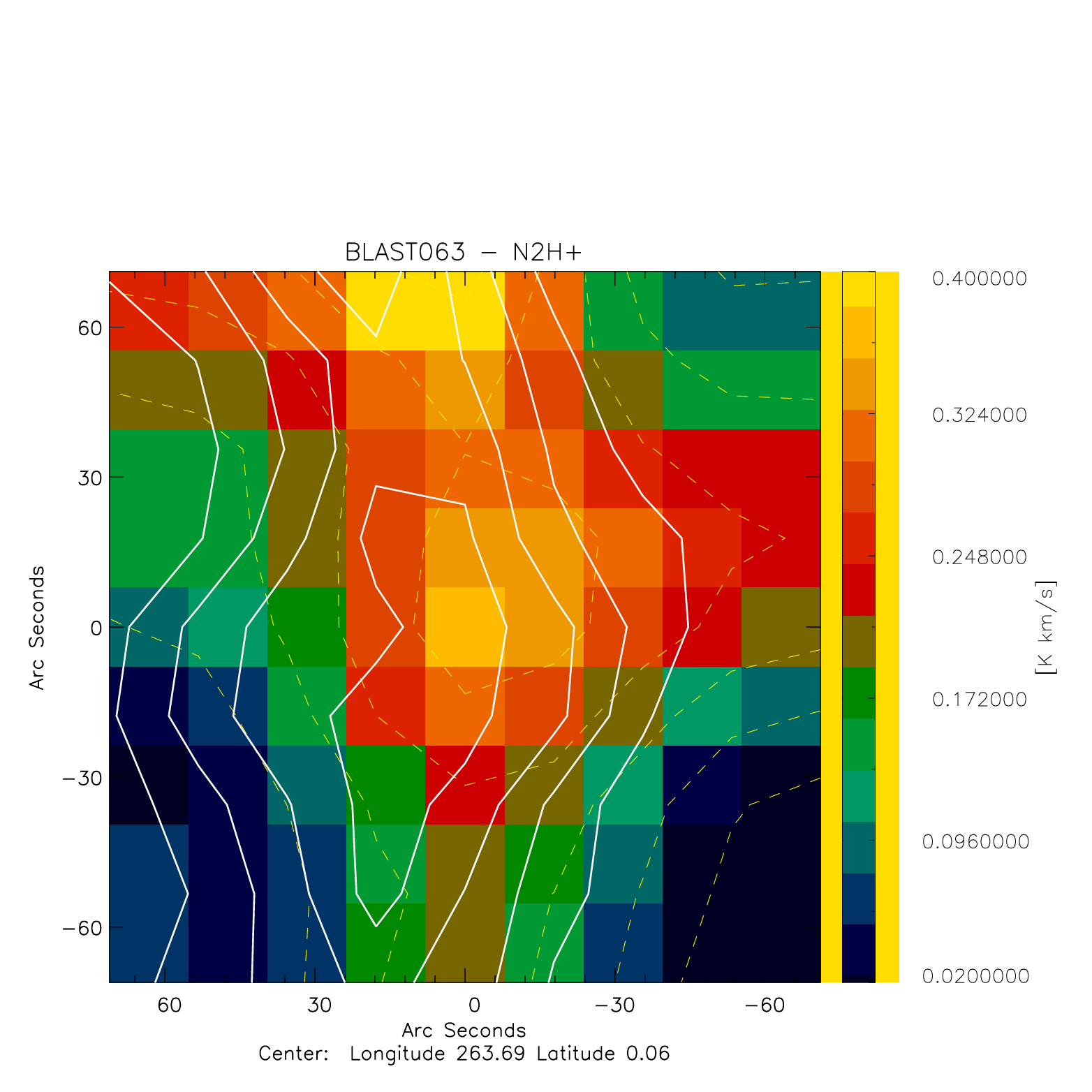}
\includegraphics[width=4.4cm,angle=0]{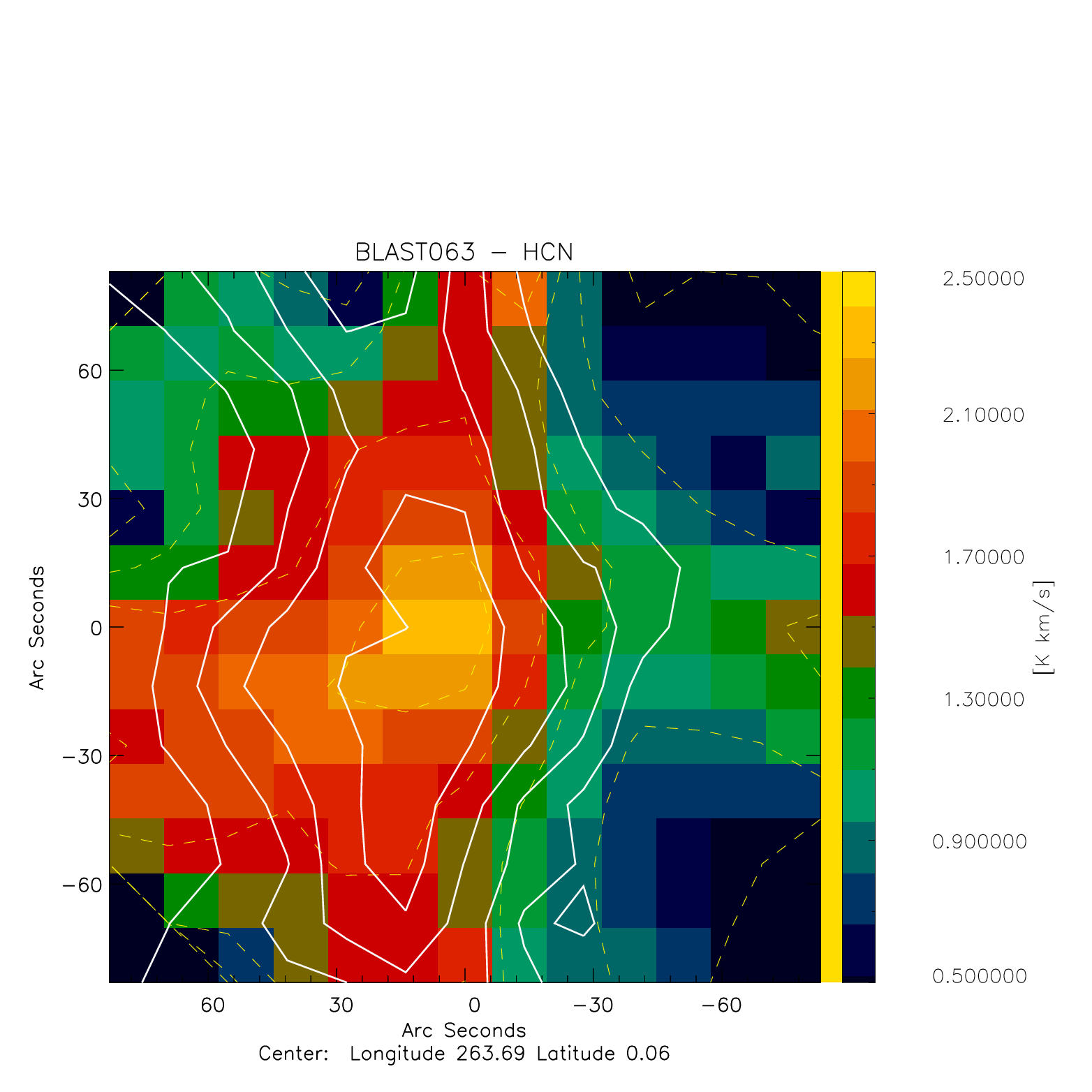}
\includegraphics[width=4.4cm,angle=0]{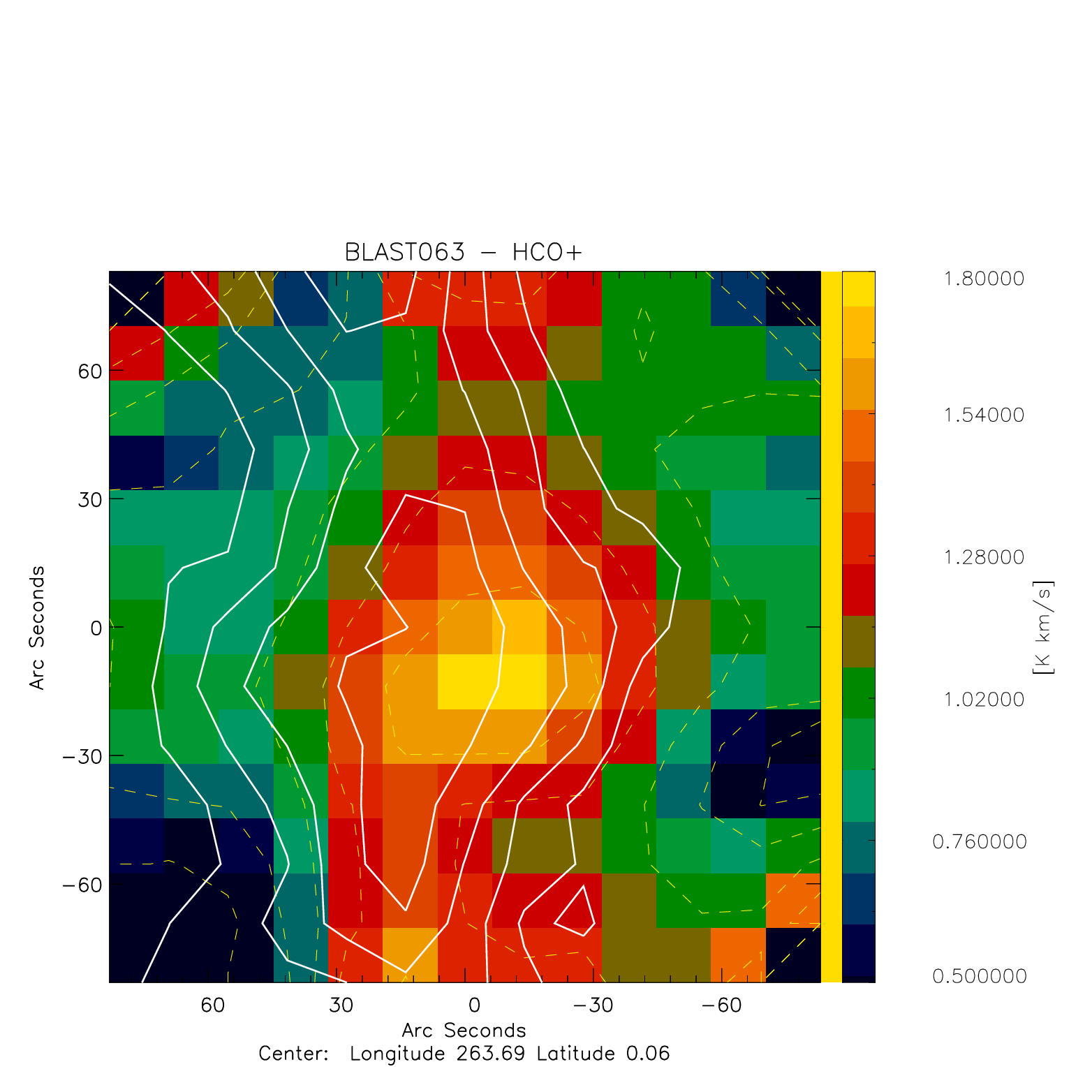}
\includegraphics[width=4.4cm,angle=0]{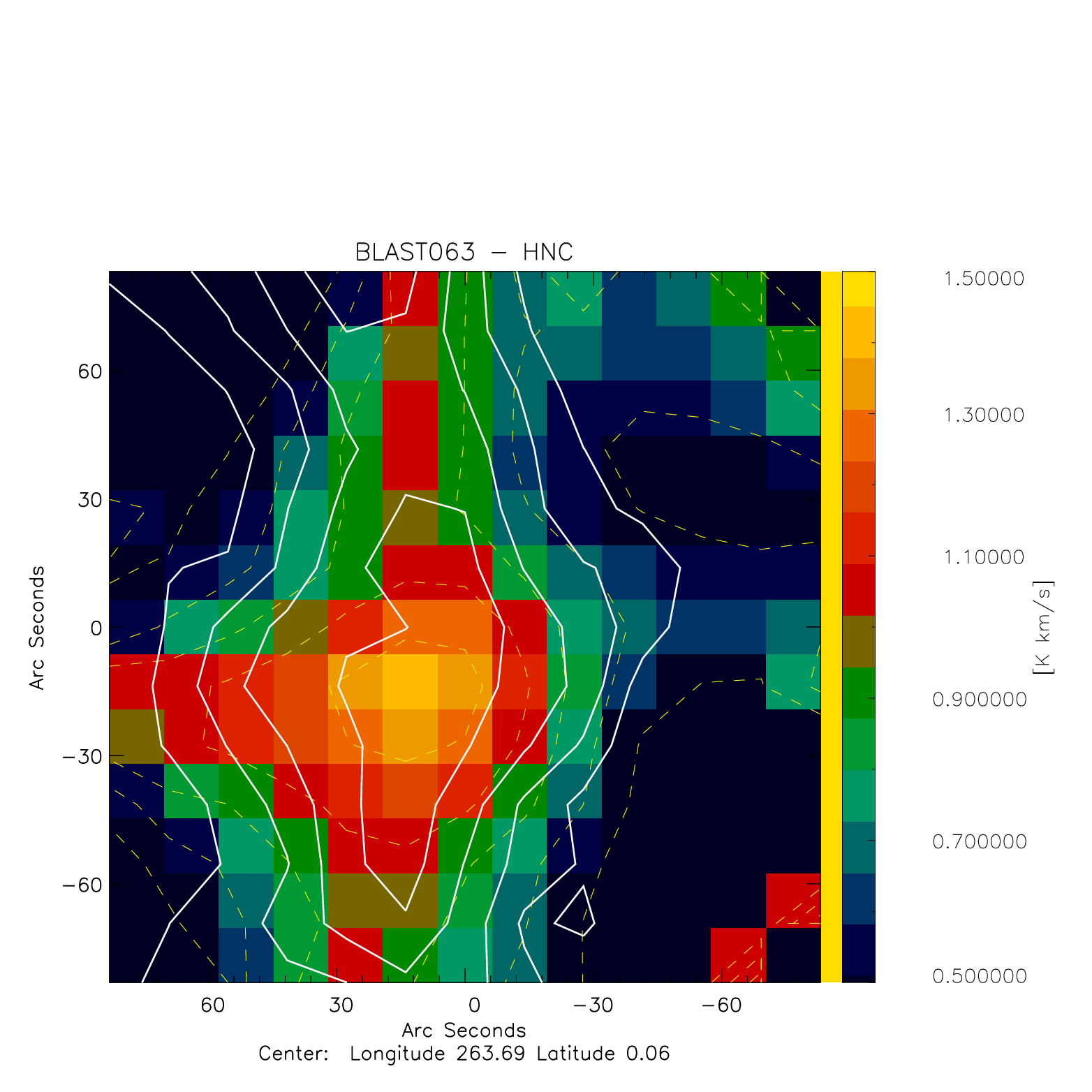} \\
\caption{
Same as Figure~\ref{fig:maps1} 
for sources BLAST050, BLAST055, BLAST056 and BLAST063.
  }
\label{fig:maps3}
\end{figure*}

\clearpage 

%
\begin{figure*}
\includegraphics[width=4.4cm,angle=0]{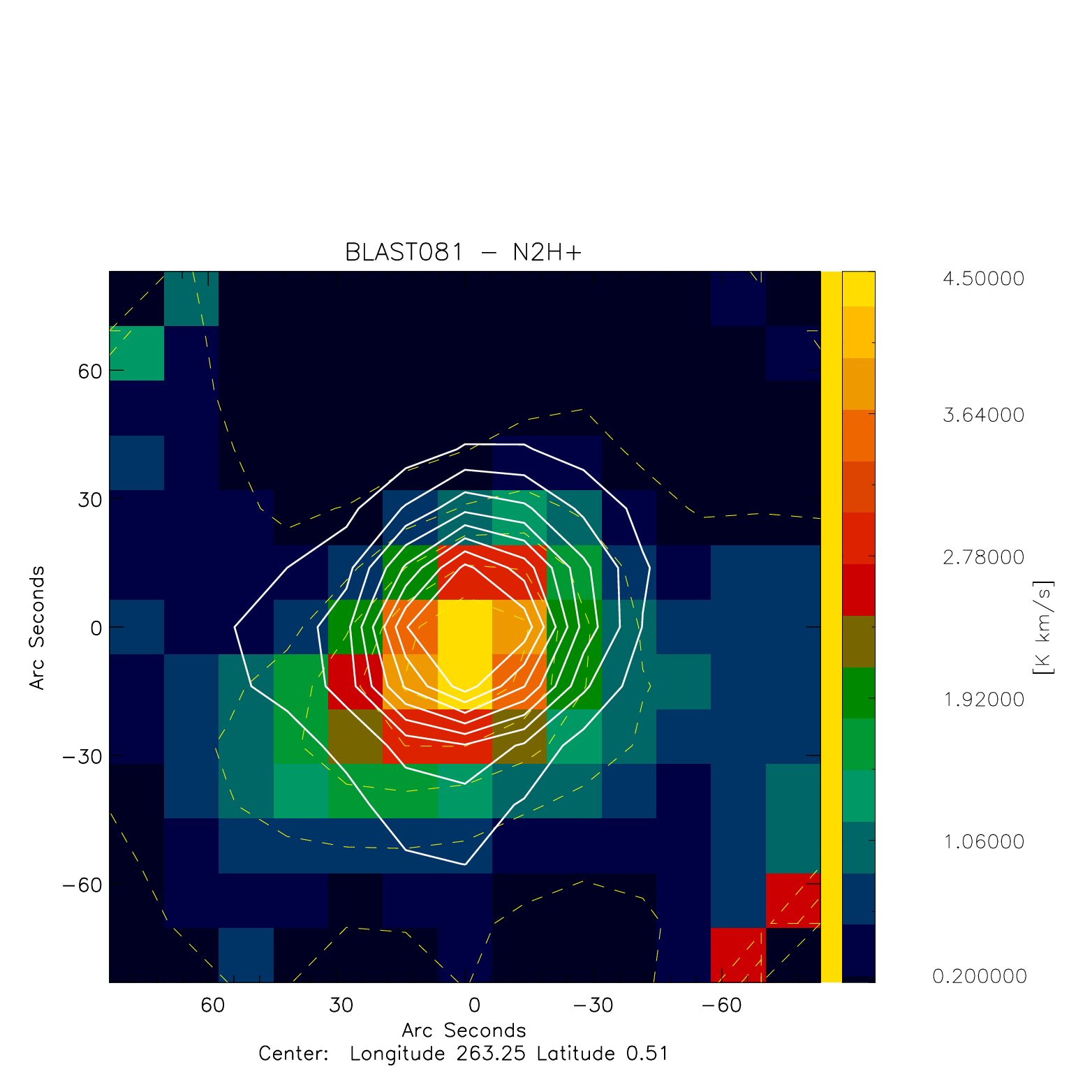}
\includegraphics[width=4.4cm,angle=0]{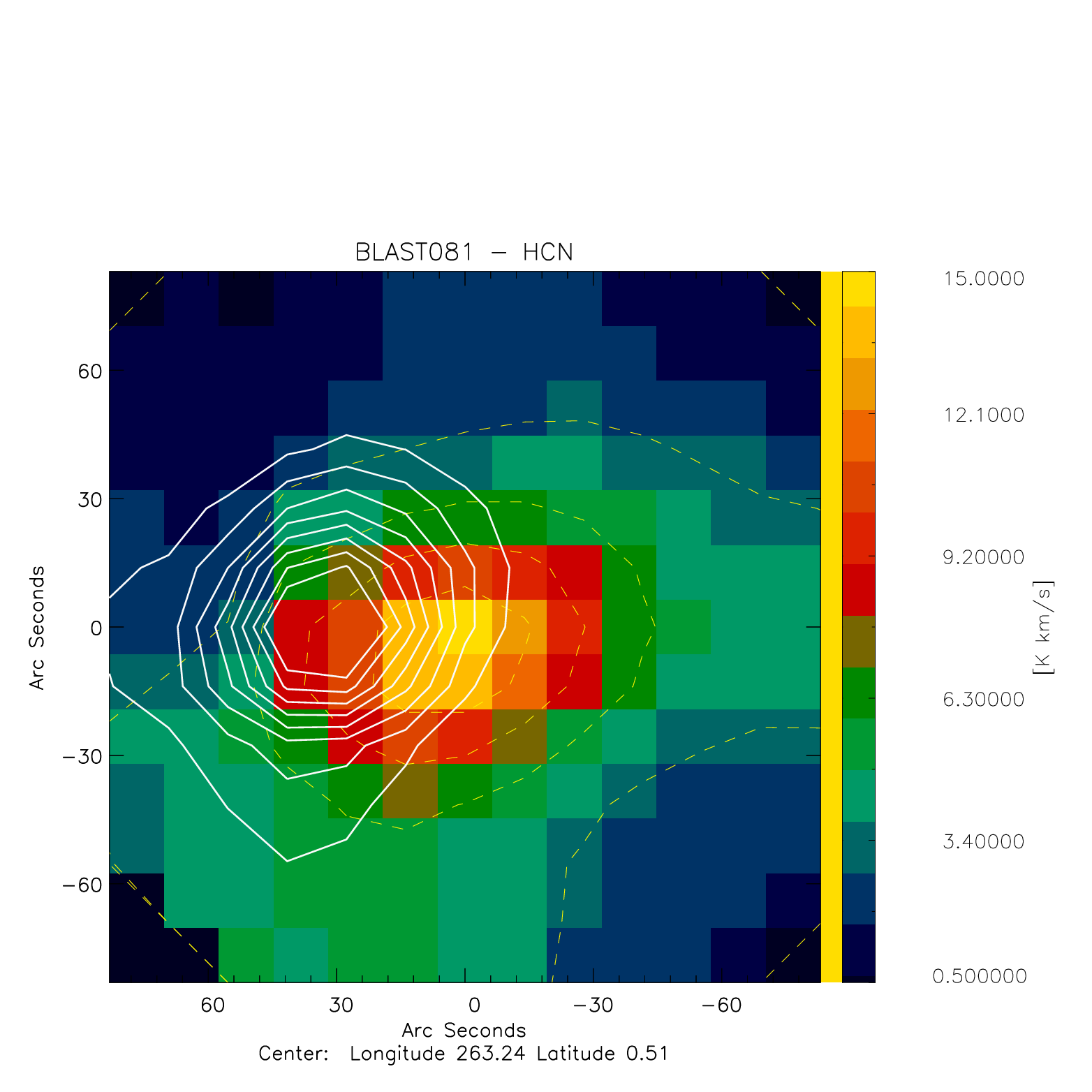}
\includegraphics[width=4.4cm,angle=0]{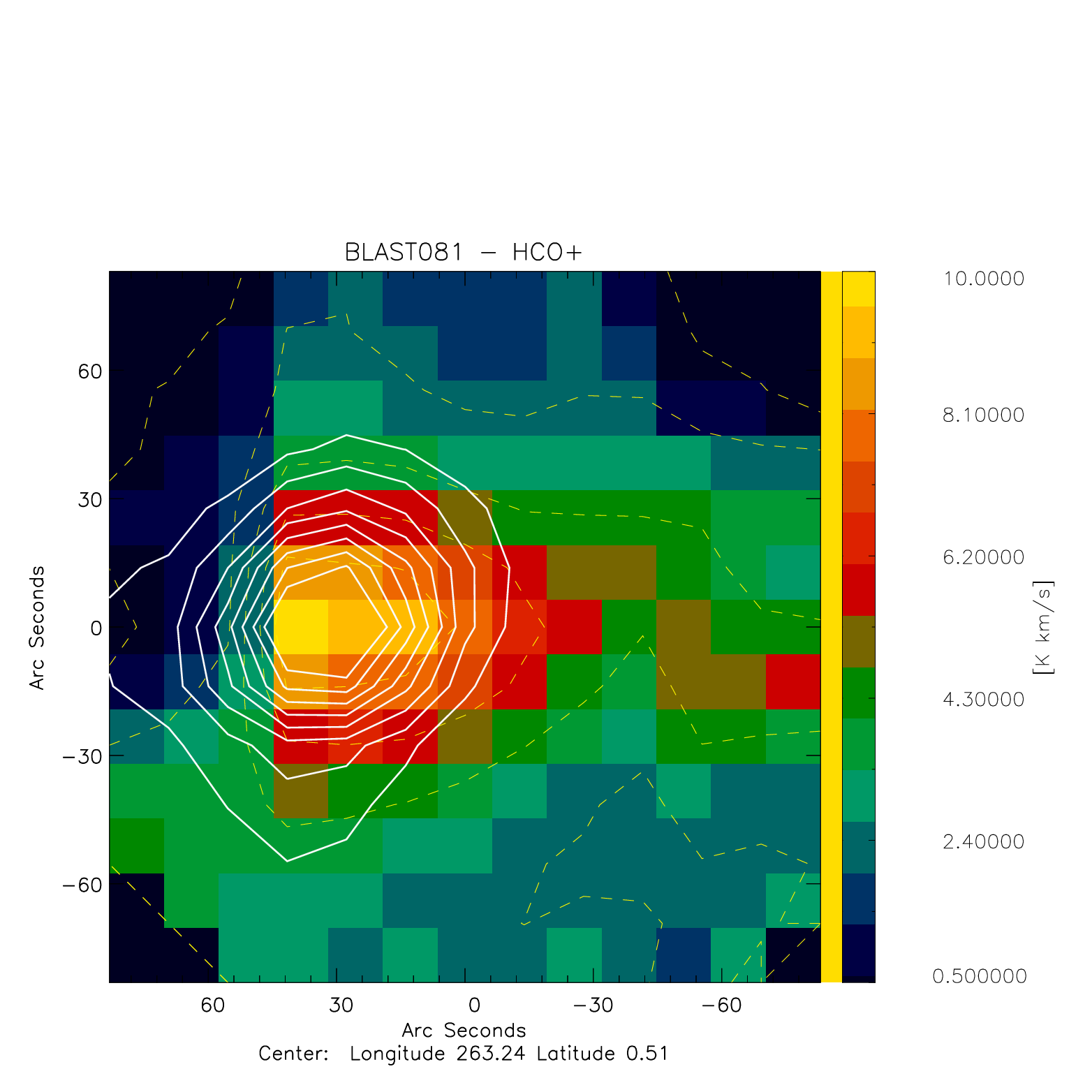}
\includegraphics[width=4.4cm,angle=0]{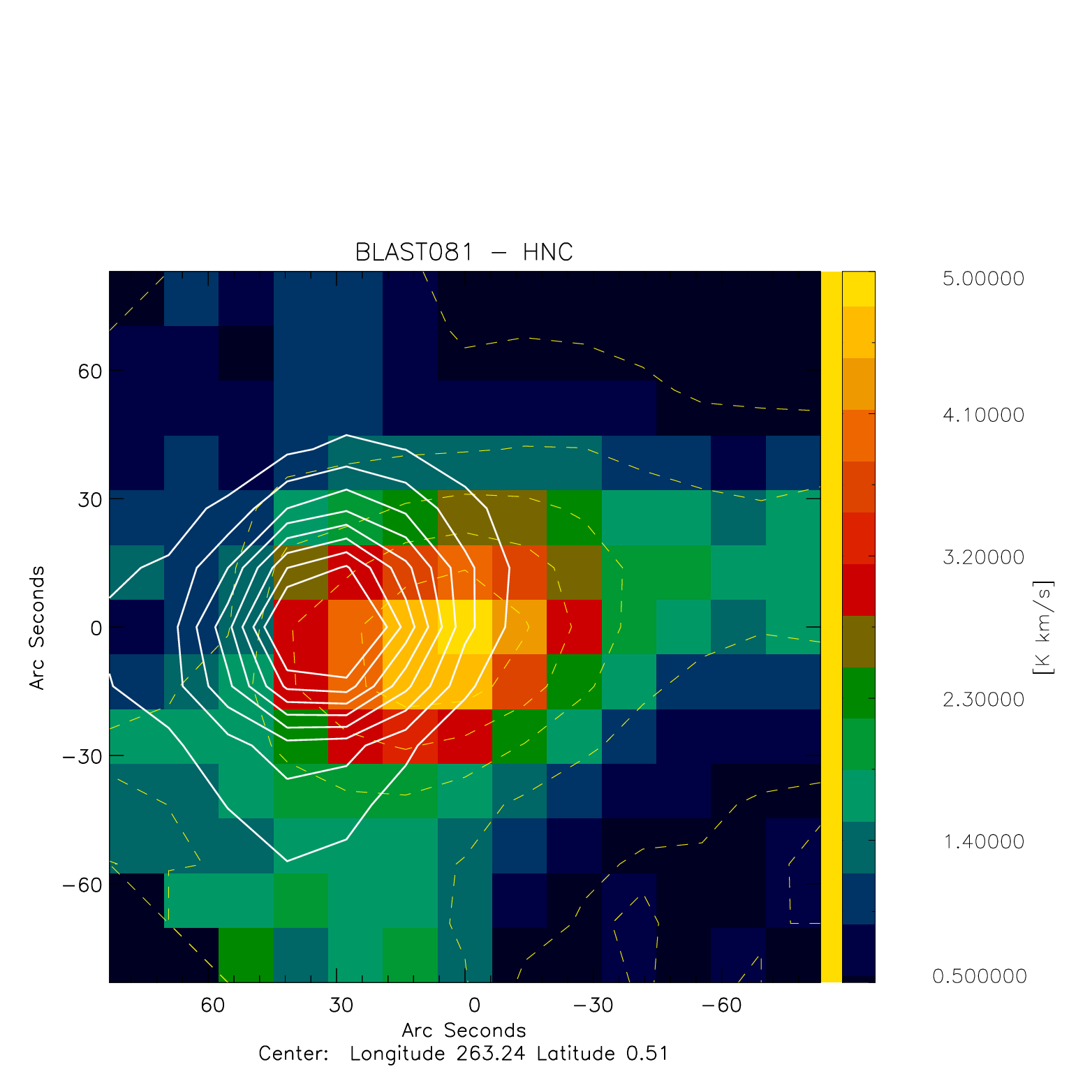} \\
\includegraphics[width=4.4cm,angle=0]{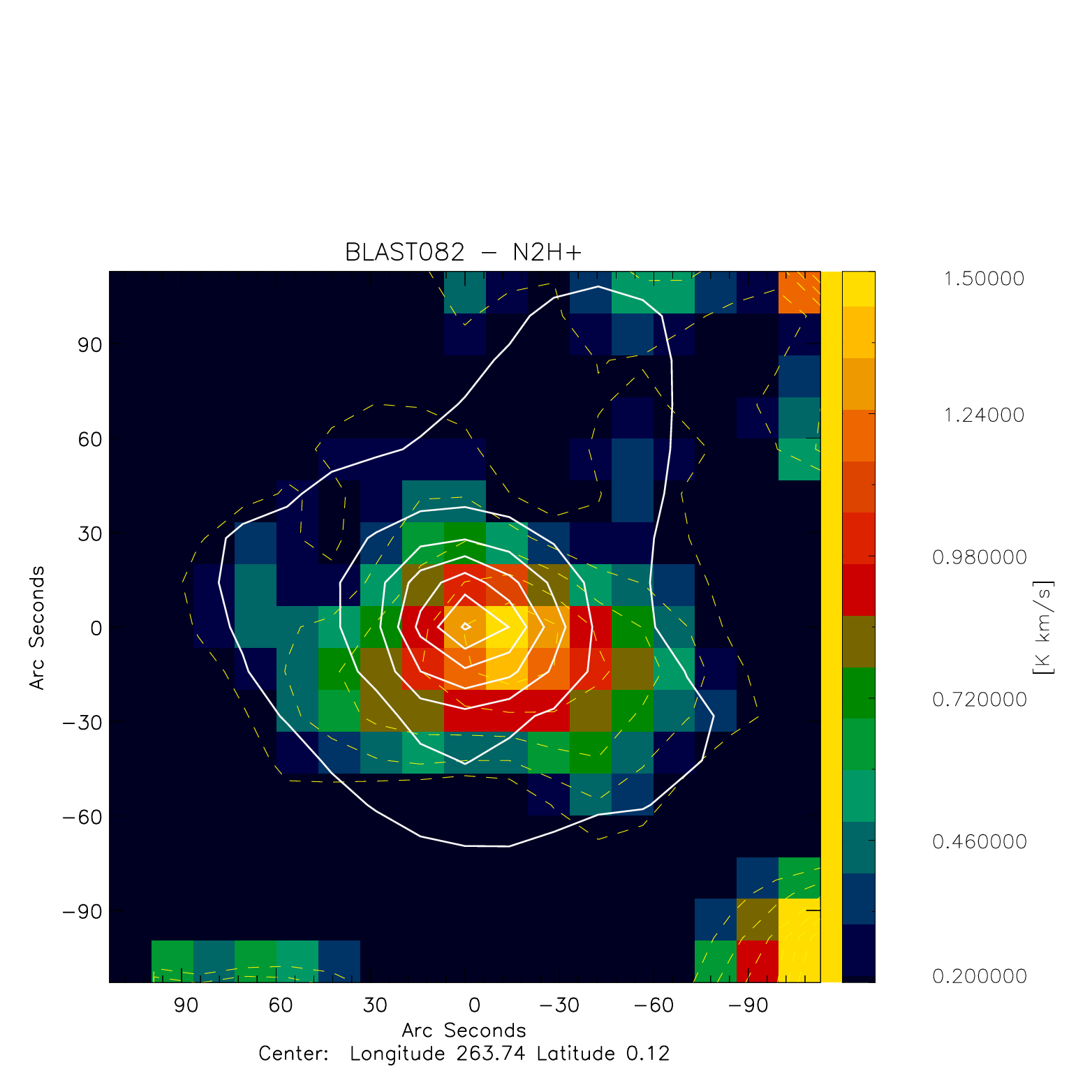}
\includegraphics[width=4.4cm,angle=0]{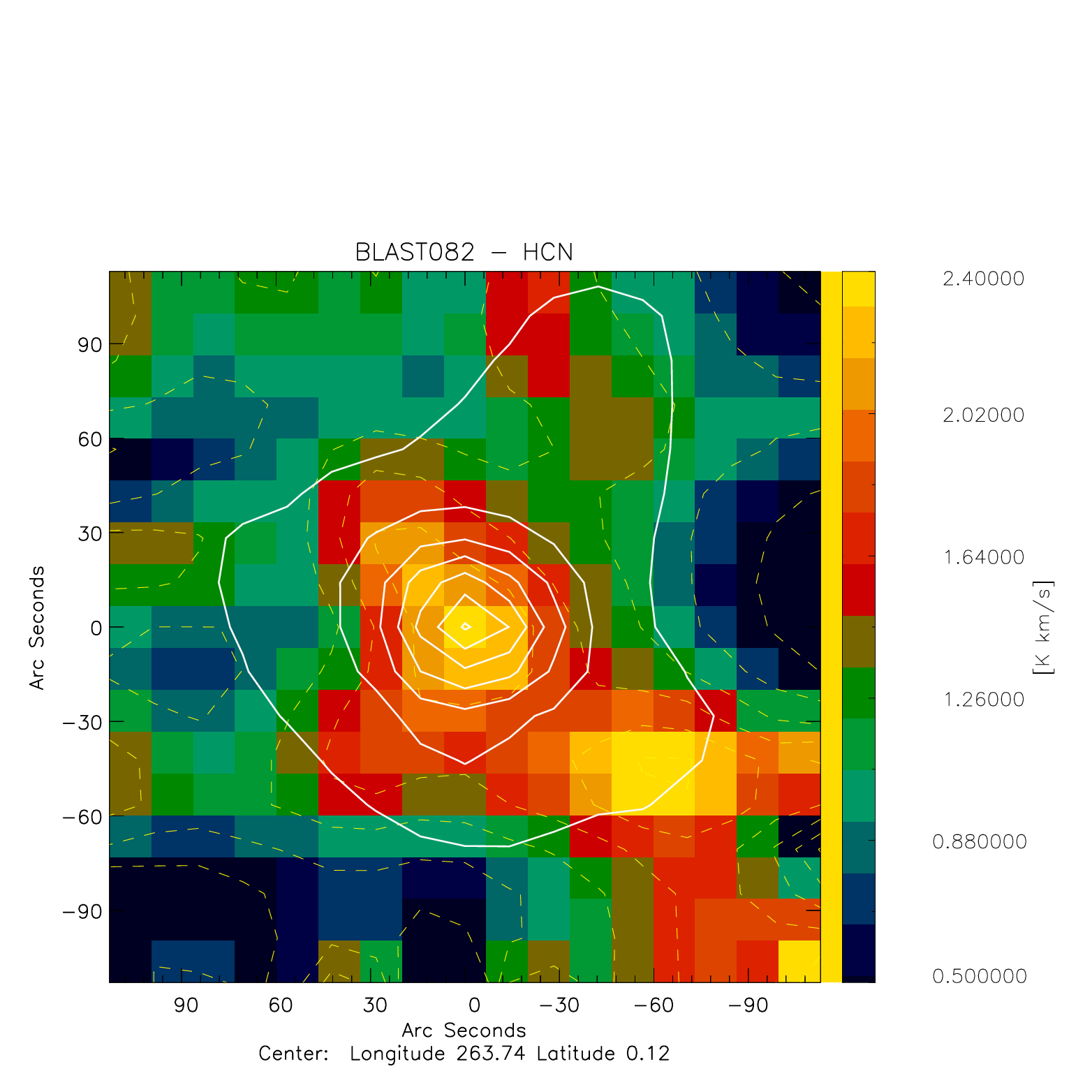}
\includegraphics[width=4.4cm,angle=0]{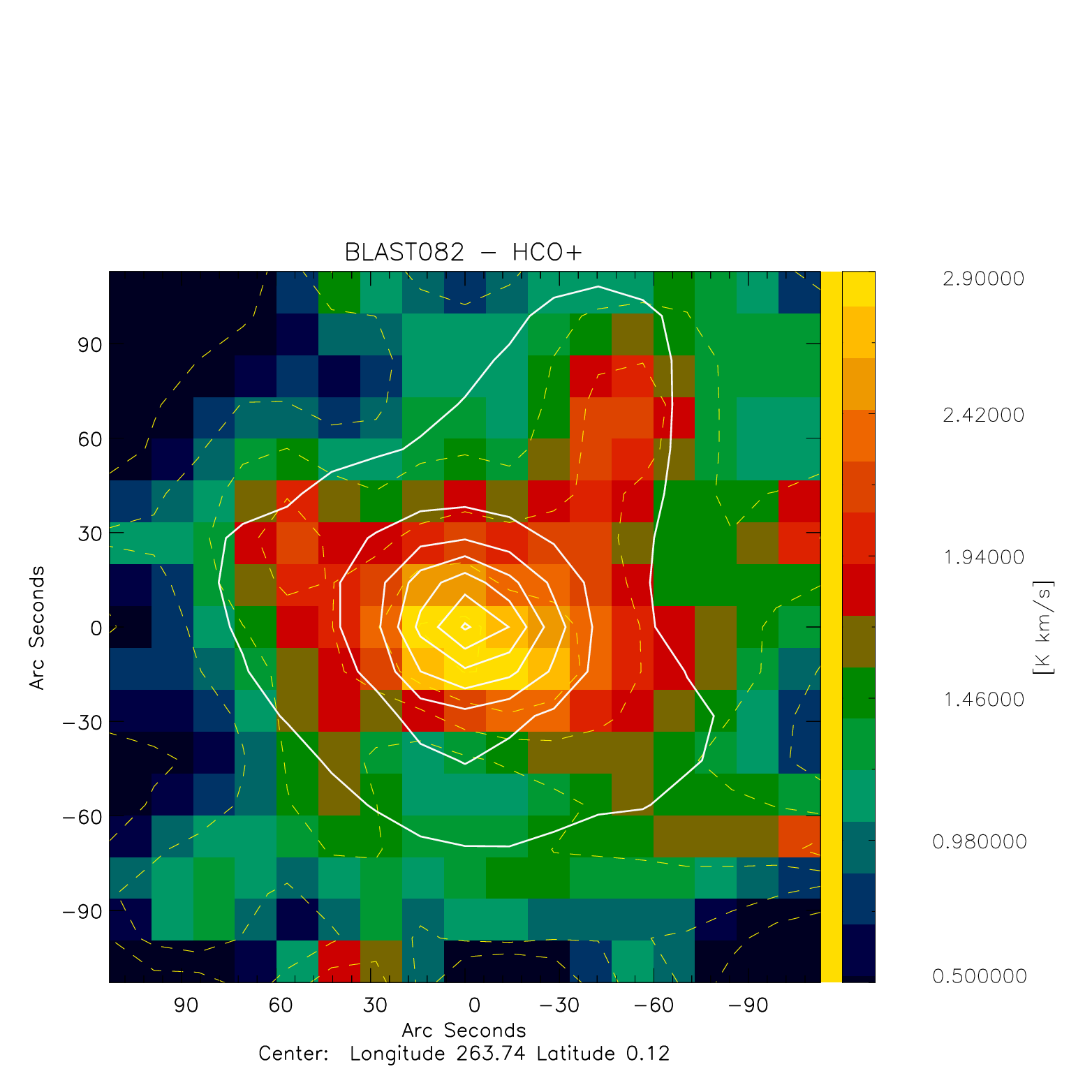}
\includegraphics[width=4.4cm,angle=0]{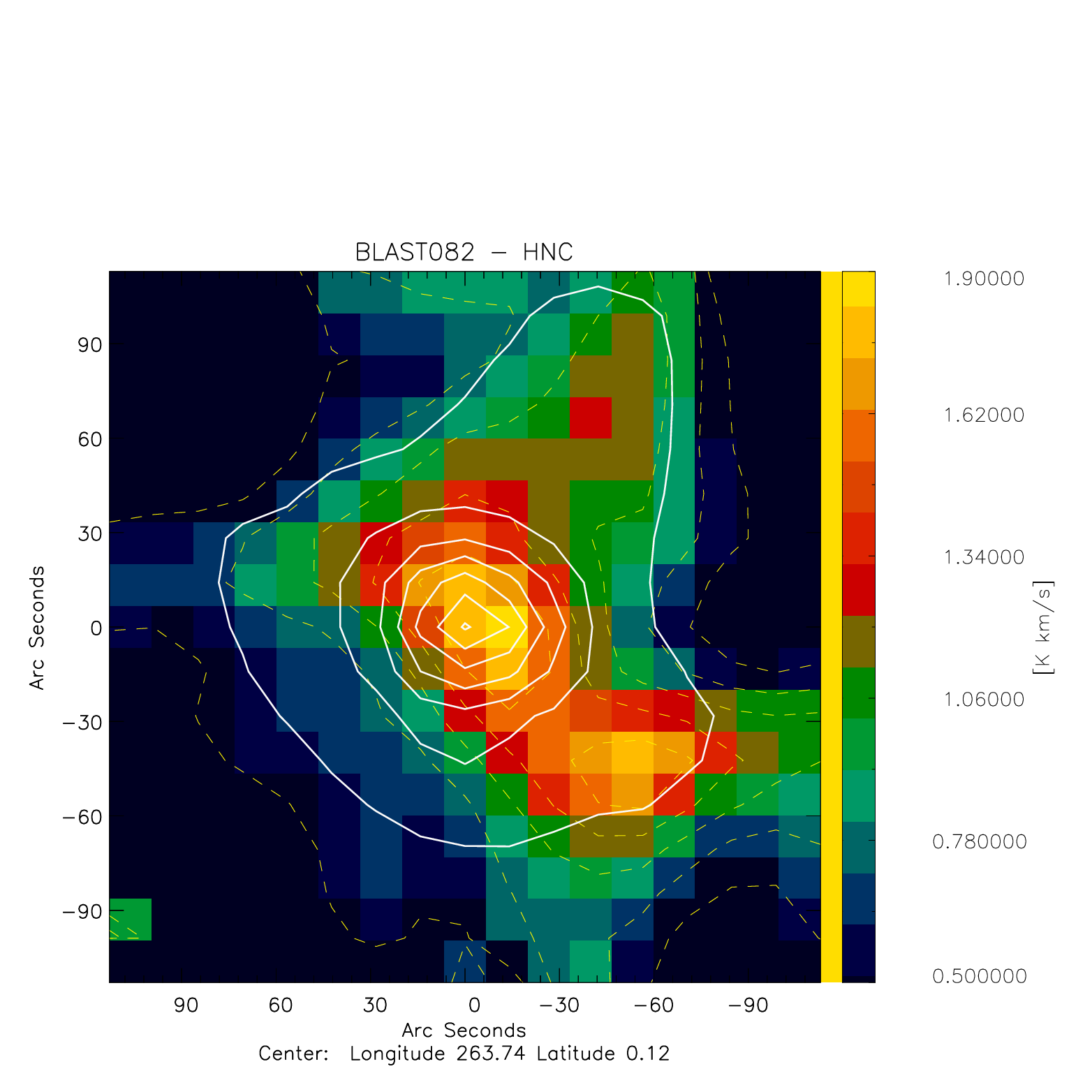} \\
\hspace*{4.4cm}
\includegraphics[width=4.4cm,angle=0]{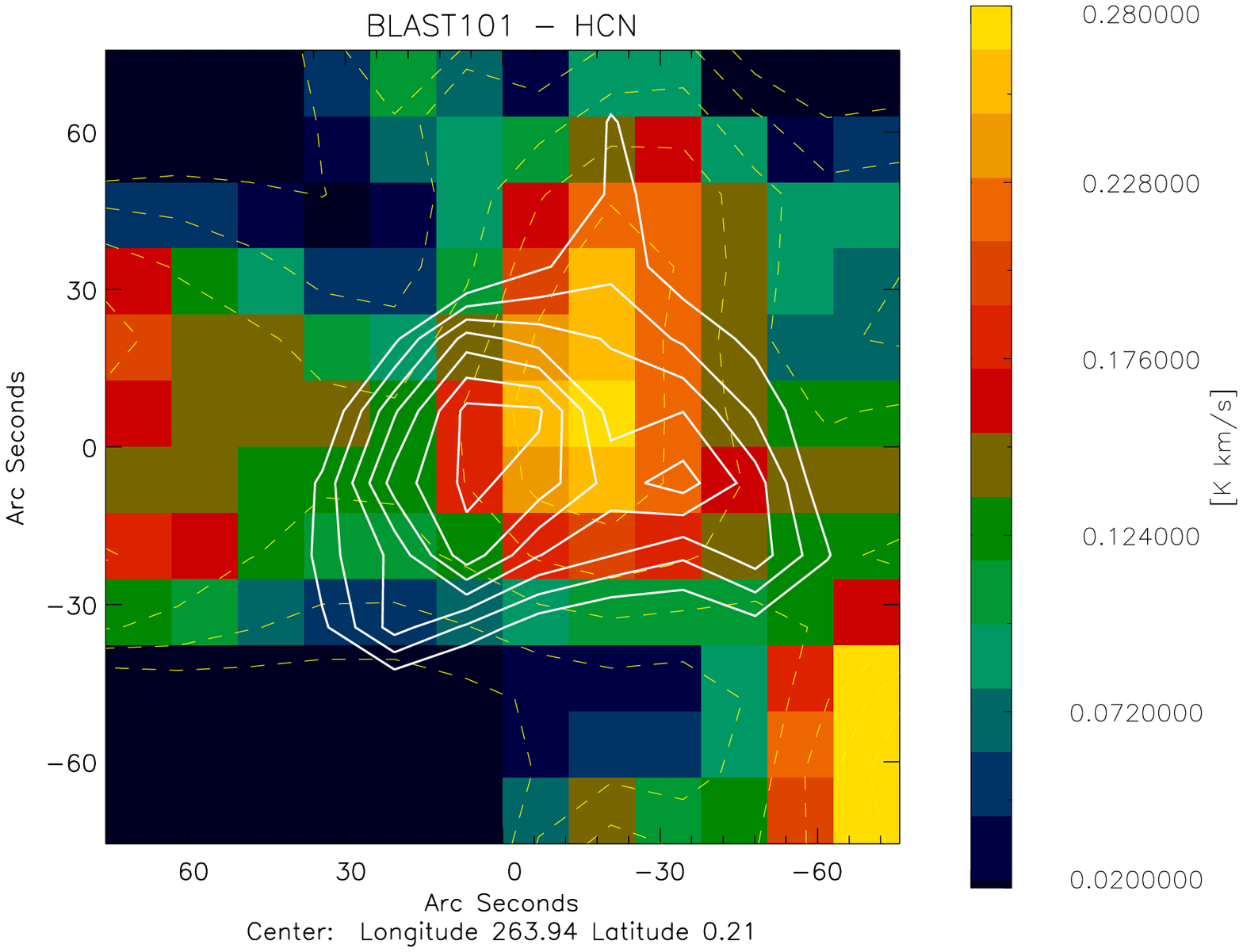}
\includegraphics[width=4.4cm,angle=0]{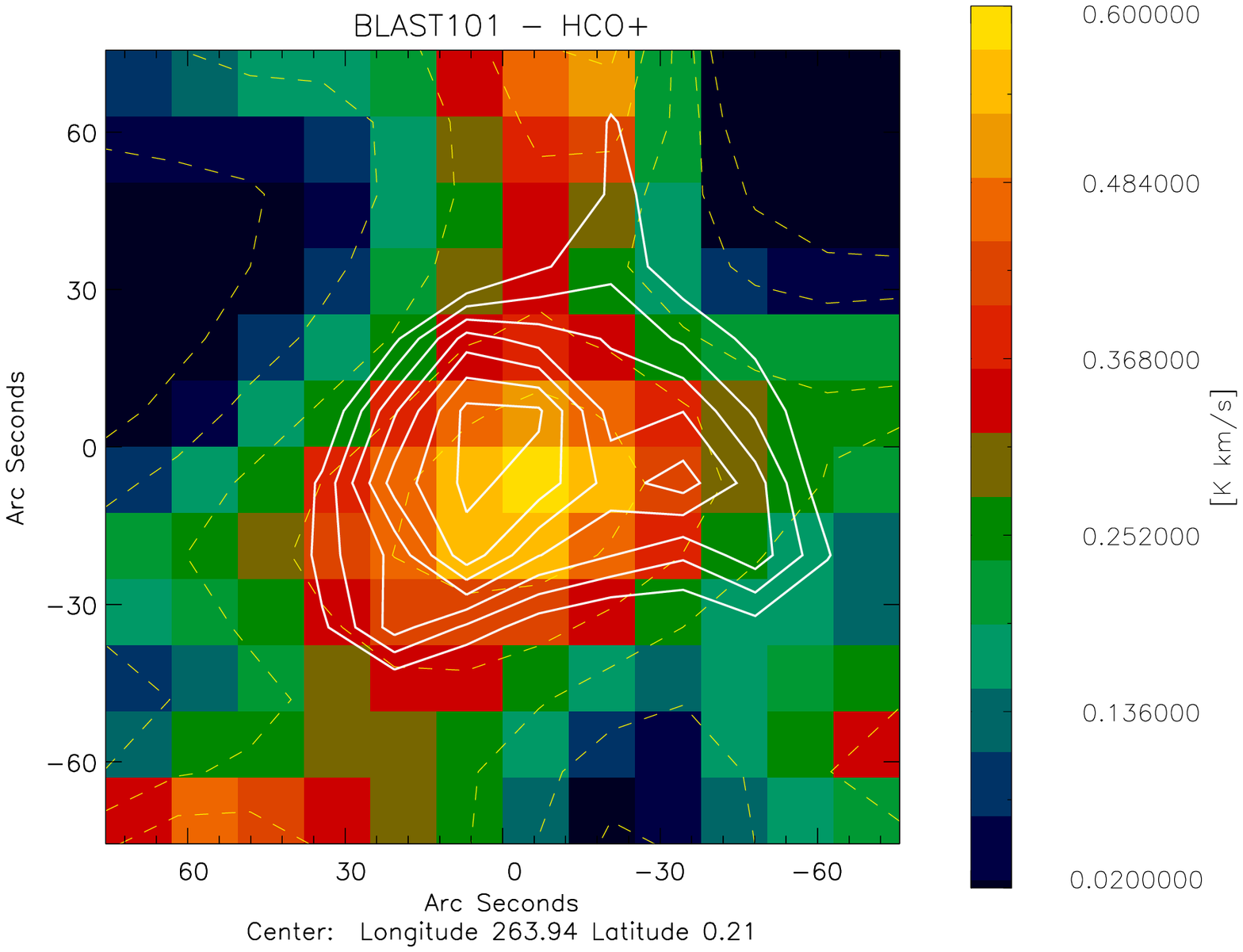}
\includegraphics[width=4.4cm,angle=0]{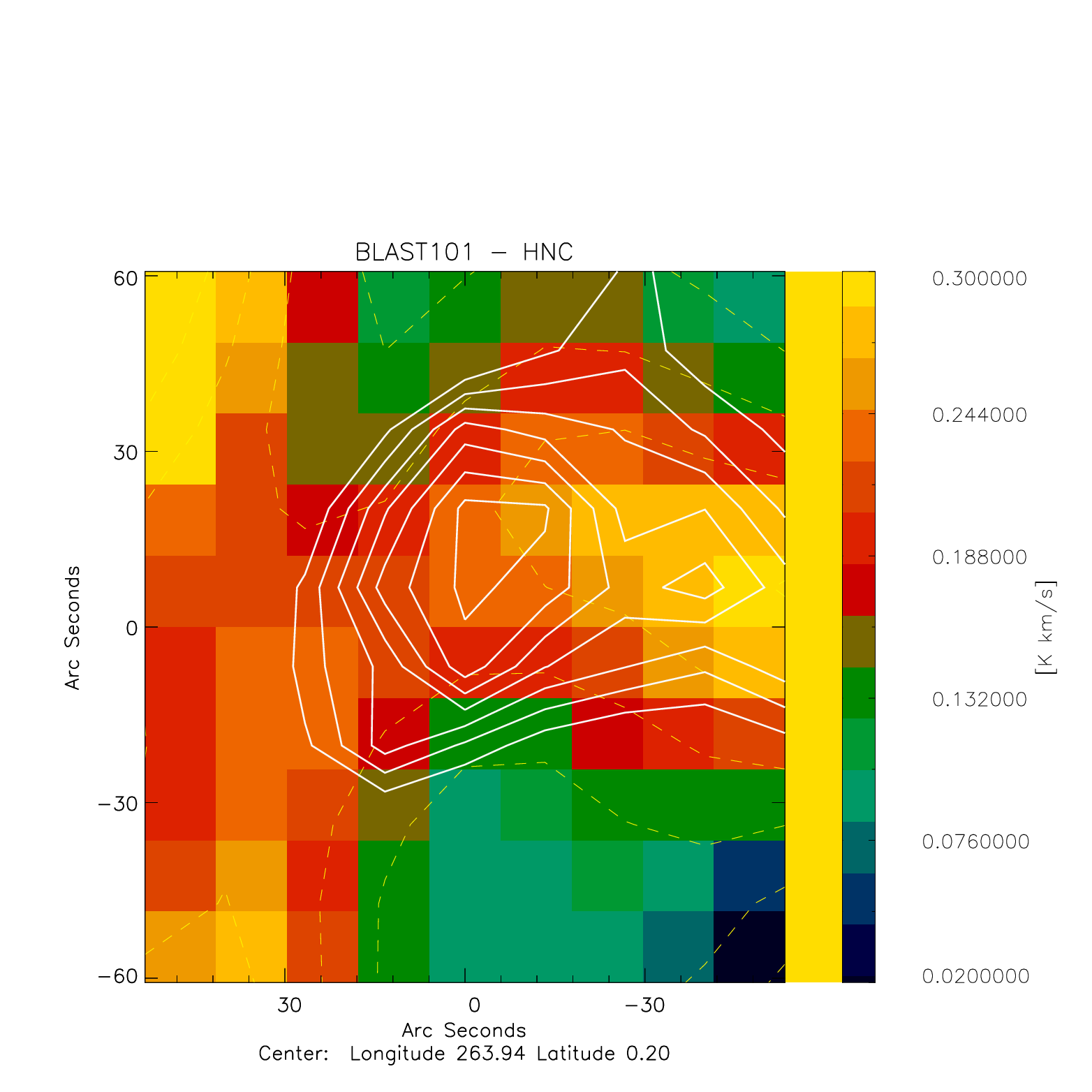} \\
\includegraphics[width=4.4cm,angle=0]{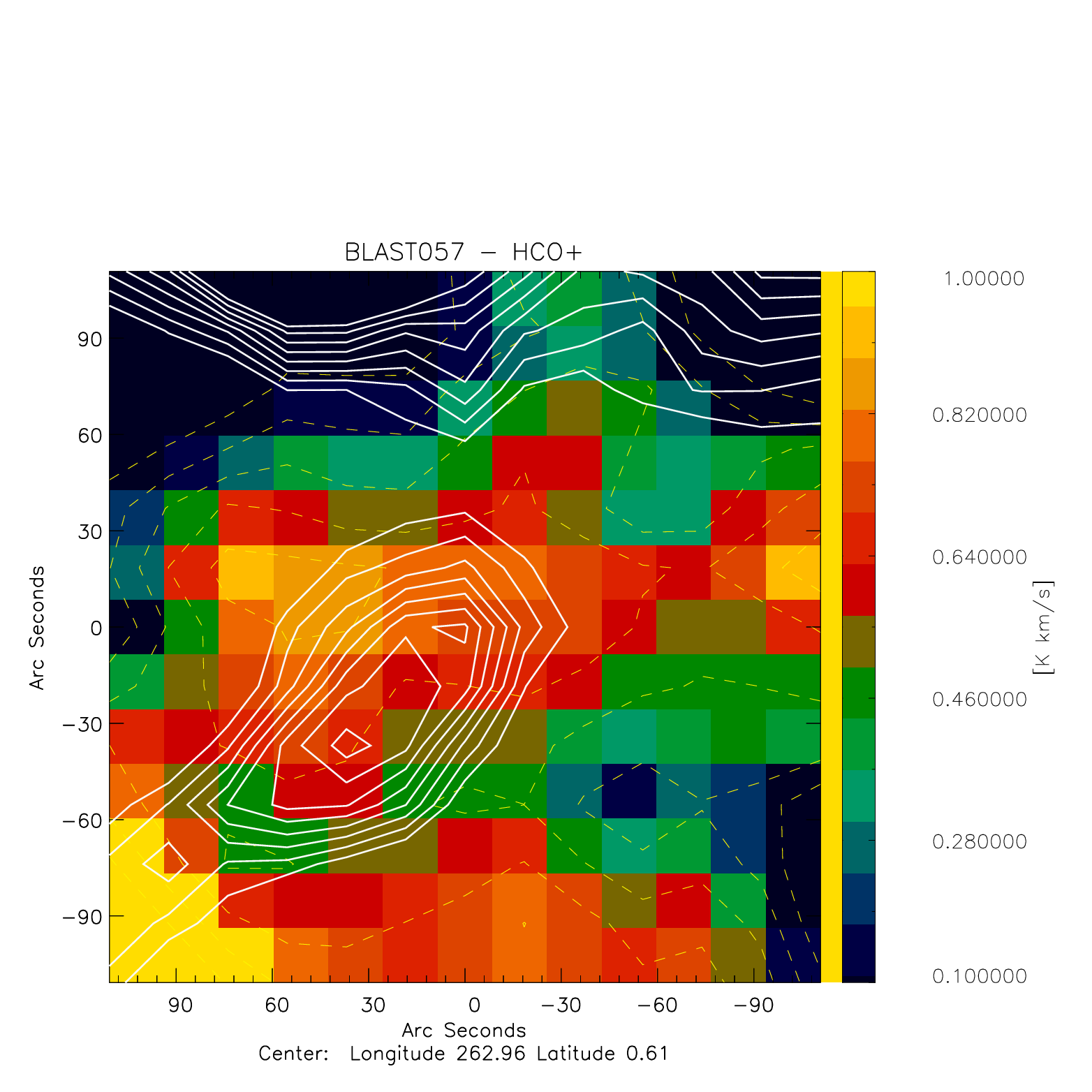}
\includegraphics[width=4.4cm,angle=0]{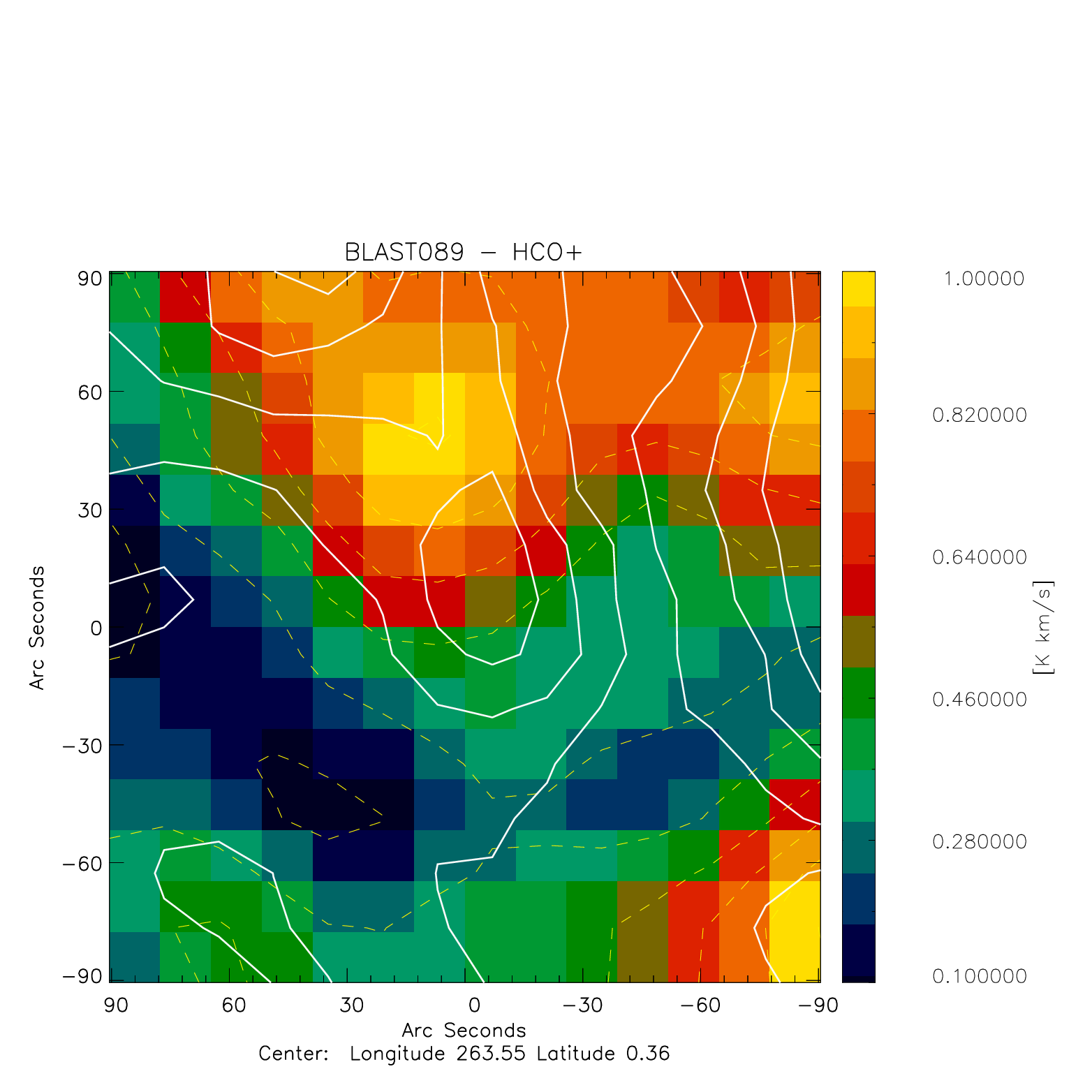}
\includegraphics[width=4.4cm,angle=0]{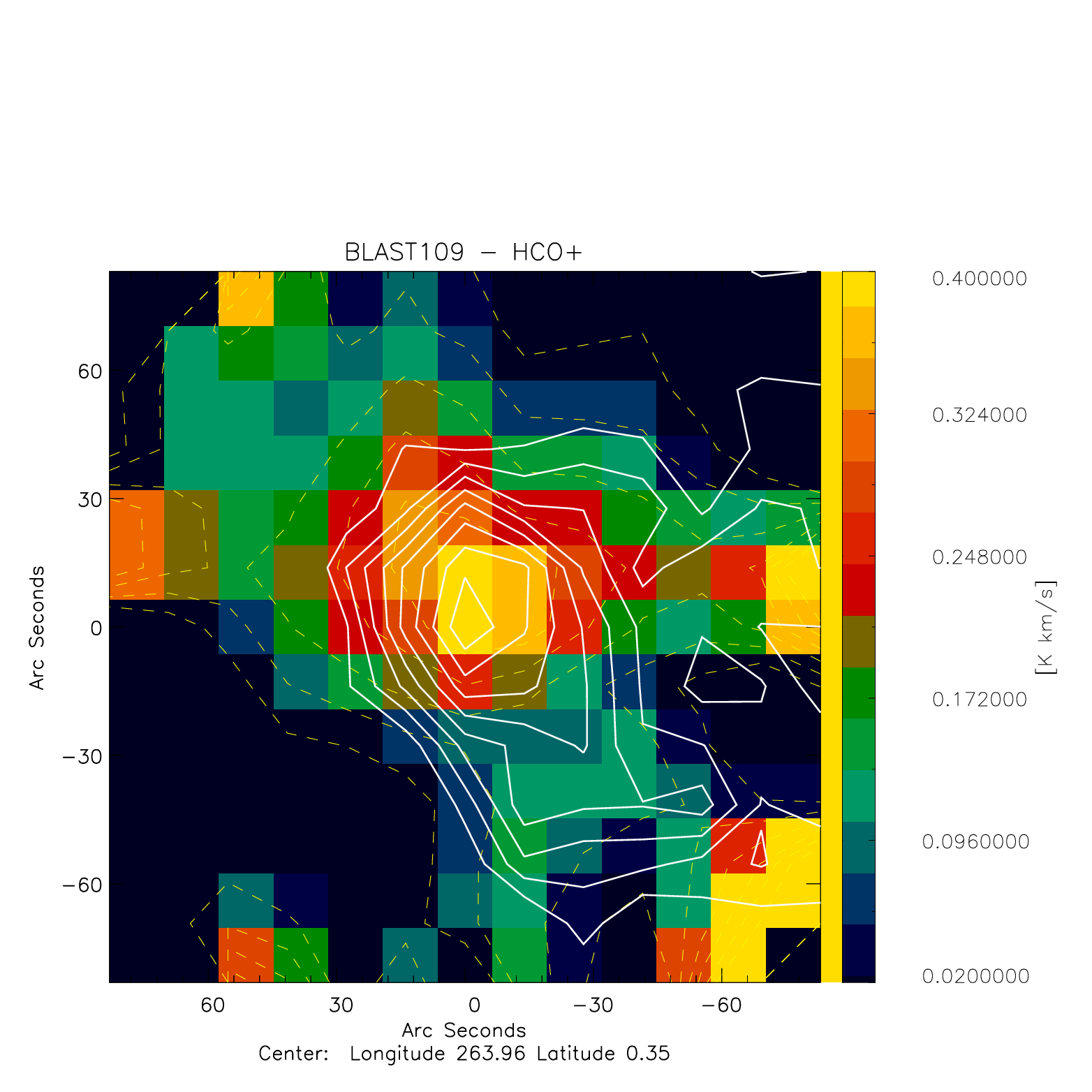} \\
\caption{
Same as Figure~\ref{fig:maps1} 
for sources BLAST081, BLAST082 and
BLAST101. The bottom row shows the HCO$^+(1-0)$ integrated intensity
maps of sources BLAST057, BLAST089 and BLAST109.
  }
\label{fig:maps4}
\end{figure*}

\clearpage

\section{Molecular line spectra}  
\label{sec:spec}


%
\begin{figure*}[h]
\includegraphics[width=10.0cm,angle=90]{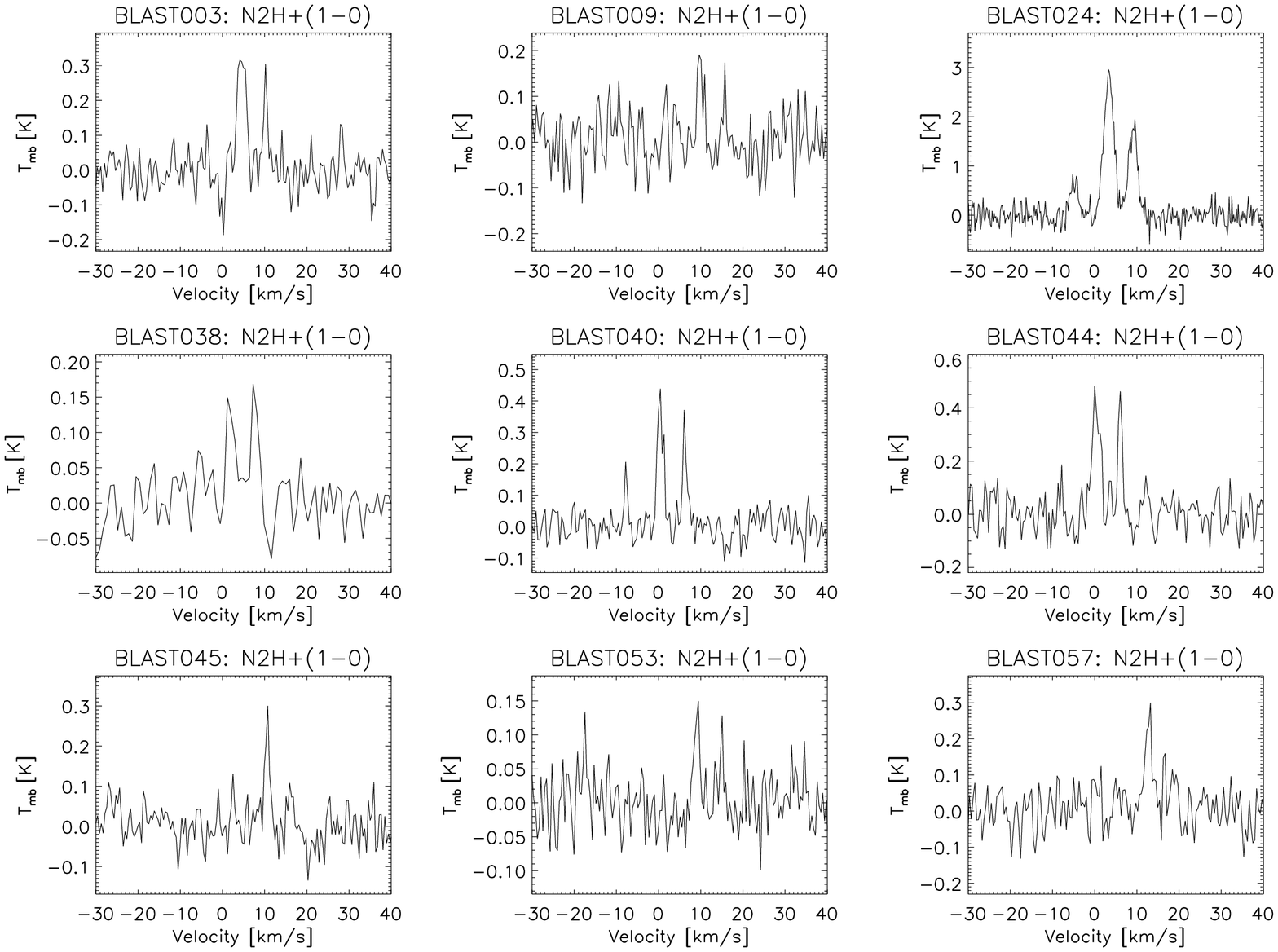}   \\
\includegraphics[width=10.0cm,angle=90]{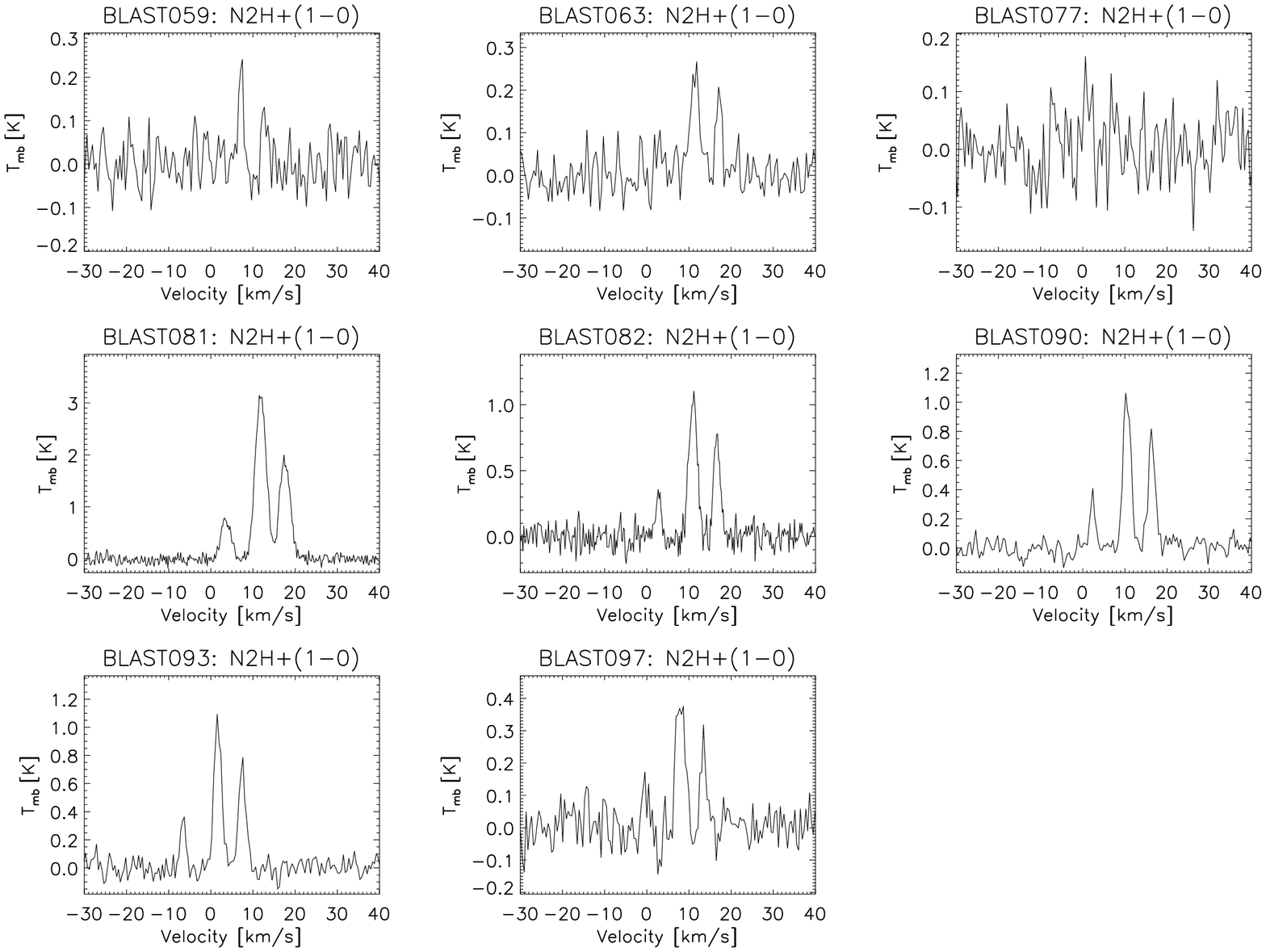} 
\caption{
Spectra of N$_2$H$^+(1-0)$ towards the Vela-D sources.
  }
\label{fig:N2Hspectra}
\end{figure*}

\clearpage

%
\begin{figure}
\includegraphics[width=10.0cm,angle=90]{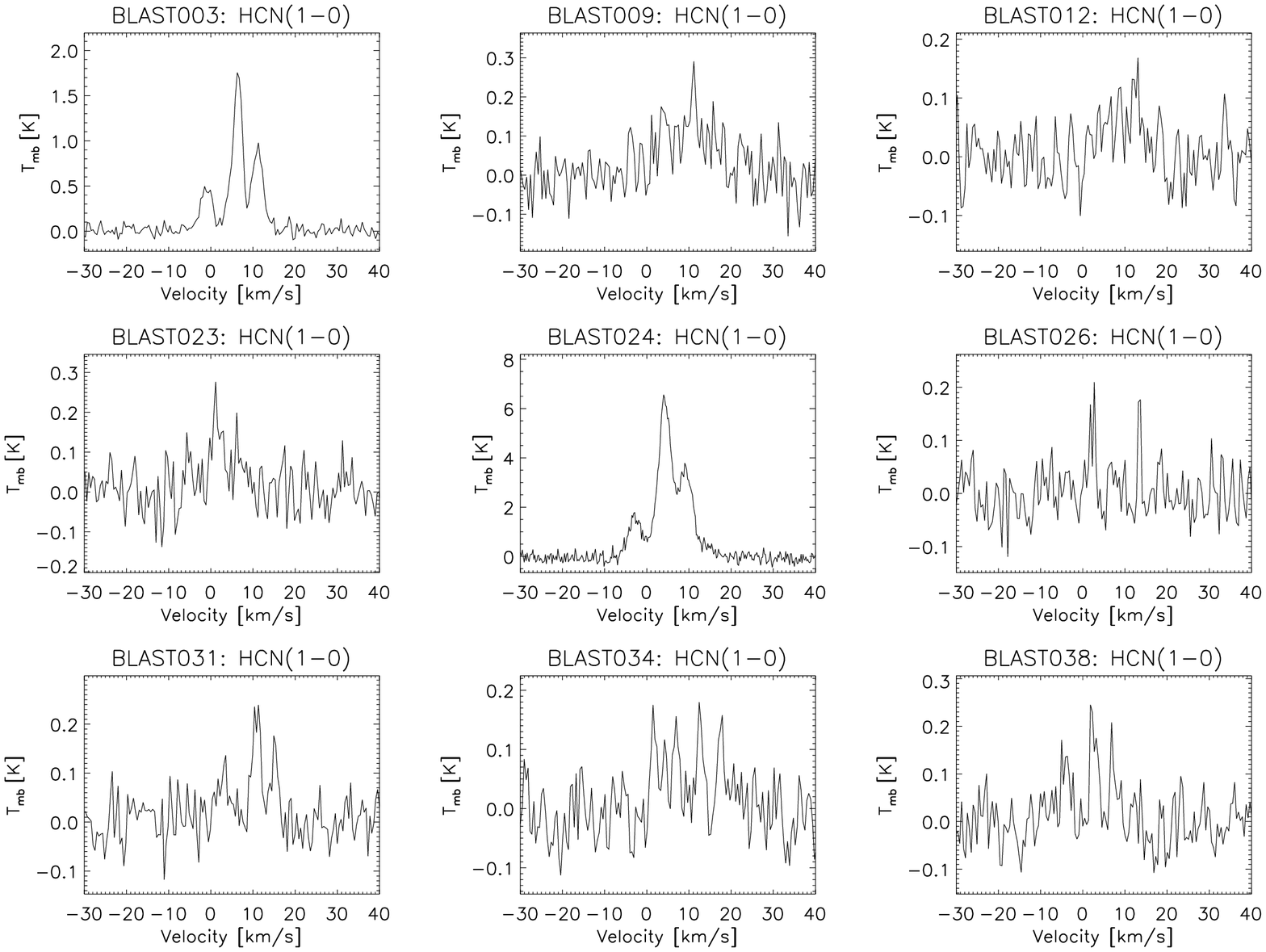}  \\
\includegraphics[width=10.0cm,angle=90]{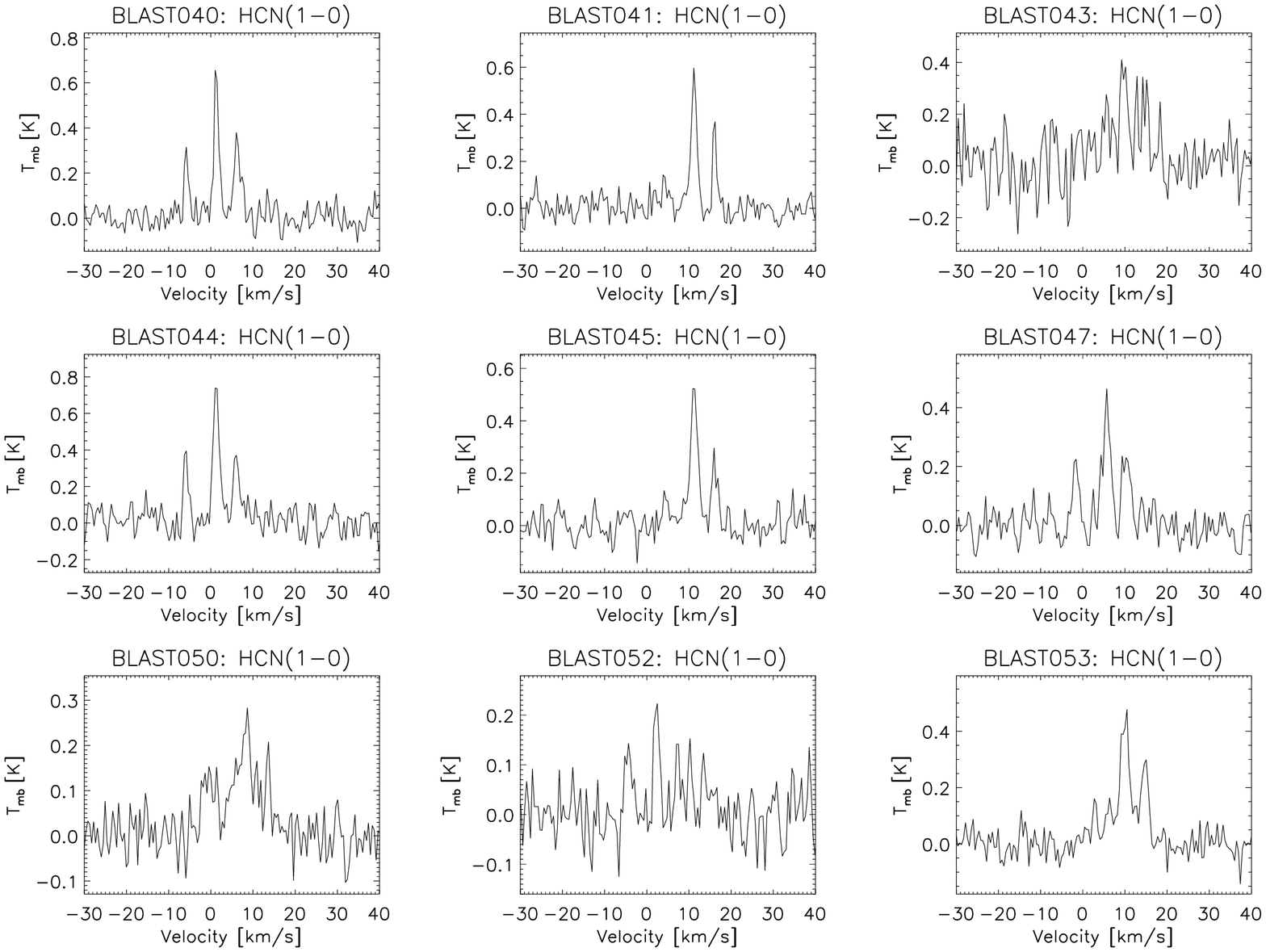} 
\caption{
Same as Figure~\ref{fig:N2Hspectra} for the HCN$(1-0)$ line
  }
\label{fig:HCNspectra1}
\end{figure}

\clearpage

%
\begin{figure*}
\includegraphics[width=10.0cm,angle=90]{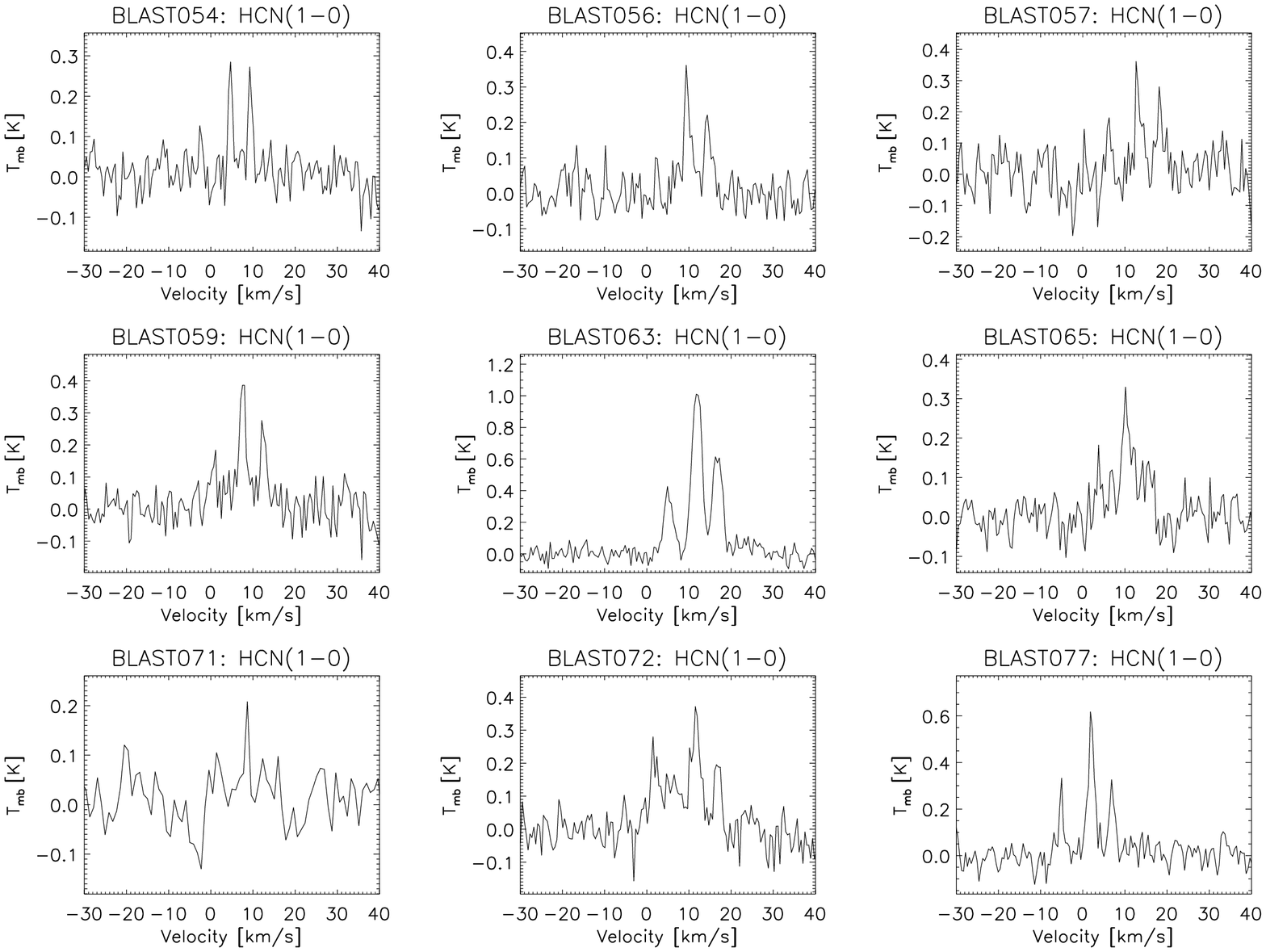} \\
\includegraphics[width=10.0cm,angle=90]{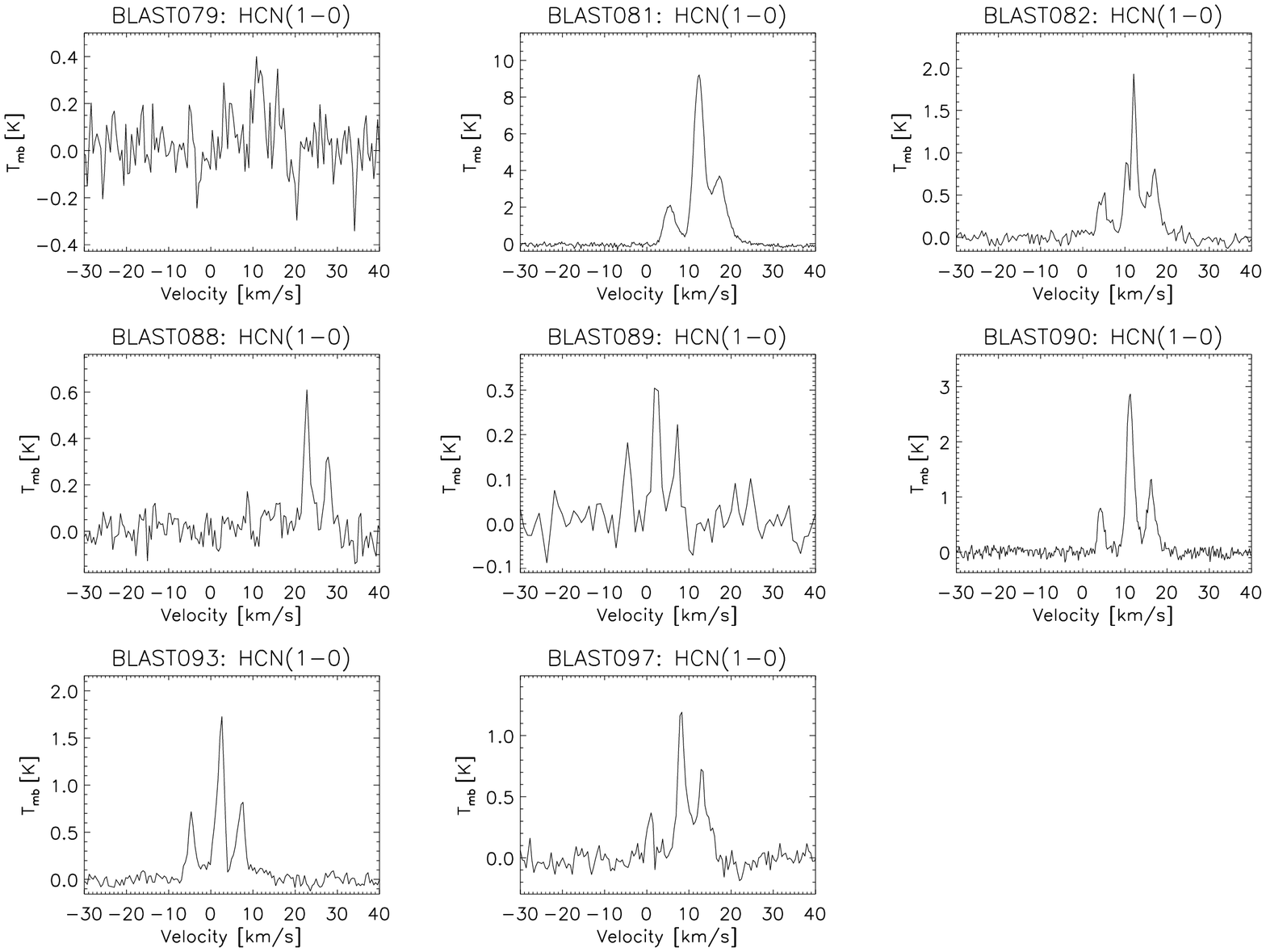} 
\caption{ 
Ctnd. from Figure~\ref{fig:HCNspectra1}
  }
\label{fig:HCNspectra2}
\end{figure*}

\clearpage

%
\begin{figure*}
\includegraphics[width=10.0cm,angle=90]{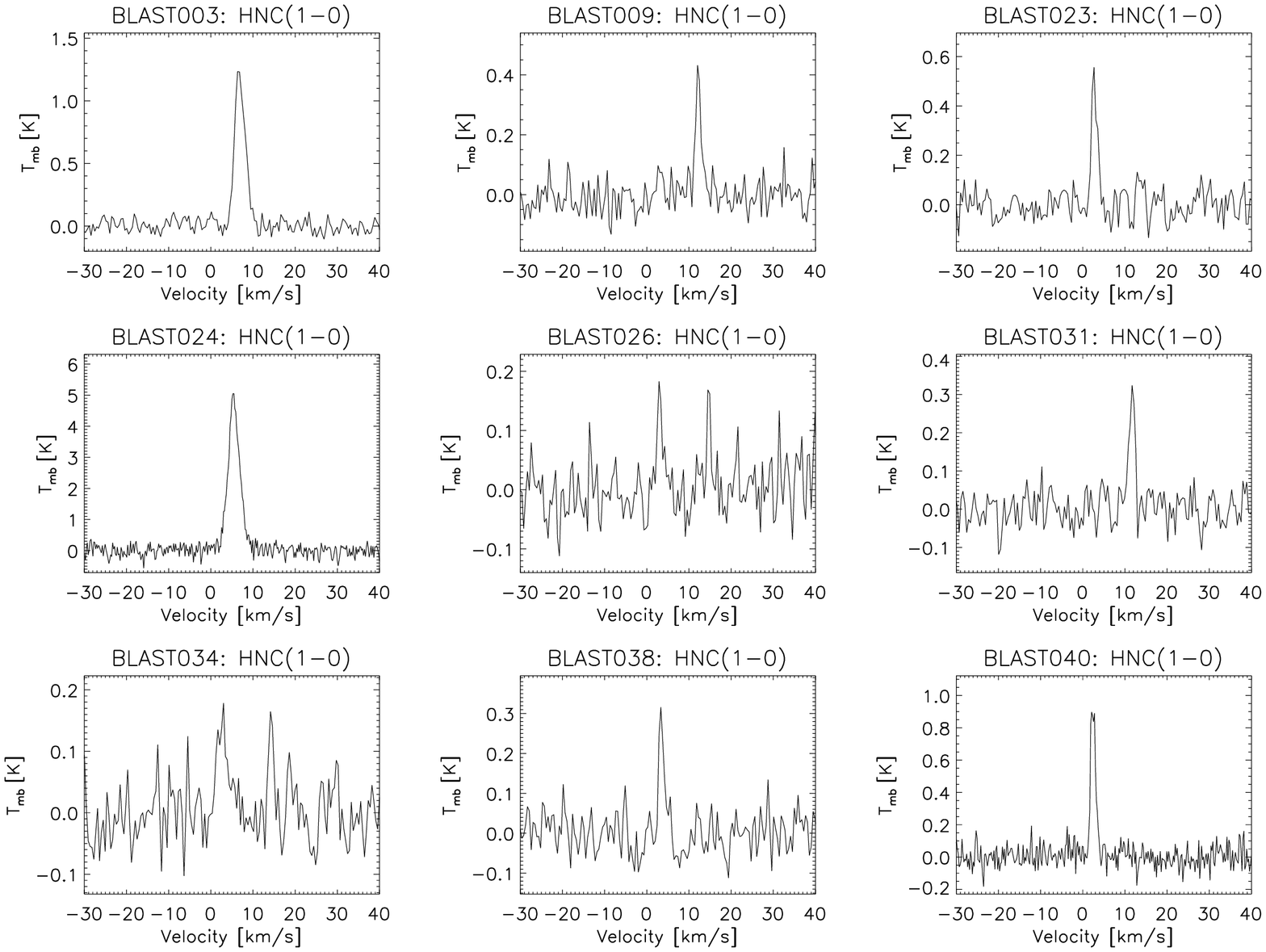} \\ 
\includegraphics[width=10.0cm,angle=90]{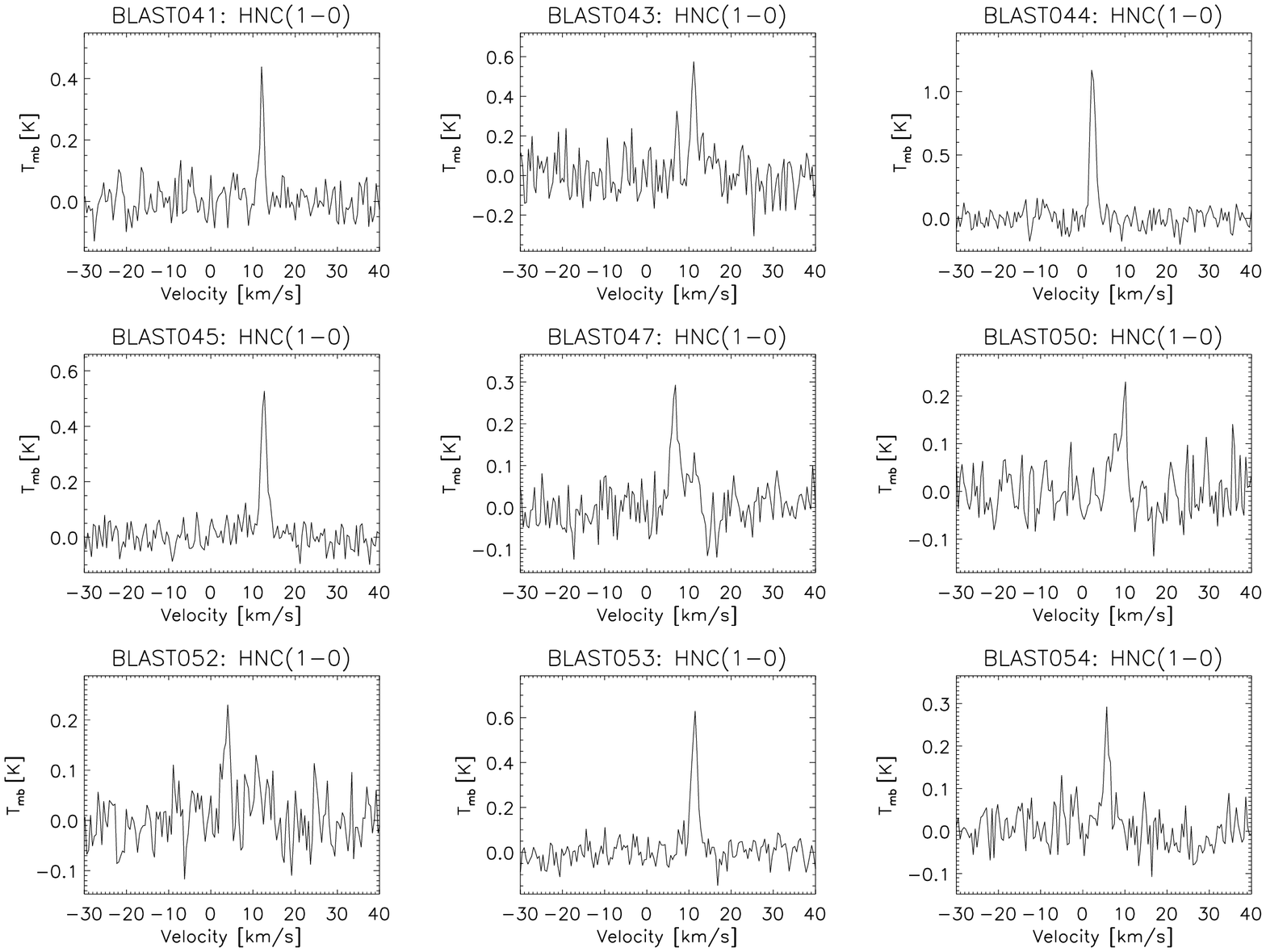} 
\caption{
Same as Figure~\ref{fig:N2Hspectra} for the HNC$(1-0)$ line
  }
\label{fig:HNCspectra1}
\end{figure*}

\clearpage

%
\begin{figure*}
\includegraphics[width=10.0cm,angle=90]{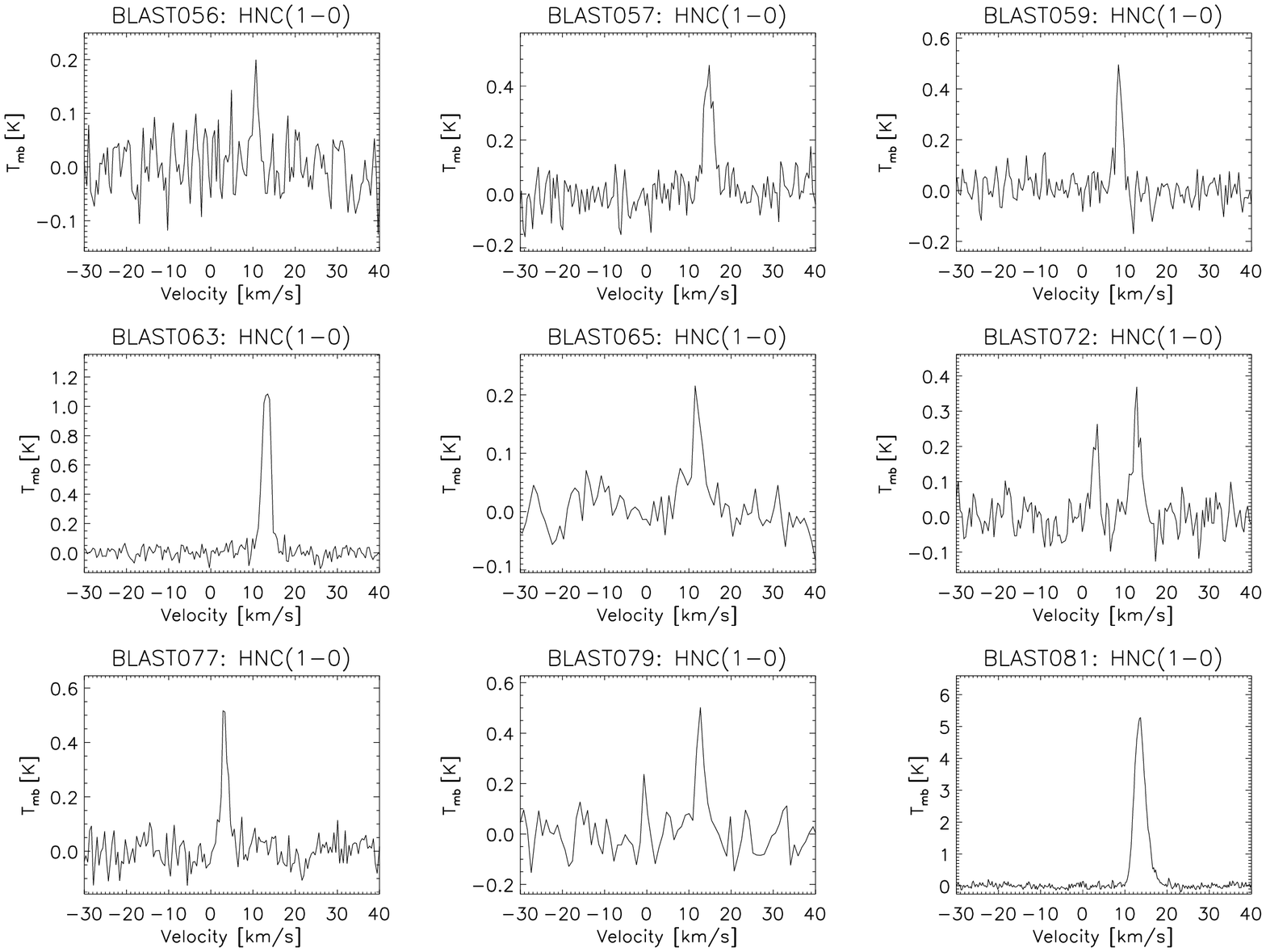} \\
\includegraphics[width=10.0cm,angle=90]{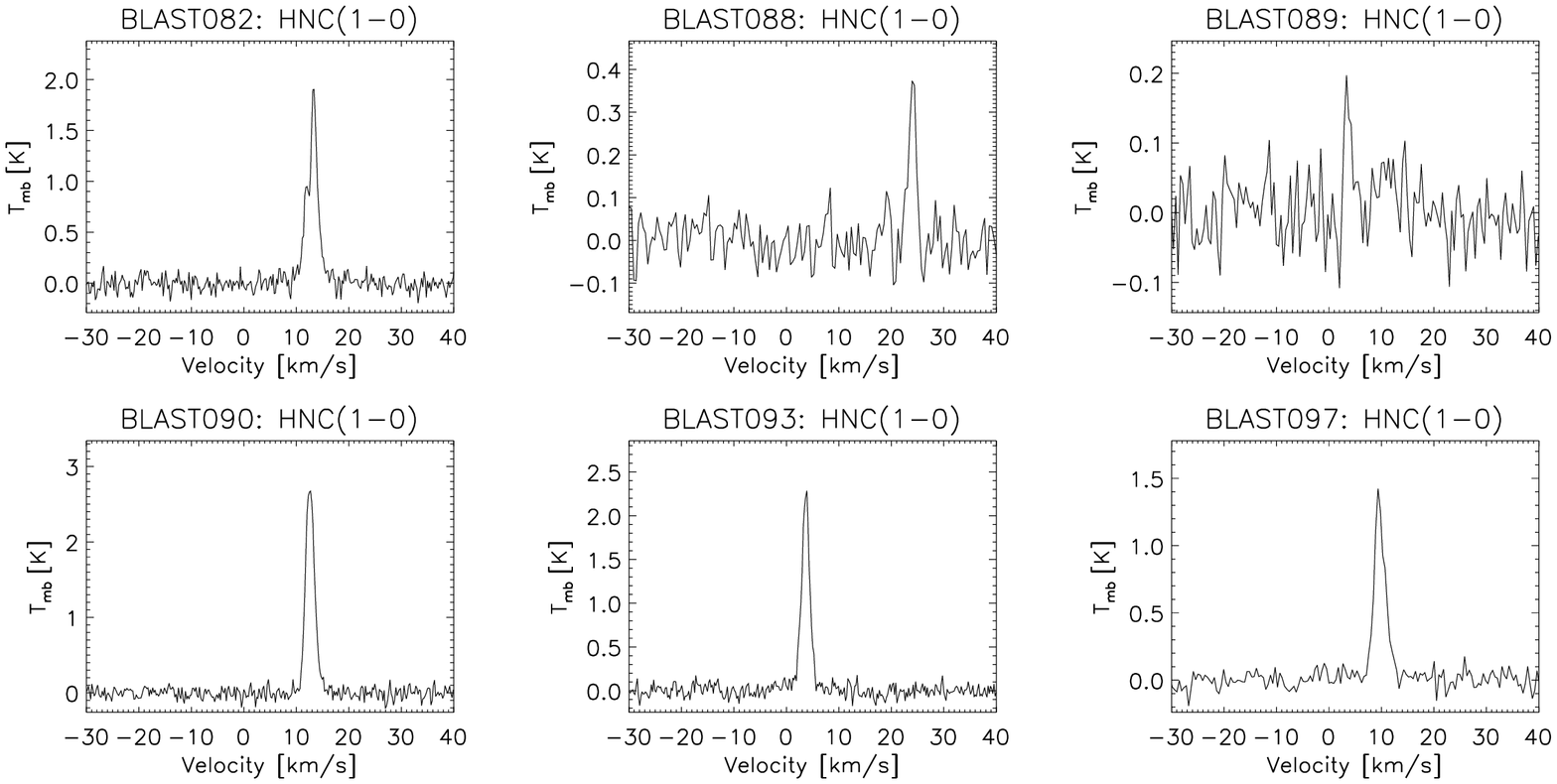} 
\caption{
Ctnd. from Figure~\ref{fig:HNCspectra1}
  }
\label{fig:HNCspectra2}
\end{figure*}

\clearpage

%
\begin{figure*}
\includegraphics[width=10.0cm,angle=90]{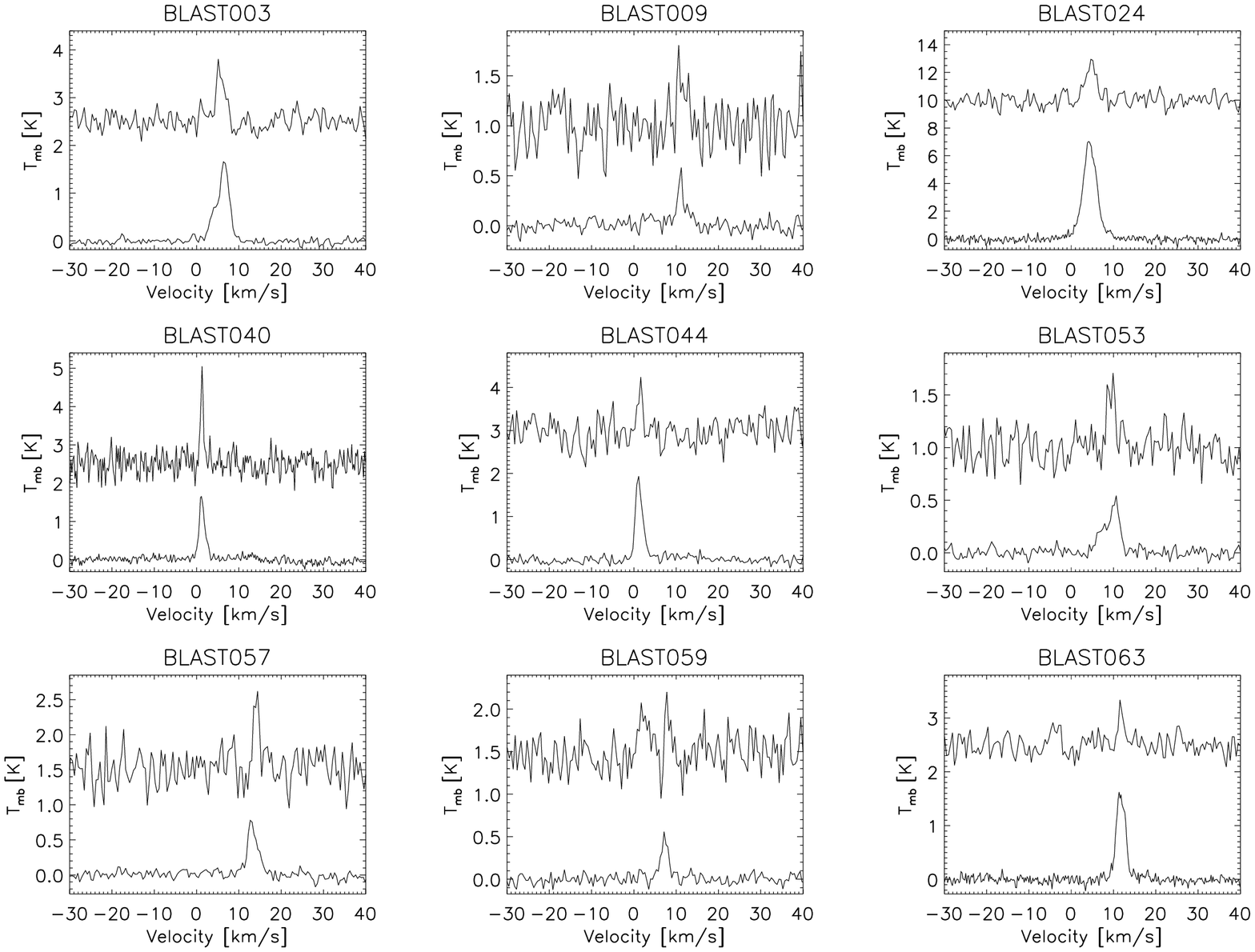} \\
\includegraphics[width=10.0cm,angle=90]{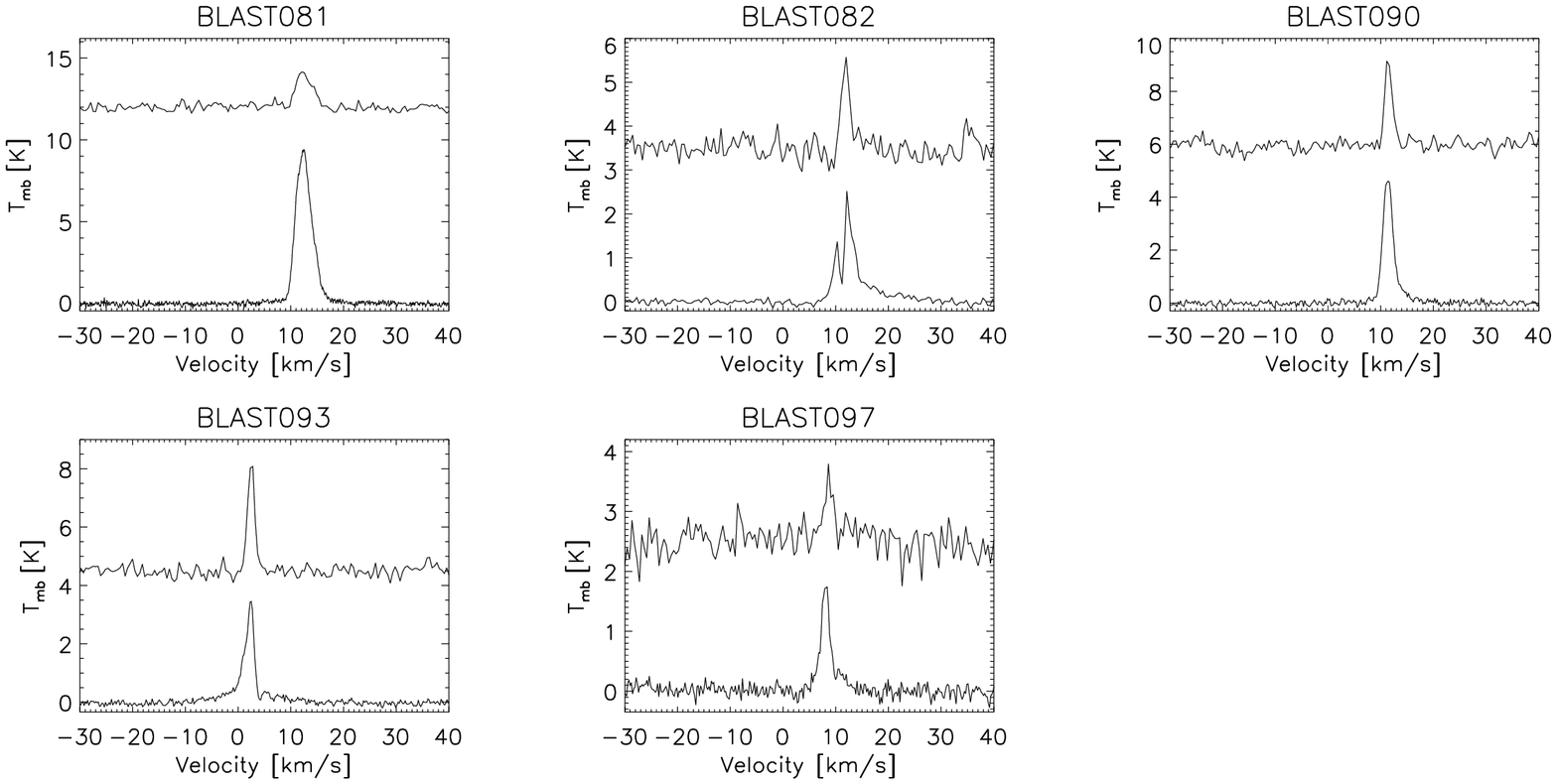} 
\caption{
Same as Figure~\ref{fig:N2Hspectra} for the
HCO$^+(1-0)$ (lower spectra) and H$^{13}$CO$^+(1-0)$ (upper spectra) lines.
Only sources where H$^{13}$CO$^+(1-0)$ was detected are shown. 
  }
\label{fig:HCOspectra}
\end{figure*}

\clearpage

%
\begin{figure*}
\includegraphics[width=9.0cm,angle=90]{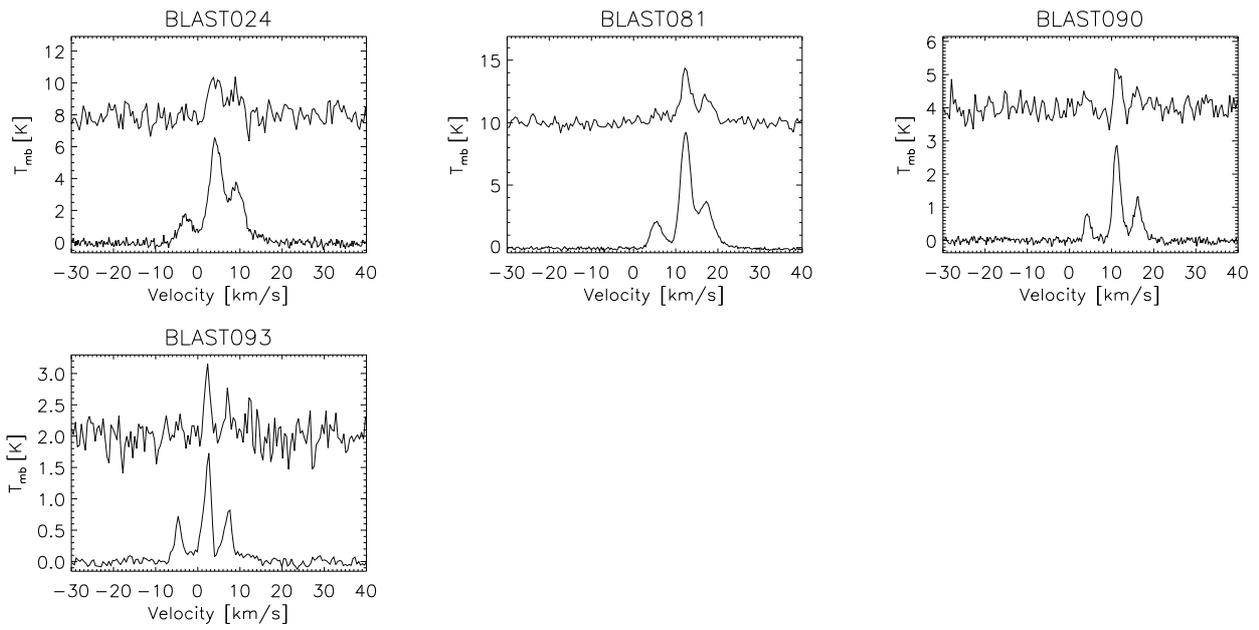} 
\caption{
Same as Figure~\ref{fig:N2Hspectra} for the H$^{13}$CN$(1-0)$ (upper spectra) lines, shown
together with the corresponding HCN$(1-0)$ (lower spectra) lines.
Only sources where H$^{13}$CN$(1-0)$ was detected are shown.
  }
\label{fig:HC13Nspectra}
\end{figure*}

%

%
\begin{figure*}
\includegraphics[width=10.cm,angle=90]{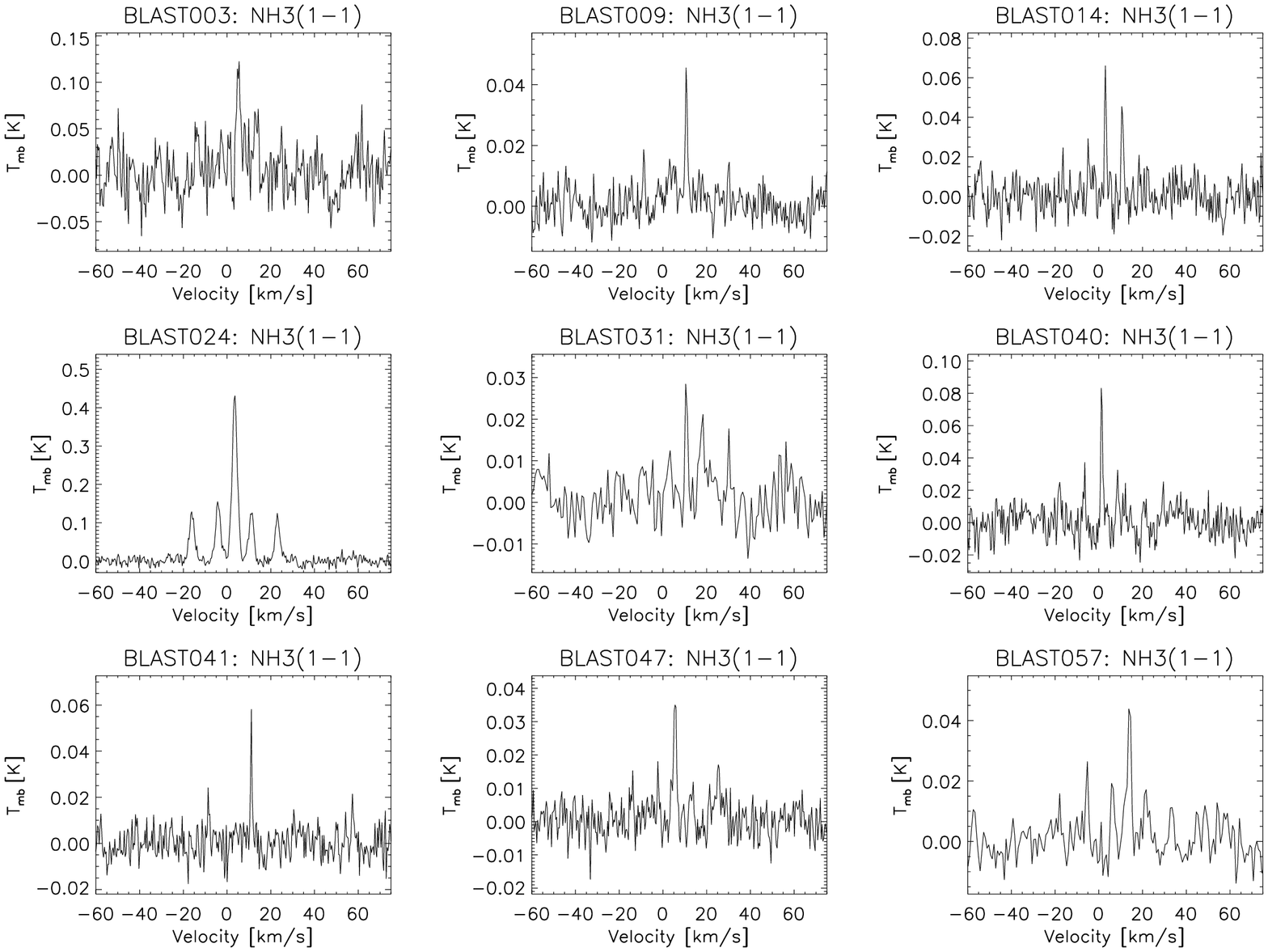} \\
\includegraphics[width=10.cm,angle=90]{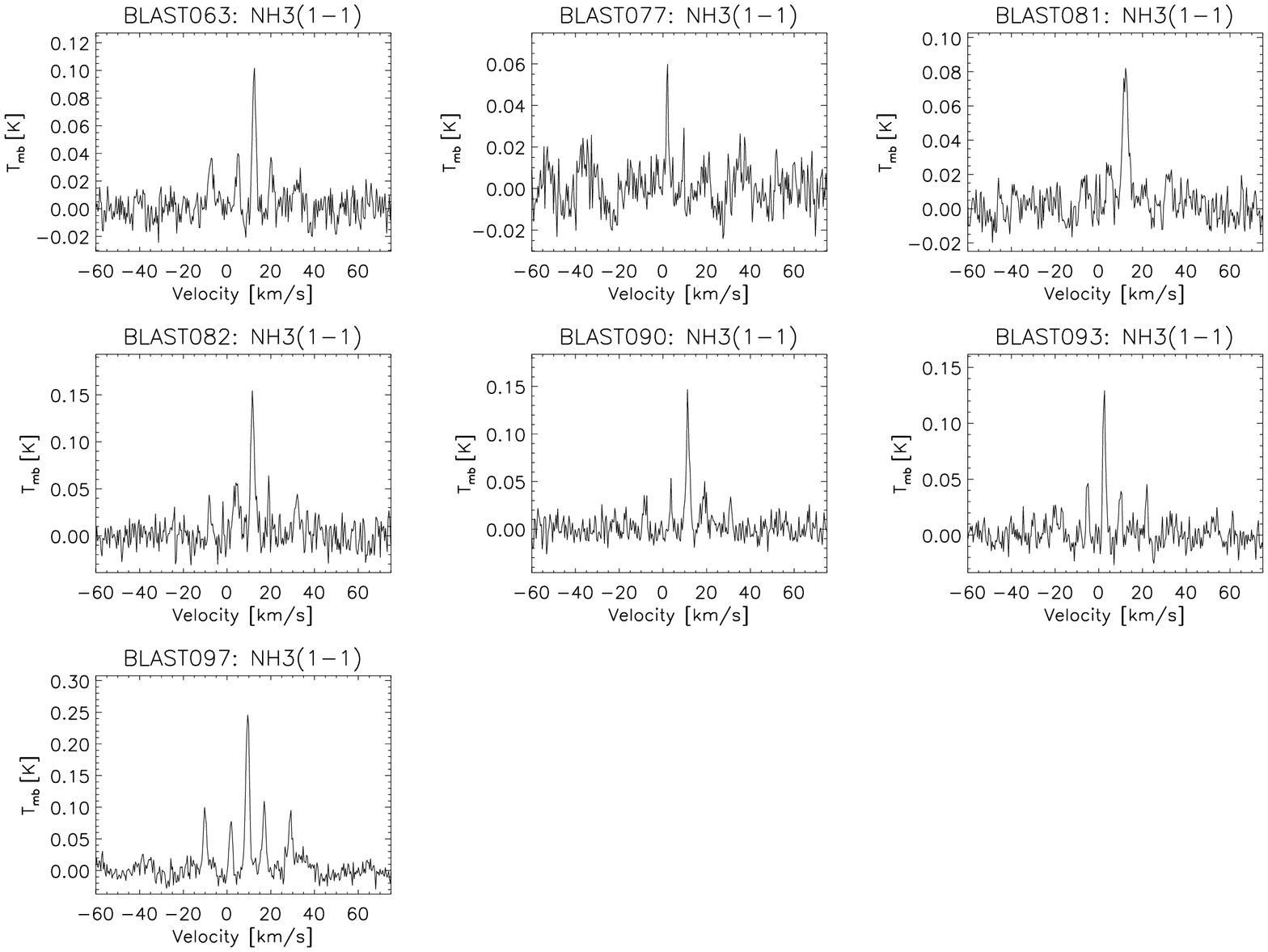}
\caption{
Spectra of the NH$_3$(1,1) line obtained with the Parkes telescope.
  }
\label{fig:NH3spectra}
\end{figure*}

\clearpage 

\section{Molecular lines parameters}     
\label{sec:lineparams}

%
\begin{table*}
\caption{N$_2$H$^+$ line parameters.}
\label{tab:N2H}
\centering
\begin{tabular}{lccccc}
\hline\hline
Source \# & \multicolumn{5}{c}{\bf N$_2$H$^+$} \\
\cline{2-6}
 & $T_A^{\star}\,\tau $  & $V_{lsr}$ & FWHM & $\tau$ & $T_{ex}$ \\
 & [K] & [km\,s$^{-1}$] & [km\,s$^{-1}$] & & [K]  \\  
%
%
%
\hline
 3     &   0.57     &   4.3      &   1.0      &   0.72     &   4.5    \\
 9     &   0.35     &   9.8      &   0.7      &   0.10     &    --    \\
24     &   2.84     &   3.2      &   2.3      &   0.10     &    --    \\
40     &   0.79     &   0.3      &   0.7      &   0.10     &    --    \\
44     &   0.79     &   0.2      &   0.9      &   0.25     &    --    \\
45     &   0.28     &  10.4      &   1.1      &   0.10     &    --    \\
63     &   0.34     &  11.2      &   1.6      &   0.78     &   3.7    \\
81     &   3.05     &  11.6      &   2.7      &   0.10     &    --    \\
82     &   1.38     &  10.8      &   1.6      &   0.65     &   7.2    \\
90     &   1.51     &  10.4      &   1.4      &   0.90     &   6.3    \\
93     &   1.40     &   1.5      &   1.5      &   0.72     &   6.9    \\
97     &   0.55     &   7.7      &   1.6      &   1.09     &   3.8    \\
\hline
\end{tabular}
\tablefoot{
Only sources with a single-point observation are listed (see Table~\ref{tab:srclist}).
Undetected sources and unreliable hfs fits are not shown.
}
\end{table*}


%
\begin{table*}
\caption{HCN and H$^{13}$CN line parameters.}
\label{tab:HCN}
\centering
\begin{tabular}{lcccccccccc}
\hline\hline
Source \# &
\multicolumn{5}{c}{\bf HCN} &
 &
\multicolumn{4}{c}{\bf H$^{13}$CN} \\
\cline{2-6}
\cline{8-11}
 & $T_A^{\star}\,\tau $  & $V_{lsr}$ & FWHM & $\tau$ & $T_{ex}$ & & $T_A^{\star}\,\tau $ & $V_{lsr}$ & FWHM & $\tau$ \\
 & [K] & [km\,s$^{-1}$] & [km\,s$^{-1}$] & & [K] & & [K] & [km\,s$^{-1}$] & [km\,s$^{-1}$] \\
%
%
\hline
 3       & 0.83    &  7.1     &  2.8     &  0.12    & --      &   & --      &  --       & --       & --     \\
24       & 3.12    &  5.1     &  3.6     &  0.16    & --      &   & 0.25    &  2.8      & 2.9      & 0.18   \\
26       & 0.06    &  3.1     &  1.7     &  0.10    & --      &   & --      &  --       & --       & --     \\
26       & 0.06    &  3.1     &  1.7     &  0.10    & --      &   & --      &  --       & --       & --     \\
31       & 0.22    & 11.4     &  2.0     &  2.06    & 3.0     &   & --      &  --       & --       & --     \\
38       & 1.17    &  3.6     &  1.2     & 17.03    & 2.9     &   & --      &  --       & --       & --     \\
40       & 0.43    &  2.2     &  1.3     &  0.83    & 3.8     &   & --      &  --       & --       & --     \\
41       & 0.29    & 12.1     &  1.4     &  0.10    & --      &   & --      &  --       & --       & --     \\
43       & 0.21    &  9.7     &  5.2     &  1.77    & 3.0     &   & --      &  --       & --       & --     \\
44       & 0.49    &  2.1     &  1.6     &  0.84    & 4.0     &   & --      &  --       & --       & --     \\
45       & 0.24    & 12.1     &  1.9     &  0.10    & --      &   & --      &  --       & --       & --     \\
47       & 0.50    &  6.4     &  1.8     &  3.08    & 3.1     &   & --      &  --       & --       & --     \\
50       & 0.93    &  7.7     &  2.9     & 13.81    & 2.9     &   & --      &  --       & --       & --     \\
52       & 0.36    &  3.3     &  1.1     &  4.13    & 2.9     &   & --      &  --       & --       & --     \\
53       & 0.22    & 10.7     &  2.6     &  0.10    & --      &   & --      &  --       & --       & --     \\
54       & 0.42    &  5.4     &  0.9     &  3.21    & 3.0     &   & --      &  --       & --       & --     \\
56       & 0.16    & 10.4     &  1.9     &  0.10    & --      &   & --      &  --       & --       & --     \\
57       & 0.38    & 14.2     &  1.5     &  3.06    & 3.0     &   & --      &  --       & --       & --     \\
59       & 0.23    &  8.4     &  1.9     &  0.53    & 3.7     &   & --      &  --       & --       & --     \\
63       & 0.78    & 12.7     &  2.4     &  1.11    & 4.2     &   & --      &  --       & --       & --     \\
77       & 0.28    &  2.9     &  1.8     &  0.10    & --      &   & --      &  --       & --       & --     \\
79       & 0.36    & 12.1     &  2.1     &  2.02    & 3.1     &   & --      &  --       & --       & --     \\
81       & 4.05    & 13.2     &  3.1     &  0.10    & --      &   & 0.38    & 11.3      & 3.1      & 0.10   \\
82       & 0.67    & 12.8     &  3.0     &  0.11    & --      &   & --      &  --       & --       & --     \\
88       & 0.27    & 23.7     &  1.8     &  0.10    & --      &   & --      &  --       & --       & --     \\
89       & 0.35    &  3.1     &  1.5     &  2.08    & 3.1     &   & --      &  --       & --       & --     \\
90       & 1.27    & 12.1     &  2.1     &  0.10    & --      &   & 0.12    & 10.2      & 1.6      & 0.18   \\
93       & 0.80    &  3.1     &  2.2     &  0.24    & --      &   & 0.11    &  1.1      & 1.5      & 0.10   \\
97       & 0.53    &  9.1     &  2.7     &  0.10    & --      &   & --      &  --       & --       & --     \\
\hline
\end{tabular}
\tablefoot{
Only sources with a single-point observation are listed (see Table~\ref{tab:srclist}).
Undetected sources and unreliable hfs fits are not shown. If two velocity components
were detected in a given source, only the more intense component is listed here.
}
\end{table*}

%
\begin{table*}
\caption{HNC line parameters.}
\label{tab:HNC}
\centering
\begin{tabular}{lccc}
\hline\hline
%
Source \# & \multicolumn{3}{c}{\bf HNC} \\
\cline{2-4}
 & $\int{T_A^{\star}\, {\rm d}V }$  & $V_{lsr}$ & FWHM \\
 & [K\,km\,s$^{-1}$] & [km\,s$^{-1}$] & [km\,s$^{-1}$] \\  
%
%
\hline
 3     &   1.79     &   6.9      &   2.9      \\
 9     &   0.31     &  12.2      &   1.5      \\
23     &   0.42     &   2.7      &   1.6      \\
24     &   7.68     &   5.5      &   3.2      \\
26     &   0.12     &   3.1      &   1.4      \\
26     &   0.08     &  14.7      &   0.8      \\
31     &   0.25     &  11.7      &   1.6      \\
34     &   0.21     &   2.7      &   2.9      \\
38     &   0.21     &   3.4      &   1.3      \\
40     &   0.66     &   2.4      &   1.4      \\
41     &   0.23     &  12.1      &   1.1      \\
43     &   0.42     &  11.1      &   1.5      \\
44     &   0.90     &   2.4      &   1.5      \\
45     &   0.44     &  12.6      &   1.7      \\
52     &   0.20     &   3.7      &   1.9      \\
53     &   0.53     &  11.3      &   1.8      \\
54     &   0.22     &   5.7      &   1.7      \\
56     &   0.14     &  10.6      &   1.6      \\
57     &   0.61     &  14.7      &   2.8      \\
59     &   0.42     &   8.6      &   1.8      \\
63     &   1.38     &  13.2      &   2.4      \\
72     &   0.36     &  12.8      &   2.4      \\
77     &   0.46     &   3.3      &   1.8      \\
79     &   0.55     &  12.7      &   2.3      \\
81     &   8.07     &  13.6      &   3.1      \\
82     &   1.97     &  13.1      &   2.5      \\
88     &   0.30     &  24.0      &   1.5      \\
89     &   0.13     &   3.6      &   1.4      \\
90     &   2.71     &  12.6      &   2.0      \\
93     &   2.01     &   3.7      &   1.8      \\
97     &   1.64     &   9.7      &   2.5      \\
\hline
\end{tabular}
\tablefoot{
Only sources with a single-point observation are listed (see Table~\ref{tab:srclist}).
Undetected sources and unreliable fits are not shown.
If two velocity components
were detected in a given source, only the more intense component is listed here.
}
\end{table*}

%
\begin{table*}
\caption{HCO$^+$ and H$^{13}$CO$^+$ Line Parameters.}
\label{tab:HCO}
\centering
\begin{tabular}{lccccccccccc}
\hline\hline
%
Source \# & \multicolumn{6}{c}{\bf HCO$^+$} & & \multicolumn{4}{c}{\bf H$^{13}$CO$^+$} \\
\cline{2-7}
\cline{9-12}
 & $T_A^{\star}\,\tau $  & $V_{lsr}$ & FWHM & $\tau$ & $T_{ex}$ & Line Wings? & & $T_A^{\star}\,\tau $ & $V_{lsr}$ & FWHM & $\tau$ \\
 & [K] & [km\,s$^{-1}$] & [km\,s$^{-1}$] & & [K] & & & [K] & [km\,s$^{-1}$] & [km\,s$^{-1}$] \\
%
\hline
 3     &   2.66     &   6.3      &   3.6      &   7.97     &   4.4     & NG  &   &   0.33     &   5.6      &   2.4      &   0.20   \\
 9     &   0.45     &  11.1      &   1.9      &  12.07     &   3.3     & NG  &   &   0.17     &  11.3      &   2.8      &   0.30   \\
23     &   0.37     &   1.2      &   0.9      &     --     &   3.7     & N  &   &     --     &     --     &     --     &     --   \\
24     &  13.00     &   4.4      &   3.8      &   4.13     &  10.1     & Y  &   &   0.99     &   4.7      &   2.9      &   0.10   \\
26     &   0.31     &   1.9      &   2.0      &     --     &   3.1     & N  &   &     --     &     --     &     --     &     --   \\
26     &   0.20     &  13.5      &   1.5      &     --     &   3.1     & N  &   &     --     &     --     &     --     &     --   \\
31     &   0.60     &  10.4      &   1.7      &     --     &   3.6     & N  &   &     --     &     --     &     --     &     --   \\
34     &   0.50     &   2.0      &   2.0      &     --     &   3.3     & N  &   &     --     &     --     &     --     &     --   \\
34     &   0.24     &  13.1      &   1.6      &     --     &   3.1     & N  &   &     --     &     --     &     --     &     --   \\
38     &   0.36     &   1.9      &   1.8      &     --     &   3.2     & N  &   &     --     &     --     &     --     &     --   \\
40     &   1.30     &   1.3      &   1.7      &     --     &   4.5     & N  &   &   0.23     &   1.3      &   0.8      &     --   \\
41     &   0.62     &  11.0      &   1.4      &     --     &   3.7     & N  &   &     --     &     --     &     --     &     --   \\
43     &   0.63     &   9.9      &   1.4      &     --     &   3.8     & N  &   &     --     &     --     &     --     &     --   \\
44     &   1.85     &   1.2      &   2.0      &   6.65     &   4.8     & N  &   &   0.16     &   1.5      &   1.1      &   0.17   \\
45     &   1.14     &  11.1      &   2.1      &     --     &   4.0     & NG  &   &     --     &     --     &     --     &     --   \\
47     &   0.37     &   5.4      &   2.5      &     --     &   3.1     & N  &   &     --     &     --     &     --     &     --   \\
47     &   0.18     &   9.2      &   1.3      &     --     &   3.1     & N  &   &     --     &     --     &     --     &     --   \\
52     &   0.44     &   2.3      &   2.3      &     --     &   3.2     & N  &   &     --     &     --     &     --     &     --   \\
54     &   0.24     &   4.7      &   1.9      &     --     &   3.1     & N  &   &     --     &     --     &     --     &     --   \\
56     &   0.35     &   9.4      &   2.0      &     --     &   3.2     & N  &   &     --     &     --     &     --     &     --   \\
57     &   1.04     &  13.2      &   2.9      &     --     &   3.6     & N  &   &   0.21     &  14.1      &   1.5      &     --   \\
59     &   0.51     &   7.3      &   2.1      &     --     &   3.3     & N  &   &   0.08     &   7.8      &   0.9      &     --   \\
63     &   2.04     &  11.7      &   2.5      &   4.99     &   4.6     & Y  &   &   0.13     &  11.8      &   1.4      &   0.12   \\
71     &   0.17     &   7.5      &   2.1      &     --     &   3.0     & N  &   &     --     &     --     &     --     &     --   \\
72     &   0.86     &   1.9      &   2.1      &     --     &   3.7     & N  &   &     --     &     --     &     --     &     --   \\
72     &   0.58     &  11.6      &   2.9      &     --     &   3.2     & N  &   &     --     &     --     &     --     &     --   \\
77     &   1.00     &   1.8      &   2.1      &     --     &   3.9     & N  &   &     --     &     --     &     --     &     --   \\
79     &   0.87     &  11.0      &   3.8      &     --     &   3.3     & N  &   &     --     &     --     &     --     &     --   \\
81     &  16.33     &  12.4      &   3.6      &   2.07     &  13.7     & Y  &   &   0.95     &  12.5      &   3.6      &   0.05   \\
82     &   0.66     &  10.0      &   1.1      &     --     &   4.1     & Y  &   &   0.44     &  11.9      &   1.8      &     --   \\
82     &   2.52     &  12.6      &   2.5      &  11.49     &   5.0     & Y  &   &   0.44     &  11.9      &   1.8      &   0.29   \\
89     &   0.57     &   2.3      &   2.4      &     --     &   3.3     & N  &   &     --     &     --     &     --     &     --   \\
90     &   5.37     &  11.4      &   2.3      &   7.57     &   7.8     & Y  &   &   0.65     &  11.5      &   1.6      &   0.19   \\
93     &   3.52     &   2.1      &   2.3      &  14.82     &   6.1     & Y  &   &   0.73     &   2.5      &   1.5      &   0.37   \\
97     &   1.81     &   8.1      &   2.2      &   6.64     &   4.6     & Y  &   &   0.28     &   8.7      &   2.2      &   0.17   \\
\hline
\end{tabular}
\tablefoot{
Only sources with a single-point observation are listed (see Table~\ref{tab:srclist}).
Undetected sources and unreliable hfs fits are not shown.
In the $7^{th}$ column ``NG'' indicates a non-Gaussian line profile.
}
\end{table*}

\clearpage


\bibliographystyle{aa}
\bibliography{refs}

\begin{thebibliography}{41}
\expandafter\ifx\csname natexlab\endcsname\relax\def\natexlab#1{#1}\fi

\bibitem[{{Aikawa} {et~al.}(2008){Aikawa}, {Wakelam}, {Garrod}, \&
  {Herbst}}]{aikawa2008}
{Aikawa}, Y., {Wakelam}, V., {Garrod}, R.~T., \& {Herbst}, E. 2008, \apj, 674,
  984

\bibitem[{{Bachiller} {et~al.}(1987){Bachiller}, {Guilloteau}, \&
  {Kahane}}]{bach1987}
{Bachiller}, R., {Guilloteau}, S., \& {Kahane}, C. 1987, \aap, 173, 324

\bibitem[{{Benson} {et~al.}(1998){Benson}, {Caselli}, \& {Myers}}]{benson1998}
{Benson}, P.~J., {Caselli}, P., \& {Myers}, P.~C. 1998, \apj, 506, 743

\bibitem[{{Blake} {et~al.}(1995){Blake}, {Sandell}, {van Dishoeck},
  {Groesbeck}, {Mundy}, \& {Aspin}}]{blake1995}
{Blake}, G.~A., {Sandell}, G., {van Dishoeck}, E.~F., {et~al.} 1995, \apj, 441,
  689

\bibitem[{{Carey} {et~al.}(2005){Carey}, {Noriega-Crespo}, {Price}, {Padgett},
  {Kraemer}, {Indebetouw}, {Mizuno}, {Ali}, {Berriman}, {Boulanger}, {Cutri},
  {Ingalls}, {Kuchar}, {Latter}, {Marleau}, {Miville-Deschenes}, {Molinari},
  {Rebull}, \& {Testi}}]{car05}
{Carey}, S.~J., {Noriega-Crespo}, A., {Price}, S.~D., {et~al.} 2005, in
  Bulletin of the American Astronomical Society, 1252

\bibitem[{{Caselli} {et~al.}(2002){Caselli}, {Benson}, {Myers}, \&
  {Tafalla}}]{caselli2002}
{Caselli}, P., {Benson}, P.~J., {Myers}, P.~C., \& {Tafalla}, M. 2002, \apj,
  572, 238

\bibitem[{{Elia} {et~al.}(2007){Elia}, {Massi}, {Strafella}, {De Luca},
  {Giannini}, {Lorenzetti}, {Nisini}, {Campeggio}, \& {Maiolo}}]{elia2007}
{Elia}, D., {Massi}, F., {Strafella}, F., {et~al.} 2007, \apj, 655, 316

\bibitem[{{Evans}(1999)}]{evans1999}
{Evans}, II, N.~J. 1999, \araa, 37, 311

\bibitem[{{Friesen} {et~al.}(2010){Friesen}, {Di Francesco}, {Shimajiri}, \&
  {Takakuwa}}]{friesen2010}
{Friesen}, R.~K., {Di Francesco}, J., {Shimajiri}, Y., \& {Takakuwa}, S. 2010,
  \apj, 708, 1002

\bibitem[{{Fuente} {et~al.}(2005){Fuente}, {Rizzo}, {Caselli}, {Bachiller}, \&
  {Henkel}}]{fuente2005}
{Fuente}, A., {Rizzo}, J.~R., {Caselli}, P., {Bachiller}, R., \& {Henkel}, C.
  2005, \aap, 433, 535

\bibitem[{{Giannini} {et~al.}(2007){Giannini}, {Lorenzetti}, {De Luca},
  {Nisini}, {Marengo}, {Allen}, {Smith}, {Fazio}, {Massi}, {Elia}, \&
  {Strafella}}]{gianni07}
{Giannini}, T., {Lorenzetti}, D., {De Luca}, M., {et~al.} 2007, \apj, 671, 470

\bibitem[{{Godard} {et~al.}(2010){Godard}, {Falgarone}, {Gerin}, {Hily-Blant},
  \& {de Luca}}]{godard2010}
{Godard}, B., {Falgarone}, E., {Gerin}, M., {Hily-Blant}, P., \& {de Luca}, M.
  2010, \aap, 520, A20

\bibitem[{{Goodman} {et~al.}(1993){Goodman}, {Benson}, {Fuller}, \&
  {Myers}}]{good1993}
{Goodman}, A.~A., {Benson}, P.~J., {Fuller}, G.~A., \& {Myers}, P.~C. 1993,
  \apj, 406, 528

\bibitem[{{Hirota} {et~al.}(1998){Hirota}, {Yamamoto}, {Mikami}, \&
  {Ohishi}}]{hirota1998}
{Hirota}, T., {Yamamoto}, S., {Mikami}, H., \& {Ohishi}, M. 1998, \apj, 503,
  717

\bibitem[{{Kruegel} \& {Walmsley}(1984)}]{krugel1984}
{Kruegel}, E. \& {Walmsley}, C.~M. 1984, \aap, 130, 5

\bibitem[{{Lada} {et~al.}(2008){Lada}, {Muench}, {Rathborne}, {Alves}, \&
  {Lombardi}}]{lada2008}
{Lada}, C.~J., {Muench}, A.~A., {Rathborne}, J., {Alves}, J.~F., \& {Lombardi},
  M. 2008, \apj, 672, 410

\bibitem[{{Ladd} {et~al.}(2005){Ladd}, {Purcell}, {Wong}, \&
  {Robertson}}]{ladd2005}
{Ladd}, N., {Purcell}, C., {Wong}, T., \& {Robertson}, S. 2005, \pasa, 22, 62

\bibitem[{{Lee} {et~al.}(2004){Lee}, {Bergin}, \& {Evans}}]{lee2004}
{Lee}, J.-E., {Bergin}, E.~A., \& {Evans}, II, N.~J. 2004, \apj, 617, 360

\bibitem[{{Liseau} {et~al.}(1992){Liseau}, {Lorenzetti}, {Nisini}, {Spinoglio},
  \& {Moneti}}]{Lis92}
{Liseau}, R., {Lorenzetti}, D., {Nisini}, B., {Spinoglio}, L., \& {Moneti}, A.
  1992, \aap, 265, 577

\bibitem[{MacLaren {et~al.}(1988)MacLaren, Richardson, \&
  Wolfendale}]{maclaren1988}
MacLaren, I., Richardson, K.~M., \& Wolfendale, A.~W. 1988, \apj, 333, 821

\bibitem[{{Mardones} {et~al.}(1997){Mardones}, {Myers}, {Tafalla}, {Wilner},
  {Bachiller}, \& {Garay}}]{mardones1997}
{Mardones}, D., {Myers}, P.~C., {Tafalla}, M., {et~al.} 1997, \apj, 489, 719

\bibitem[{{Massi} {et~al.}(2007){Massi}, {De Luca}, {Elia}, {Giannini},
  {Lorenzetti}, \& {Nisini}}]{massi07}
{Massi}, F., {De Luca}, M., {Elia}, D., {et~al.} 2007, \aap, 466, 1013

\bibitem[{{Molinari} {et~al.}(2010){Molinari}, {Swinyard}, {Bally}, {Barlow},
  {Bernard}, {Martin}, {Moore}, {Noriega-Crespo}, {Plume}, {Testi}, {Zavagno},
  {Abergel}, {Ali}, {Anderson}, {Andr{\'e}}, {Baluteau}, {Battersby},
  {Beltr{\'a}n}, {Benedettini}, {Billot}, {Blommaert}, {Bontemps}, {Boulanger},
  {Brand}, {Brunt}, {Burton}, {Calzoletti}, {Carey}, {Caselli}, {Cesaroni},
  {Cernicharo}, {Chakrabarti}, {Chrysostomou}, {Cohen}, {Compiegne}, {de
  Bernardis}, {de Gasperis}, {di Giorgio}, {Elia}, {Faustini}, {Flagey},
  {Fukui}, {Fuller}, {Ganga}, {Garcia-Lario}, {Glenn}, {Goldsmith}, {Griffin},
  {Hoare}, {Huang}, {Ikhenaode}, {Joblin}, {Joncas}, {Juvela}, {Kirk},
  {Lagache}, {Li}, {Lim}, {Lord}, {Marengo}, {Marshall}, {Masi}, {Massi},
  {Matsuura}, {Minier}, {Miville-Desch{\^e}nes}, {Montier}, {Morgan}, {Motte},
  {Mottram}, {M{\"u}ller}, {Natoli}, {Neves}, {Olmi}, {Paladini}, {Paradis},
  {Parsons}, {Peretto}, {Pestalozzi}, {Pezzuto}, {Piacentini}, {Piazzo},
  {Polychroni}, {Pomar{\`e}s}, {Popescu}, {Reach}, {Ristorcelli}, {Robitaille},
  {Robitaille}, {Rod{\'o}n}, {Roy}, {Royer}, {Russeil}, {Saraceno}, {Sauvage},
  {Schilke}, {Schisano}, {Schneider}, {Schuller}, {Schulz}, {Sibthorpe},
  {Smith}, {Smith}, {Spinoglio}, {Stamatellos}, {Strafella}, {Stringfellow},
  {Sturm}, {Taylor}, {Thompson}, {Traficante}, {Tuffs}, {Umana}, {Valenziano},
  {Vavrek}, {Veneziani}, {Viti}, {Waelkens}, {Ward-Thompson}, {White},
  {Wilcock}, {Wyrowski}, {Yorke}, \& {Zhang}}]{molinari2010}
{Molinari}, S., {Swinyard}, B., {Bally}, J., {et~al.} 2010, \aap, 518, L100

\bibitem[{{Motte} {et~al.}(1998){Motte}, {Andr\'e}, \& {Neri}}]{motte1998}
{Motte}, F., {Andr\'e}, P., \& {Neri}, R. 1998, \aap, 336, 150

\bibitem[{{Murphy} \& {May}(1991)}]{mur:may}
{Murphy}, D.~C. \& {May}, J. 1991, \aap, 247, 202

\bibitem[{{Netterfield} {et~al.}(2009){Netterfield}, {Ade}, {Bock}, {Chapin},
  {Devlin}, {Griffin}, {Gundersen}, {Halpern}, {Hargrave}, {Hughes}, {Klein},
  {Marsden}, {Martin}, {Mauskopf}, {Olmi}, {Pascale}, {Patanchon}, {Rex},
  {Roy}, {Scott}, {Semisch}, {Thomas}, {Truch}, {Tucker}, {Tucker}, {Viero}, \&
  {Wiebe}}]{netterfield2009}
{Netterfield}, C.~B., {Ade}, P.~A.~R., {Bock}, J.~J., {et~al.} 2009, \apj, 707,
  1824

\bibitem[{{Nikoli{\'c}} {et~al.}(2003){Nikoli{\'c}}, {Johansson}, \&
  {Harju}}]{nikolic2003}
{Nikoli{\'c}}, S., {Johansson}, L.~E.~B., \& {Harju}, J. 2003, \aap, 409, 941

\bibitem[{{Olmi} {et~al.}(2009){Olmi}, {Ade}, {Angl{\'e}s-Alc{\'a}zar}, {Bock},
  {Chapin}, {De Luca}, {Devlin}, {Dicker}, {Elia}, {Fazio}, {Giannini},
  {Griffin}, {Gundersen}, {Halpern}, {Hargrave}, {Hughes}, {Klein},
  {Lorenzetti}, {Marengo}, {Marsden}, {Martin}, {Massi}, {Mauskopf},
  {Netterfield}, {Patanchon}, {Rex}, {Salama}, {Scott}, {Semisch}, {Smith},
  {Strafella}, {Thomas}, {Truch}, {Tucker}, {Tucker}, {Viero}, \&
  {Wiebe}}]{olmi2009}
{Olmi}, L., {Ade}, P.~A.~R., {Angl{\'e}s-Alc{\'a}zar}, D., {et~al.} 2009, \apj,
  707, 1836

\bibitem[{{Olmi} {et~al.}(2010){Olmi}, {Araya}, {Chapin}, {Gibb}, {Hofner},
  {Martin}, \& {Poventud}}]{olmi2010}
{Olmi}, L., {Araya}, E.~D., {Chapin}, E.~L., {et~al.} 2010, \apj, 715, 1132

\bibitem[{{Olmi} \& {Testi}(2002)}]{olmi2002}
{Olmi}, L. \& {Testi}, L. 2002, \aap, 392, 1053

\bibitem[{{Pascale} {et~al.}(2008){Pascale}, {Ade}, {Bock}, {Chapin}, {Chung},
  {Devlin}, {Dicker}, {Griffin}, {Gundersen}, {Halpern}, {Hargrave}, {Hughes},
  {Klein}, {MacTavish}, {Marsden}, {Martin}, {Martin}, {Mauskopf},
  {Netterfield}, {Olmi}, {Patanchon}, {Rex}, {Scott}, {Semisch}, {Thomas},
  {Truch}, {Tucker}, {Tucker}, {Viero}, \& {Wiebe}}]{pascale2008}
{Pascale}, E., {Ade}, P.~A.~R., {Bock}, J.~J., {et~al.} 2008, \apj, 681, 400

\bibitem[{{Patanchon} {et~al.}(2008)}]{patanchon2008}
{Patanchon}, G. {et~al.} 2008, \apj, 681, 708

\bibitem[{{Pirogov} \& {Zinchenko}(1993)}]{pirogov1993}
{Pirogov}, L.~E. \& {Zinchenko}, I.~I. 1993, Astronomy Reports, 37, 484

\bibitem[{{Saito} {et~al.}(2008){Saito}, {Saito}, {Yonekura}, \&
  {Nakamura}}]{saito2008}
{Saito}, H., {Saito}, M., {Yonekura}, Y., \& {Nakamura}, F. 2008, \apjs, 178,
  302

\bibitem[{{Tafalla} {et~al.}(2004){Tafalla}, {Myers}, {Caselli}, \&
  {Walmsley}}]{tafalla2004}
{Tafalla}, M., {Myers}, P.~C., {Caselli}, P., \& {Walmsley}, C.~M. 2004, \aap,
  416, 191

\bibitem[{{Tafalla} {et~al.}(2002){Tafalla}, {Myers}, {Caselli}, {Walmsley}, \&
  {Comito}}]{tafalla2002}
{Tafalla}, M., {Myers}, P.~C., {Caselli}, P., {Walmsley}, C.~M., \& {Comito},
  C. 2002, \apj, 569, 815

\bibitem[{{Truch} {et~al.}(2008)}]{truch2008}
{Truch}, M.~D.~P. {et~al.} 2008, \apj, 681, 415

\bibitem[{{Ungerechts} {et~al.}(1986){Ungerechts}, {Winnewisser}, \&
  {Walmsley}}]{ung1986}
{Ungerechts}, H., {Winnewisser}, G., \& {Walmsley}, C.~M. 1986, \aap, 157, 207

\bibitem[{{van der Tak} {et~al.}(2007){van der Tak}, {Black}, {Sch{\"o}ier},
  {Jansen}, \& {van Dishoeck}}]{vanderTak2007}
{van der Tak}, F.~F.~S., {Black}, J.~H., {Sch{\"o}ier}, F.~L., {Jansen}, D.~J.,
  \& {van Dishoeck}, E.~F. 2007, \aap, 468, 627

\bibitem[{{Womack} {et~al.}(1992){Womack}, {Ziurys}, \& {Wyckoff}}]{womack1992}
{Womack}, M., {Ziurys}, L.~M., \& {Wyckoff}, S. 1992, \apj, 387, 417

\bibitem[{{Zinchenko} {et~al.}(2009){Zinchenko}, {Caselli}, \&
  {Pirogov}}]{zinchenko2009}
{Zinchenko}, I., {Caselli}, P., \& {Pirogov}, L. 2009, \mnras, 395, 2234

\end{thebibliography}

\end{document}